%% file: paper.tex
\documentclass[preprint,3p]{elsarticle}

\usepackage{graphicx}

\usepackage[breaklinks]{hyperref}
\graphicspath{%
  {./figures/}}

\usepackage{amssymb}

\usepackage{abbrevs}
\commenttrue

\usepackage{alphalph}
\renewcommand{\theaffn}{\alphalph{\value{affn}}}

\journal{Nuclear Instruments and Methods A}

\title{\textit{b}-Jet Identification in the \Dzero\ Experiment}

\date{February 23, 2010}

\begin{document}


\hfill FERMILAB-PUB-10-037-E

\begin{frontmatter}

 \input{authorlist}

 \begin{abstract}
    Algorithms distinguishing jets originating from \bquark\ quarks
    from other jet flavors are important tools in the physics program of the
    \Dzero\ experiment at the Fermilab Tevatron \ppbar\ collider. This article
    describes the methods that have been used to identify \bquark-quark jets,
    exploiting in particular the long lifetimes of \bquark-flavored hadrons, and
    the calibration of the performance of these algorithms based on collider
    data.
  \end{abstract}

  \begin{keyword}
    b-jet identification \sep b-tagging \sep D0 \sep Tevatron \sep Collider

    \PACS 29.85.+c
  \end{keyword}
\end{frontmatter}


\tableofcontents


\setcounter{footnote}{0}
\renewcommand{\thefootnote}{\fnsymbol{footnote}}


\input{introduction}

\input{reconstruction}

\input{preliminaries}

\input{svt}

\input{jlip}

\input{csip}

\input{nn}

\input{performance}

\input{fakes}

\input{conclusion}

\section*{Acknowledgement}
\input{acknowledgement_paragraph_r2}
\bibliographystyle{elsarticle-num}
\bibliography{paper}







\end{document}

%% file: authorlist.tex
%
\author[aff36]{V.M.~Abazov}
\author[aff74]{B.~Abbott}
\author[aff63]{M.~Abolins}
\author[aff29]{B.S.~Acharya}
\author[aff49]{M.~Adams}
\author[aff47]{T.~Adams}
\author[aff6]{E.~Aguilo}
\author[aff36]{G.D.~Alexeev}
\author[aff40]{G.~Alkhazov}
\author[aff62]{A.~Alton\fnref{fna}}
% ...

\address{\vspace{0.1 in}(The D\O\ Collaboration)\vspace{0.1 in}}
% ...
\address[aff6]{Simon Fraser University, Burnaby, British Columbia, Canada;
                and York University, Toronto, Ontario, Canada}
% ...
\address[aff29]{Tata Institute of Fundamental Research, Mumbai, India}
% ...
\address[aff36]{Joint Institute for Nuclear Research, Dubna, Russia}
% ...
\address[aff40]{Petersburg Nuclear Physics Institute,
                St. Petersburg, Russia}
% ...
\address[aff47]{Florida State University, Tallahassee, Florida 32306, USA}
% ...
\address[aff49]{University of Illinois at Chicago,
                Chicago, Illinois 60607, USA}
% ...
\address[aff62]{University of Michigan, Ann Arbor, Michigan 48109, USA}
\address[aff63]{Michigan State University,
                East Lansing, Michigan 48824, USA}
% ...
\address[aff74]{University of Oklahoma, Norman, Oklahoma 73019, USA}
% ...
%
\fntext[fna]{
Visitor from Augustana College, Sioux Falls, SD, USA.}
\fntext[fnb]{
Visitor from The University of Liverpool, Liverpool, UK.}
\fntext[fnc]{
Visitor from SLAC, Menlo Park, CA, USA.}
\fntext[fnd]{
Visitor from ICREA/IFAE, Barcelona, Spain.}
\fntext[fne]{
Visitor from Centro de Investigacion en Computacion - IPN,
  Mexico City, Mexico.}
\fntext[fnf]{
Visitor from ECFM, Universidad Autonoma de Sinaloa, Culiac\'an, Mexico.}
\fntext[fng]{
Visitor from Universit{\"a}t Bern, Bern, Switzerland.}
\fntext[fnh]{
Formerly with Physikalisches Institut, Universit{\"a}t Bonn, Bonn, Germany.}

%
\vskip 0.25cm

%% file: introduction.tex
\section{Introduction}
\label{sec:intro}

The bottom quark occupies a special place among the
fundamental fermions: on the one hand, its
mass (of the order of 5 \GeVcc~\cite{PDG2008}) is substantially larger than
that of the (next lightest) charm quark. On the other hand, it is light enough
to be produced copiously at present-day high energy colliders.

In particular, unlike the top quark, the bottom quark is lighter than the $W$
boson, preventing decays to on-shell $W$ bosons. As a result, it lives long enough
for hadronization to occur before its decay. The average lifetime of \bquark-flavored
hadrons (referred to as \bquark\ hadrons in the following) has been measured to be
about 1.5 ps~\cite{PDG2008}: this is sufficiently long for \bquark\ hadrons,
even of moderate momentum, to travel distances of the order of at least a
mm. Combined with the relatively large mass of \bquark\ hadrons, the use of precise
tracking information allows the detection of the presence of \bquark\ hadrons
through their charged decay products. In addition, \bquark\ hadron decays often lead to
the production of high momentum leptons; especially
at hadron colliders, the observation of such leptons provides 
easy access to samples with enhanced \bquark-jet content. The identification of jets
originating from the hadronization of bottom quarks (referred to as \bquark-jet
identification or \bquark-tagging in the following) in the \Dzero\ experiment is the
subject of this publication.

\subsection{The upgraded \Dzero\ detector}
\label{sec:d0upgrade}

The \Dzero\ experiment is one of the two experiments operating at the Tevatron
$p\bar{p}$ Collider at Fermilab. After a successful Tevatron Run I, which led to
the discovery of the top quark~\cite{Abe:1995hr,Abachi:1995iq}, the Tevatron
was upgraded to provide both a higher center-of-mass energy (from 1.8 TeV to
1.96 TeV) and a significant increase in luminosity. Run~II started in 2001, and
the Tevatron delivered 1.6 fb$^{-1}$ of integrated luminosity to the
experiments by March 2006, at which time another detector upgrade was
commissioned. This publication refers to the Run~II data taken before March
2006, commonly denoted as the Run~IIa period.

\begin{figure}[bth]
  \begin{center}
    \includegraphics[width=0.50\textwidth]{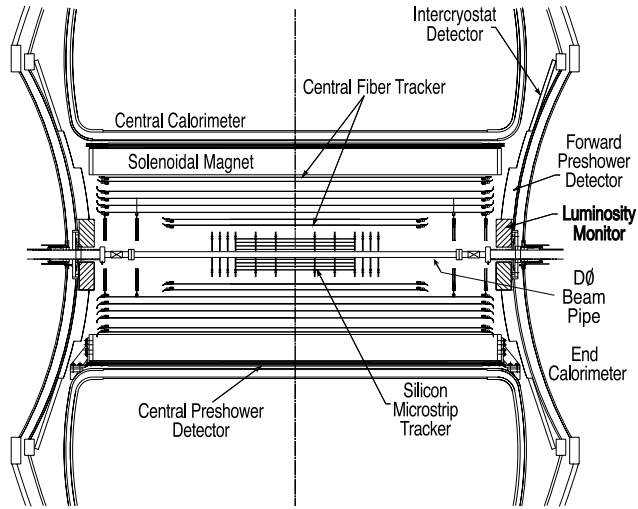}
  \end{center}
  \caption[] {The Run~IIa central tracking detectors.}
  \label{fig:tracker}
\end{figure}

To cope with the increased luminosity and decreased bunch spacing (from 3.6
$\mu$s to 396 ns) in Run~II, the \Dzero\ detector also underwent a significant
upgrade, described in detail elsewhere~\cite{run2nim}. In
particular, a 2~T central solenoid was installed to provide an axial magnetic
field used to measure the momentum of charged particles. Correspondingly, the
existing tracking detectors were removed and replaced with two new detectors,
shown in Fig.~\ref{fig:tracker}:
\begin{itemize}
 \item the central fiber tracker (CFT), consisting of about 77,000 axial and
  small-angle stereo scintillating fibers arranged in eight concentric layers,
  and covering the pseudorapidity region $|\eta| \lesssim 1.7$\footnote{The
    coordinate system used in this article is a cylindrical one with the $z$
    axis chosen along the proton beam direction, and with polar and azimuthal
    angles $\theta$ and $\phi$ (measured with respect to the selected primary
    vertex, as explained in Sec.~\ref{sec:pv}). Pseudorapidity $\eta$ is defined
    as $\eta \equiv -\ln(\tan\theta/2)$, and approximates rapidity $y \equiv
    \frac{1}{2}\ln\left(\frac{E+p_{z}}{E-p_{z}}\right)$ (rapidity differences
    are invariant under Lorentz transformations along the beam axis).};
 \item and the silicon microstrip tracker (SMT)~\cite{SMT}, a detector featuring 912
  silicon strip sensor modules arranged in six barrel and sixteen disk
  structures, allowing tracking up to $|\eta| \lesssim 3$.
  Of particular interest is the innermost layer of SMT barrel sensors: its
  proximity to the beam line (at a radius of 2.7 cm) results in a relatively small
  uncertainty in the extrapolation of tracks to the beam line, and hence in good
  vertex reconstruction capabilities.
\end{itemize}

The ability to identify efficiently the \bquark\ quarks\footnote{In this article,
charge conjugated states are implied as well.} in an event considerably
broadens the range of physics topics that can be studied by the \Dzero\ experiment
in Run~II. While the analysis leading to the observation of the top quark by \Dzero\
in Run I employed only semimuonic decays $\bquark\rightarrow\mu X$,
the use of lifetime tagging allows for a more precise determination of the top
quark properties (see \emph{e.g.}~\cite{rbpaper}).
The search for electroweak production of top quarks ($\ppbar\to \tquark\bquark+X,
\tquark q\bquark+X$) 
relies heavily as well on the efficiency to identify \bquark\ jets and reject
light and \cquark\ jets, as demonstrated in the recent observation
of this production process~\cite{Abazov:2009ii}.
The search for the standard model Higgs boson also depends on \bquark-jet
identification: a relatively light ($m_{H}\lesssim 135\GeVcc$) Higgs
boson will decay predominantly to $\bbbar$ quark pairs. Finally, various regions
of the minimal supersymmetric standard model parameter space lead to final
states containing \bquark\ quarks (from gluino or stop quark decays),
and efficient \bquark\ tagging greatly increases the sensitivity of the search
for those final states.

This article is subdivided as follows. Section~\ref{sec:reconstruction}
describes the objects that serve as input to the \bquark-tagging
algorithms. Section~\ref{sec:preliminaries} introduces the steps taken
before applying the tagging algorithms proper. Sections~\ref{sec:svt}, \ref{sec:jlip},
and~\ref{sec:csip} describe the basic ways in which lifetime-correlated
variables are extracted. Section~\ref{sec:nn} describes the combination of these
variables in an artificial neural network to obtain an optimal tagging
performance. Finally, Sections~\ref{sec:efficiency} and~\ref{sec:fake_rate}
detail how Tevatron data are used to calibrate the performance of the resulting
tagging algorithm.


%% file: reconstruction.tex
\section{Object Reconstruction}
\label{sec:reconstruction}

Besides the charged particle tracks, which are reconstructed from hits (clustered
energy deposits) in the CFT and SMT  detectors, 
the input for lifetime identification of \bquark-quark jets consists of two
kinds of reconstructed objects:
\begin{itemize}
\item primary vertices, which are built from two or more
  charged particle tracks that originate from a common point in space;
\item hadron jets, which are reconstructed primarily from their energy
  deposition in the calorimeter.
\end{itemize}
These objects are described below.

\subsection{Primary vertex reconstruction}
\label{sec:pv}
 
The knowledge of the $\ppbar$ interaction point or \emph{primary vertex} of an
interaction is important to provide the most precise reference point for the
lifetime-based tagging algorithms described in subsequent sections. In addition,
multiple interactions can occur during a single bunch crossing. It is therefore
necessary to select the primary vertex associated to the interaction of interest.
The reconstruction and identification of the primary vertex at \Dzero\ consists of
the following steps: (i) track selection; (ii) vertex fitting using a Kalman
filter algorithm to obtain a list of candidate vertices; (iii) a second vertex
fitting iteration using an adaptive algorithm to reduce the effect of outlier
tracks;
and (iv) primary vertex selection. 

In the first stage, tracks are selected if their momentum component in the plane
perpendicular to the beam line, \pt,  exceeds 0.5\GeVc, and they have two or
more hits in the SMT (counted as the number of hit ladders or disks) if the track
is within the SMT geometric acceptance as measured in the $(\eta,z)$ plane. The
selected tracks are then clustered along the $z$ direction in 2~cm regions to
separate groups of tracks coming from different interactions, as evidenced by the
tracks' $z$ coordinates at their distance of closest approach to the beam line.

In the second stage, the tracks in each of the $z$ clusters are used to
reconstruct a vertex. This is done in two passes. In the first pass, the
selected tracks in each cluster are fitted to a common vertex. A
Kalman filter vertex fitter is used for this step where tracks with the highest
$\chi^2$ contribution to the vertex are removed in turn, until the total vertex
$\chi^2$ per degree of freedom is less than 10.
In the second pass, the track selection in each $z$ cluster is refined based on
the track's distance of closest approach in the transverse plane, $d$, to the
vertex position computed in the first pass, as well on its uncertainty $\sigma_{d}$:
only tracks with \emph{impact parameter significance} $\mathcal{S}_d \equiv
d/\sigma_{d}$ satisfying $|\mathcal{S}_d| <5$ are retained.

Once the outliers with respect to the beam position
have been removed from the selected tracks, an adaptive vertex
algorithm~\cite{Fruhwirth:1991pm,D'Hondt:2004dy} is used 
to fit the selected tracks to a common vertex in each cluster. 
This algorithm differs from the Kalman filter vertex fitter in that all tracks
remaining after the Kalman filter selection procedure are
allowed to contribute to the final vertex fit instead of rejecting those tracks
whose $\chi^2$ contribution to the vertex fit exceeds a certain value.
It is especially suited to reducing the contribution of distant tracks to the vertex
fit, thus obtaining a better separation between primary and secondary vertices.
The algorithm proceeds in three iterative stages:
(i) the track candidates in a $z$-cluster are fitted using a Kalman filter; (ii)
each track is weighted according to its $\chi^2$ contribution to the vertex
found in the previous step, and if the weight is less than $10^{-6}$, the track
is eliminated; and (iii) steps (i) and (ii) are repeated until either the weights
converge (the maximum change in track weights between consecutive iterations is
less than $10^{-3}$) or more than 100 iterations have been performed. 

\begin{figure}[htb]
  \begin{center}
    \includegraphics[width=0.45\textwidth]{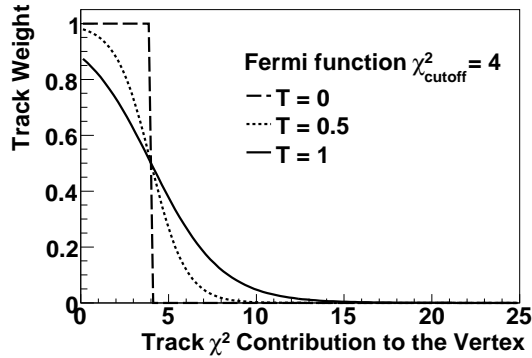}
  \end{center}
  \caption[]{Weight given to tracks as a function of the $\chi^2$ distance to the
    vertex for different T parameters. In the adaptive fitting, all tracks
    contribute to the fit when $T>0$. The value used in the reconstruction algorithm
    is $T=1$.}
  \label{fig:apv}
\end{figure}

The weights are \emph{adapted} in each iteration: a track $i$ associated with
a small weight in one iteration will affect the weights of all other tracks in
the next iteration because they are derived with respect to the new vertex position
obtained with a down-weighted contribution from track $i$. 
The tracks are weighted in step ii) according to their $\chi^2$
contribution to the vertex by a sigmoidal function:
\begin{equation}
  w_i = \frac{1}{1+e^{(\chi^2_i-\chi^2_{\mbox{\scriptsize cutoff}})/2T}}.
\end{equation}
Here, $\chi^2_i$ is the $\chi^2$ contribution of track $i$ to the vertex;
$\chi^2_{\mbox{\scriptsize cutoff}}$ is the $\chi^{2}$ value at which the weight
function drops to 0.5; and $T$, like the temperature in the Fermi function in
statistical thermodynamics, is a parameter that controls the sharpness of the
function. Figure~\ref{fig:apv} shows the weight function used at \Dzero\ with
$\chi^2_{\mbox{\scriptsize cutoff}}=4$ and $T=1$. 

\begin{figure}[htb]
  \begin{center}
    \includegraphics[width=0.45\textwidth]{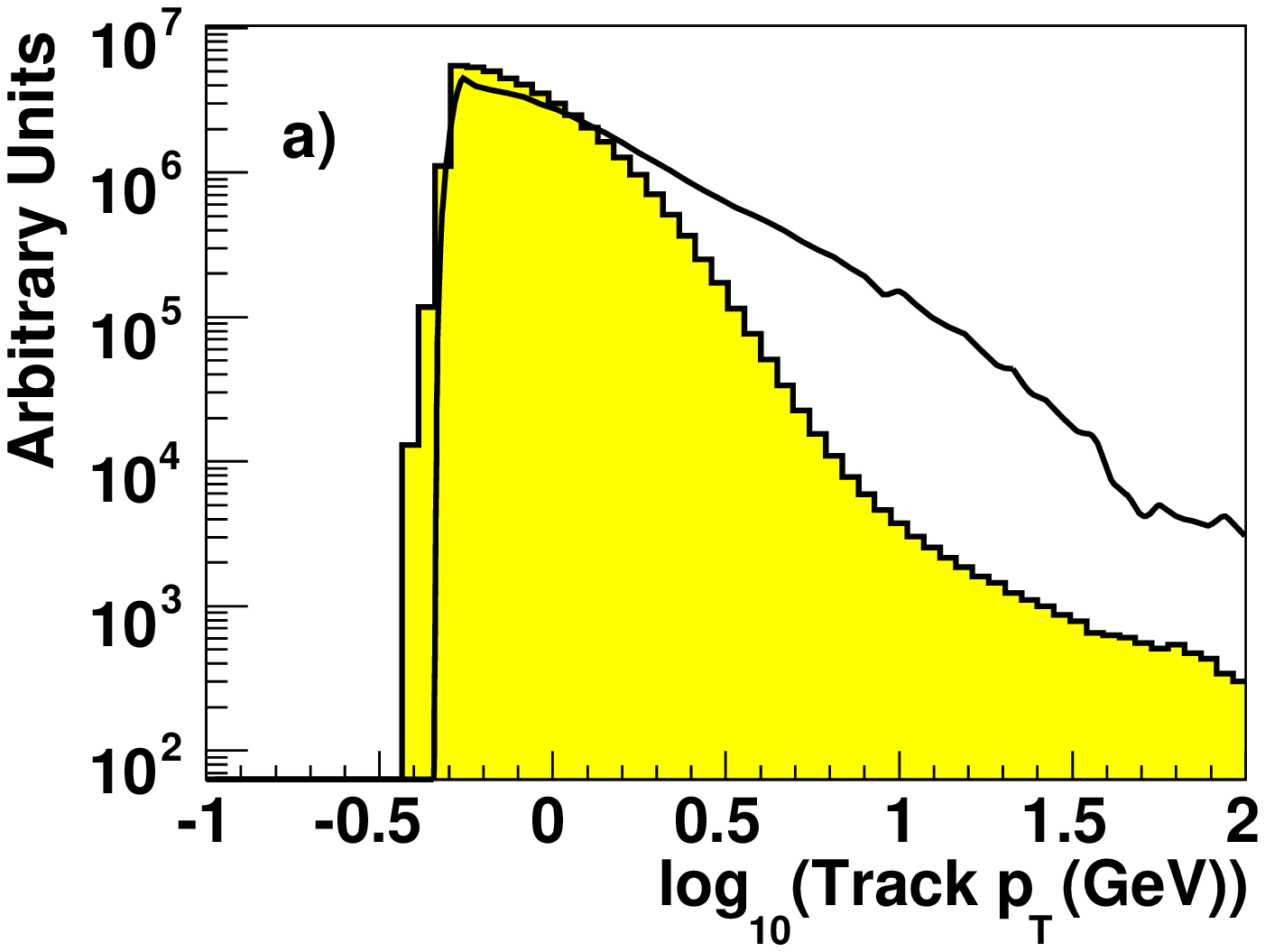}
    \includegraphics[width=0.45\textwidth]{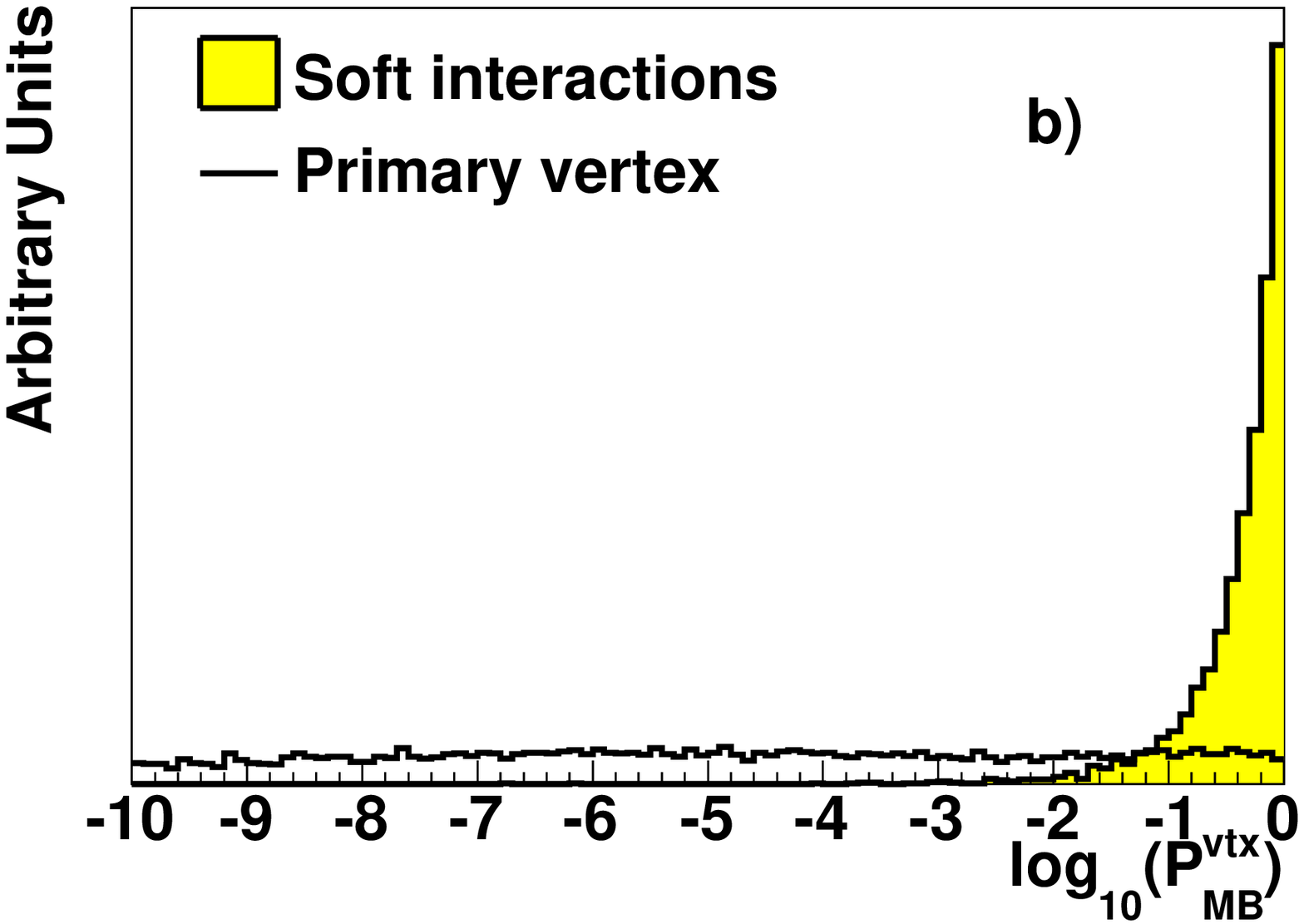}
  \end{center}
  \caption[]{(a): The track \pt\ spectrum for soft-interaction vertices in the
    \Dzero\ data selected as described in the text. For comparison, the
    corresponding distribution is also shown for the primary vertices in one
    data run selected on the basis of their
    $\mathcal{P}_{\mbox{\scriptsize MB}}^{\mbox{\scriptsize vtx}}$ values, as
    also described in the text. The soft-interaction vertex distribution is
    $\mathcal{F}(\pt)$ in Eq.~\ref{eq:PMBtrk}.
    (b): The $\mathcal{P}_{\mbox{\scriptsize MB}}^{\mbox{\scriptsize vtx}}$
    distribution for soft-interaction vertices and multijet (hard scatter) primary
    vertices.}
  \label{fig:PVsel}
\end{figure}

Once all possible vertices in the event are accurately reconstructed, the
fourth and final step consists of selecting which of the fitted vertices is the
result of the hard scatter interaction. The hard scatter vertex is distinguished
from soft-interaction vertices by the higher average \pt\ of its tracks, as
shown in Fig.~\ref{fig:PVsel}.
The probability $\mathcal{P}_{\mbox{\scriptsize MB}}^{\mbox{\scriptsize trk}}(\pt)$
that the observed \pt\ of a given track is compatible with the track originating
from a soft interaction is computed as
\begin{equation}
  \mathcal{P}_{\mbox{\scriptsize MB}}^{\mbox{\scriptsize trk}}(\pt) = 
  \frac{\int_{\pt}^{\infty}\mathcal{F}(\pt^{\prime})d\pt^{\prime}}
  {\int_{\pt^{\min}}^{\infty}\mathcal{F}(\pt^{\prime})d\pt^{\prime}}.
  \label{eq:PMBtrk}
\end{equation}
Here, $\mathcal{F}(\pt)$ is the \pt\ distribution of tracks attached to soft
interaction vertices in $Z\rightarrow\mu^{+}\mu^{-}$ candidate events requesting
that they be separated from the $Z\rightarrow\mu^{+}\mu^{-}$ interaction vertex
by more than 10 cm. $\mathcal{F}(\pt)$ is depicted in Fig.~\ref{fig:PVsel}a.
Only tracks with $\pt>\pt^{\min}=0.5\GeVc$ are used in this calculation. For each
reconstructed vertex, the probability that it is consistent with a minimum bias
interaction is formed as
\begin{eqnarray}
  \mathcal{P}_{\mbox{\scriptsize MB}}^{\mbox{\scriptsize vtx}} & = &
  \Pi \cdot \sum_{j=0}^{N_{\mbox{\scriptsize trk}}-1}
  \frac{(-\ln\Pi)^j}{j!} \;\;\;
  \mbox{with} \;\;\; \\
  \Pi & = & \prod_{i=1}^{N_{\mbox{\scriptsize trk}}}
  \mathcal{P}_{\mbox{\scriptsize MB}}^{\mbox{\scriptsize trk}}(p_{T,i})\;,\nonumber
\end{eqnarray}
where $N_{\mbox{\scriptsize trk}}$ is the number of tracks attached to the
vertex; a motivation for this expression is provided in Sec.~\ref{sec:jlip}.
The selected primary vertex is the one with the lowest
$\mathcal{P}_{\mbox{\scriptsize MB}}^{\mbox{\scriptsize vtx}}$; the resulting 
$\mathcal{P}_{\mbox{\scriptsize MB}}^{\mbox{\scriptsize vtx}}$ distribution is
shown in Fig.~\ref{fig:PVsel}b.

The reconstruction and identification efficiency in data is between 97\% and 100\%
for primary vertices reconstructed up to $|z|$ = 100~cm, as measured on
the $Z\rightarrow\mu^{+}\mu^{-}$ candidate event sample. For multijet events,
the position resolution of the selected primary vertex in the transverse plane
can be determined by subtracting the known beam width quadratically from the
width of the observed vertex position distribution. This resolution improves with
increasing track multiplicity and is better than the beam width (around 30~$\mu$m)
for events with at least 10 tracks attached to the primary vertex.

\subsection{Jet reconstruction and calibration}
\label{sec:jets}

The vast majority of data analyses in \Dzero\ make use of so-called cone jets,
which collect all calorimeter energy deposits within a fixed angular distance
$\mathcal{R}\equiv\sqrt{(\Delta y)^{2}+(\Delta\phi)^{2}}$ in ($y$,$\phi$)
space. Specifically, the cone jet reconstruction algorithm used within \Dzero\ is
the \emph{Run~II cone jet} algorithm~\cite{Blazey:2000qt}. This algorithm is
insensitive to the presence of soft or collinear radiation off partons, thus
allowing for detailed comparisons of jet distributions in the \Dzero\ data with
theoretical predictions. The cone radii used in analyses in \Dzero\ are
$\mathcal{R}=0.5$ and $\mathcal{R}=0.7$, but for most high \pt\ physics the
$\mathcal{R}=0.5$ cone jets are used. It is only these jets that are
described in this article.

As the jets are reconstructed on the basis of calorimetric information, 
comparisons between the data and jets simulated using Monte Carlo (MC) methods
require corrections for various effects:
\begin{itemize}
\item energy deposits not from the hard interaction (either from the remnant of
  the original \ppbar\ system or from additional soft interactions or noise);
\item the (energy dependent) calorimeter response to incident high-energy
  particles;
\item the net energy flow through the jet cone because of the finite size of
  showers in the calorimeter and the bending of charged particle trajectories by
  the magnetic field.
\end{itemize}
The topic of the determination of the \emph{jet energy scale} (JES) is described
in a separate paper~\cite{JES}. The resulting JES, by itself, does not yet
account for neutrinos from decays of \bquark- or \cquark-flavored hadrons
escaping undetected. While such additional corrections may be important for
physics analyses, \bquark\ tagging is only sensitive to it because the tagging
performance obtained in data (Sec.~\ref{sec:efficiency}) is parametrized in
terms of jet \et\ (and $\eta$) and applied to simulated jets. For this purpose,
corrections for undetected neutrinos (and for the energy not deposited in the
calorimeter, in the case of muons associated with jets) do not need to be
applied, provided that data and simulated jets are treated identically.


%% file: preliminaries.tex
\section{Tagging Prerequisites}
\label{sec:preliminaries}

In order to evaluate the performance of the \bquark-tagging algorithms described
in the following sections, it is necessary to define properly the meaning of
``\bquark\ jet''. Also, several steps are carried out before proceeding with
the tagging proper. These steps are discussed below.

\subsection{Flavor assignment in simulated events}
\label{sec:signal}

The \bquark-tagging algorithms used within the
\Dzero\ experiment are jet based rather than event based. This choice makes sense
especially for high-luminosity hadron colliders, where pile-up (overlapping
electronics signals) from previous interactions, as well as multiple
interactions in the same bunch crossing, may lead to other reconstructed tracks
and jets in the event besides those of the ``interesting'' high-\pt\ interaction.

However, this choice introduces an ambiguity for jets in simulated events: In
order to estimate the performance of a \bquark-tagging algorithm, it is first
necessary to specify precisely how a jet's flavor is determined. The following
choice has been made:
\begin{itemize}
\item if at the particle level (\emph{i.e.}, after the hadronization of the
  partonic final state), a hadron containing a \bquark\ quark (denoted in
  the following as \bquark\ hadron) is found within a $\mathcal{R}=0.5$ radius
  of the jet direction\footnote{Apart from the jet cone definition, as discussed in
    Sec.~\ref{sec:jets}, angular distances in this article are determined in
    ($\eta$, $\phi$) space.}, the jet is considered to be a  \bquark\ jet;
\item if no \bquark\ hadron is found, but a hadron containing a \cquark\ quark
 (henceforth denoted as a \cquark\ hadron) is found instead, the jet is considered
 to be a \cquark\ jet;
\item if no \cquark\ hadron is found either, the jet is considered to be a
 light-flavor jet.
\end{itemize}
This choice is preferred over the association with a parton level \bquark\ or
\cquark\ quark, as in the latter case, parton showering may lead to a large
distance $\Delta\mathcal{R}$ between the original quark direction and that of the
corresponding jet(s).

\subsection{Taggability}
\label{sec:taggability}

The jet tagging algorithms described in the following sections are based entirely
on tracking and vertexing of charged particles. Therefore, a very basic requirement
is that there should be charged particle tracks associated with the
(calorimeter) jet. Rather than incorporating such basic requirements in the
tagging algorithms themselves, they are implemented as a separate step.

The reason for this is that the tagging algorithm's performance must be
evaluated on the data, as detailed in Sec.~\ref{sec:efficiency}. It is
parametrized in terms of the jet kinematics (\et\ and $|\eta|$). This
parametrization presupposes that there are no further dependences. However, the
interaction region at the \Dzero\ detector is quite long, $\sigma_{z}\approx 25$
cm, and the detector acceptance affects the track reconstruction efficiency
dependence on $\eta$ differently for different values of the interaction point's
$z$ coordinate; hence the above parametrization is only possible once this $z$
dependence is accounted for.

The requirement for a jet to be \emph{taggable}, \emph{i.e.}, for it to be
considered for further application of the tagging algorithms, is that it should
be within $\Delta\mathcal{R}=0.5$ from a so-called \emph{track jet}. Track
jets are reconstructed starting from tracks having at least one hit in the SMT, a
distance to the selected primary vertex less than 2~mm in the transverse plane
and less than 4~mm in the $z$ direction, and $\pt > 0.5 \GeVc$. Starting with
``seed'' tracks having $\pt > 1~\GeVc$, the Snowmass jet
algorithm~\cite{Berger:1992zh} is used to cluster the tracks within cones of
radius $\mathcal{R}=0.5$.

\begin{figure*}[htb]
  \centering
  \includegraphics[height=0.24\textheight]{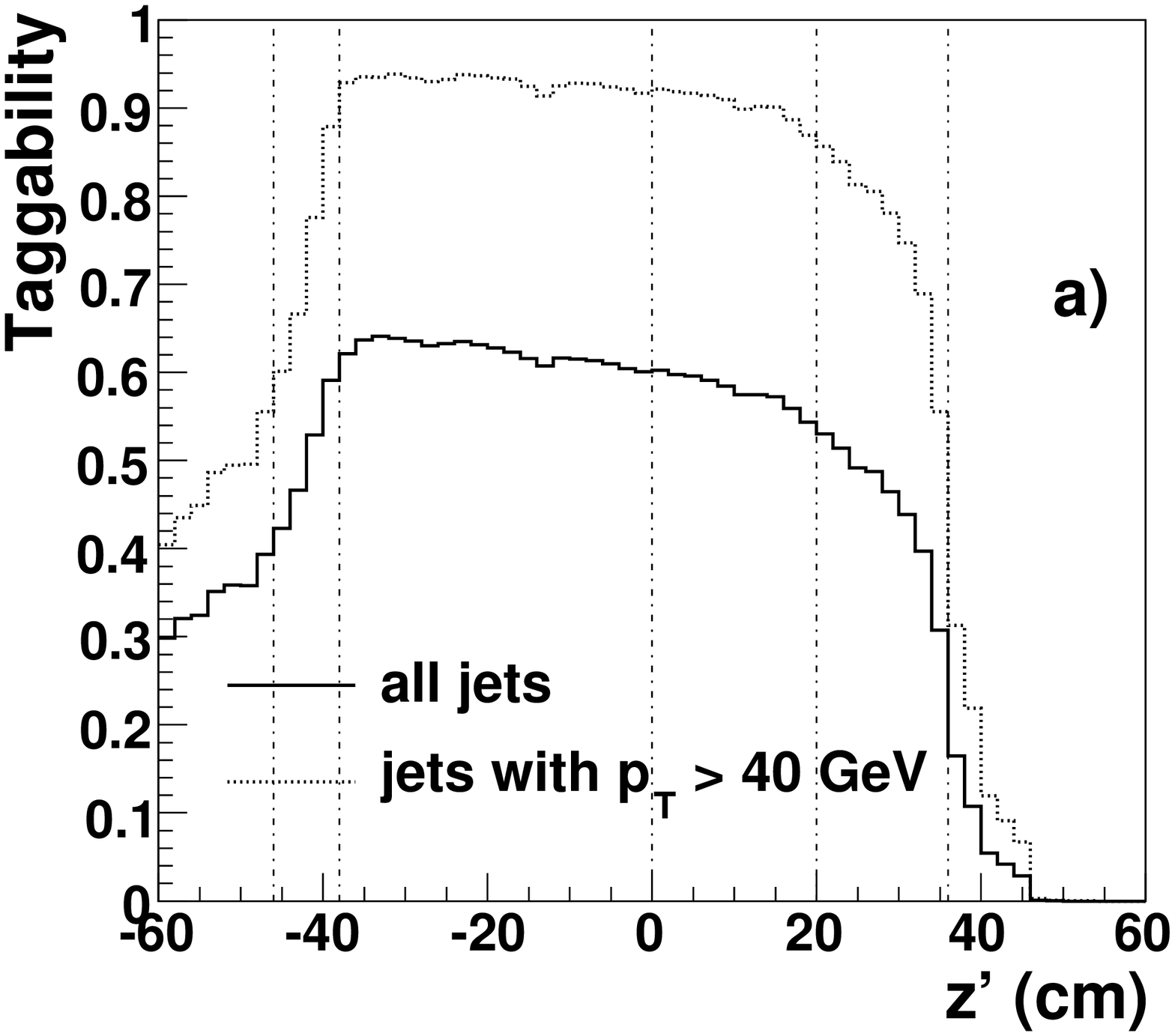}
  \includegraphics[height=0.24\textheight]{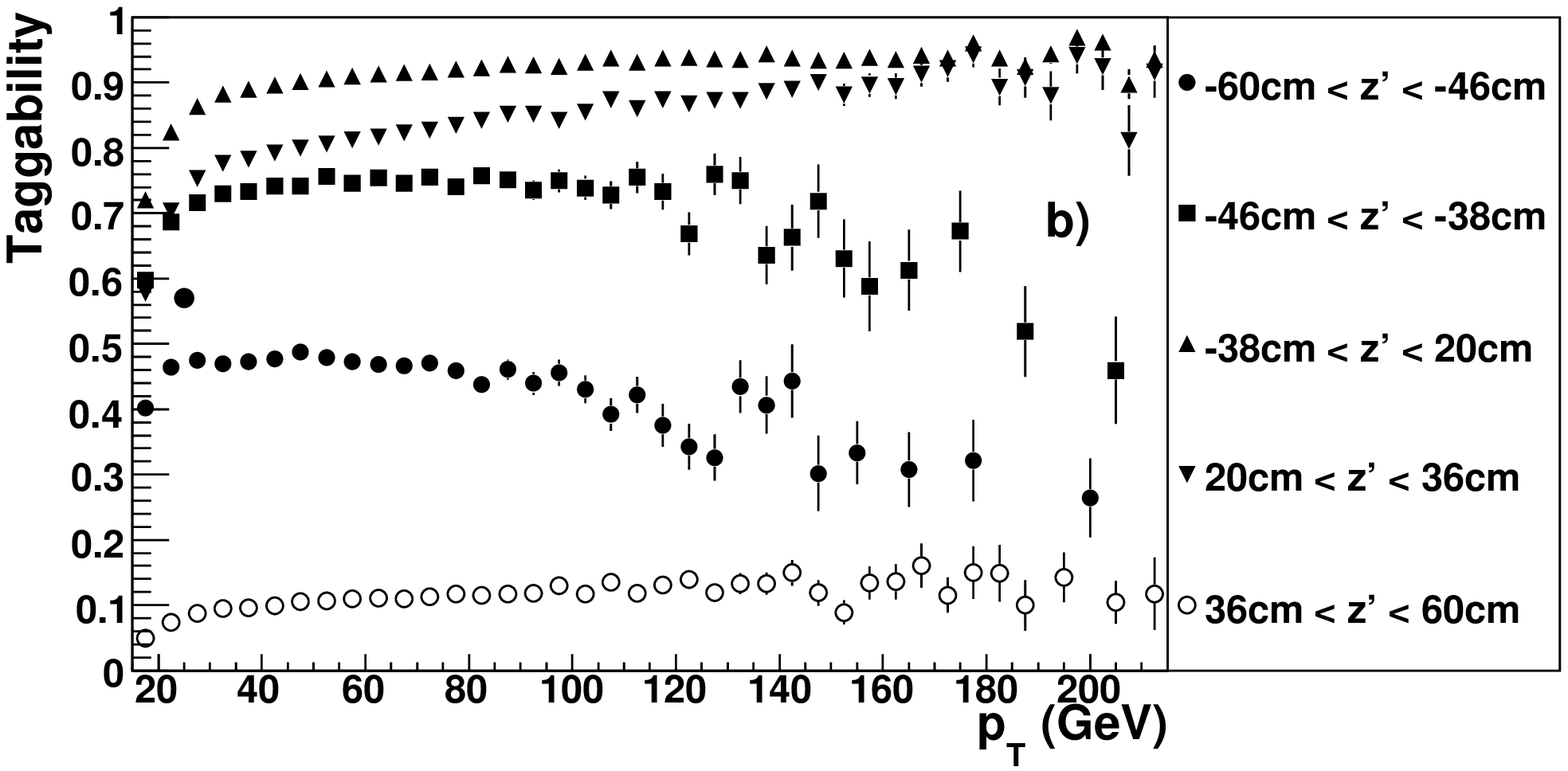}
  \caption{(a): taggability as a function of
    $z^{\prime}\equiv |z|\cdot\mbox{sign}(\eta\cdot z)$. The vertical lines
    denote the boundaries chosen for the parametrization in \et\ and $|\eta|$.
    (b): taggability as a function of jet \et, in different bins of
    $z^{\prime}$. The curves for the two central bins are very similar and have
    been combined.}
  \label{fig:taggability}
\end{figure*}

As an example, Fig.~\ref{fig:taggability} shows the taggability for the sample
used to determine the \bquark-jet efficiency in Sec.~\ref{sec:efficiency}.
The taggability is determined as a function of both the jet kinematics (\et\ and
$\eta$) and the $z$ coordinate of the selected primary
vertex, as detailed above. In detail, the coordinate used in
Fig.~\ref{fig:taggability} is
$z^{\prime}\equiv |z|\cdot\mbox{sign}(\eta\cdot z)$, as this variable 
makes optimal use of the geometrical correlations between $\eta$ and $z$. In
different $z^{\prime}$ regions, as shown in
Fig.~\ref{fig:taggability}a, only $|\eta|$ and \et\ are used as
further independent variables. The taggability parametrization in each
$z^{\prime}$ bin is two dimensional, even though only the projection
onto \et\ is shown.
The tagging algorithms discussed in subsequent chapters are applied only to jets
retained by the taggability criterion, and also the quoted performances do not
account for the inefficiency incurred by this criterion.

\subsection{$V^{0}$ rejection}
\label{sec:v0}

By construction, the lifetime tagging algorithms assume that any measurable
lifetime is indicative of heavy flavor jets. However, light strange hadrons also
decay weakly, with long lifetimes. Particularly pernicious backgrounds
arise from \PKzS\ and \PgL, as their lifetimes (90~ps and 263~ps, respectively)
do not differ vastly from those of \bquark\ hadrons.
In addition, $\gamma\rightarrow\epem$ conversions may occur in the detector
material at large distances from the beam line.

\begin{figure}[bth]
  \begin{center}
    \includegraphics[width=0.48\textwidth]{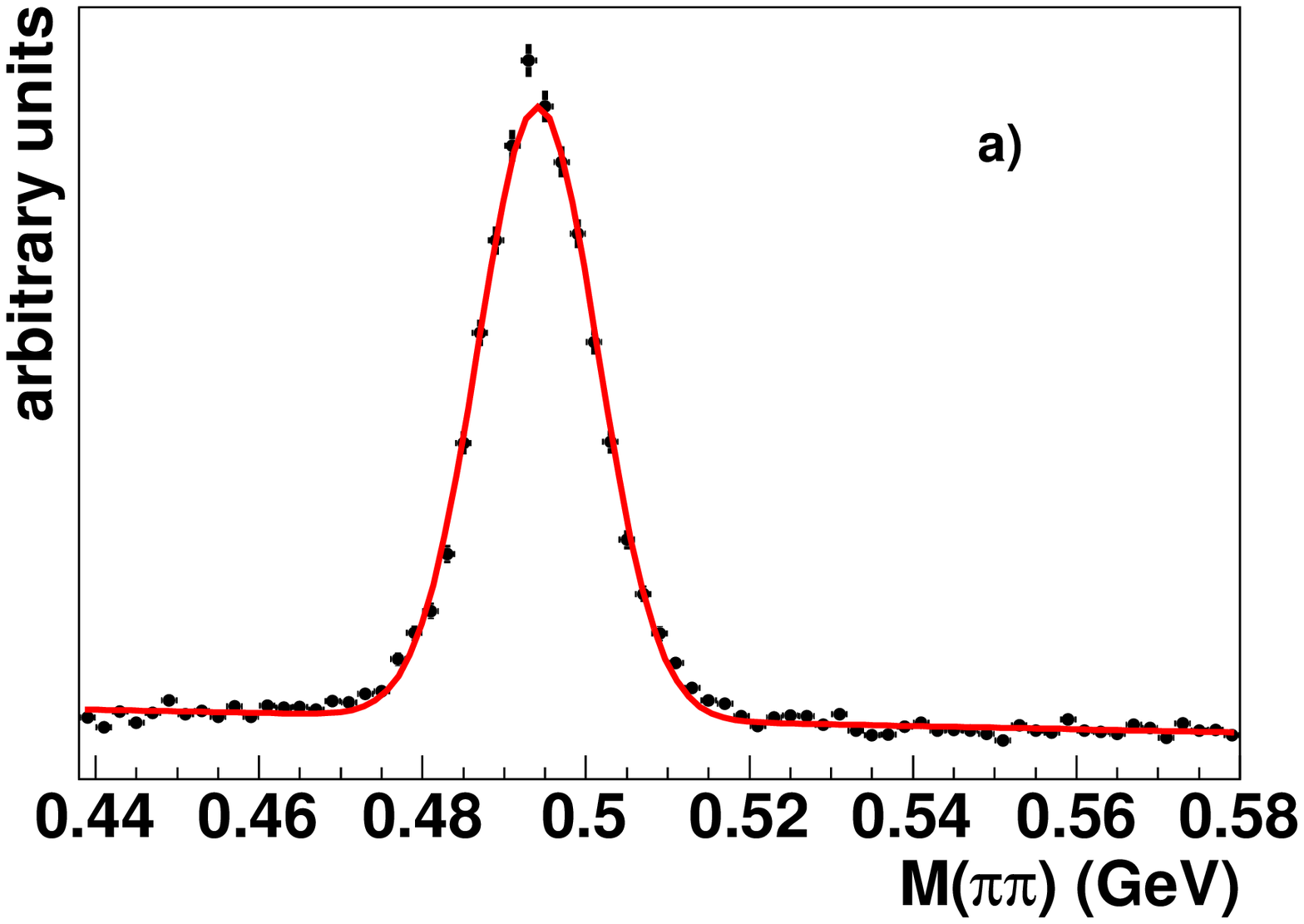}
    \includegraphics[width=0.48\textwidth]{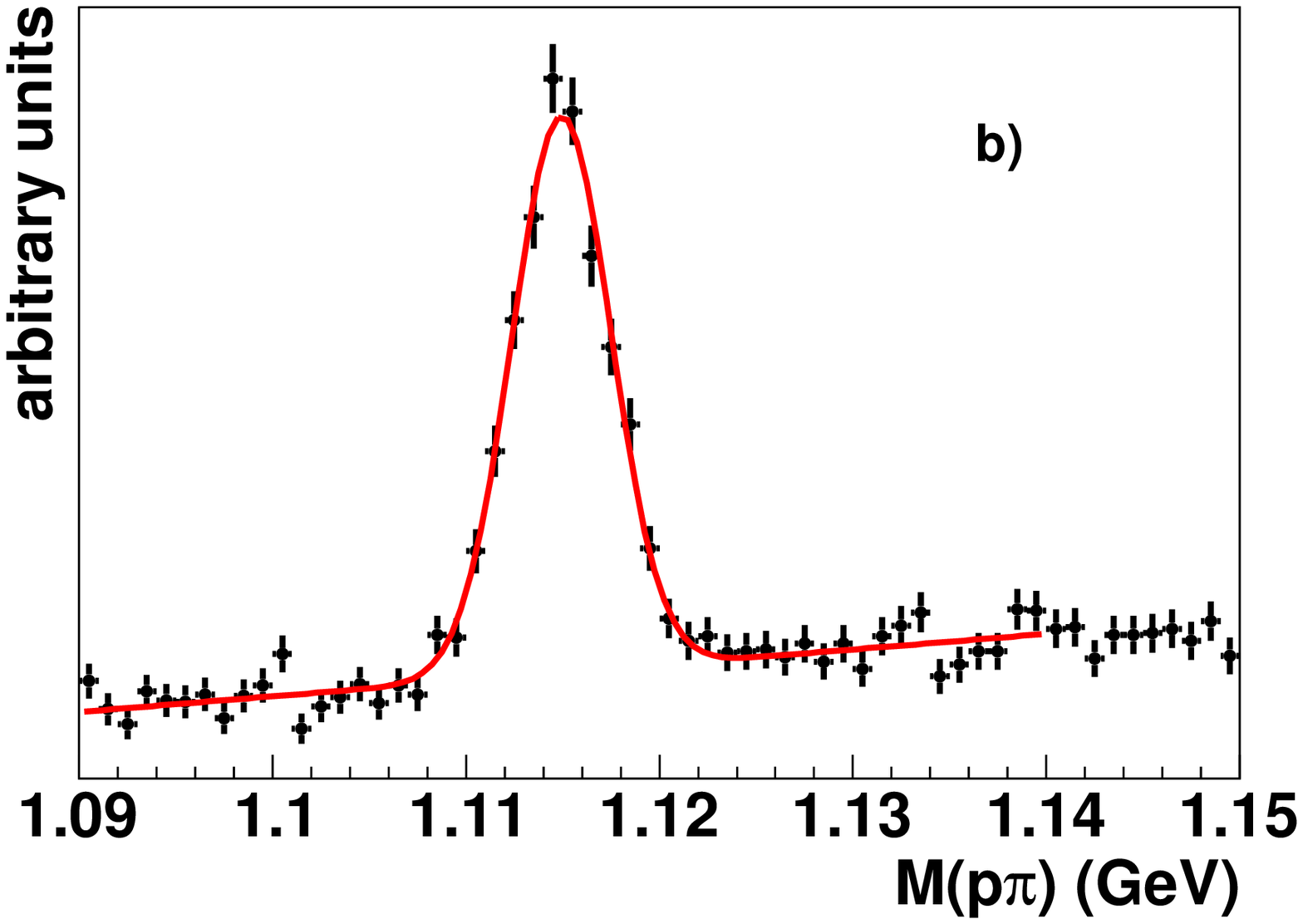}
  \end{center}
  \vspace{-0.5cm}
  \caption[] {Reconstructed \PKzS\ (a) and \PgL\ (b) mass peaks.}
  \label{fig:v0plot}
\end{figure}

\PKzS\ and \PgL\ candidates, commonly denoted as $V^{0}$s, are identified through
two oppositely charged tracks satisfying the following criteria:
\begin{itemize}
\item The significance of the distance of closest approach (DCA) to the selected
  primary vertex in the transverse plane, $\mathcal{S}_{d}$ (see
  Sec.~\ref{sec:pv}), of both tracks must satisfy $|\mathcal{S}_{d}| > 3$.
\item The tracks' $z$ coordinates at the point of closest approach in the
  transverse plane must be displaced from the primary vertex less than 1 cm, to
  suppress misreconstructed tracks.
\item The resulting $V^{0}$ candidate must have a distance of closest approach to the
  primary vertex of less than 200 $\mu$m. This requirement is intended to select
  only those $V^{0}$ candidates originating from the primary vertex, while
  candidates originating from heavy flavor decays may be taken into account
  during the tagging.
\item The reconstructed mass should satisfy $472\MeVcc < m < 516\MeVcc$ for
  \PKzS\ candidates, and $1108\MeVcc < m < 1122\MeVcc$ for \PgL\ candidates (in
  the latter case, the higher \pt\ tracks are considered to be protons; the
  other tracks, or both tracks in the case of \PKzS\ reconstruction, are assumed
  to be charged pions). The invariant mass distributions of reconstructed \PKzS\ and
  \PgL\ candidates are shown in Fig.~\ref{fig:v0plot}.
\end{itemize}

\begin{figure}[bth]
  \begin{center}
    \includegraphics[width=0.48\textwidth]{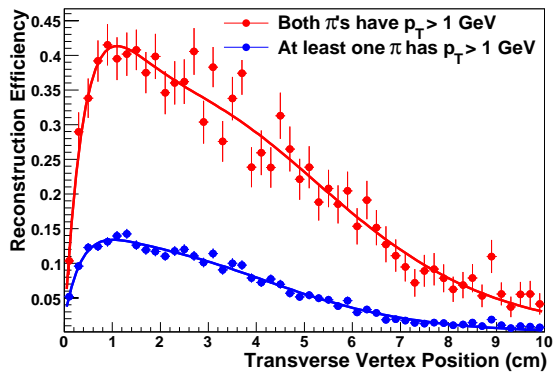}
  \end{center}
  \vspace{-0.5cm}
  \caption[] {The $\PKzS\rightarrow\pi^{+}\pi^{-}$ finding efficiency as a
    function of the transverse position of the decay vertex as determined in
    simulated multijet events.}
  \label{fig:v0eff}
\end{figure}

The $V^{0}$ finding efficiency depends on the transverse
momentum of the $V^{0}$ decay products, as well as on the position of the
decay vertex inside the tracker volume. As an example, Fig.~\ref{fig:v0eff}
shows the \PKzS\ finding efficiency as a function of the transverse position
of the decay vertex for the cases when both or at least one of the decay pions
have a transverse momentum greater than 1~GeV. This efficiency has been
determined using simulated multijet events.

\begin{figure}[bth]
  \begin{center}
    \includegraphics[width=0.48\textwidth]{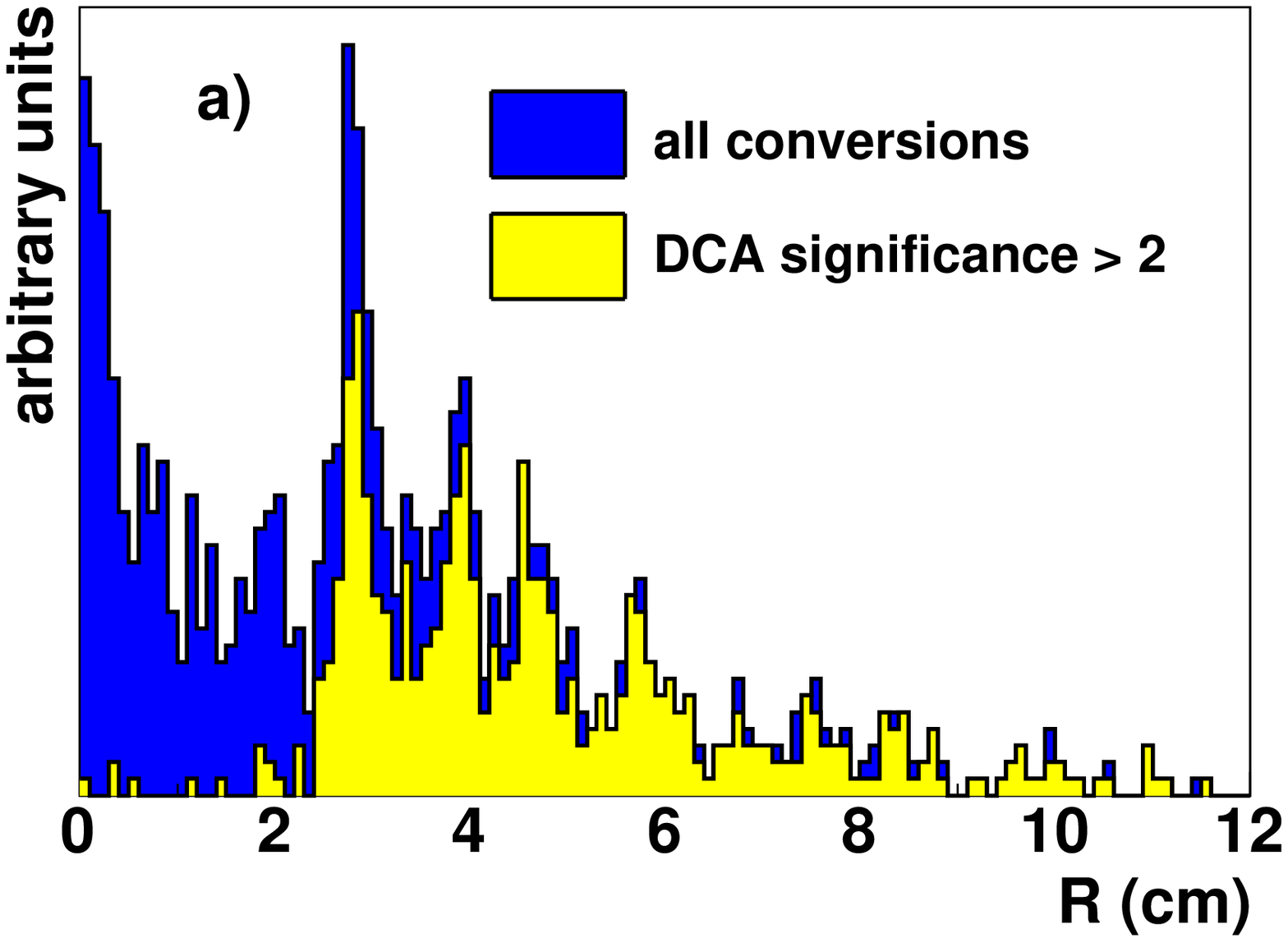}
    \includegraphics[width=0.48\textwidth]{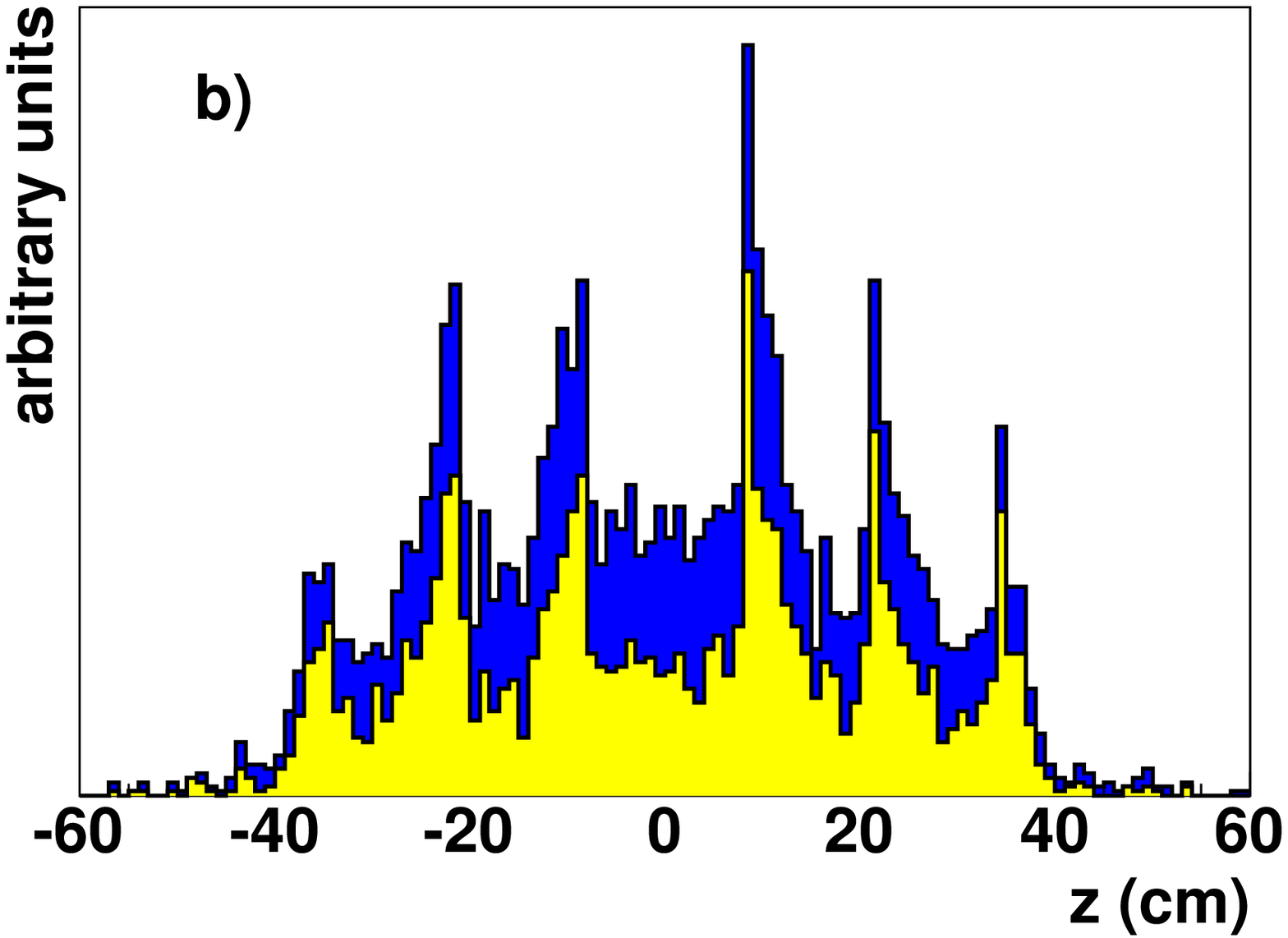}
  \end{center}
  \vspace{-0.5cm}
  \caption[] {Reconstructed radial (a) and $z$ coordinate (b) of candidate
    conversion vertices. The peaks correspond to the radial positions of the SMT
    barrels and $z$ positions of the SMT disk structures, respectively.}
  \label{fig:conversions}
\end{figure}

Photon conversions are most easily recognized by the fact that the opening angle
between the electron and positron is negligibly small. In the plane
perpendicular to the beam line, this is exploited by requiring that the tracks
be less than 30 $\mu$m apart at the location where their trajectories are
parallel to each other. In addition, they must again be oppositely charged,
and their invariant mass is required to be less than 25\MeVcc. Since
conversions happen inside material, the locations of their vertices
reflect the distribution of material inside the detector,
as illustrated by Fig.~\ref{fig:conversions}: the locations of the SMT barrel
ladders and the disks are clearly visible in the distribution of the radial and $z$
coordinates, respectively, of reconstructed photon conversions. 

%% file: svt.tex
\section{The Secondary Vertex Tagger}
\label{sec:svt}

The vast majority of \bquark-hadron decays give rise to multiple charged particles
emanating from the \bquark-hadron's decay point. The most intuitive tagging method
is therefore to attempt to reconstruct this decay point explicitly and to require
the presence of a displaced or \emph{secondary} vertex. The requirement that a
number of tracks all be extrapolated to the same point in three dimensions is
expected to lead to an algorithm, the Secondary Vertex Tagger
(SVT)~\cite{schwartzmanthesis}, which is robust even in the presence of
misreconstructed tracks.

After the identification and selection of the primary interaction vertex,
the reconstruction of secondary vertices starts from the track jet associated
with each (taggable) calorimeter jet (see Sec.~\ref{sec:taggability}).
The tracks considered are those associated with the track jet, subject to
additional selection criteria: 
they should have at least two SMT hits, transverse momentum exceeding
$0.5\GeVc$, transverse impact parameter with respect to the primary vertex $|d|
<$ 1.5~mm, and a separation in the $z$ direction between the point of closest
approach to the beam line and the primary vertex less than 4~mm.
Subsequently, tracks associated with identified $V^{0}$ vertices (see
Sec.~\ref{sec:v0}) are removed from consideration.
All remaining tracks are used in a so-called \emph{build-up} vertex finding
algorithm. In detail, the algorithm consists of the following steps:

\begin{enumerate}
\item Tracks within track jets with large transverse impact parameter
  significance, $|\mathcal{S}_{d}| > 3$, are selected.

\item Vertices are reconstructed from all pairs of tracks using a Kalman vertex
  fitting technique~\cite{Fruhwirth:1987fm}, and are retained if the vertex fit
  yields a goodness-of-fit $\chi^{2}<\chi^{2}_{\max} = 100$.

\item Additional tracks pointing to these seed vertices are added one by one,
  according to the resulting $\chi^{2}$ contribution to the vertex fit. The
  combination yielding the smallest increase in fit $\chi^{2}$ is retained.

\item This procedure is repeated until the increase in fit $\chi^{2}$ exceeds a
  set maximum, $\Delta\chi^{2}_{\max} = 15$, or the total fit $\chi^{2}$ exceeds
  $\chi^{2}_{\max}$.

\item The resulting vertex is selected if in addition, the angle $\zeta$ between
  the reconstructed momentum of the displaced vertex (computed as the sum of the
  constituent tracks' momenta) and the direction from the
  primary to the displaced vertex (in the transverse plane) satisfies $\cos\zeta
  > 0.9$, and the vertex decay length in the transverse direction $L_{xy}<2.6$
  cm.

\item Many displaced vertex candidates may result, with individual tracks
  possibly contributing to multiple candidates. Duplicate displaced vertex
  candidates are removed until no two candidates are associated with identical
  sets of tracks.

\item Secondary vertices are associated with the nearest calorimeter jets if
  $\Delta \mathcal{R}(\mbox{vertex},\mbox{jet}) < 0.5$. Here, the vertex
  direction is computed as the difference of the secondary and primary vertex
  positions.

\end{enumerate}

\begin{figure}[bth]
  \begin{center}
    \includegraphics[width=0.4\textwidth]{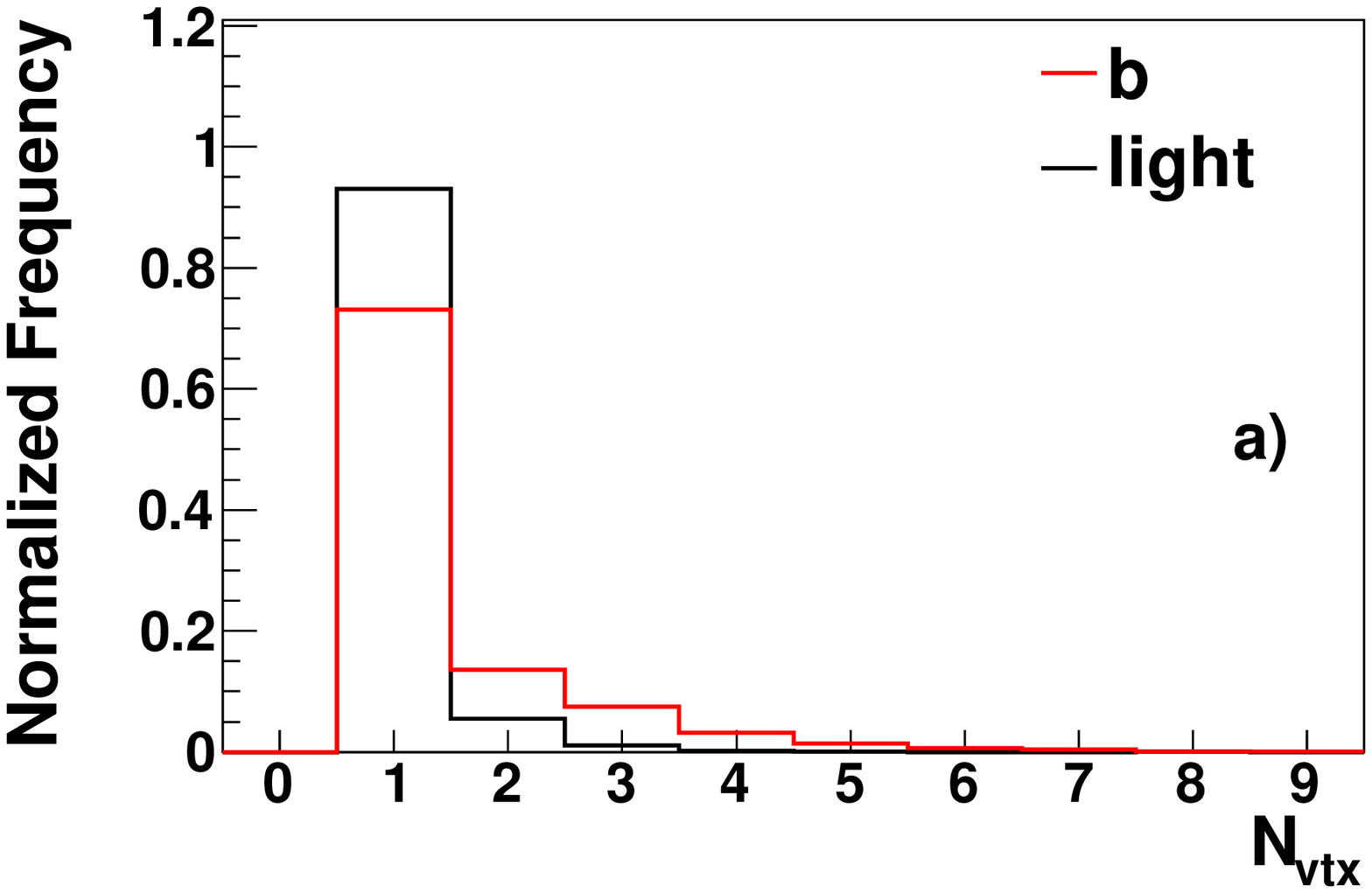}
    \includegraphics[width=0.4\textwidth]{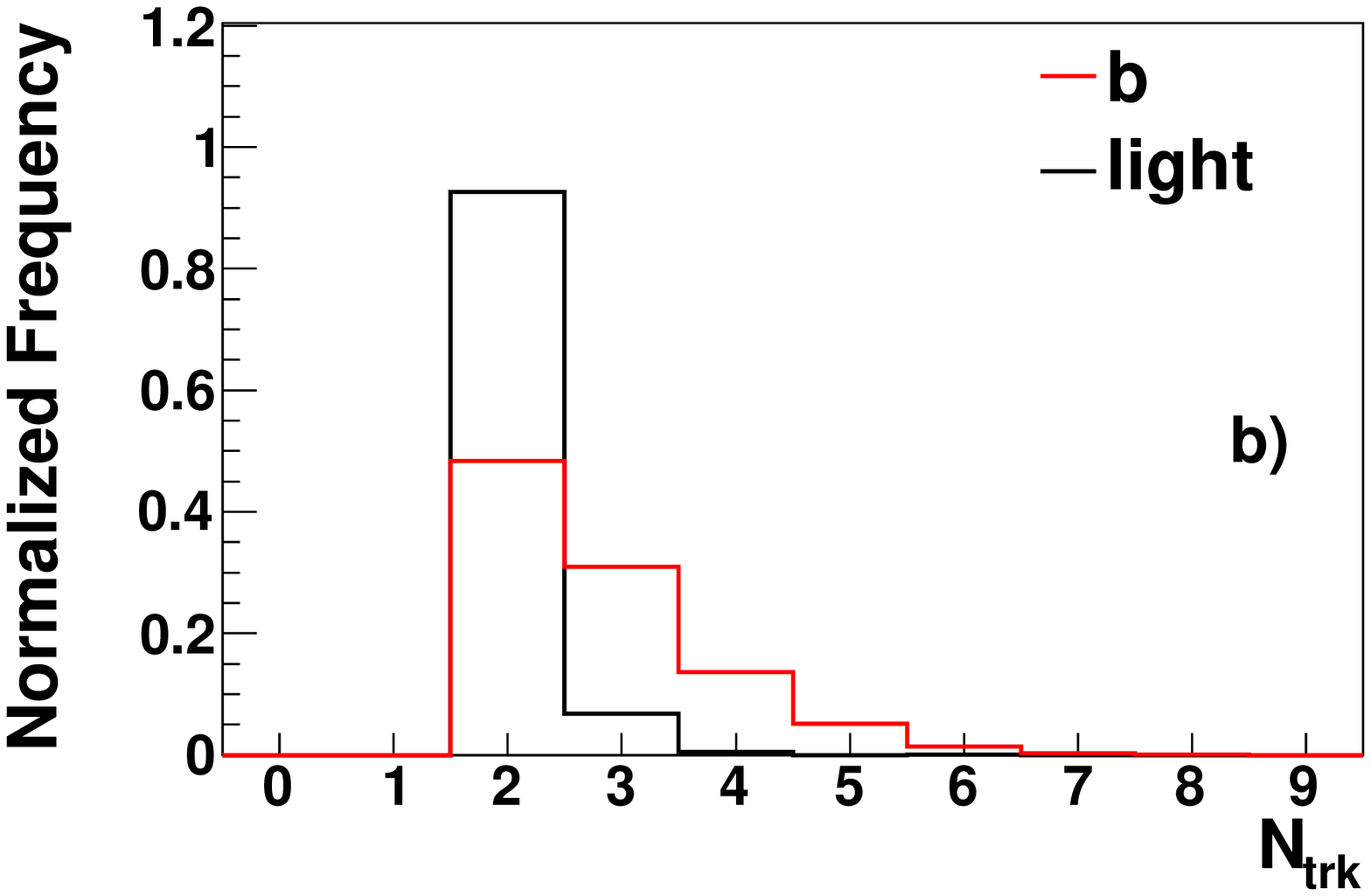}
    \includegraphics[width=0.4\textwidth]{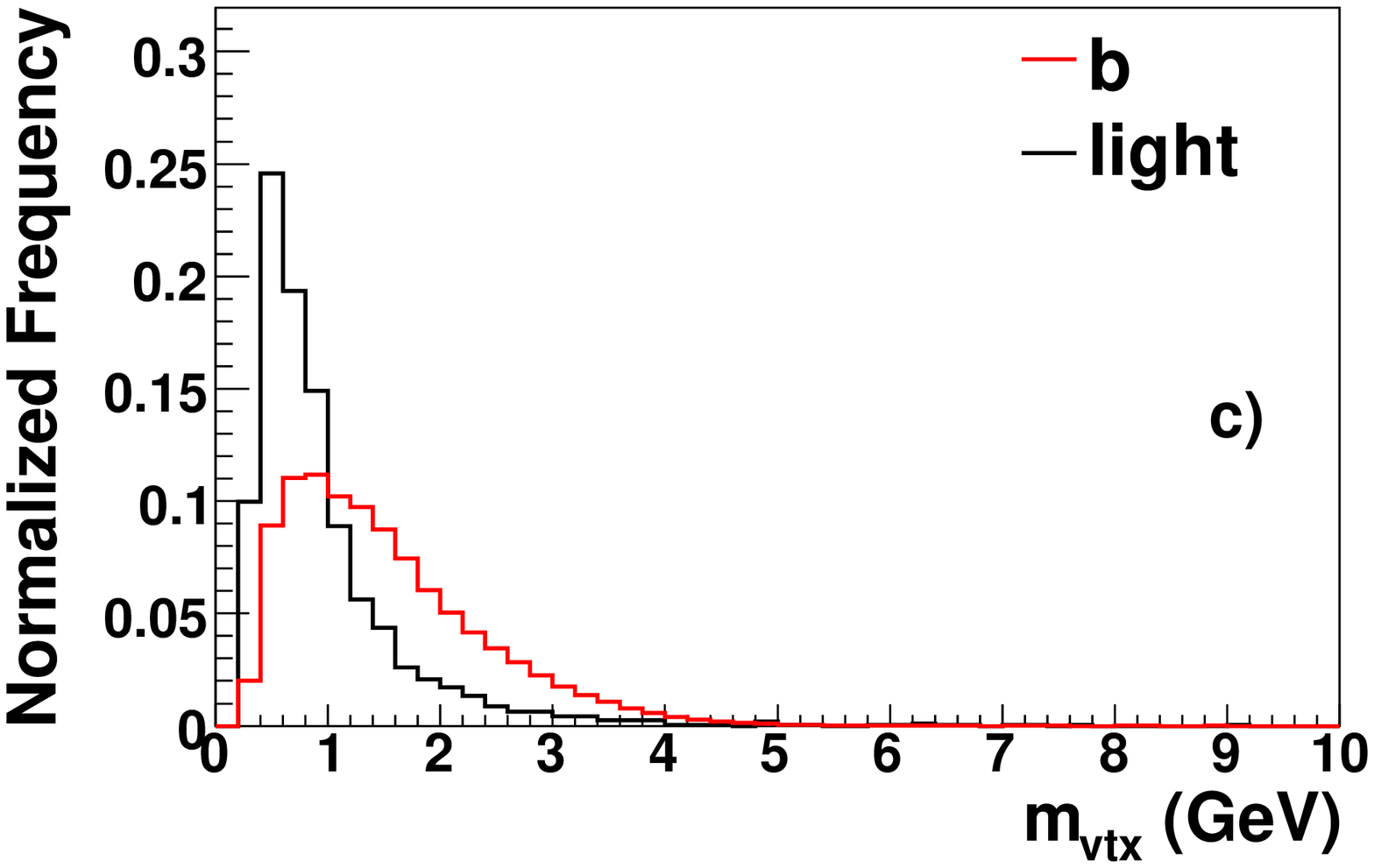}
    \includegraphics[width=0.4\textwidth]{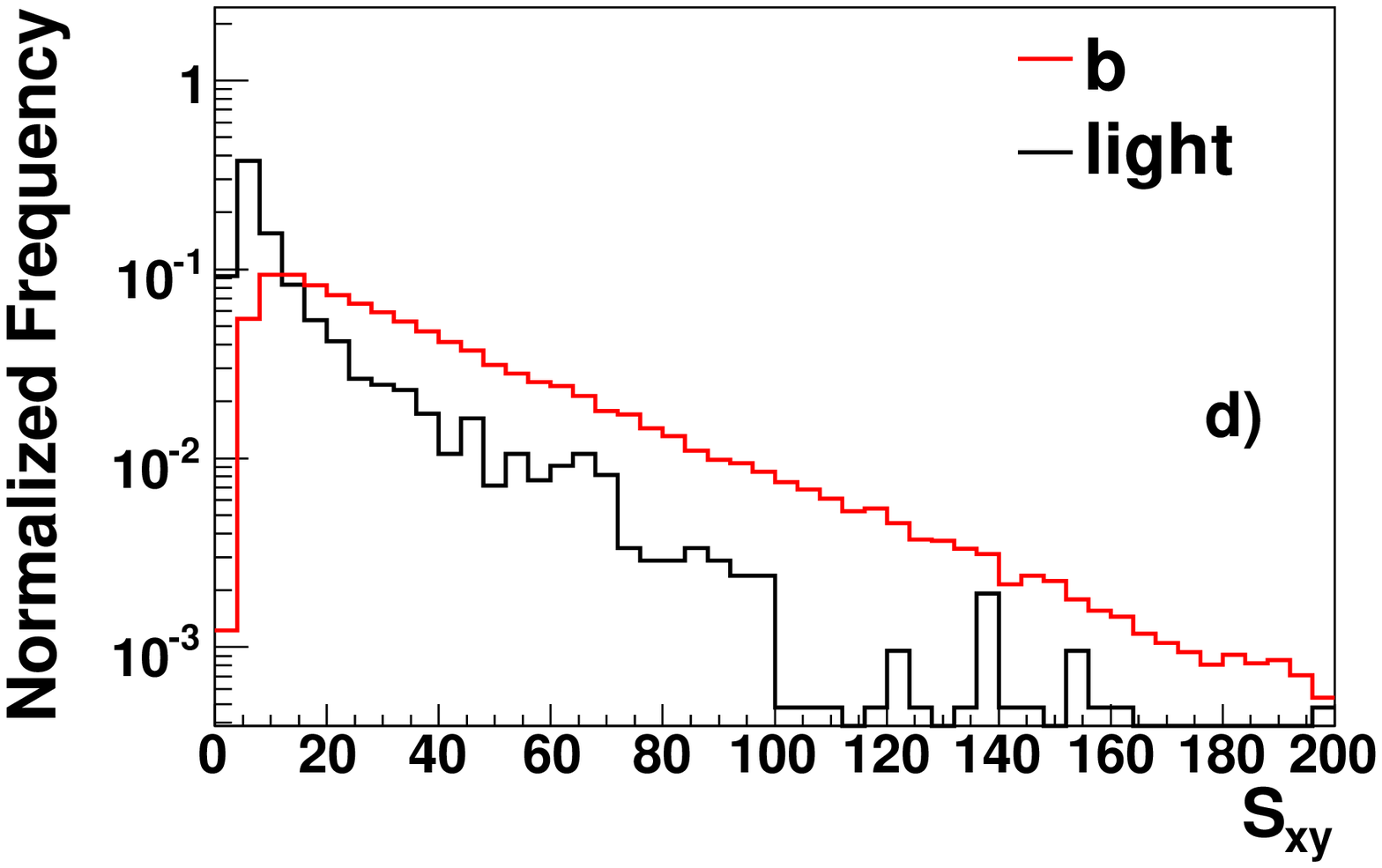}
  \end{center}
  \caption[] {Properties of the secondary vertices for tagged \bquark\ and light
    jets in \ttbar\ MC:
    multiplicity of vertices found in a track jet (a), 
    the number of tracks associated with the vertex (b), 
    the mass of the vertex (c), and  
    the decay length significance (d).}
  \label{fig:svt}
\end{figure}

Figure~\ref{fig:svt} shows distributions that characterize the 
properties of \bquark-jet and light-flavor secondary vertices reconstructed in
\ttbar\ events simulated using the \textsc{Alpgen}~\cite{Alpgen} event
generator:
the multiplicity of vertices found in a track jet ($N_{\mbox{\scriptsize vtx}}$), 
the number of tracks associated with the vertex ($N_{\mbox{\scriptsize trk}}$), 
the mass of the vertex ($m_{\mbox{\scriptsize vtx}}$)
calculated as the invariant mass of all track four-momentum vectors assuming
that all particles are pions, and the largest decay length
significance, $\mathcal{S}_{xy} \equiv L_{xy}/\sigma(L_{xy})$, where
$\sigma(L_{xy})$ represents the uncertainty on $L_{xy}$.


A calorimeter jet is tagged as a \bquark\ jet if it has at least one secondary
vertex with decay length significance
greater than a nominal value.
In order to characterize the performance of this tagging algorithm,
various versions have been studied, differing in the requirements
used in the selection procedure.

The algorithm as described above is referred to as the Loose algorithm.
In Sec.~\ref{sec:nn-variables} a high-efficiency version, labeled SuperLoose,
with a correspondingly high tagging rate for light-flavor jets is also used. The
high efficiency is obtained by placing no requirement on the impact parameter 
significance of the tracks used to reconstruct the secondary vertices.


%% file: jlip.tex
\section{The Jet Lifetime Probability Tagger}
\label{sec:jlip}

The impact parameters of all tracks associated with a calorimeter jet can be
combined into a single variable, the Jet Lifetime Probability (JLIP)
$\mathcal{P}_{\mbox{\scriptsize JLIP}}$~\cite{grederthesis,ref:jlip1,ref:jlip2},
which can be interpreted as the confidence level that all tracks in a jet
originate from the (selected) primary interaction point.
Jets from light quark fragmentation are expected to present a uniform
$\mathcal{P}_{\mbox{\scriptsize JLIP}}$ distribution between 0 and 1, whereas
jets from \cquark\ and \bquark\ quarks will exhibit a peak at a very low
$\mathcal{P}_{\mbox{\scriptsize JLIP}}$ value.
It is thus easy to select \bquark-quark jets by requiring this probability not
to exceed a given threshold, the value of which depends on the signal efficiency
and background rejection desired for a given physics analysis.

Using the impact parameters of reconstructed tracks allows control
of their resolution by using data, minimizing the need for simulated samples.
For this purpose, the impact parameter is signed by using the
coordinates of the track at the point of closest approach to the fitted primary
vertex, ${\vec d}$, and the jet momentum vector, ${\vec \pt}$(jet).
In the plane transverse to the beam axis, the distance of closest approach to
the primary vertex ($d = |{\vec d}|$) is given the same sign as the scalar
product ${\vec d} \cdot {\vec \pt}$(jet).
The signed $d$ distribution for tracks from light quark fragmentation is almost
symmetric, whereas the distribution for tracks from \bquark-hadron decay exhibits a
long tail at positive values. Therefore, provided that the sign of
${\vec d} \cdot {\vec \pt}$(jet) is correctly determined,
the negative part of the $d$ distribution allows the $d$ resolution function to
be parametrized.

\subsection{Calibration of the impact parameter resolution}
 
In order to tune the impact parameter uncertainty, $\sigma_{d}$, computed from
the track fit, the following variable is introduced:
$p_{\mbox{\scriptsize scat}} = p (\sin \theta)^{3/2}$, where $p$ is the particle momentum and
$\theta$ its polar angle relative to the beam axis.
In the plane transverse to the beam axis, the smearing due to multiple
scattering is inversely proportional to $\pt = p \sin\theta$ and proportional to
the square root of the distance traveled by the track. Assuming the detector
material to be distributed along cylinders aligned with the beam, this distance
is also inversely proportional to $\sin\theta$.
The $d$ distributions are then computed in sixteen different
$p_{\mbox{\scriptsize scat}}$ intervals.
 
In order to parametrize the $d$ resolution, five track categories are 
considered:
\begin{itemize}
\item at most six CFT hits (each doublet layer may give rise to one hit) and at
  least one SMT hit in the innermost layer, for tracks with $|\eta|>1.6$;
\item at least seven CFT hits and 1, 2, 3, or 4 SMT superlayer hits (the eight
  SMT barrel layers are grouped into four superlayers; the two neighboring
  layers constituting one superlayer provide full azimuthal acceptance).
\end{itemize}
The first category includes forward tracks outside the full CFT acceptance, 
the latter are central tracks with different numbers of SMT hits.
 
\begin{figure*}[bth]
  \begin{center}
   \includegraphics[width=0.45\textwidth]{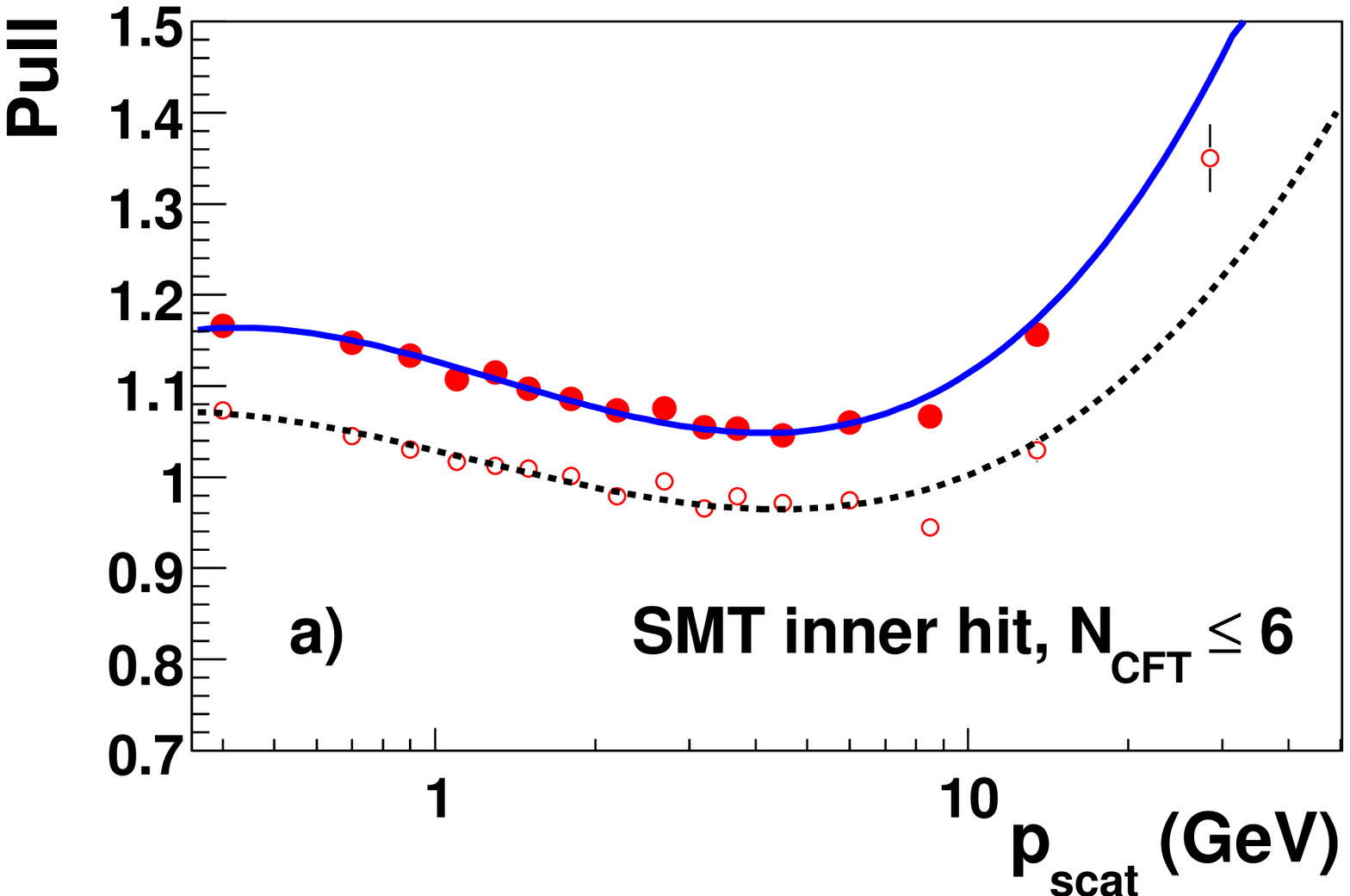}
    \includegraphics[width=0.45\textwidth]{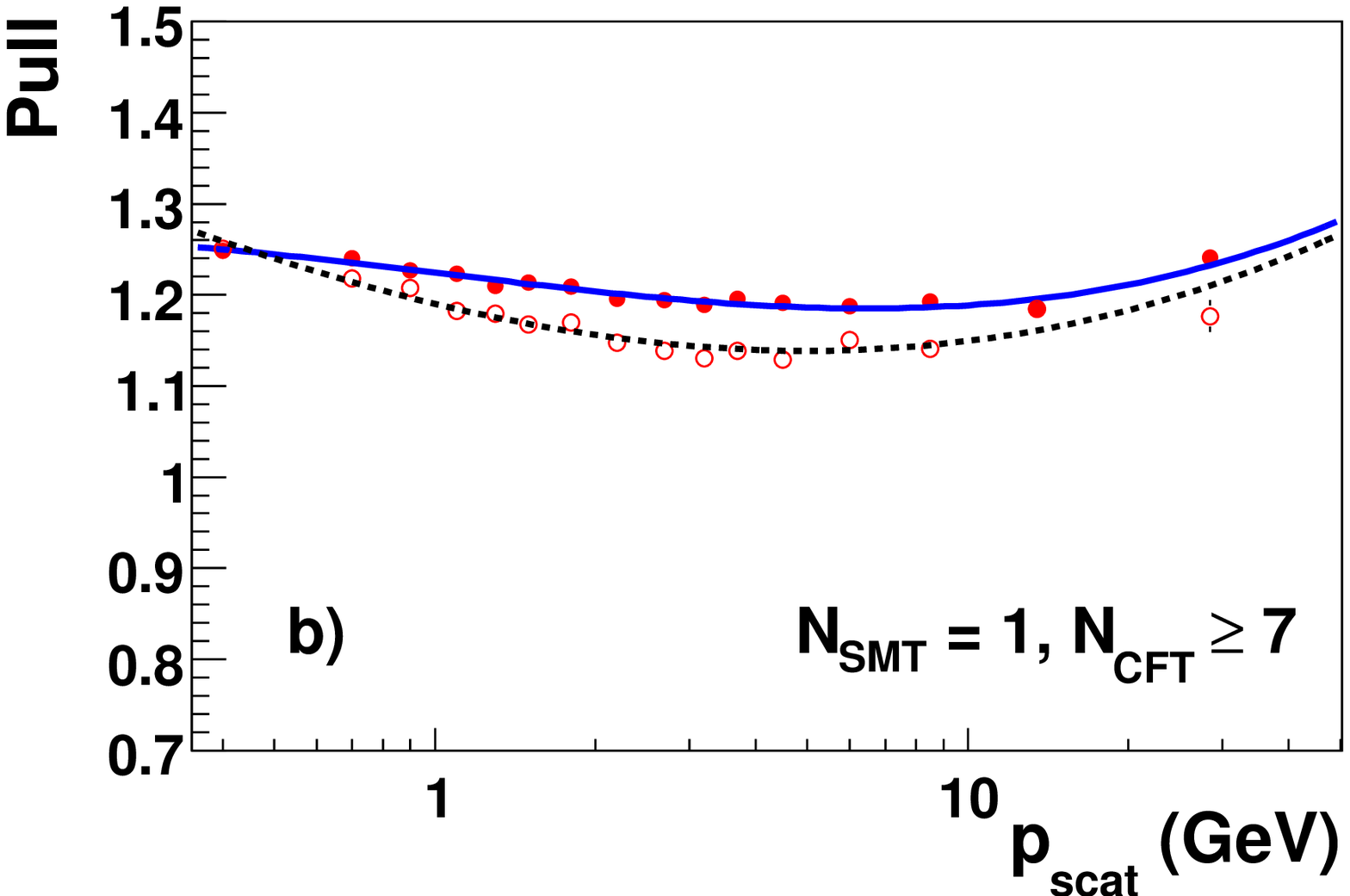}
    \includegraphics[width=0.45\textwidth]{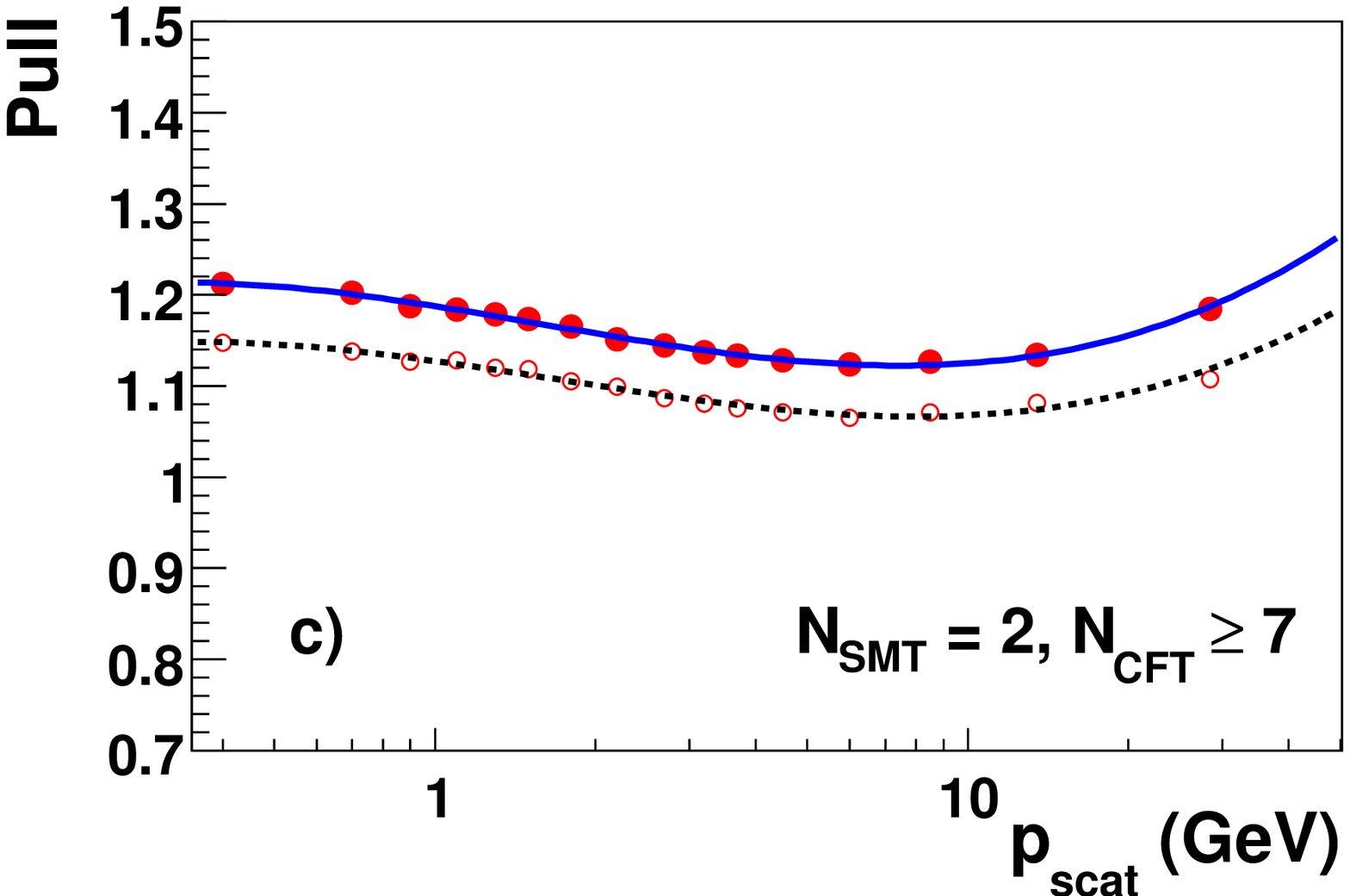}
    \includegraphics[width=0.45\textwidth]{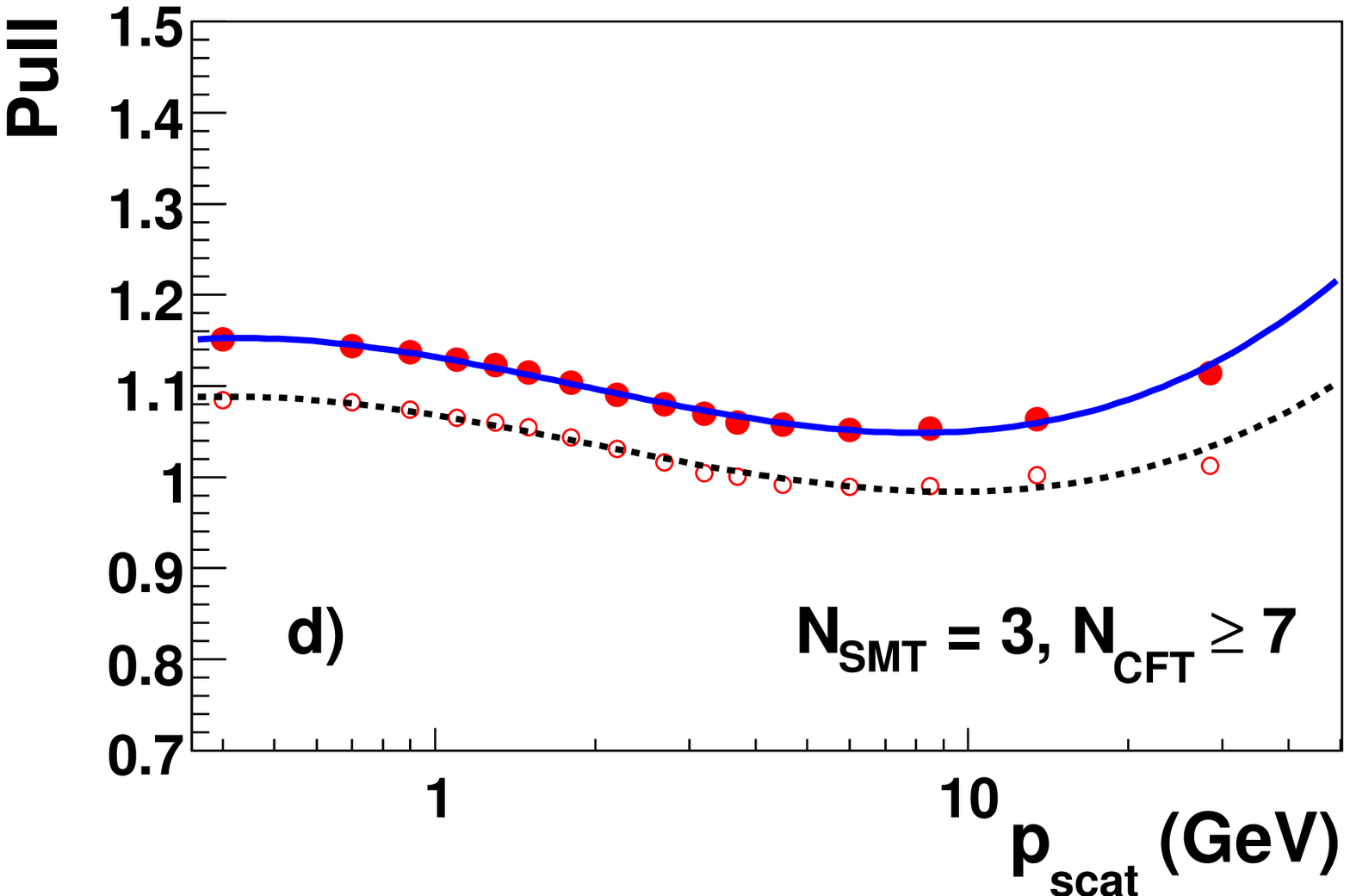}
  \end{center}
  \caption[] {
    Fitted pull values of the track impact parameters as a function of
    $p_{\mbox{\scriptsize scat}} = p (\sin\theta)^{3/2}$, for different 
    track categories. The superimposed solid (dashed) curves are empirical
    parametrizations of the data (QCD MC).}
  \label{fig:ip_pull}
\end{figure*}

The calibration is performed starting from the impact parameter significance
$\mathcal{S}_{d} = d/\sigma_{d}$, with the sign of $d$ determined as described
above. In each $p_{\mbox{\scriptsize scat}}$ interval and for each category, the
$\mathcal{S}_{d}$ distribution is fitted using a Gaussian function to describe the
$d$ resolution. The fitted pull values (the variance of the Gaussian) are presented in
Fig.~\ref{fig:ip_pull} for multijet data. The same calibration
procedure is carried out for simulated QCD events, and the corresponding results
are also shown in Fig.~~\ref{fig:ip_pull}.
The superimposed curves are empirical parametrizations of the data and the
simulation. The pull values are found to go up to 1.2 in data,
while they are closer to 1 in the simulation.

\begin{figure}[bth]
  \begin{center}
    \includegraphics[width=0.45\textwidth]{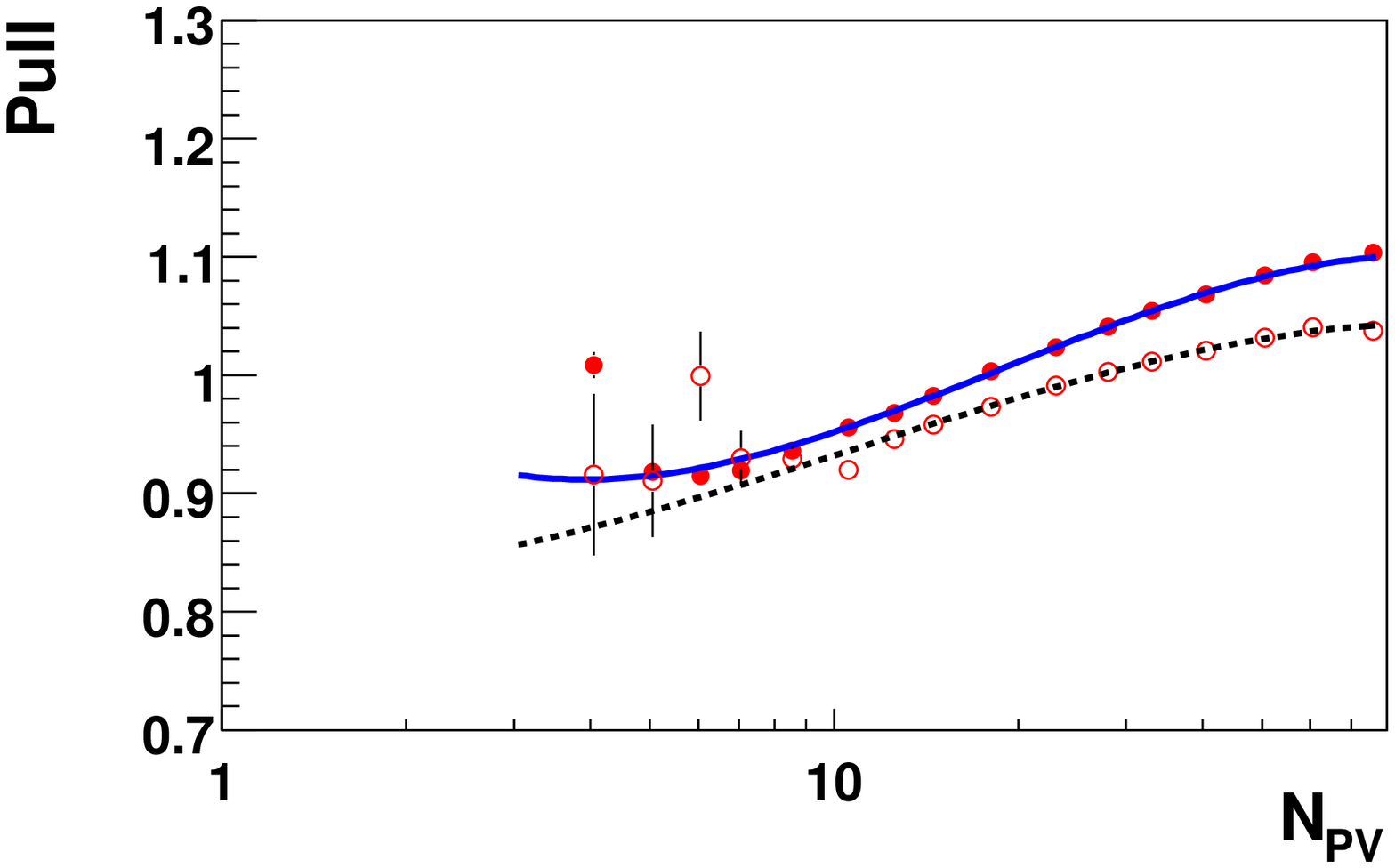}
  \end{center}
  \caption[] {
    Fitted pull values of the track impact parameters, corrected here for their
    $p_{\mbox{\scriptsize scat}}$ and category dependence,
    as a function of the number of tracks attached
    to the primary vertex, combining all track categories.
    The superimposed solid (dashed) curves are empirical
    parametrizations of the data (QCD MC).}
  \label{fig:ip_pull_pvNtr}
\end{figure}

As the impact parameter resolution may be sensitive to the primary vertex
resolution, the $d$ significance is also fitted separately for events with
different numbers of tracks, $N_{\mbox{\scriptsize PV}}$, attached to the primary vertex.
As shown in Fig.~\ref{fig:ip_pull_pvNtr} for multijet data and simulated QCD
events, the pull value increases significantly with $N_{\mbox{\scriptsize PV}}$
(here the $p_{\mbox{\scriptsize scat}}$ and category dependence of the pull
value is already corrected for).

\begin{figure*}[bth]
  \begin{center}
    \includegraphics[width=0.45\textwidth]{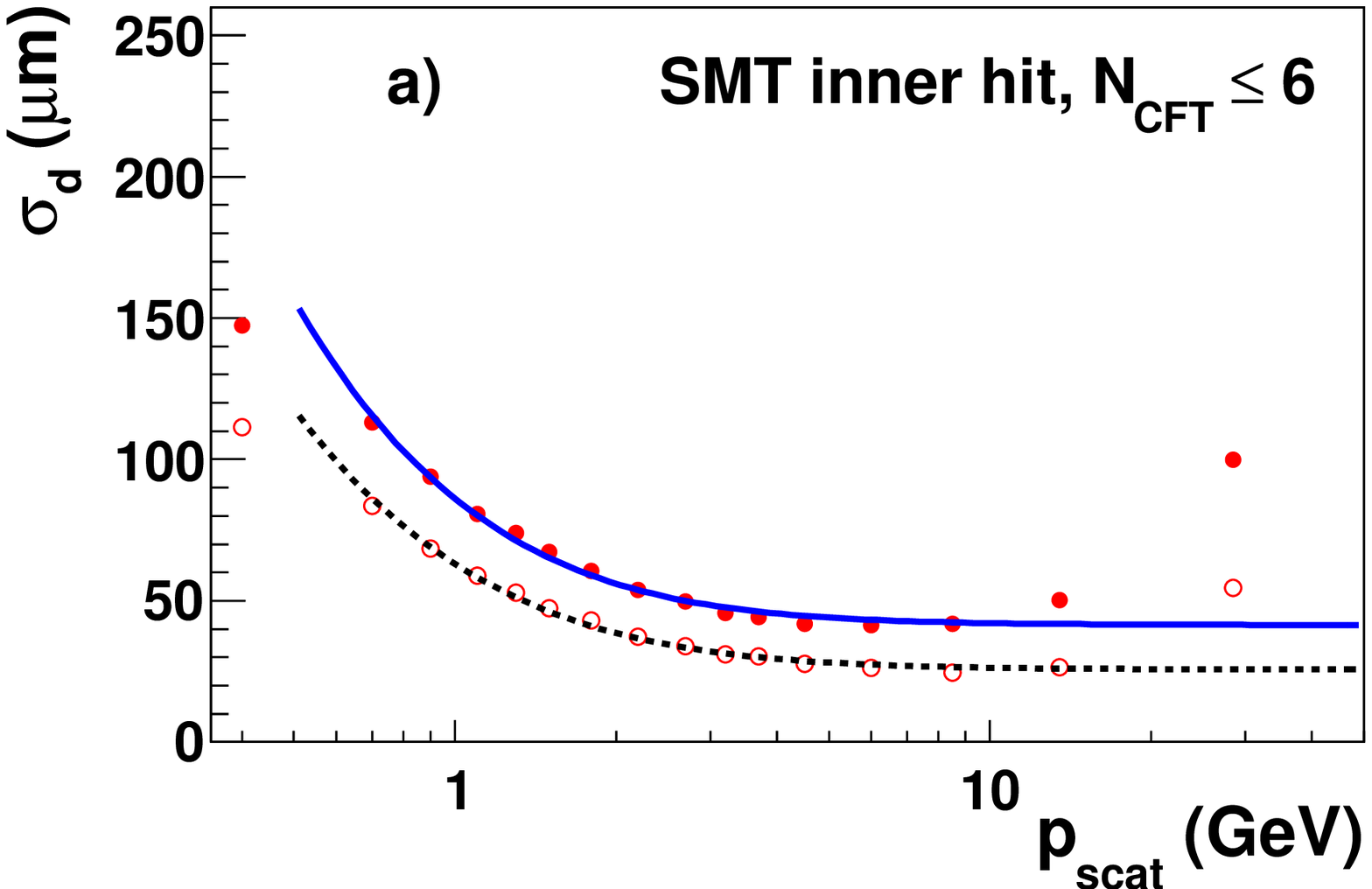}
    \includegraphics[width=0.45\textwidth]{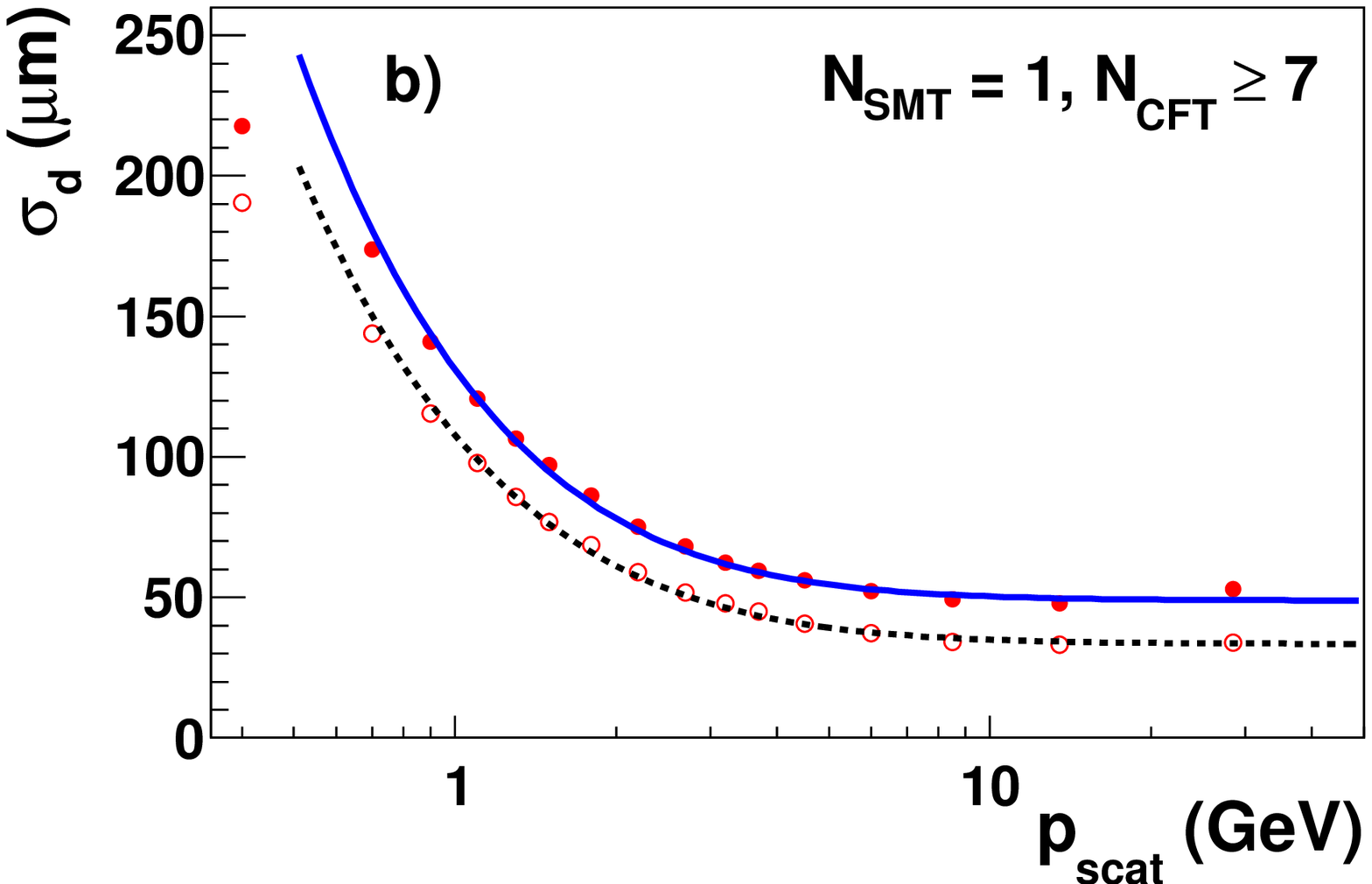}
    \includegraphics[width=0.45\textwidth]{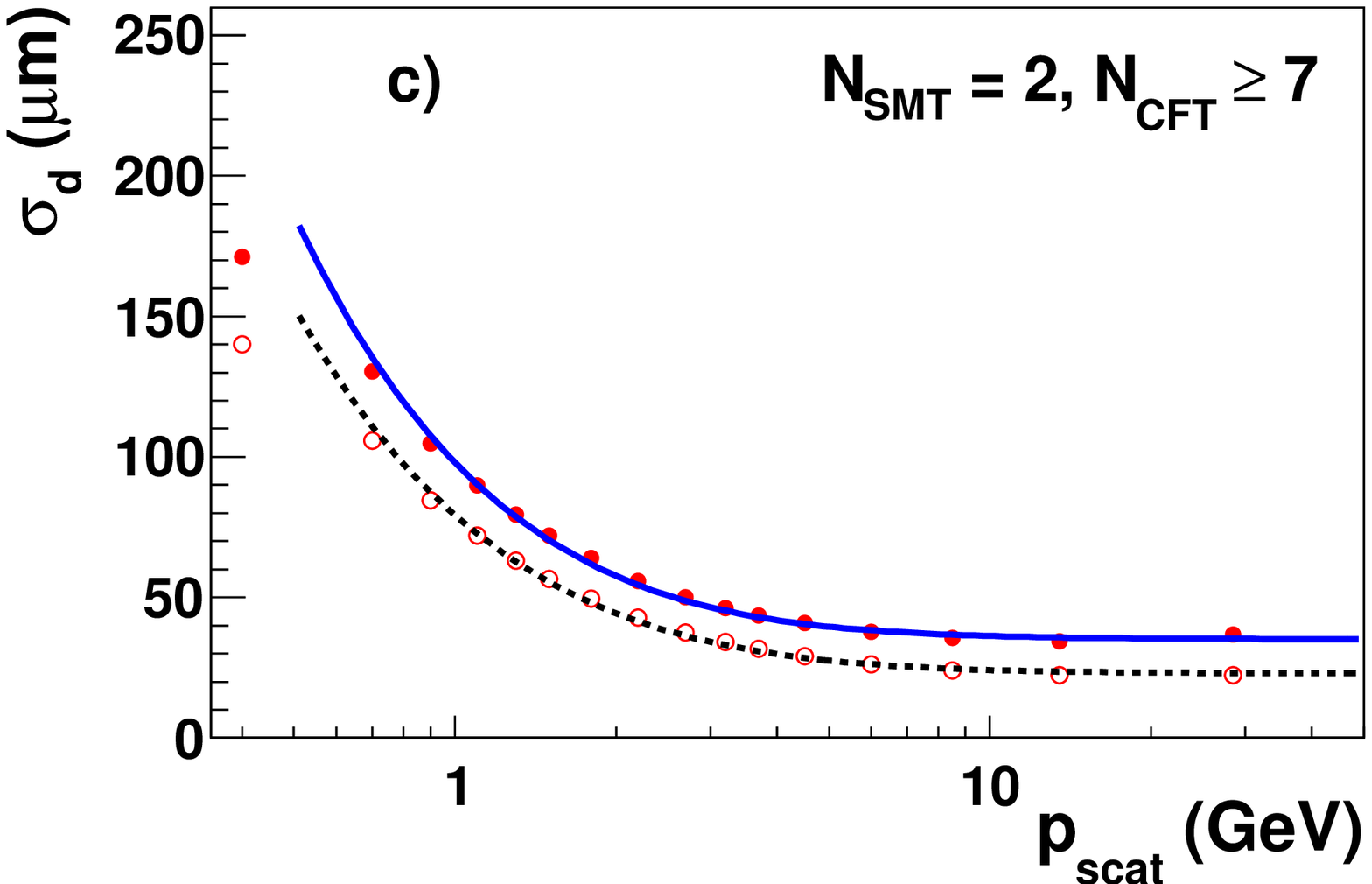}
    \includegraphics[width=0.45\textwidth]{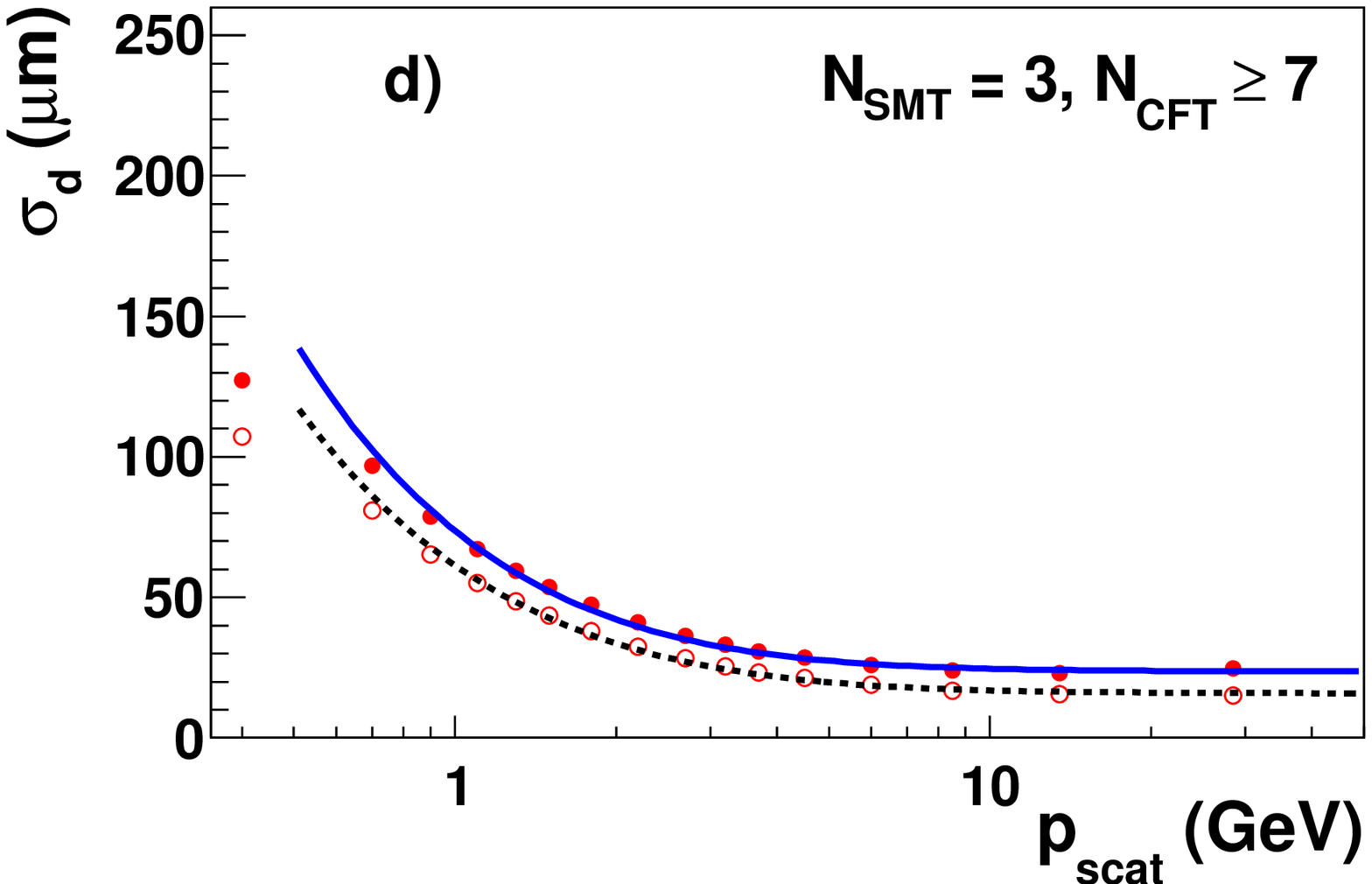}
 \end{center}
 \caption[] {
    Corrected track impact parameter resolution as a function of
    $p_{\mbox{\scriptsize scat}}$, for different track categories. The solid
    (dashed) curve is a fit to the data (QCD MC).}
  \label{fig:ip_res}
\end{figure*}

Then for each track, the impact parameter uncertainty and associated
significance are corrected according to the track's measured
$p_{\mbox{\scriptsize scat}}$ value, category $i$ and number of tracks
$N_{\mbox{\scriptsize PV}}$ attached to the primary vertex:
\begin{eqnarray}
  \sigma_{d}^{\mbox{\scriptsize corr}} &=& \mbox{pull}(p_{\mbox{\scriptsize
      scat}},i,N_{\mbox{\scriptsize PV}}) \cdot \sigma_{d} \\
  \mathcal{S}_{d}^{\mbox{\scriptsize corr}} &=& \mathcal{S} /
  \mbox{pull}(p_{\mbox{\scriptsize scat}},i,N_{\mbox{\scriptsize PV}}) \; \nonumber
\end{eqnarray}
The corrected $\sigma_{d}^{\mbox{\scriptsize corr}}$ resolutions are shown in
Fig.~\ref{fig:ip_res} for multijet data and simulated QCD events.
In the approximation of small angles~\cite{PDG2008}, they can be parametrized as
\begin{equation}
  \sigma^{\mbox{\scriptsize corr}}_{d} = \frac{a}{p \; (\sin \theta)^{3/2}} \oplus b,
\end{equation}
where $a$ describes multiple scattering effects, and $b$ is the asymptotic
resolution (which is sensitive to the primary vertex resolution, detector
alignment, SMT intrinsic resolution, \emph{etc.}); the symbol $\oplus$ denotes their
summing in quadrature.
This parametrization is superimposed in Fig.~\ref{fig:ip_res}.

The impact parameter resolution is observed to be better in the simulation than
in data.
For forward tracks with fewer than 7~CFT hits and with
$p_{\mbox{\scriptsize scat}} >10\GeVc$, the multiple-scattering small angles 
approximation is no longer valid. This can lead to misreconstructed track
momenta, and consequently the measured $d$ resolution is larger than its
asymptotic fitted value (see Fig.~\ref{fig:ip_res}a).
The parametrization is also imperfect at the lowest $p_{\mbox{\scriptsize
    scat}}$ values, although the reason for this has not been ascertained.

\subsection{Lifetime probability}
 
\begin{figure}[bth]
  \begin{center}
    \includegraphics[width=0.45\textwidth]{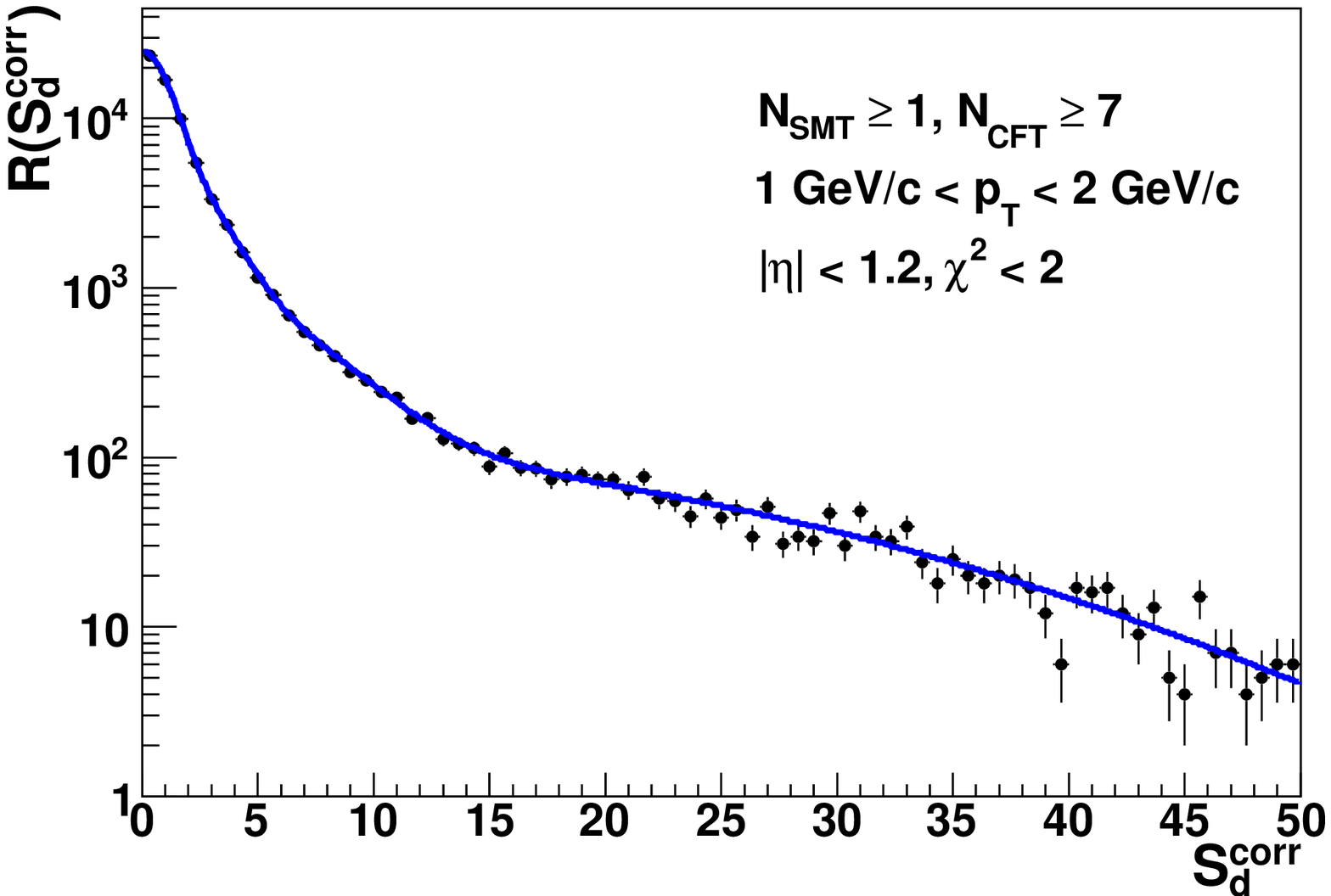}
  \end{center}
  \caption[] {
    The impact parameter resolution functions, shown here for one of the 29 track
    categories described in Table~\ref{tab:categories}, are parametrized as a sum 
    of four Gaussian functions.}
  \label{fig:ip_resfunc}
\end{figure}

The data are used to calibrate the impact parameter significance.
For multijet data or QCD MC, the negative part of the $d$
significance distribution, denoted impact parameter resolution function
$\mathcal{R}(\mathcal{S}_{d}^{\mbox{\scriptsize corr}})$, is parametrized as
the sum of four Gaussian functions, as illustrated in Fig.~\ref{fig:ip_resfunc}.
After removing tracks originating from $V^{0}$ candidates (see
Sec.~\ref{sec:preliminaries}), the track categories used in the previous section
are extended to take into account the number of SMT and CFT hits, $|\eta|$,
fit $\chi^2$, and \pt\ values of the tracks, as listed in Table~\ref{tab:categories}.
The category ranges are adjusted to describe as much as possible
geometric and tracking effects.
This further refinement results in 29 track categories.

\begin{table*}[bth]
  \begin{center}
    \begin{tabular}{|c|c|c|c|c|}
      \hline
      SMT hits                & CFT hits & $|\eta|$     & fit $\chi^2$   & \pt\ ($\!\!\!\GeVc$) \\
      \hline
      $\ge 1$ inner layer hit & $\le 6$  & 1.6--2, $>2$ & $>0$       & $>1$ \\
      \hline
      1 superlayer            & $\ge 7$  & $<1.2$       & 0--2, $>2$ & $>1$ \\
            "                 &    "     & $>1.2$       & $>0$       & $>1$ \\
      \hline
      2, 3, 4 superlayers      & $\ge 7$  & $<1.2$       & 0--2       & 1--2, 2--4, $>4$ \\
                "             &    "     &     "        & 2--4, $>4$ & $>1$             \\
                "             &    "     & 1.2--1.6     & 0--2, $>2$ & $>1$             \\
                "             &    "     & $>1.6$       & $>0$       & $>1$             \\
     \hline
    \end{tabular}
    \caption{Track categories used for the parametrization of the impact
      parameter resolution functions. The ``$\ge 1$ inner layer hit'' line denotes the
      requirement of at least one hit in the innermost SMT superlayer.}
    \label{tab:categories}
  \end{center}
\end{table*}

For tracks with positive $d$, the parametrized resolution function is then
converted into a probability for this track to originate from the primary
interaction point
\begin{equation}
  \mathcal{P}_{\mbox{\scriptsize trk}}(\mathcal{S}_{d}^{\mbox{\scriptsize corr}}) =
  \frac{\int_{-\infty}^{-|\mathcal{S}_{d}^{\mbox{\scriptsize corr}}|} \mathcal{R}(s) ds}
  {\int_{-\infty}^{0}\mathcal{R}(s) ds} \; .
 \label{eq:jlip_ptrk}
\end{equation}

\begin{figure*}[bth]
  \begin{center}
   \includegraphics[width=0.40\textwidth]{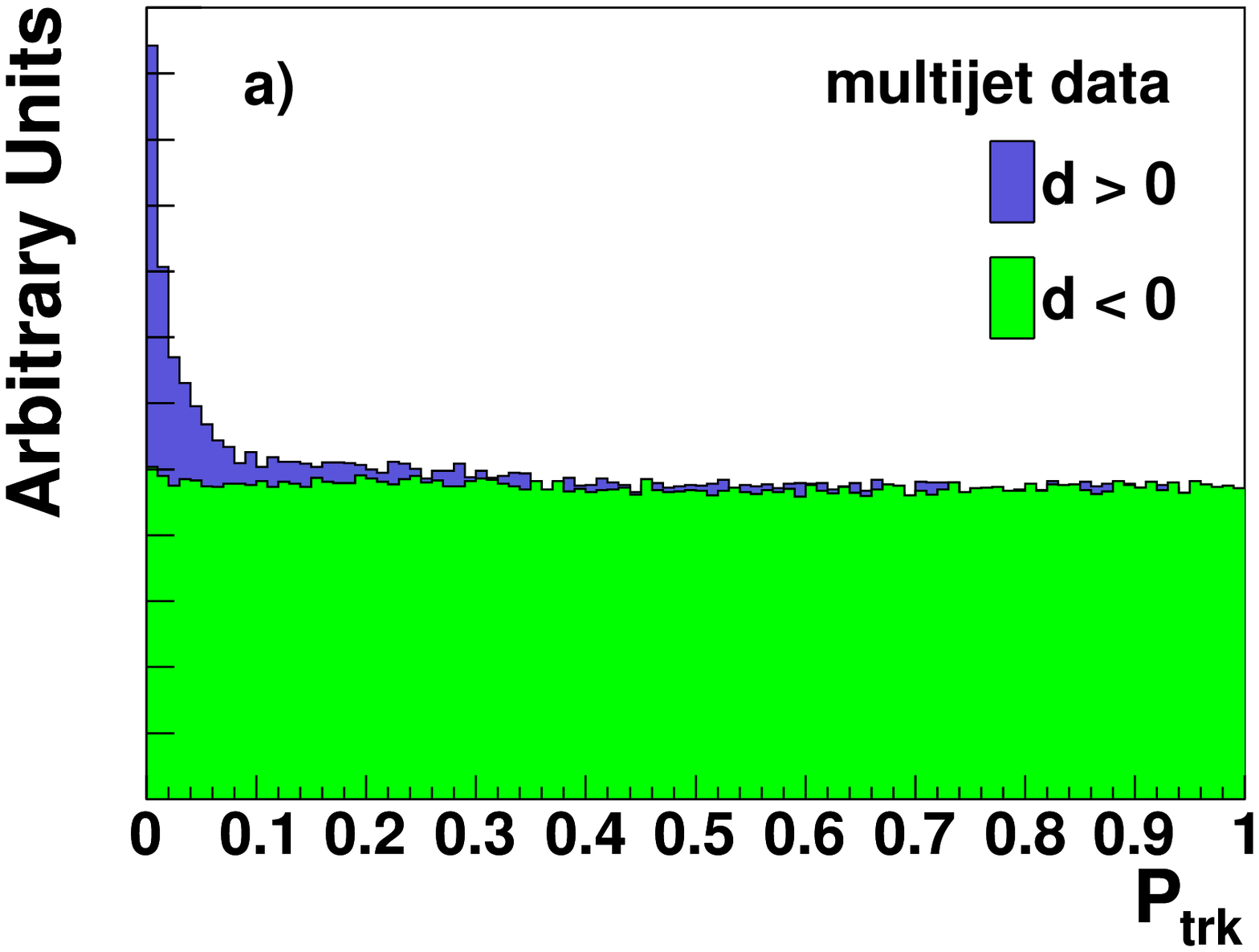}
   \includegraphics[width=0.40\textwidth]{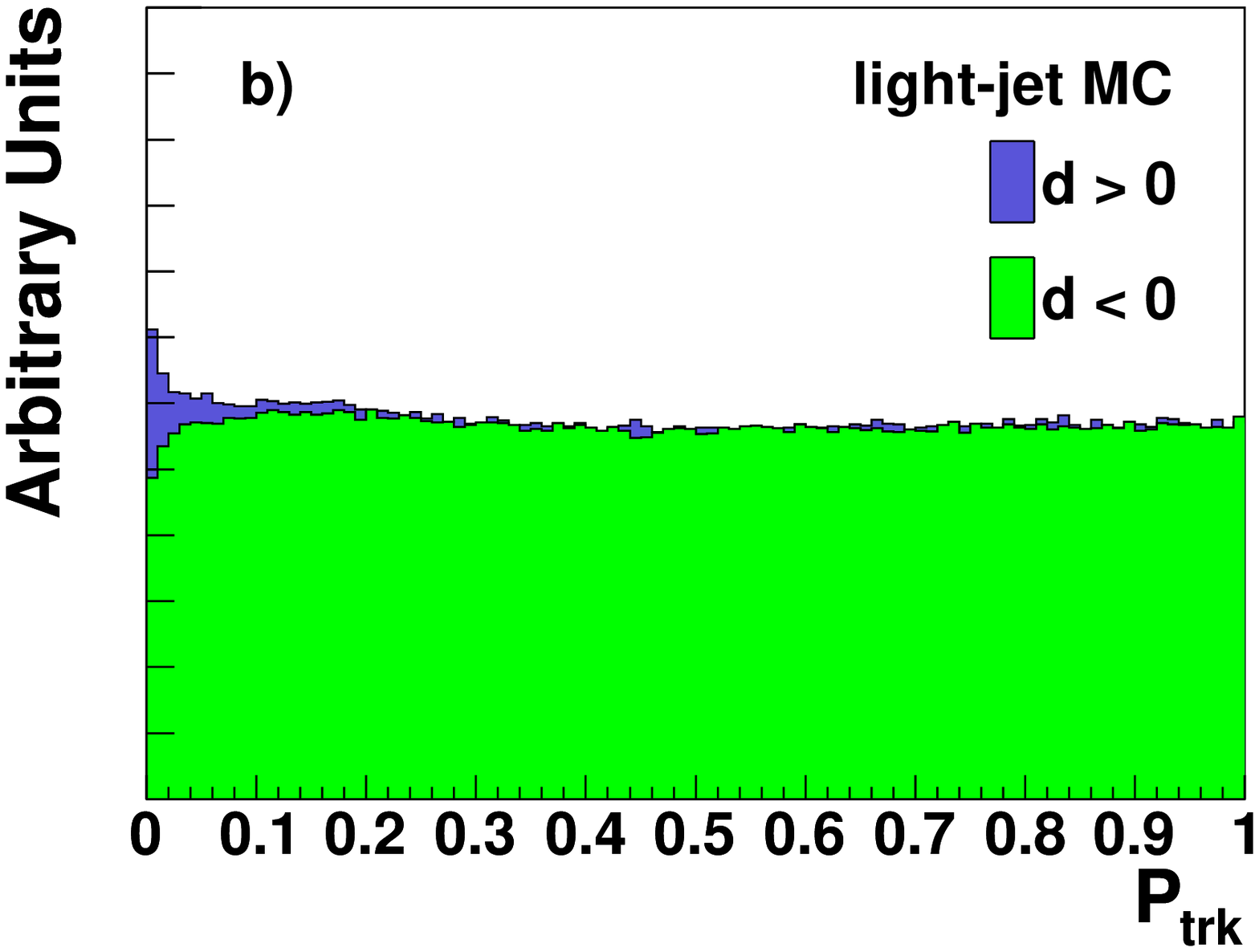}
   \includegraphics[width=0.40\textwidth]{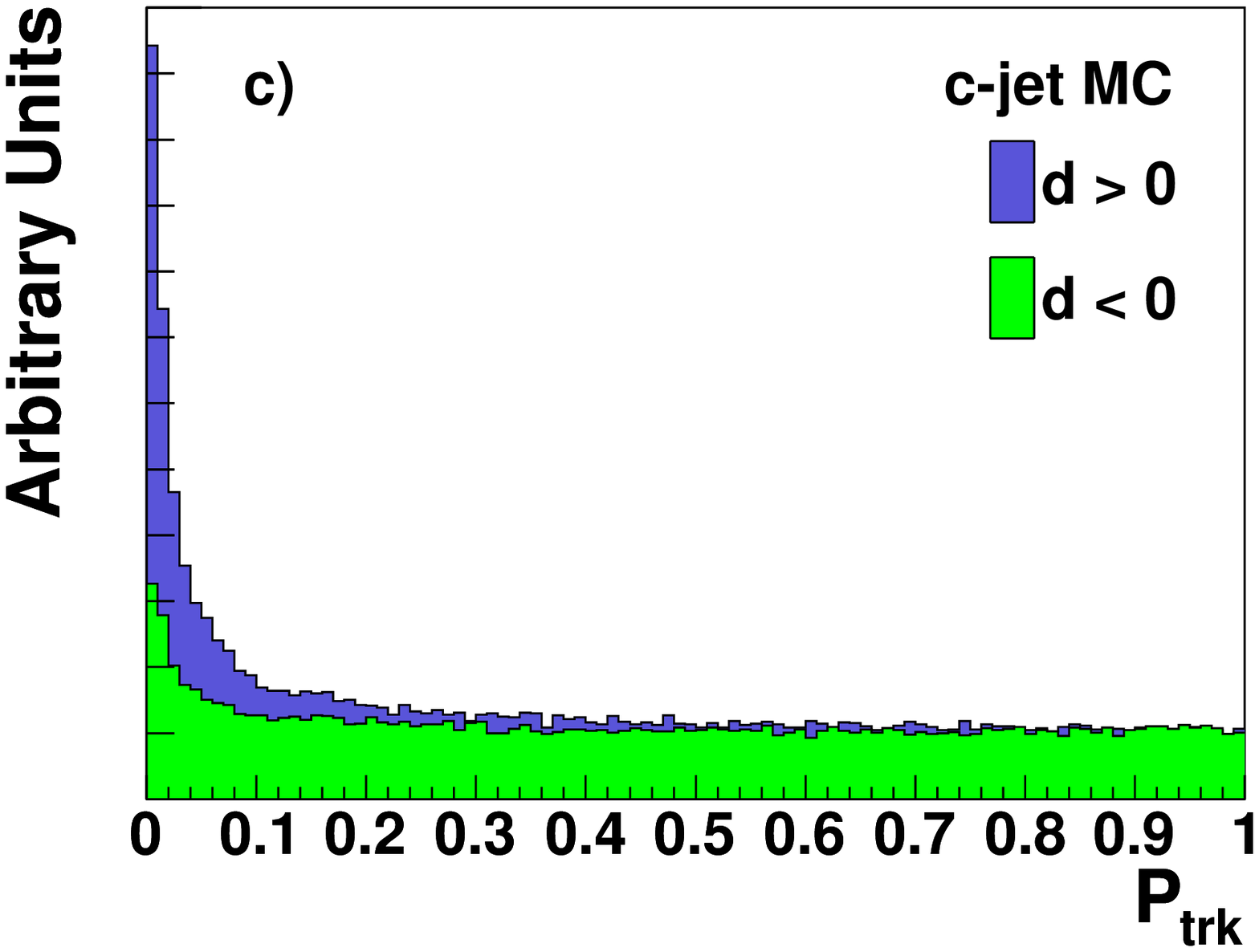}
   \includegraphics[width=0.40\textwidth]{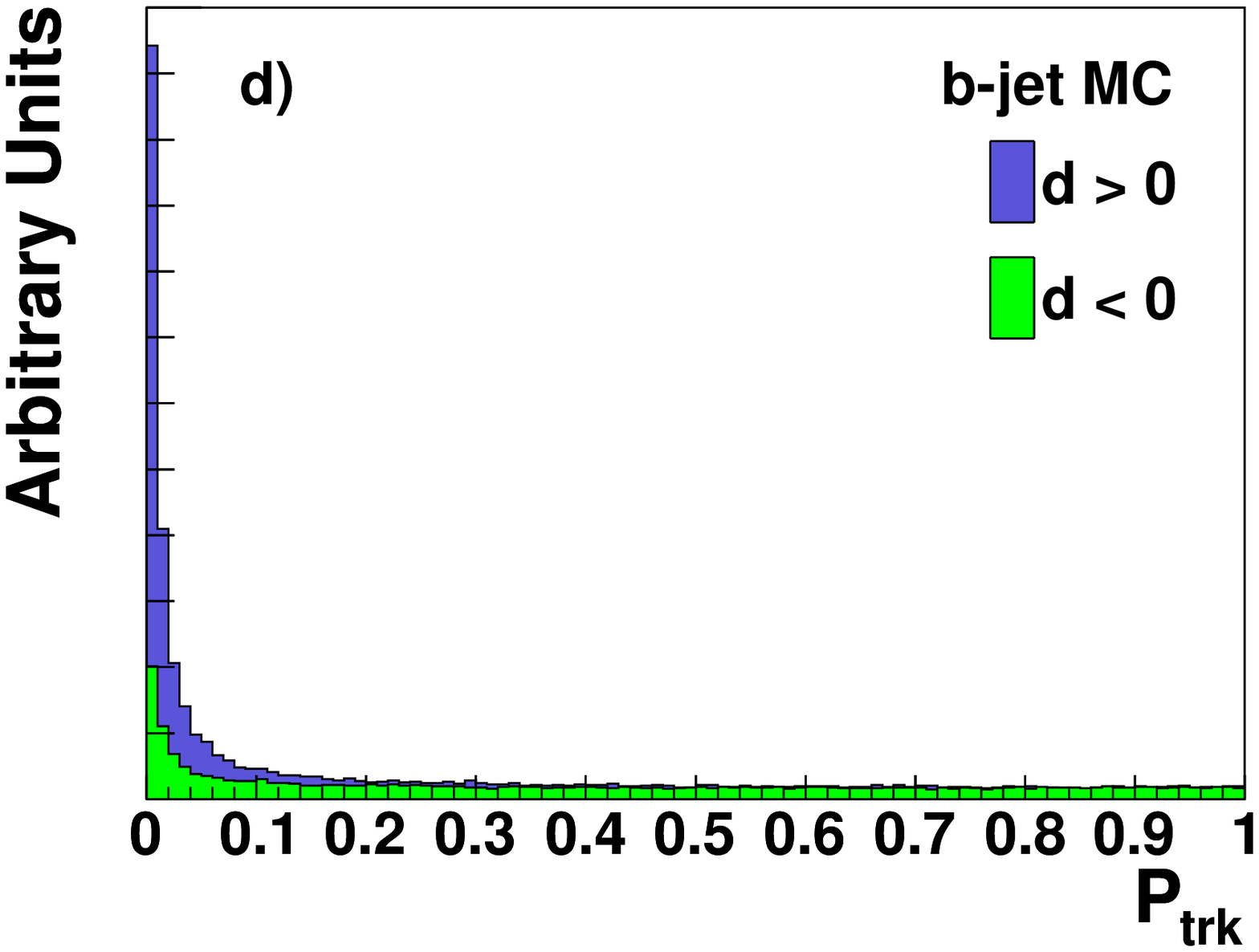}
  \end{center}
 \caption[] {
    Track probability ($\mathcal{P}_{\mbox{\scriptsize trk}}$, see
    Eq.~\ref{eq:jlip_ptrk}) distribution in multijet data (a) and QCD MC
    simulation of light-flavor (b), \cquark\ (c), and \bquark\ (d) jets,
    for positive (dark histograms) and negative (light histograms) $d$ values.}
  \label{fig:prob_trk}
\end{figure*}

The corresponding track probabilities are shown in Fig.~\ref{fig:prob_trk} for
multijet data and simulated jets of different flavors, and for positive and
negative $d$ values.
Tracks with negative $d$ values in multijet data and in simulated light quark
jets are used to define the $d$ resolution functions, thus ensuring uniform
$\mathcal{P}_{\mbox{\scriptsize trk}}(\mathcal{S}_{d}^{\mbox{\scriptsize corr}}<0)$
probability distributions. For positive $d$, a significant peak at low
$\mathcal{P}_{\mbox{\scriptsize trk}}(\mathcal{S}_{d}^{\mbox{\scriptsize corr}}>0)$
probability is present in simulated \cquark\ and \bquark\ jets.
In multijet data, a peak is also observed at low values which is partly due to
the presence of $V^{0}$s (which are not all removed, see Sec.~\ref{sec:v0}), but
also to tracks from charm and \bquark-hadron decays. 
Note that for simulated \cquark\ and \bquark\ jets, a slight peak remains at
negative $d$ due to a flip of the $d$ sign, mainly due to tracks very close to the
jet axis direction (see also Sec.~\ref{sec:csip}). In addition, a small dip is
observed at low
$\mathcal{P}_{\mbox{\scriptsize trk}}(\mathcal{S}_{d}^{\mbox{\scriptsize
    corr}}<0)$. This dip is due to the fact that, like for the data, the
resolution function for the simulated sample has been derived \emph{without}
removal of the heavy flavor component.
 
Finally, the selected $N_{\mbox{\scriptsize trk}}$ tracks with positive $d$
significance are used to compute the \emph{jet probability}
$\mathcal{P}_{\mbox{\scriptsize JLIP}}$ as
\begin{eqnarray}
  \mathcal{P}_{\mbox{\scriptsize JLIP}} \equiv
  \mathcal{P}^{+}_{\mbox{\scriptsize JLIP}} & = & \Pi \times
  \sum_{j=0}^{N_{\mbox{\scriptsize trk}}-1}
  \frac{(-\ln\Pi)^j}{j!} \\
  \mbox{with} \;\;\; \Pi & = & \prod_{i=1}^{N_{\mbox{\scriptsize trk}}}
  \mathcal{P}_{\mbox{\scriptsize trk}}(\mathcal{S}^{\mbox{\scriptsize
      corr}}_{d,i})\; . \nonumber
\end{eqnarray}
For tracks with negative $d$, a jet probability $\mathcal{P}^{-}_{\mbox{\scriptsize JLIP}}$
can be computed analogously.

\begin{figure*}[bth]
  \begin{center}
   \includegraphics[width=0.40\textwidth]{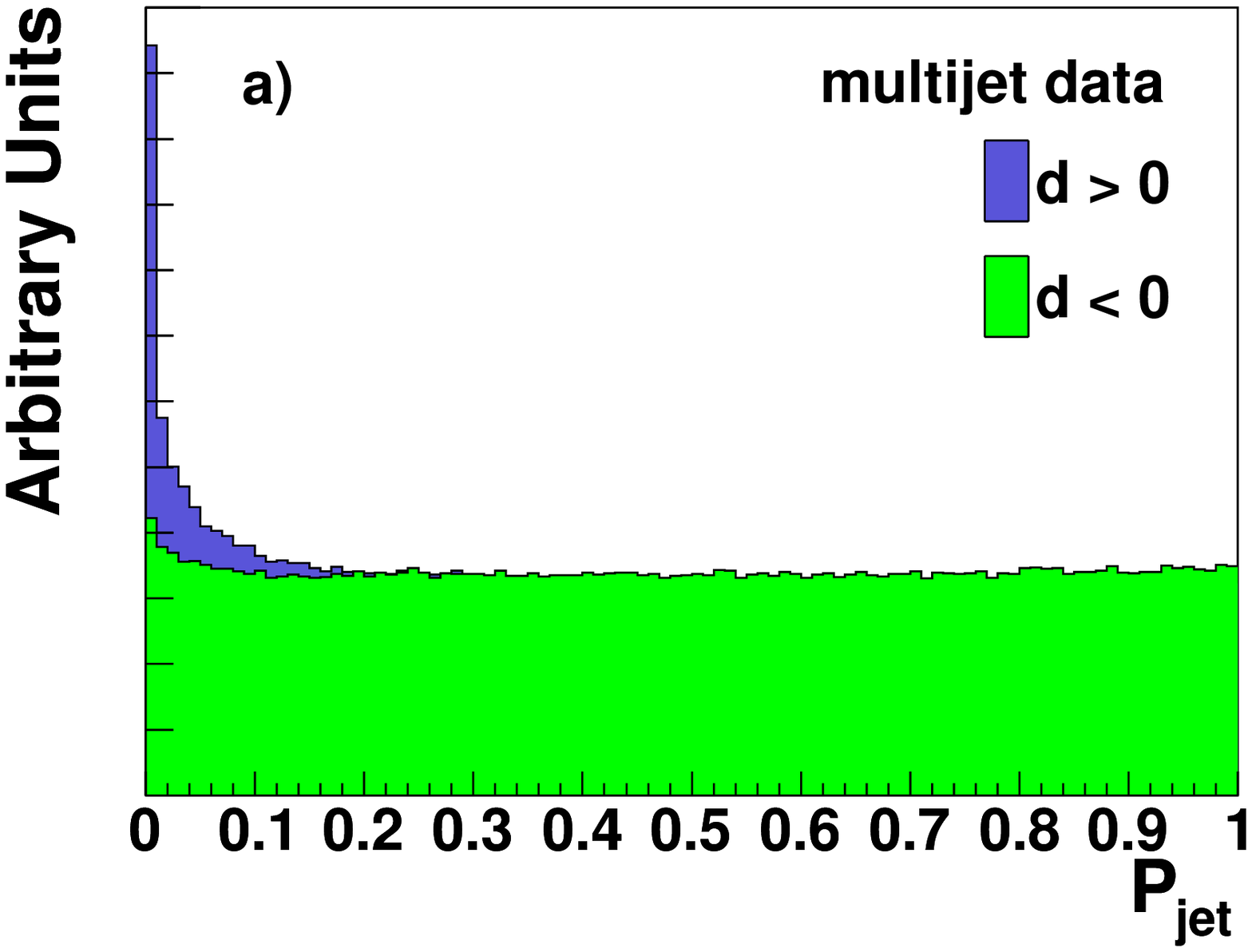}
    \includegraphics[width=0.40\textwidth]{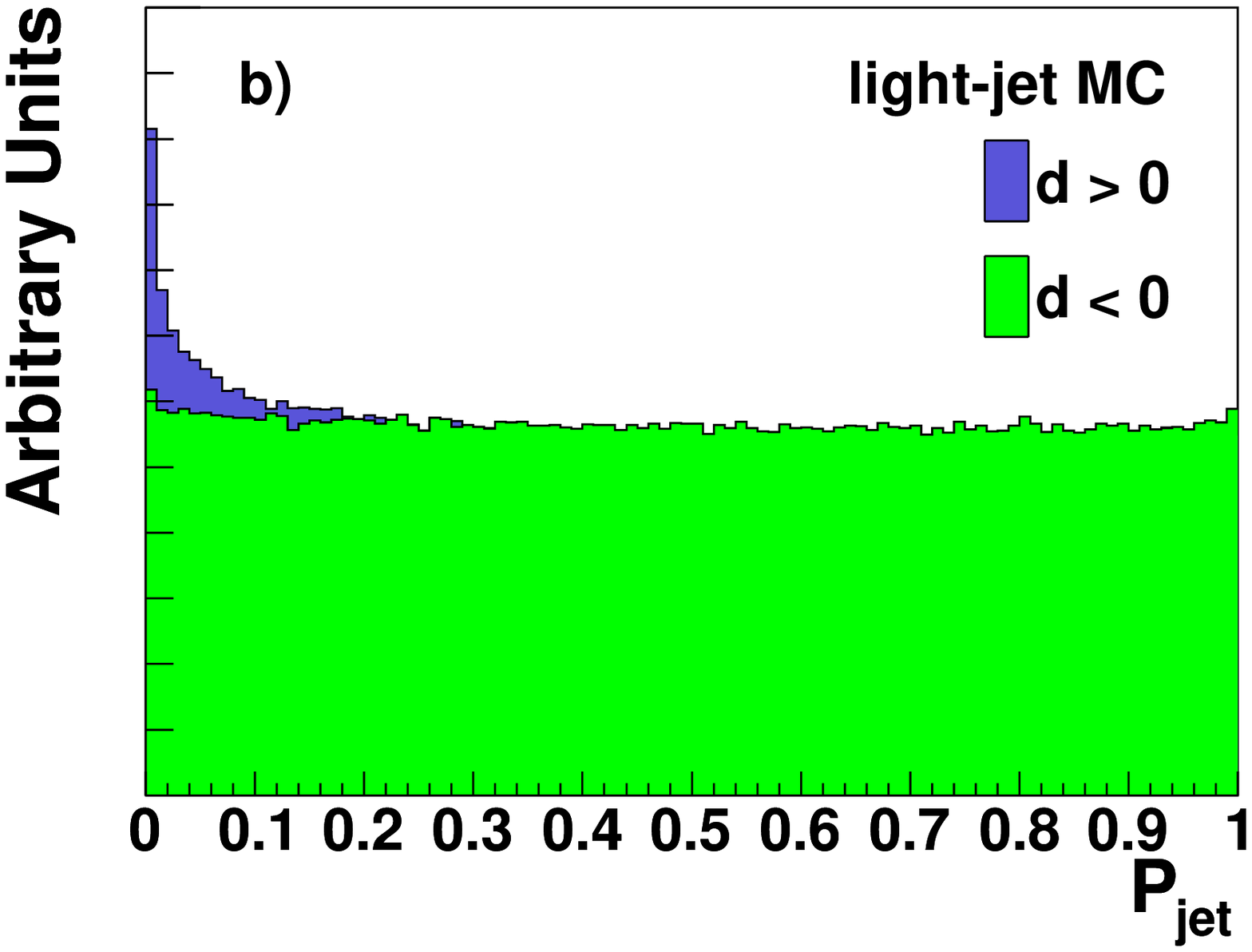}
    \includegraphics[width=0.40\textwidth]{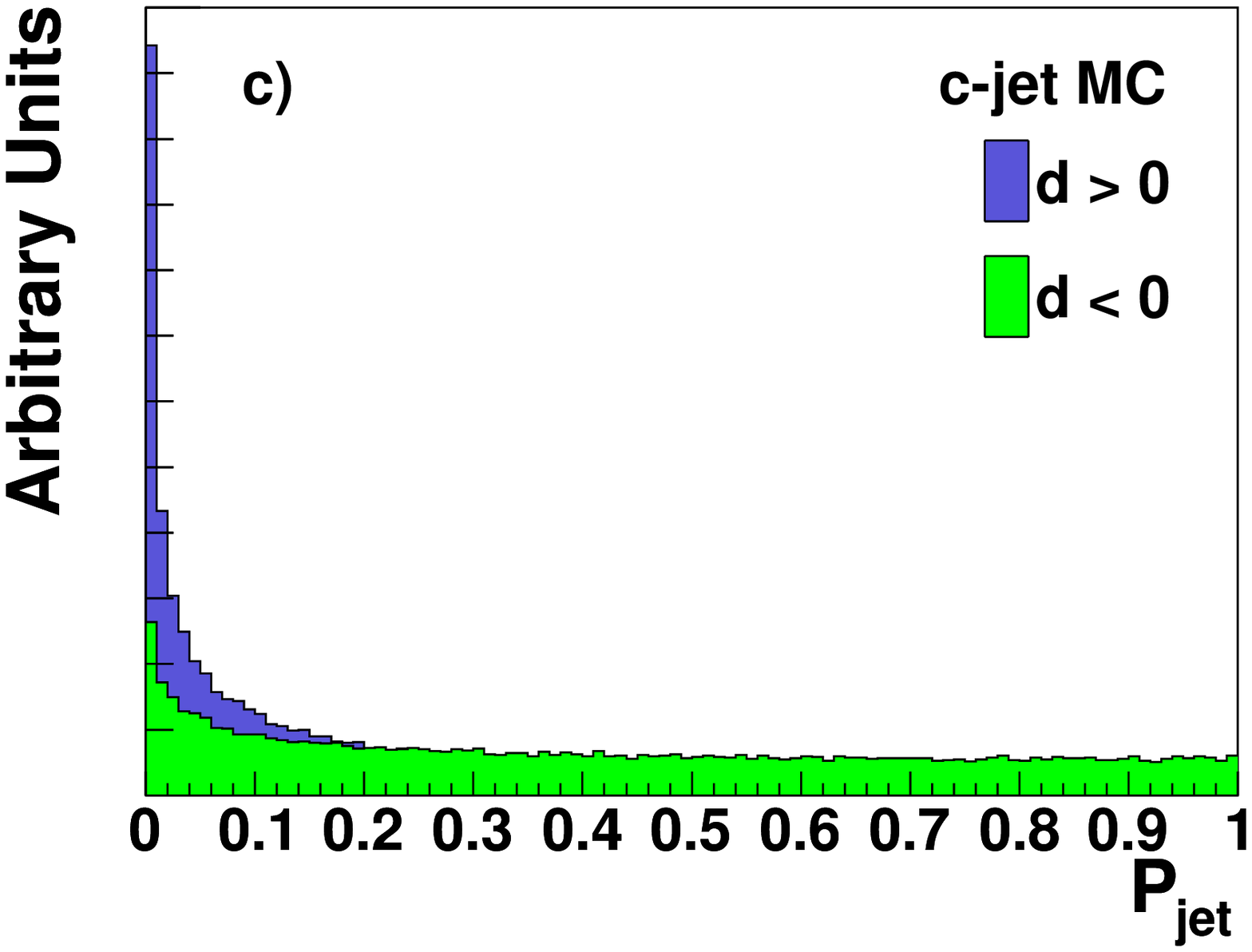}
    \includegraphics[width=0.40\textwidth]{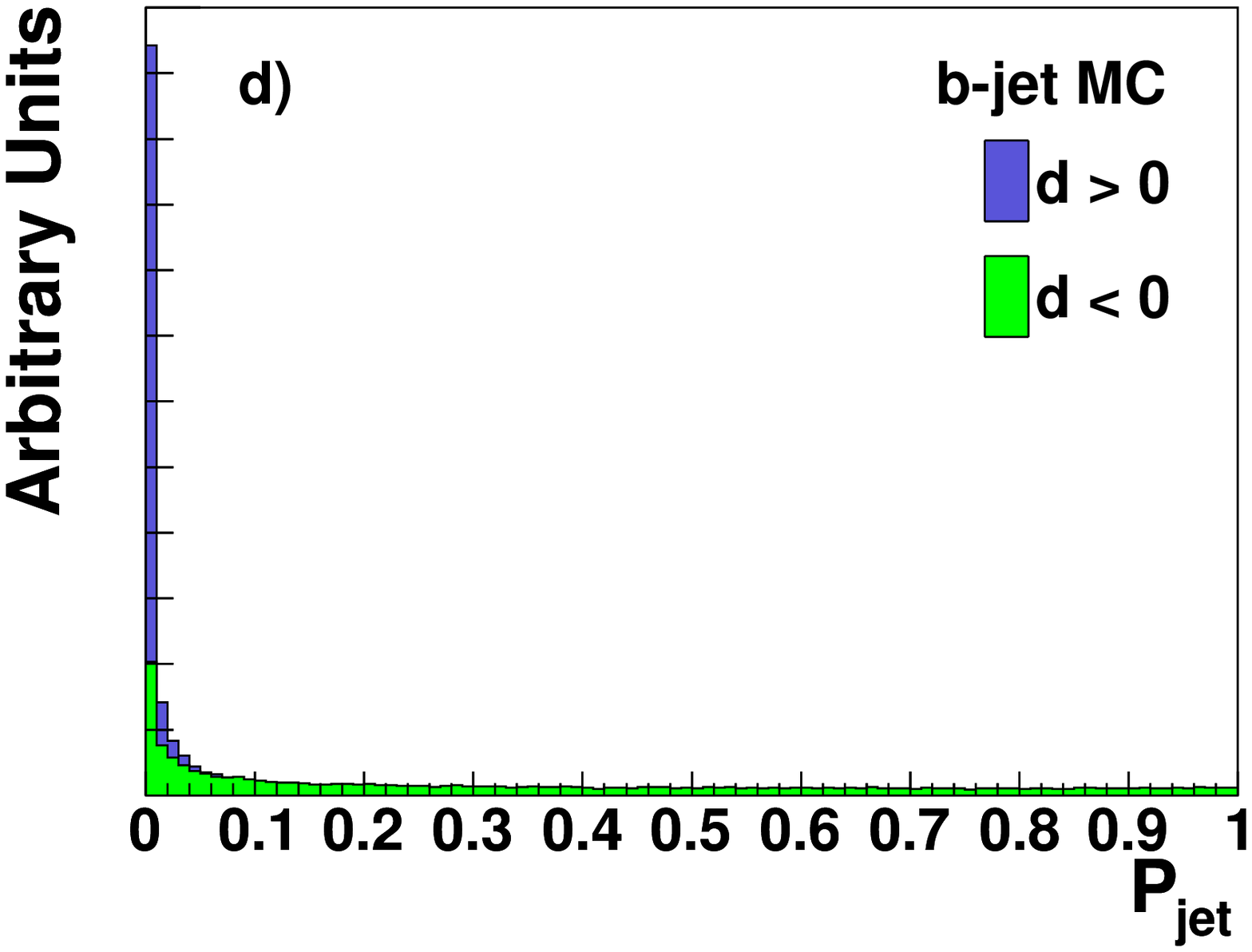}
  \end{center}
  \caption{
    Jet probability ($\mathcal{P}_{\mbox{\scriptsize JLIP}}$) distribution in
    multijet data (a) and QCD MC simulation of light-flavor (b),
    \cquark\ (c), and \bquark\ (d) jets, for positive (dark histograms) and
    negative (light histograms) $d$ values.}
  \label{fig:prob_jet}
\end{figure*}

By construction, if the $\mathcal{P}_{\mbox{\scriptsize trk}}$ are
uniformly distributed and uncorrelated, $\mathcal{P}_{\mbox{\scriptsize JLIP}}$
will also be uniformly distributed, independent of
$N_{\mbox{\scriptsize trk}}$~\cite{knuth}. Therefore, apart from wrongly assigned negative
$d$ in the case of tracks originating from the decay of long-lived particles,
and from any correlations that are induced by the common primary vertex (which
is reconstructed from the tracks under consideration, among others), the
resulting $\mathcal{P}_{\mbox{\scriptsize JLIP}}$ distribution is indeed
expected to be uniform for negative $d$ tracks in multijet data.
These distributions are shown in Fig.~\ref{fig:prob_jet} for multijet data
and simulated jets of different flavors, and for positive and negative $d$
values.
Applying this tagging algorithm simply entails requiring that a
jet's $\mathcal{P}_{\mbox{\scriptsize JLIP}}$ value does not exceed some given
maximum value.

%% file: csip.tex
\section{The Counting Signed Impact Parameter Tagger}
\label{sec:csip}

In this method~\cite{khanovthesis}, as in Sec.~\ref{sec:jlip}, there is no
attempt to use reconstructed secondary vertices. Instead, the signed impact
parameter significance $S_{d}$ is calculated for all good tracks located within
a $\mathcal{R}=0.5$ cone around the jet axis. For the present purpose, the
definition of a good track is as follows:
\begin{itemize}
\item the track should be associated with the hard interaction (the difference
  between the $z$ coordinates of the DCA point and the primary vertex should be
  less than 1~cm);
\item the track DCA must not be too large, $|d|< 2$~mm;
\item the track transverse momentum should satisfy
  $\pt > 1 \GeVc$;
\item the track fit should be of good quality: its $\chi^{2}$ per degree of
  freedom $\chi^{2}_{\mbox{\scriptsize dof}}$ should satisfy
  $\chi^{2}_{\mbox{\scriptsize dof}} < 9$;
\item tracks with $\chi^{2}_{\mbox{\scriptsize dof}} < 3$
  are required to have at least 2 SMT hits;
\item tracks with $3 \le \chi^{2}_{\mbox{\scriptsize dof}} < 9$
  are required to have either 4 SMT hits and at least 13 CFT hits,
  or at least 5 SMT hits and either 0 or at least 11 CFT hits.
\end{itemize}
The effect of the last criterion is to impose more stringent quality requirements
on tracks having $|\eta| \approx 1.5$. This is done as this region suffers from a
higher fake track rate, leading to track candidates with a small but non-zero
number of CFT hits.
Finally, tracks originating from a $V^{0}$ candidate, as detailed in
Sec.~\ref{sec:preliminaries}, are also excluded.

A jet is considered to be tagged by the Counting Signed Impact Parameter (CSIP)
tagger if there are at least two good tracks with $\mathcal{S}_{d}/a>3$ or at
least three good tracks with $\mathcal{S}_{d}/a>2$, where $a$ is a scaling
parameter.
The choice of $a$ determines the operating point (\bquark-tagging efficiency and
mistag rate) of the algorithm.
Alternatively, if a jet has at least two good tracks with positive
$\mathcal{S}_{d}$, then the minimum value of $a$ at which
there are at least two good tracks with
$\mathcal{S}_{d}/a>3$ or at least three good tracks with
$\mathcal{S}_{d}/a>2$ can be used as a continuous output
variable of the tagger. Here, $a$ is set to 1.2, as suggested by optimization
studies using simulated data.

\begin{figure}[bth]
  \begin{center}
    \includegraphics[width=0.48\textwidth]{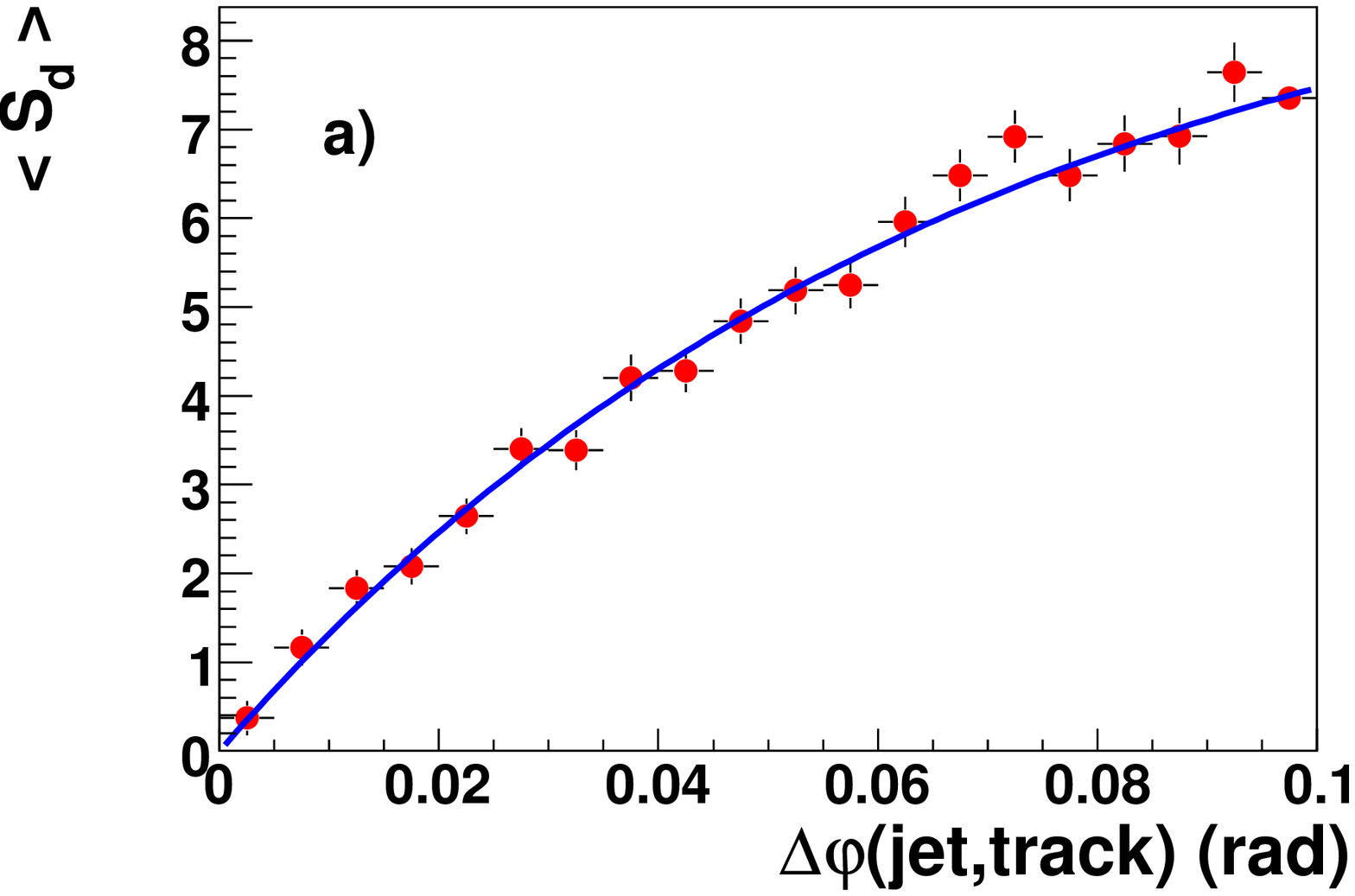}
    \includegraphics[width=0.48\textwidth]{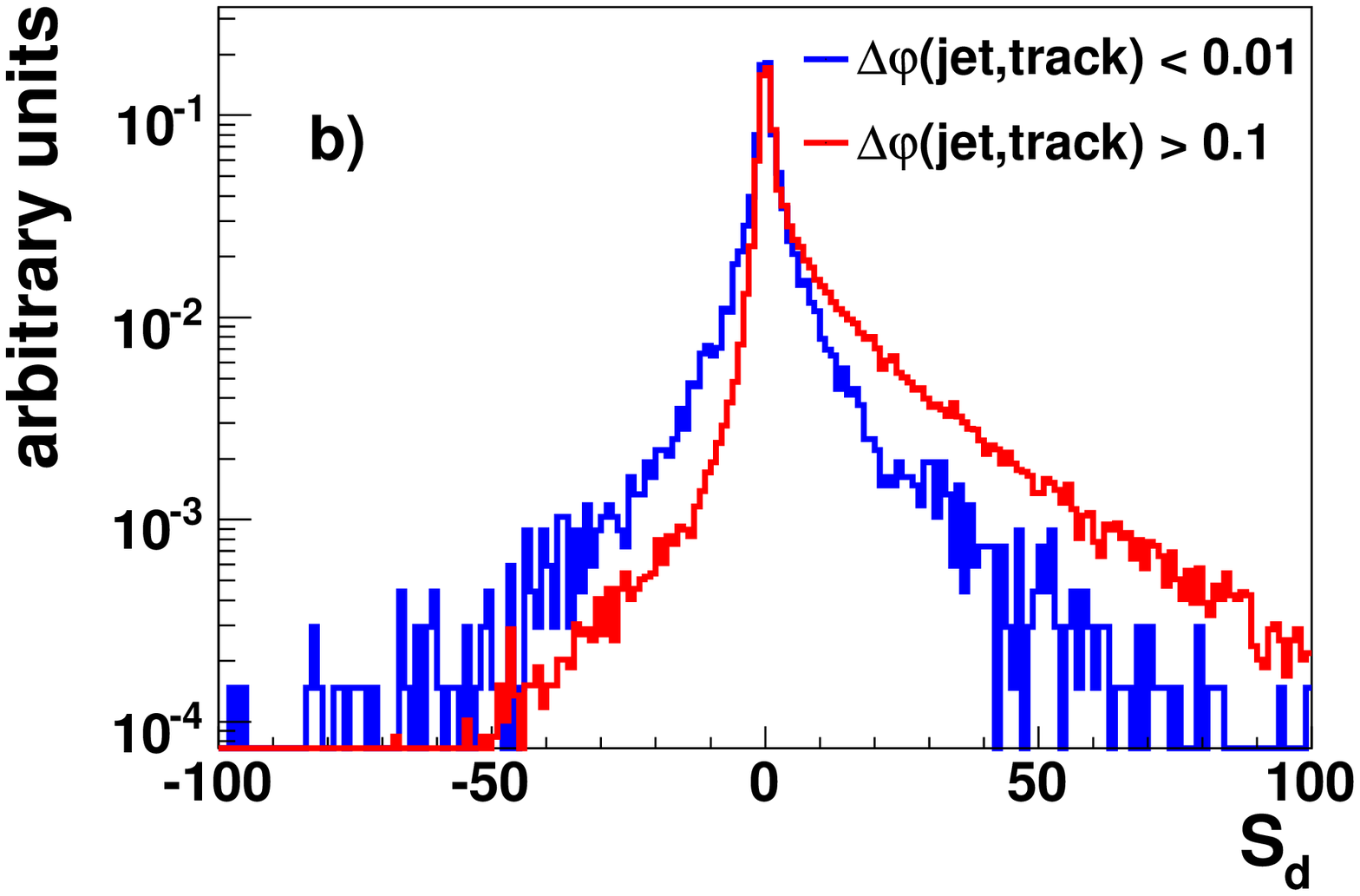}
  \end{center}
  \vspace{-0.5cm}
  \caption[] {Mean signed impact parameter significance $\mathcal{S}_{d}$
    (averaged over all tracks in all $b$-jets in simulated events)
    as a function of the azimuthal angle between the track and $b$-jet directions
    $\Delta\varphi$ (a) and $\mathcal{S}_{d}$ distributions for tracks
    at low and high $\Delta\varphi$ (b).}
  \label{fig:csip_dphi}
\end{figure}

In the actual implementation of the algorithm, there is an additional
condition related to the fact that the sign of
$\mathcal{S}_{d}$ cannot be determined 
accurately for tracks that are very close to the jet axis, as
illustrated by Fig.~\ref{fig:csip_dphi}. The
criterion of closeness is empirically chosen as the difference
in the azimuthal angle between the track and jet directions, $\Delta\varphi$,
being less than $\Delta\varphi_0=$ 20~mrad. This value
has been optimized by comparing algorithm performance for various values of
$\Delta\varphi_0$.
Four categories of tracks are counted separately:
\begin{itemize}
\item tracks with $\mathcal{S}_{d}/a>3$,
 $|\Delta\varphi|>\Delta\varphi_0$ (``$3\sigma$-strong'' tracks, their total number to
 be denoted as $N_{3s}$),
\item tracks with $2<\mathcal{S}_{d}/a<3$,
 $|\Delta\varphi|>\Delta\varphi_0$ (``$2\sigma$-strong'' tracks, $N_{2s}$),
\item tracks with $|\mathcal{S}_{d}/a|>3$,
 $|\Delta\varphi|<\Delta\varphi_0$ (``$3\sigma$-weak'' tracks, $N_{3w}$),
\item tracks with $2<|\mathcal{S}_{d}/a|<3$,
 $|\Delta\varphi|<\Delta\varphi_0$ (``$2\sigma$-weak'' tracks, $N_{2w}$).
\end{itemize}
If CSIP is used as a stand-alone algorithm,
the jet is considered tagged if $N_{2s}+N_{3s}+N_{2w}+N_{3w}\ge 3$
and $N_{2s}+N_{3s}\ge 1$, or $N_{3s}+N_{3w}\ge 2$ and $N_{3s}\ge 1$.
In other words, in addition to the original tagging condition
(at least two good tracks with $\mathcal{S}_{d}/a>3$ or at
least three good tracks with $\mathcal{S}_{d}/a>2$),
at least one of the tagging tracks is required to be strong.
In the present implementation, the four numbers
($N_{3s}$, $N_{2s}$, $N_{3w}$, $N_{2w}$)
are packed in a single variable which is used in the
combined algorithm, as explained in Sec.~\ref{sec:nn}.


%% file: nn.tex
\section{The Neural Network Tagger}
\label{sec:nn}

Artificial neural networks (see \emph{e.g.}~\cite{nn}) are modeled after the synaptic
processes in the brain, and have proved to be a versatile machine learning
approach to the general problem of separating samples of events characterized by
many event variables. In particular, the potential of neural networks to exploit
the correlations between variables, and the possibility to \emph{train} the
network to recognize such correlations, make their use in high energy physics
attractive.

The neural network (NN) tagger attempts to discriminate between \bquark\ jets
and other jet flavors by combining input variables from the
SVT, JLIP, and CSIP tagging algorithms~\cite{scanlonthesis}. The NN
implementation chosen is the \textsc{TMultiLayerPerceptron} from the 
\textsc{root}~\cite{ROOT} framework.

\subsection{Optimization}

The following NN parameters were optimized: input variables (number
and type), NN structure, number of training epochs, and jet selection
criteria. The choice of input variables is crucial for the performance
of the NN and so was optimized first.

The NN parameters were optimized by minimizing the light-flavor tagging
efficiency or \emph{fake rate} for
fixed benchmark \bquark-tagging efficiencies. The optimization plots were
produced from a high \pt\ \textsc{Alpgen} \ttbar{} sample and cross
checked with a \textsc{Pythia}~\cite{Pythia} \bbbar\ sample to ensure there
was no sample, \pt, or MC generator dependence in the optimization.

The NN was trained on simulated multijet light-flavor and \bbbar\ samples.
To avoid overtraining,
\emph{i.e.}, a focus on features that are too event specific, the signal
sample of 270~000 \bbbar\ events and the background sample of
470~000 light-flavor events were each split in half, with one half
used as the training sample and the other half
to evaluate the network's performance. It was verified that for the parameter
settings as described below, no overtraining occurs.

\subsubsection{Input variables} \label{sec:nn-variables}

Initially, a large number of lifetime-related variables were investigated (\emph{e.g.}
variables for different SVT versions, or various combinations of the numbers of
tracks in the different CSIP track categories).
After a first assessment of their performance, nine input variables were
selected for the final input variable optimization due to their good
discrimination between \bquark\ jets and light-flavor jets. Six of the variables
are based on the secondary vertices reconstructed using the SVT algorithm. The
remaining three summarize information from the JLIP and CSIP algorithms.
The input variables are:
\begin{description}
\item[SVT $\mathcal{S}_{xy}$:] the decay length significance (the decay length
  in the transverse plane divided by its uncertainty) of the
  secondary vertex with respect to the primary vertex.
\item[SVT $\chi^{2}_{\mbox{\scriptsize dof}}$:] the $\chi^{2}$ per degree of
  freedom of the secondary vertex fit.
\item[SVT $N_{\mbox{\scriptsize trk}}$:] the number of tracks used to
  reconstruct the secondary vertex.
\item[SVT $m_{\mbox{\scriptsize vtx}}$:] the mass of the secondary vertex.
\item[SVT $N_{\mbox{\scriptsize vtx}}$:] the number of secondary vertices
  reconstructed in the jet.
\item[SVT $\Delta\mathcal{R}$:] the distance in ($\eta$,$\phi$) space between
  the jet axis and the difference between the secondary and primary vertex
  positions.
\item[JLIP $\mathcal{P}_{\mbox{\scriptsize JLIP}}$:]  the ``jet lifetime
  probability'' computed in Sec.~\ref{sec:jlip}.
\item[JLIP $\mathcal{P}_{\mbox{\scriptsize RedJLIP}}$:]  JLIP
 $\mathcal{P}_{\mbox{\scriptsize JLIP}}$ re-calculated with the track with the
 highest significance removed from the calculation.
\item[CSIP $\mathcal{N}_{\mbox{\scriptsize CSIP}}$:] a combined variable based
 on the number of tracks with an impact parameter significance
 greater than an optimized value. This variable is discussed in
 more detail below.
\end{description}

Since more than one secondary vertex can be found for each jet,
vertex variables are ranked in order of the most powerful
discriminator, the decay length significance ($\mathcal{S}_{xy}$).
The secondary vertex with the largest $\mathcal{S}_{xy}$ in a jet is
used to provide the input variables. If no secondary vertex is
found, the SVT values are set to 0, apart from the SVT
$\chi^{2}_{\mbox{\scriptsize dof}}$ which is set to 75 corresponding
to the upper bound of $\chi^{2}_{\mbox{\scriptsize dof}}$ values.

\begin{figure}[htbp]
  \centering
  \includegraphics[width=0.49\textwidth]{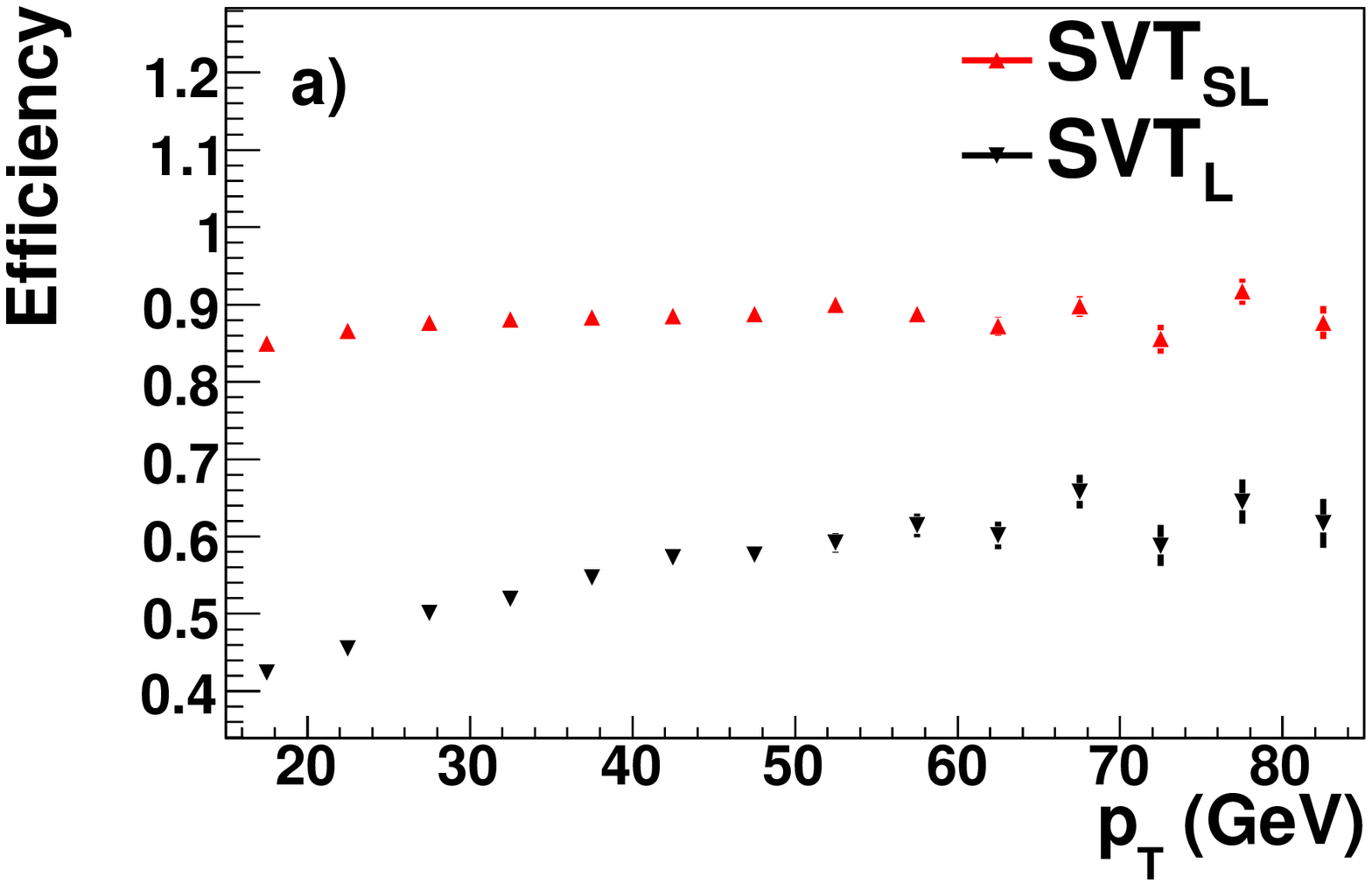}
  \includegraphics[width=0.49\textwidth]{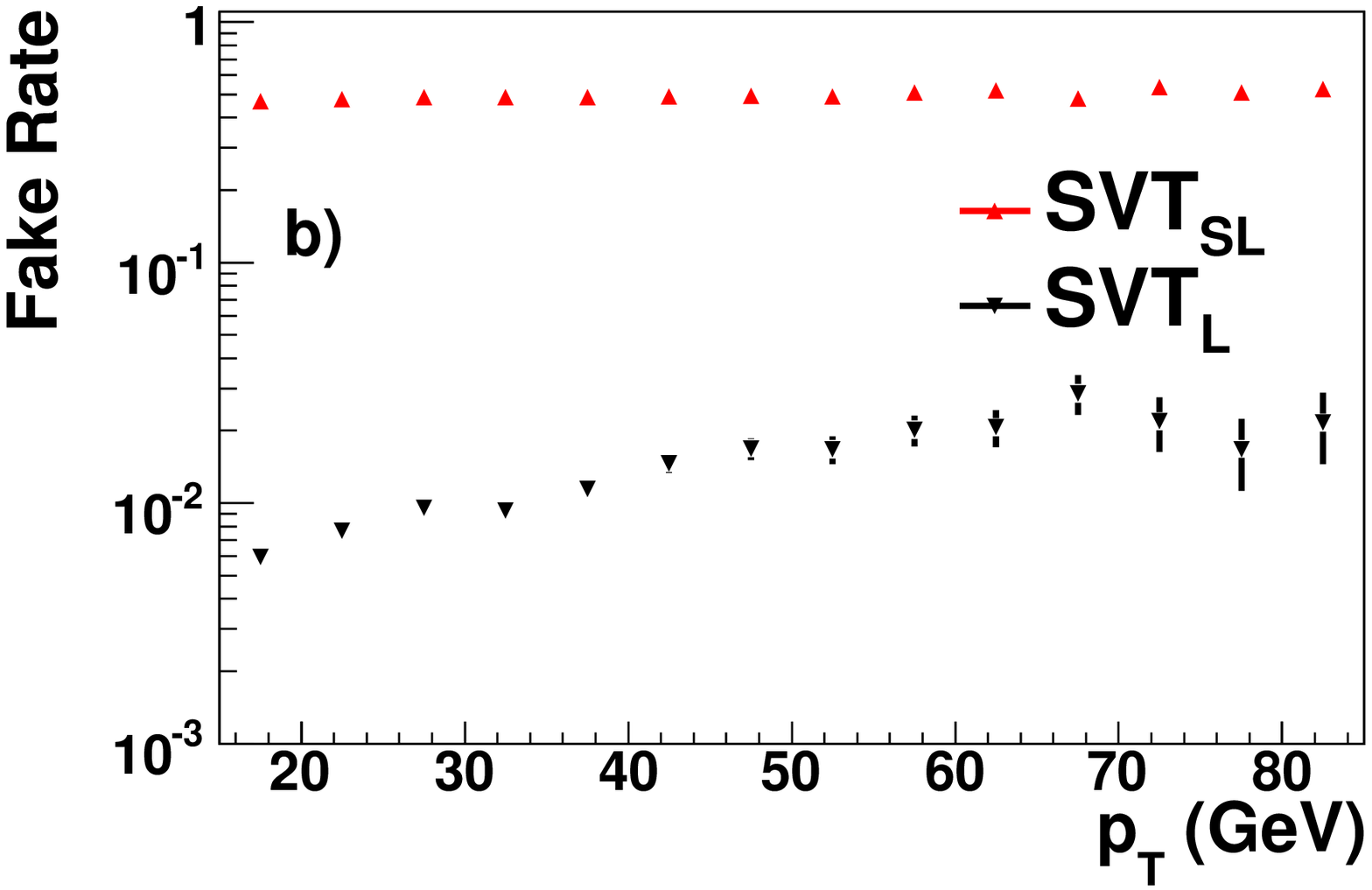}
  \caption{Efficiencies of the SuperLoose (SVT$_{SL}$, up triangles) and
    Loose (SVT$_{L}$, down triangles) SVT taggers for $b\bar{b}$ (a) and
    light-flavor (b) MC jets.}
  \label{fig:svt_eff}
\end{figure}

The standard, Loose, implementation of the SVT
algorithm requires a displaced vertex constructed from significantly
displaced tracks. While such an approach helps to isolate a pure
sample of heavy flavor decays, it typically results in a low
efficiency. In the context of an NN optimization, this is undesirable
as any vertex-related information is only available if a displaced
vertex is found. For this reason, the SuperLoose SVT algorithm
described in Sec.~\ref{sec:svt} is used: even if the vertex
candidates it finds are a less pure sample, it finds significantly
more displaced vertices and they provide additional discrimination
between \bquark\ jets and other flavors. Figure~\ref{fig:svt_eff}
shows the efficiency, as a function of jet \et, for both algorithm choices for
light-flavor and \bquark\ jets. The NN tagger is found to 
perform best if information from both the SuperLoose and the 
Loose SVT algorithms is used: the $N_{\mbox{\scriptsize trk}}$
variable is taken from the latter, and all other SVT variables from
the former.

The CSIP $\mathcal{N}_{\mbox{\scriptsize CSIP}}$ variable is based
on the four CSIP variables $N_{3s}$, $N_{2s}$, $N_{3w}$, and $N_{2w}$
described in Sec.~\ref{sec:csip}. Neural networks tend to perform
best when provided with continuous values.
Since the CSIP variables have small integer values which are not very good
as inputs, they are combined in one variable which brings the
advantage of reducing the number of input variables, hence
simplifying the NN:

\begin{equation}
  \mathcal{N}_{\mbox{\scriptsize CSIP}} =
  6\times N_{3s} + 4\times N_{2s} + 3\times N_{3w} + 2\times N_{2w}.
\end{equation}

The weights were determined in an empirical manner to give optimum
performance for this variable alone.

First, the input variables were optimized. At this stage, the other NN
parameters were set to the following values: NN structure
$N$:2$N$:1, where $N$ is the number of input variables; 500 training
epochs; and selection criteria SVT$_{SL}$ $\mathcal{S}_{xy}>2$ or CSIP
$\mathcal{N}_{\mbox{\scriptsize CSIP}}>8$ or JLIP
$\mathcal{P}_{\mbox{\scriptsize JLIP}}<0.02$. (The subscript $SL$ serves to
clarify that this variable is obtained from the SuperLoose SVT algorithm, as
described above.)

The input variables were optimized by first identifying the two most powerful
input variables by testing every possible combination in a two-input NN,
resulting in the choice of $\mathcal{S}_{xy}$ and
$\mathcal{N}_{\mbox{\scriptsize CSIP}}$ as initial variables.
Starting with an initial NN with these $n=2$ variables and a list of $m$
variables, the remaining variables were then ranked in order of power using the
following procedure:

\begin{enumerate}
  \item each of the $m$ variables to be tested was added individually to the
   initial $n$ variable NN, resulting in $m$ NNs with $n+1$ variables each;
  \item the variable whose addition yielded the largest improvement in NN
    performance was identified. This variable was added permanently to the $n$
    variable NN;
  \item the above steps were repeated, testing each of the remaining $m-1$
   variables with the new $n+1$ variable NN.
\end{enumerate}

\begin{figure}[htbp]
  \centering
  \includegraphics[width=0.48\textwidth]{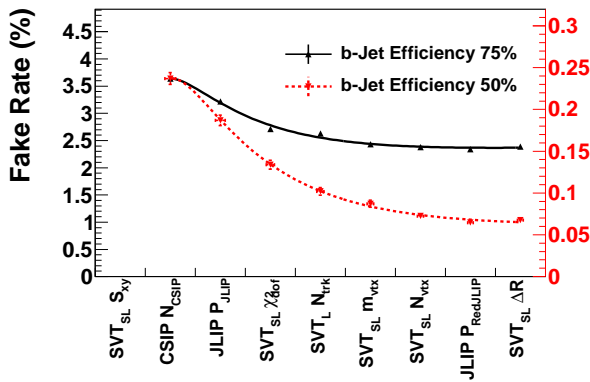}
  \caption{Fake rate for fixed signal efficiencies of $75\%$ (up triangles, left
    axis) and $50\%$ (down triangles, right axis) as a function of additional NN
    variables. The NN variables were added to the NN in order of performance. The
    lines are intended to guide the eye only. The errors are statistical only.}
  \label{fig:var_opt}
\end{figure}

The fake rate of each NN at a 70\% \bquark-jet efficiency benchmark scenario 
was used to select the optimal variable. As a cross check, the procedure was
repeated for signal efficiencies ranging from 50\% to 75\% in 5\% steps.
The same set of variables was found to give the greatest reduction in fake rate
in each case. As an example, two benchmark scenarios are shown in
Fig.~\ref{fig:var_opt}.

\begin{figure*}[htbp]\centering
  \includegraphics[width=0.45\textwidth]{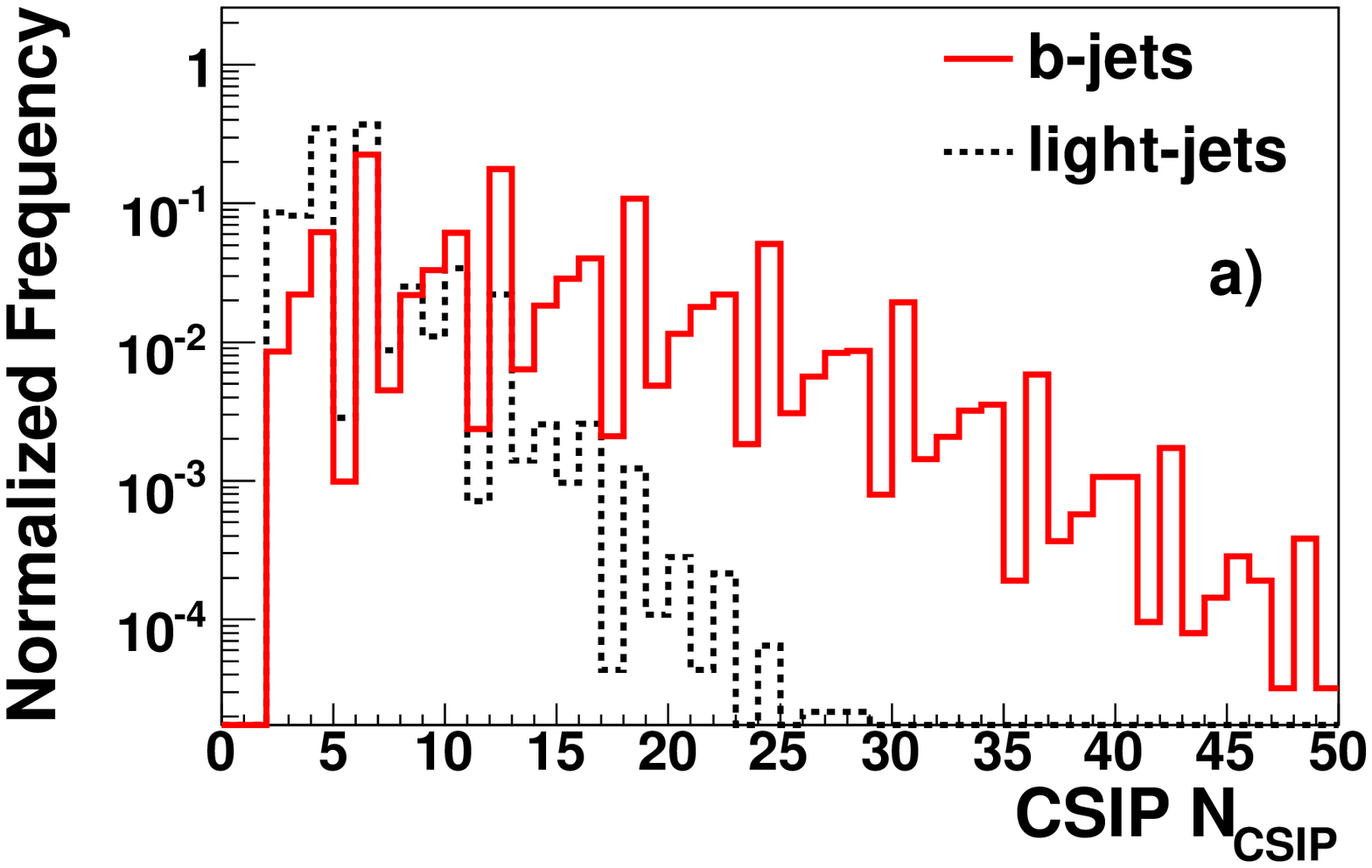}
  \includegraphics[width=0.45\textwidth]{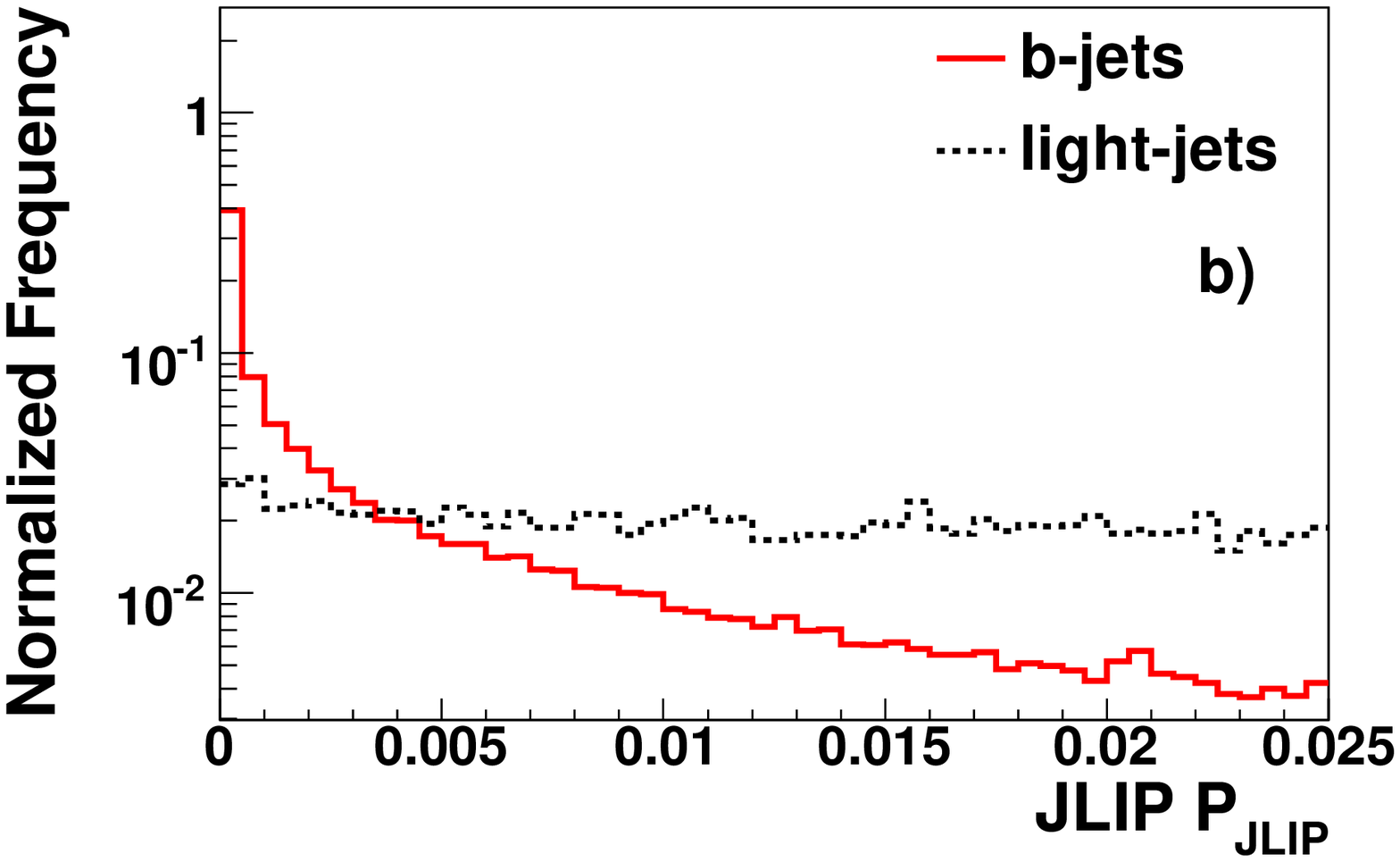}
  \includegraphics[width=0.45\textwidth]{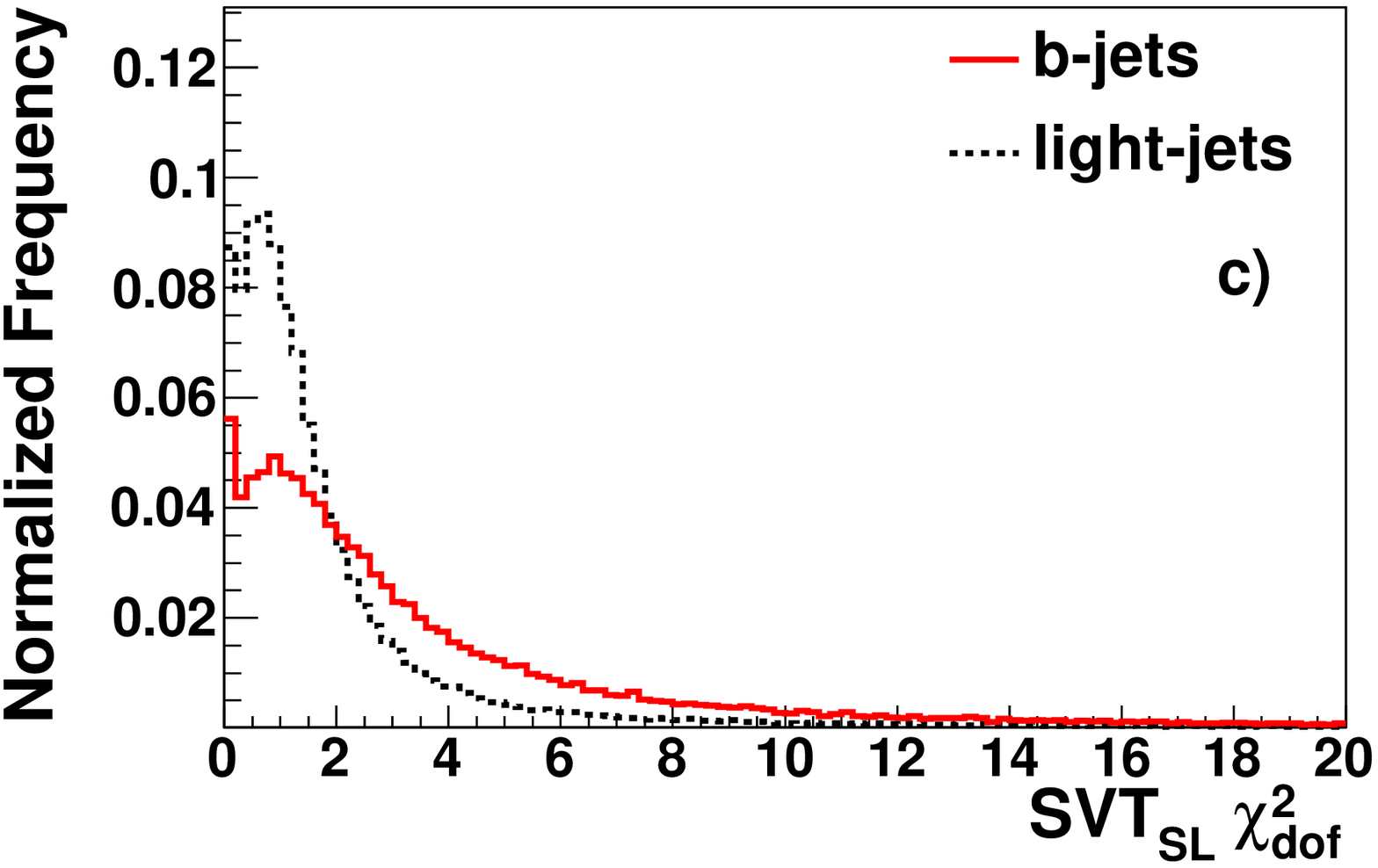}
  \includegraphics[width=0.45\textwidth]{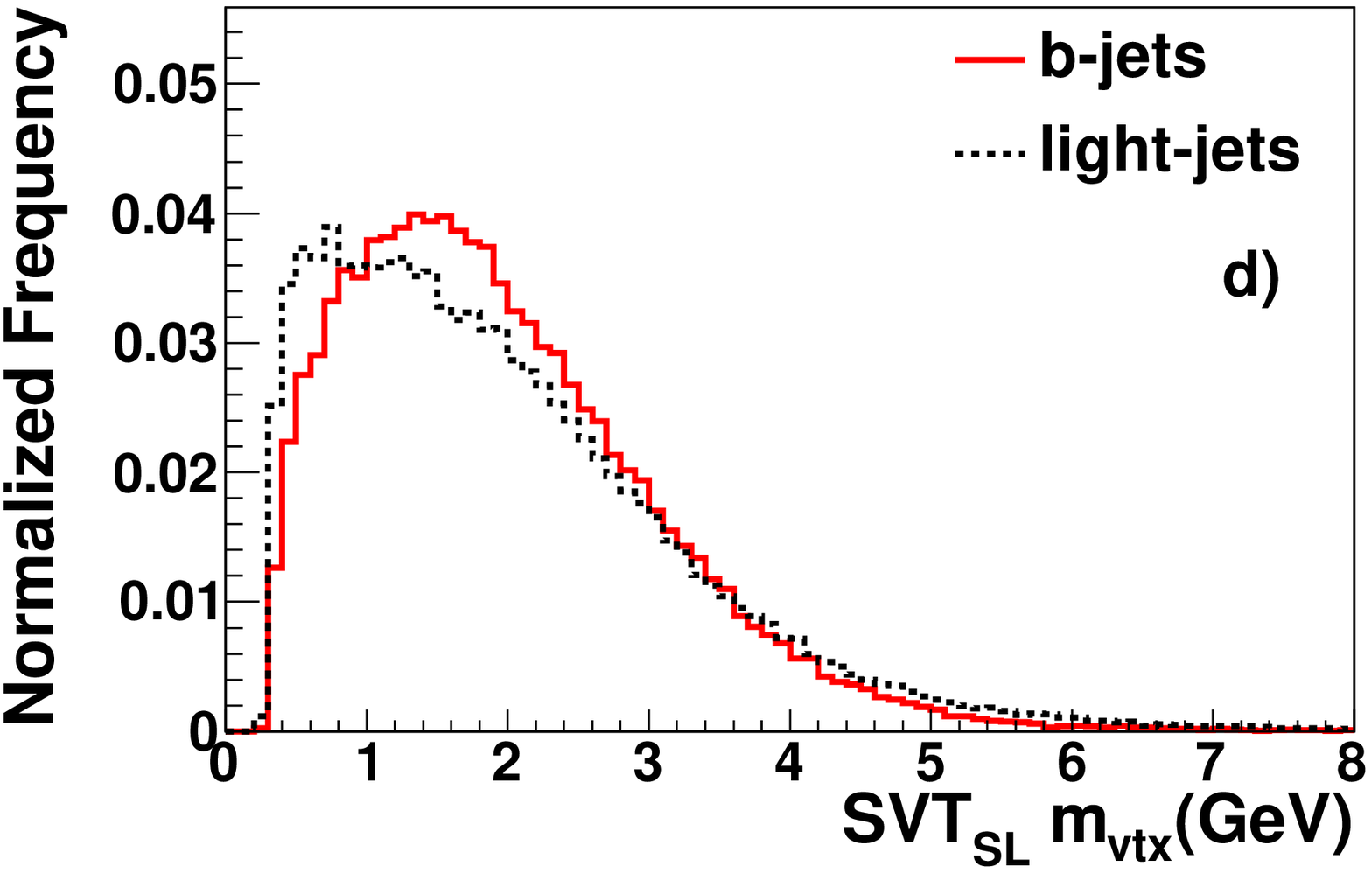}
  \includegraphics[width=0.45\textwidth]{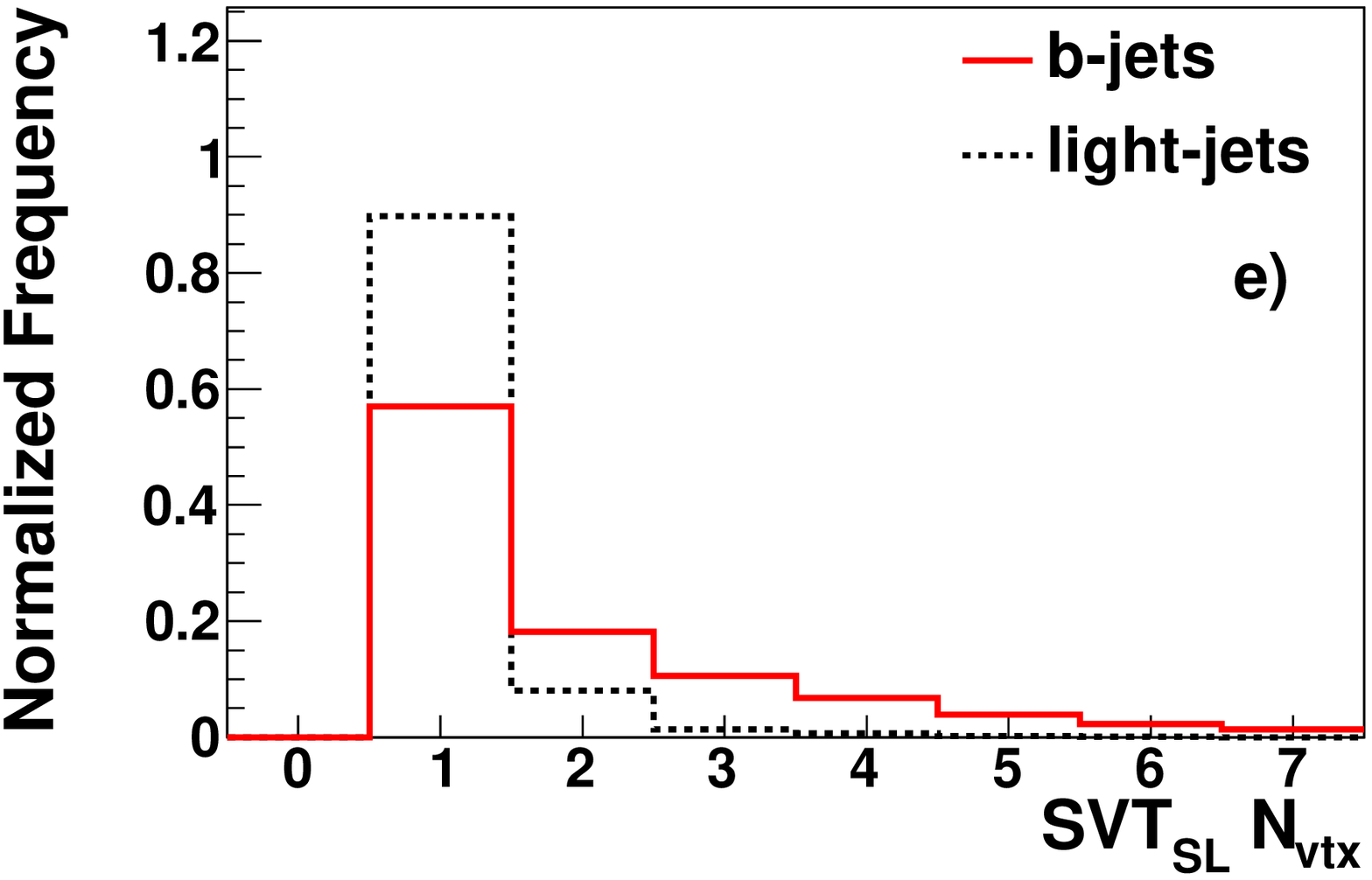}
  \includegraphics[width=0.45\textwidth]{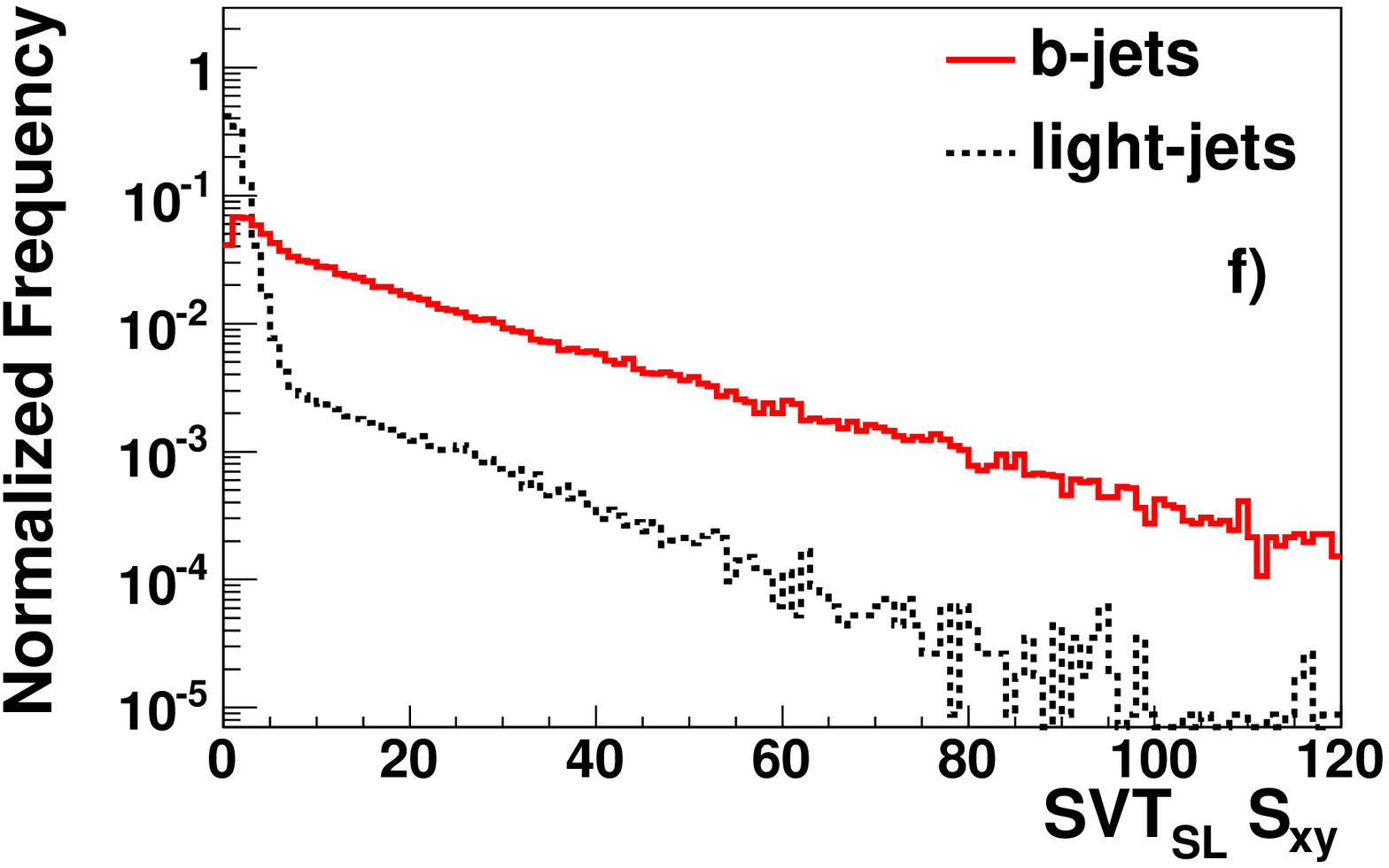}
  \includegraphics[width=0.45\textwidth]{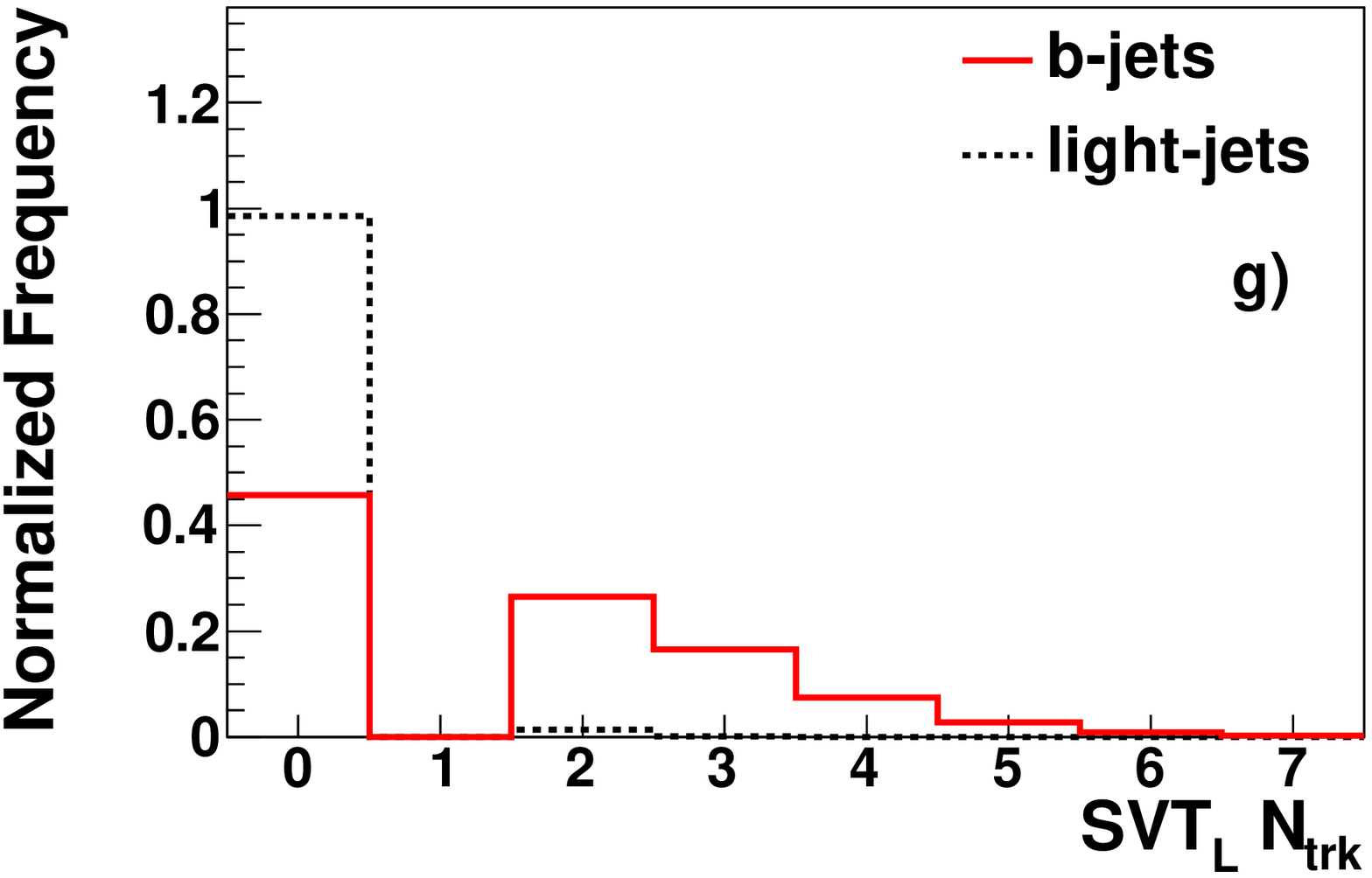}
  \caption{The NN variables CSIP $\mathcal{N}_{\mbox{\scriptsize CSIP}}$ (a), JLIP
    $\mathcal{P}_{\mbox{\scriptsize JLIP}}$ (b),
    SVT$_{SL}$ $\chi^{2}_{\mbox{\scriptsize dof}}$ (c),
    $m_{\mbox{\scriptsize vtx}}$ (d), $N_{\mbox{\scriptsize vtx}}$ (e), $S_{xy}$
    (f), and SVT$_{L}$ $N_{\mbox{\scriptsize trk}}$ (g) for QCD \bbbar\ MC (continuous
    lines), and light jet QCD MC (dashed lines). All histograms are
    normalized to unit area.}
\label{fig:input_variables}
\end{figure*}

The NNs with seven to nine variables have the best performance.
Therefore the seven variable NN is chosen as the optimal solution,
keeping the NN as simple as reasonably possible.
The final selected variables, ranked in order of performance,
are SVT$_{SL}$ $S_{xy}$, CSIP $\mathcal{N}_{\mbox{\scriptsize CSIP}}$, JLIP
$\mathcal{P}_{\mbox{\scriptsize JLIP}}$, SVT$_{SL}$ $\chi^{2}_{\mbox{\scriptsize dof}}$,
SVT$_{L}$ $N_{\mbox{\scriptsize trk}}$, SVT$_{SL}$ $m_{\mbox{\scriptsize vtx}}$, and 
SVT$_{SL}$ $N_{\mbox{\scriptsize vtx}}$. The distributions of these
variables in simulated QCD samples are shown in Fig.~\ref{fig:input_variables}.

\subsubsection{Number of training epochs and neural network structure}

The number of training epochs was varied from 50 up to 2000.
For each of the benchmark scenarios the
majority of the minimization is reached by $\approx 400$ epochs, with only small
further improvement thereafter.

The number of hidden layers was set to one, as one layer should be
sufficient to model any continuous function \cite{nntheory} and this
minimizes CPU usage. The number of hidden nodes was optimized by
varying their number from seven through thirty-four.
Twenty-four was chosen as the optimal number of hidden nodes.

\subsubsection{Input selection criteria}

Another important attribute of the NN is the selection of the jets
which are used to train the NN. A selection too loose can cause a
loss of performance as the NN training is dominated by signal and background
jets which could have been separated with a simple requirement, causing a loss
of resolution. A selection which is too tight will cause a significant loss of
\bquark\ jets and therefore limit the maximum possible efficiency.

\begin{figure}[htbp]\centering
  \includegraphics[width=2.67in]{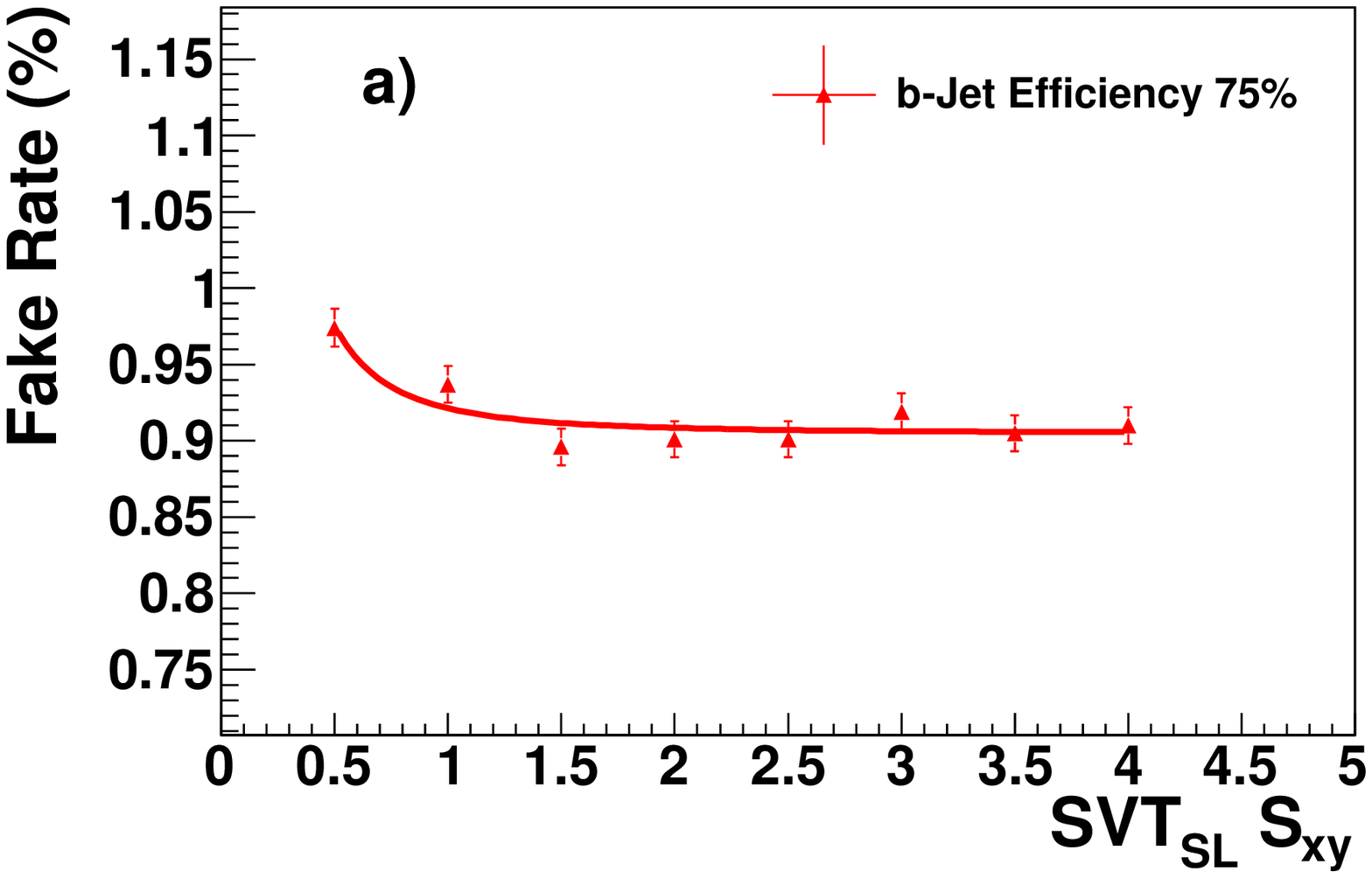}
  \includegraphics[width=2.67in]{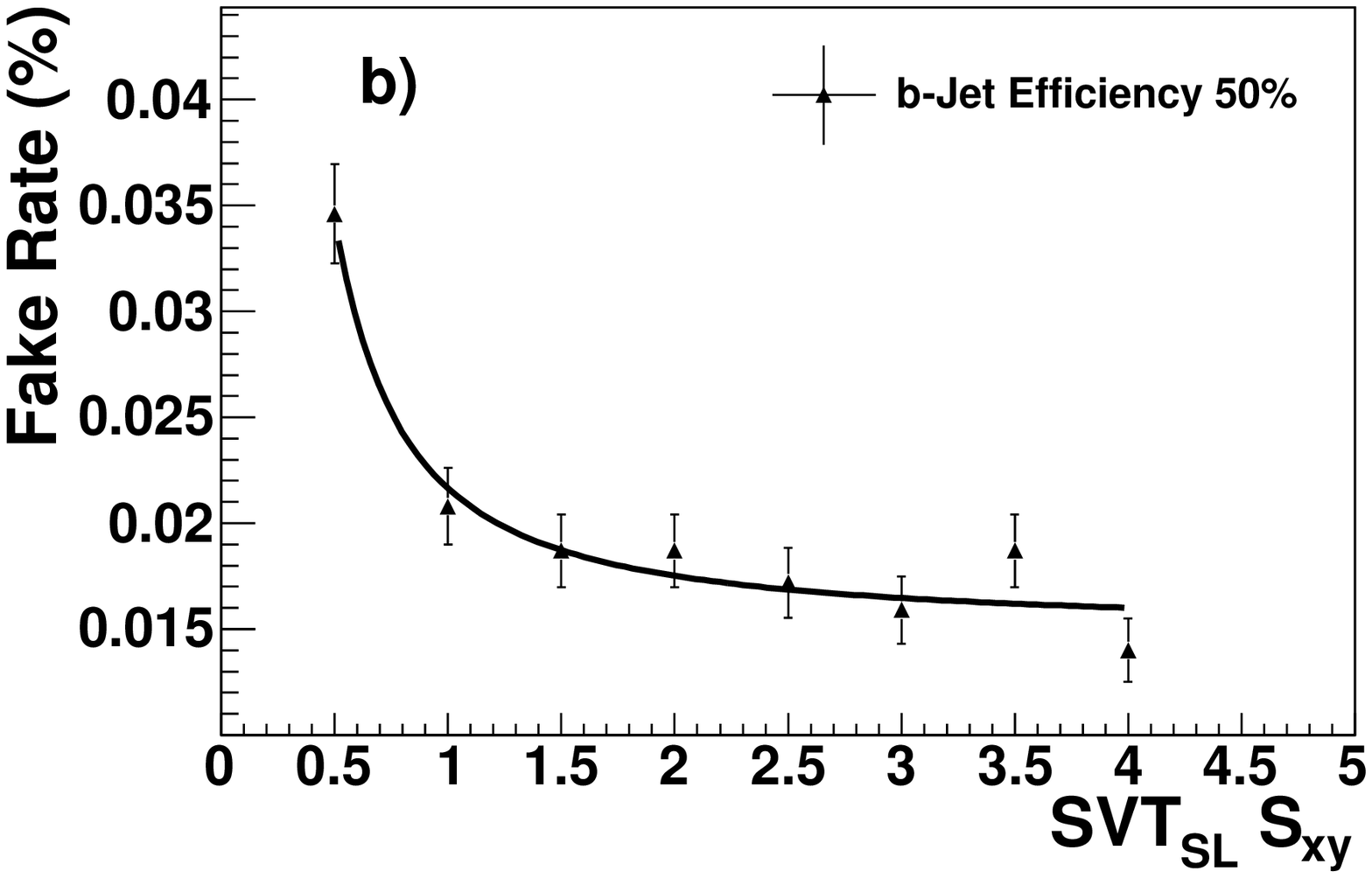}
  \caption{Fake rate for fixed signal efficiencies of $75\%$ (a)
    and $50\%$ (b) as a function of the SVT$_{SL}$ $\mathcal{S}_{xy}$
    requirement on the input jets. The lines are intended to guide the eye
    only.}
  \label{fig:input_variables_opt}
\end{figure}

The input selection criteria were optimized by considering each variable
in turn, starting with the most important variable,
SVT$_{SL}$~$\mathcal{S}_{xy}$, then JLIP~$\mathcal{P}_{\mbox{\scriptsize
JLIP}}$, and finally CSIP~$\mathcal{N}_{\mbox{\scriptsize CSIP}}$ (at this
stage, a requirement on JLIP~$\mathcal{P}_{\mbox{\scriptsize JLIP}}$ performs better
than one on CSIP~$\mathcal{N}_{\mbox{\scriptsize CSIP}}$).
The optimal values were chosen as SVT$_{SL}$ $\mathcal{S}_{xy}>$~2.5, JLIP
$\mathcal{P}_{\mbox{\scriptsize JLIP}} <~0.02$, and CSIP
$\mathcal{N}_{\mbox{\scriptsize CSIP}}>~8$. The results for
SVT$_{SL}$~$\mathcal{S}_{xy}$ are shown in Fig.~\ref{fig:input_variables_opt} (in this
case, the requirement is fixed at an~$\mathcal{S}_{xy}$ value of 2.5 since for
the loosest operating points, the performance degrades for even larger
$\mathcal{S}_{xy}$ values).

\subsubsection{Optimized NN parameters}

The optimized parameter values for the NN tagger are summarized in Table
\ref{tab:nnparameters}.

\begin{table*}[htbp]
  \begin{center}\begin{tabular}{|l|c|}
      \hline
      Parameter & Value \\ \hline
      NN structure & 7 input nodes:24 hidden nodes:1 output node\\ \hline
      Input variables & (1) SVT$_{SL}$ $\mathcal{S}_{xy}$ (2) CSIP
      $\mathcal{N}_{\mbox{\scriptsize CSIP}}$ (3) JLIP
      $\mathcal{P}_{\mbox{\scriptsize JLIP}}$ (4) 
      SVT$_{SL}$ $\chi^{2}_{\mbox{\scriptsize dof}}$ \\ 
      (performance ranked) & (5) SVT$_{L}$ $N_{\mbox{\scriptsize trk}}$
      (6) SVT$_{SL}$ $m_{\mbox{\scriptsize vtx}}$  (7) SVT$_{SL}$ $N_{\mbox{\scriptsize vtx}}$ \\
      \hline
      Input selection criteria & SVT$_{SL}$ $\mathcal{S}_{xy}>$~2.5 or JLIP
      $\mathcal{P}_{\mbox{\scriptsize JLIP}} <~0.02$  or CSIP
      $\mathcal{N}_{\mbox{\scriptsize CSIP}}>~8$ \\
      (failure results in NN output of 0) &  \\
      \hline
      Number of training epochs & 400 \\  \hline
    \end{tabular}
    \caption{Optimized NN parameters.}
    \label{tab:nnparameters}
  \end{center}
\end{table*}

\subsection{Performance}

\begin{figure}[p]
  \begin{center}
    \includegraphics[width=0.48\textwidth]{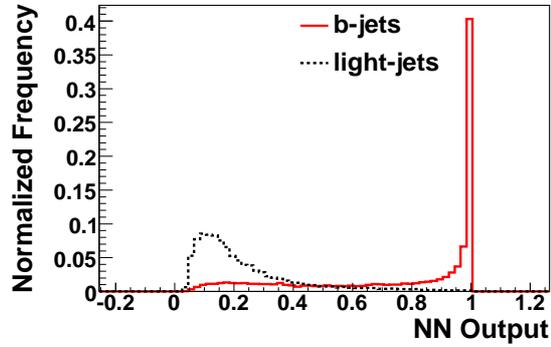}
  \end{center}
  \caption{The NN output for light-flavor (dashed line) and \bquark\
    (continuous line) jets (with $\pt > 15 \GeVc$ and $|\eta| < 2.5$) in
    simulated QCD events. Both distributions are normalized to unity.}
  \label{fig:NN_output}
\end{figure}

The output from the optimized NN \bquark\ tagger on \bbbar\ and light-flavor
simulated jets is shown in Fig.~\ref{fig:NN_output}. There is 
a significant separation between the signal and background samples.
It should be noted that the light-flavor jets in the distribution have all
passed the loose tagging input selection criteria listed in
Table~\ref{tab:nnparameters}.

\begin{figure}[htbp]
  \centering
  \includegraphics[width=0.48\textwidth]{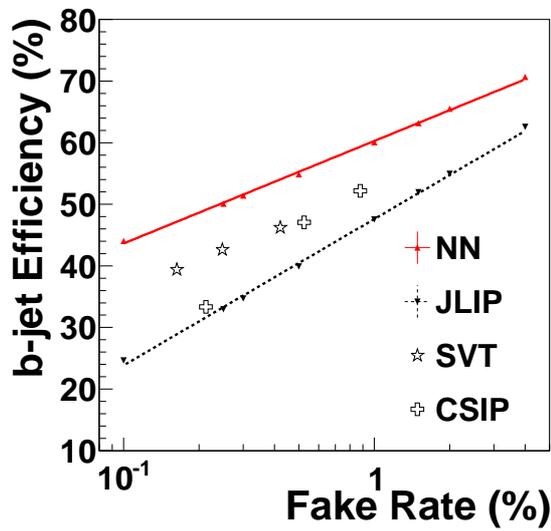}
  \caption{Performance of the NN (up triangles, continuous line),
    JLIP (down triangles, dotted line), CSIP (hollow crosses) and Loose SVT
    (hollow stars) taggers computed on simulated $\PZz\rightarrow \bbbar$ and
    $\PZz\rightarrow \qqbar$ jet samples. Due to the discrete nature of the CSIP
    and SVT taggers' outputs a continuous performance curve is not shown.}
  \label{fig:effrej_jlipnn}
\end{figure}

The advantage of combining the input variables from several
taggers in an NN is shown in Fig.~\ref{fig:effrej_jlipnn},
which compares the NN \bquark-tagging performance to the JLIP, SVT and CSIP taggers.
There is a substantial improvement,
with relative efficiency increases of $\approx$ 20 -- 50\% for a fake
rate of $0.2\%$ and $\approx 15\%$ for a fake rate of $4\%$. The fake
rate is reduced by a factor of between two or three for fixed signal efficiencies.

The NN tagger performance in data is evaluated in the following sections for
twelve operating points, corresponding to NN output discriminant threshold
values ranging from 0.1 to 0.925. For illustrative purposes, detailed results
will be provided for threshold values of 0.325 and 0.775, referred to as L2 and
Tight, respectively.


%% file: performance.tex
\section{Efficiency Estimation}
\label{sec:efficiency}

The performance of the tagging algorithm cannot simply be inferred from
simulated samples. Several effects cause differences between the data and
these simulated samples:
\begin{itemize}
\item Simulated hit resolutions, both in the CFT and in the SMT, have been
  tuned to reproduce those in the data. However, the tuning cannot be expected
  to be perfect as the observed resolutions in the data are also affected by
  poorly understood geometrical effects which are not modeled in the simulation.
\item A small but non-negligible fraction of the detector elements, in
  particular in the SMT, are disabled from time to time.
\end{itemize}
These effects lead to different effective resolutions and efficiencies in data
and simulated samples. A
calibration is therefore required. This section describes the estimation of the
\bquark- and \cquark-jet tagging efficiencies, both of which are dominated by
genuine heavy flavor decays. Jets 
originating from \uquark, \dquark, or \squark\ quarks or from gluons are jointly
referred to as light ($l$) jets. The light-jet tag rate estimation is described
in Sec.~\ref{sec:fake_rate}.

\subsection{MC and data samples}
\label{sec:samples}

The data used in the performance measurements were collected from July 2002 to
February 2006 and correspond to an integrated luminosity of
$\approx 1~\mbox{fb}^{-1}$.
The tagging efficiency in simulated events is measured using several processes
simulated using the \textsc{pythia} ($\PZz\rightarrow \bbbar$, $\PZz\rightarrow
\ccbar$, $\PZz\rightarrow \qqbar$, QCD) and \textsc{alpgen} (inclusive $\ttbar$)
event generators.
Large samples of simulated \bquark\ and \cquark\ jets are created by combining the
appropriate flavor jets from the individual samples.

\subsection{The \textit{SystemD} method}
\label{subsec:math}

The {\it SystemD} method~\cite{clementthesis} was developed to determine
identification efficiencies using almost exclusively the data.
Simulated samples are used only to estimate correction factors.
The method involves several, essentially uncorrelated, identification criteria
which are applied to the same data sample.
Combining these criteria allows the definition of a system of equations
which can be solved to extract the efficiency of each criterion.

The data sample is assumed to be composed of a signal and $n$ backgrounds.
Denoting by $f_0$ the fraction of signal events, and by $f_{i=1\ldots n}$ the
fraction of each considered background, these fractions must satisfy:
\begin{equation}
  \label{eq:total}
  \sum_{i=0}^{n}f_i=1.
\end{equation}
Subsequently, $m$ uncorrelated identification criteria are considered with
different selection efficiencies $\varepsilon_{i=0\ldots n}^{k=1\ldots m}$ for the
signal and backgrounds.
Only a fraction $Q^{k}$ of the total number of events will pass the $k$-th
identification criterion.
Then a new set of equations can be added for each selection:
\begin{equation}
  \label{eq:1tag}
  \sum_{i=0}^{n}\varepsilon_{i}^{k}f_{i} = Q^{k} .
\end{equation}
If the selection criteria are uncorrelated, the total efficiency
$\varepsilon_{i=0\ldots n}^{k_{1},\ldots,k_{r}}$ ($r\leq m$) of applying several
of them successively can be factorized in terms of the individual efficiencies
$\varepsilon_{i}^{k}$:
\begin{equation}
  \label{eq:factorization}
  \varepsilon_{i}^{k_1,\ldots,k_r} = \prod_{v=1}^{r}\varepsilon_i^{k_v} .
\end{equation}
A generalization of Eq.~(\ref{eq:1tag}) can then be obtained for a combination
of several criteria:
\begin{equation}
  \label{eq:ntag}
  Q^{k_1,\ldots,k_r}=\sum_{i=0}^{n} \Big(\prod_{v=1}^{r}\varepsilon_i^{k_v}\Big)f_i .
\end{equation}
The signal and background fractions are $n+1$ unknown parameters and each
identification criterion introduces $n+1$ new unknowns in the form of selection
efficiencies.
The number of equations of the form of Eq.~(\ref{eq:ntag}) depends on the number
of combinations of the $m$ criteria, leading to a total of
$\sum_{r=0}^{m}{r\choose m}=2^{m}$ equations.
To obtain a system of equations which can be solved, $n$ and $m$ must satisfy
\begin{equation}
  (1+m)\times(1+n) \leq 2^{m} .
\end{equation}
The simplest non-trivial solutions are
\begin{itemize}
\item $m=3$, $n=1$: 8 equations with 8 unknowns;
\item $m=4$, $n=2$: 16 equations with 15 unknowns.
\end{itemize}
These systems of equations are nonlinear and have several solutions.
Only the simplest case of 8 equations will be considered in the following.
This system has two solutions, which differ by the interchange of efficiencies
assigned to the signal and background samples. As will be detailed in
Sec.~\ref{sec:system8-equations}, further \emph{a priori} knowledge of at
least one of the unknown parameters is required to resolve the ambiguity.
The input parameters are the fractions of events, $Q_i^{k_0,..,k_r}$,
which are determined directly from the data.
There is therefore no input from simulated events.
Solving the system gives access to the signal and background fractions
and to the various efficiencies.

\subsection{Application to \bquark-tagging efficiency measurements}
\label{sec:system8-application}

The {\it SystemD} method is used here in order to extract the \bquark-tagging
efficiencies of the NN tagger.
The method is applied to a data jet sample that is expected to have a
significantly higher heavy flavor content than generic QCD events, but is not
biased by lifetime requirements and provides large statistics.
In detail, taggable jets are selected based on the following additional criteria:
\begin{itemize}
\item the jet must have $\et>15\GeV$ and $|\eta|<2.5$;
\item the jet must contain a muon with $\pt^{\mu}>4\GeVc$, within a distance
 $\Delta\mathcal{R}=0.5$ from the jet axis.
\end{itemize}

The lifetime composition of the resulting sample could be biased by 
trigger requirements applying impact parameter or secondary vertex
requirements. To avoid such biases, events are required to have passed at least
one lifetime-unbiased trigger. These requirements result in a sample of
141$\times 10^{6}$ events.

This sample contains a mixture of \bquark, \cquark, and light-flavor jets.
The first two are mostly due to semimuonic decays of \bquark\
and \cquark\ hadrons; muons in light-flavor jets arise mainly from in-flight decays of
$\pi^{\pm}$ and $K^{\pm}$ mesons. To apply the \textit{SystemD} method
as described above, with $m=3$ criteria and $n=1$ backgrounds, however, only a
single source of background can be dealt with. The \cquark\ and light-flavor
backgrounds are therefore combined and denoted as ``cl''. An important
consequence of this combination is the fact
that the use of the \textit{SystemD} method only allows the efficiency to be
determined for a specific mixture of \cquark\ and light-flavor jets; it is
therefore not useful to extract efficiencies for the separate background
sources.

\subsubsection{{\it SystemD} selection criteria}
\label{sec:system8-criteria}

The three identification criteria used are:
\begin{enumerate}
 \item The NN tagger operating point under study.
 \item A requirement on the transverse momentum of the muon relative
  to the direction obtained by adding the muon and jet momenta, \ptrel. This
  criterion is chosen because the high \ptrel\ values in \bquark-hadron decays are
  due to the high mass of the \bquark\ quark, and as such are in principle
  expected to be independent of the lifetime criterion.
 \item The requirement that the event contain another jet satisfying
  $\mathcal{P}_{\mbox{\scriptsize JLIP}}<0.005$, referred to below as
  \emph{away-side tag}. As \bquark\ quarks are usually produced in pairs, this
  selection criterion allows an increase in the fraction of \bquark\ jets
  \emph{without} being applied directly to the muon jet itself, and hence no
  correlation with the two other criteria is \emph{a priori} expected.
\end{enumerate}

\subsubsection{Correction factors}
\label{sec:system8-corrections}

In practice, correlations between these selection criteria are not altogether
absent, and they need to be accounted for. As insufficient information is
available in the data to estimate these correlations, they are evaluated on the
simulated samples instead. Even though this approach entails some dependence on
MC, the dependence can be expressed in terms of
\emph{correction factors}, as detailed below, and the efficiencies themselves are
determined on the data only.

\begin{figure}[htbp]
  \begin{center}
    \includegraphics[width=0.45\textwidth]{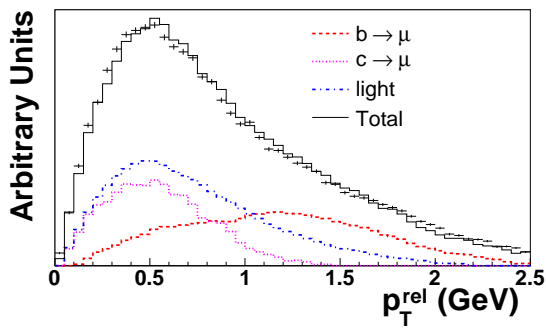}
  \end{center}
  \vspace{-0.5cm}
  \caption[] {
    \ptrel\ distribution of muons in taggable jets.
    The superimposed histograms represent
    the contribution of simulated $b\rightarrow \mu X$ (dashed),
    $c\rightarrow \mu X$ (dotted),
    light quark jets (dash-dotted), and their sum (solid histogram). The
    template fit employed the method described in Ref.~\cite{barlow}.}
\label{fig:ptrel}
\end{figure}

The first correlation studied is that between the NN tagger and the \ptrel\ requirement.
The \ptrel\ distribution of the data sample is shown in Fig.~\ref{fig:ptrel} for
all taggable jets. The data are fitted as the sum of MC distributions for
each quark flavor, with a free normalization for each flavor.
Reasonable agreement is obtained. Although such a fitting procedure could in
principle be used to estimate \bquark-jet tagging efficiencies, it relies on more
assumptions than the \textit{SystemD} method and is therefore not used for this
purpose.

\begin{figure}[htbp]
  \begin{center}
    \includegraphics[width=2.67in]{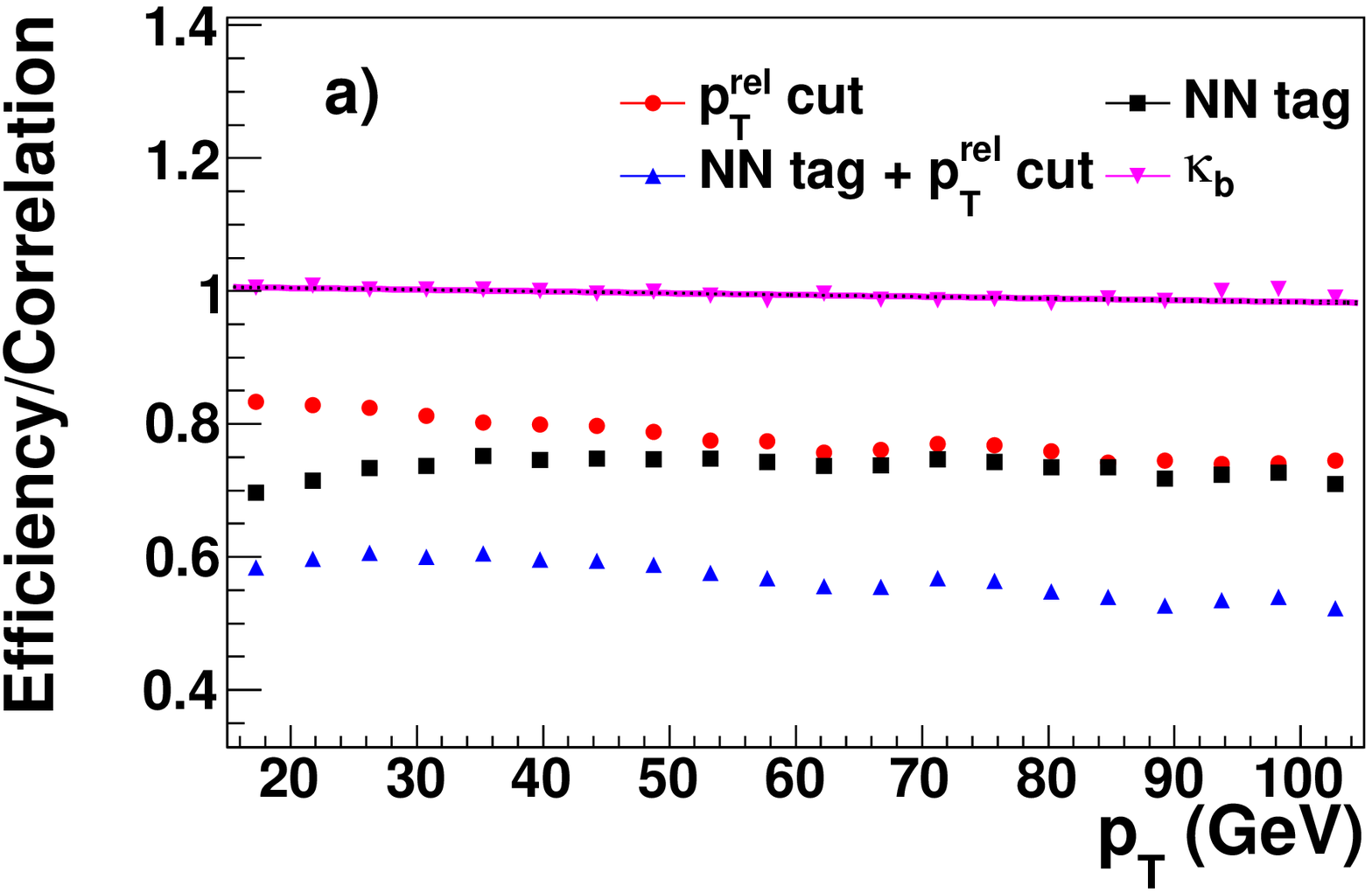}
    \includegraphics[width=2.67in]{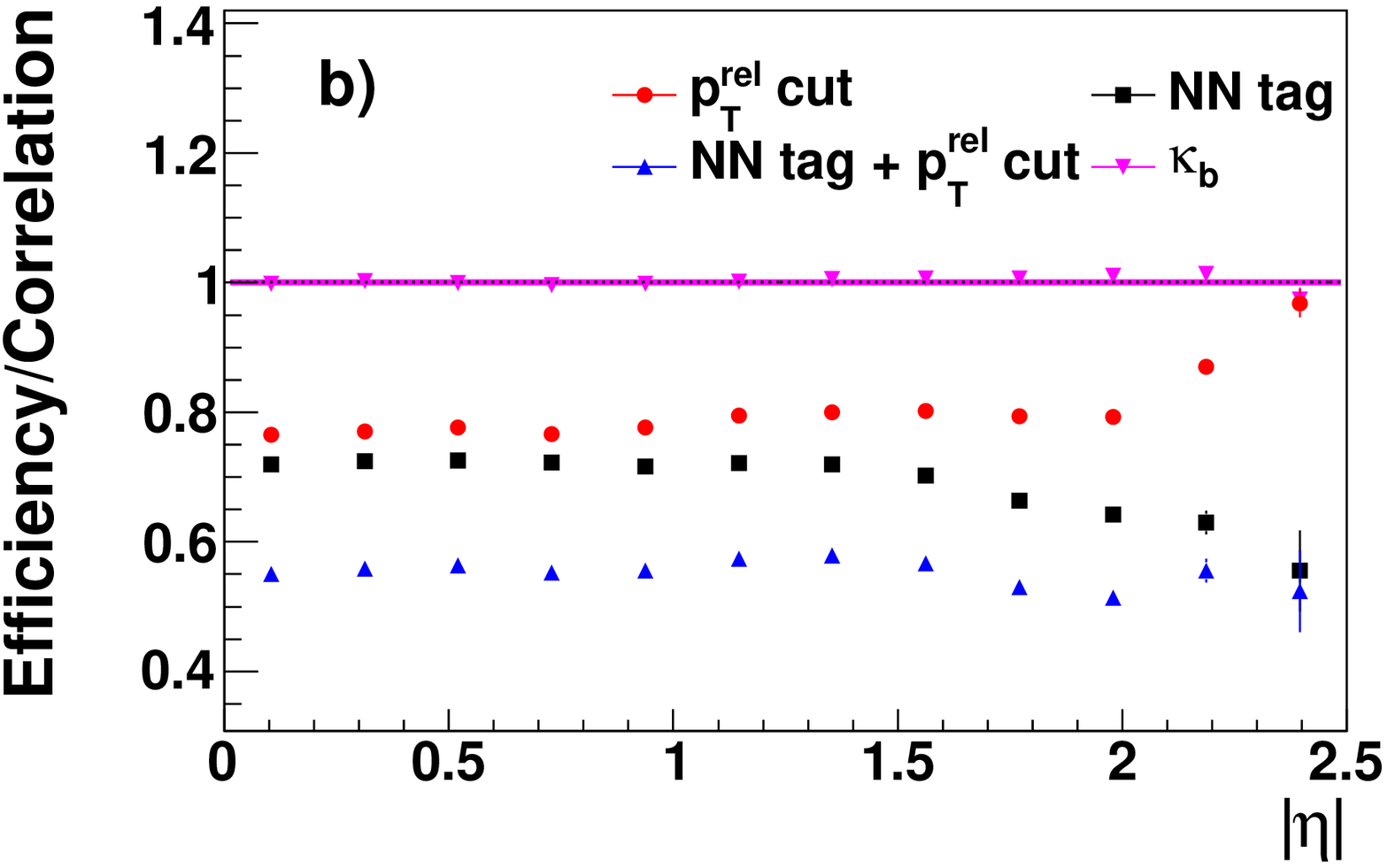}
    \includegraphics[width=2.67in]{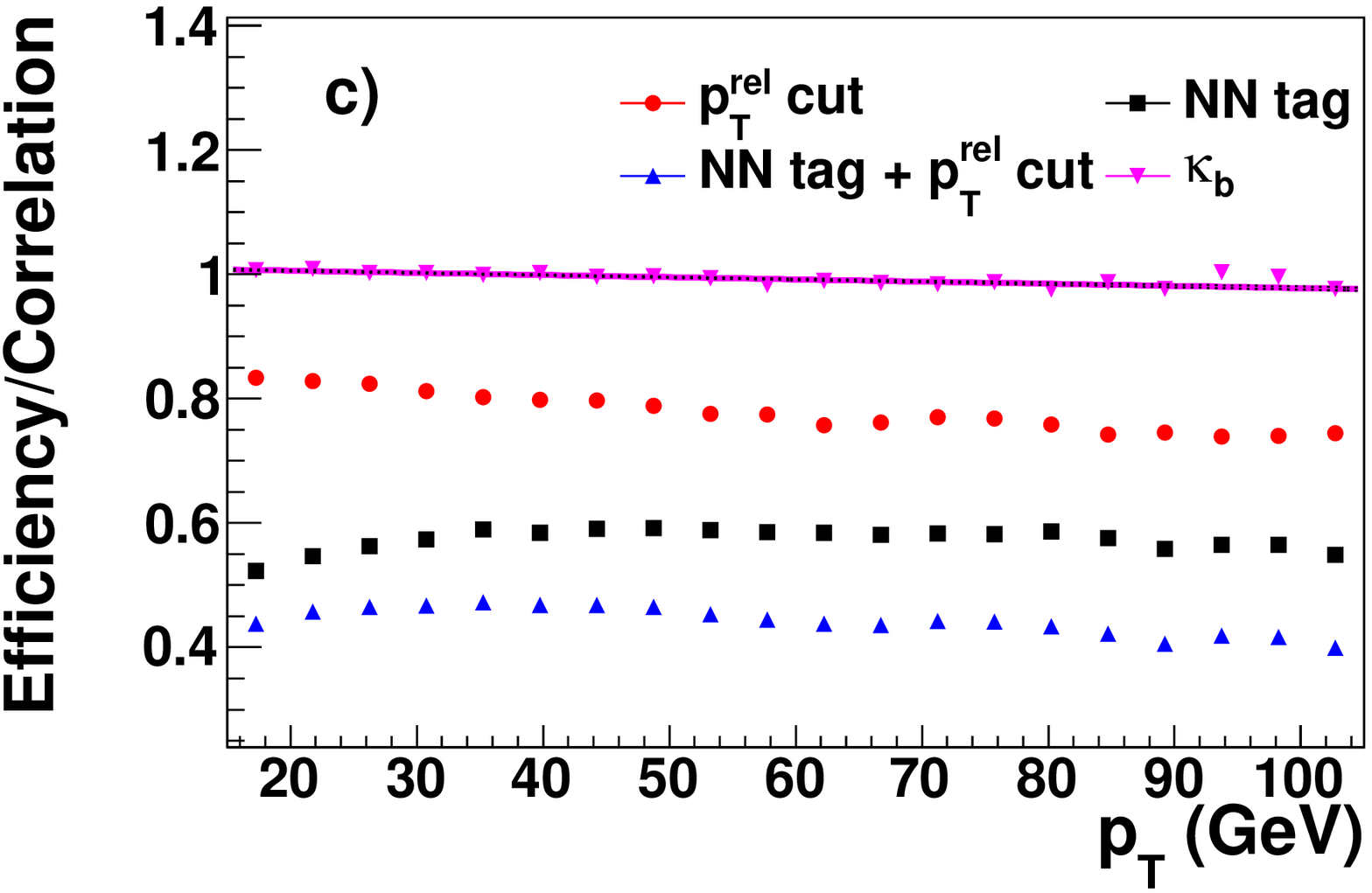}
    \includegraphics[width=2.67in]{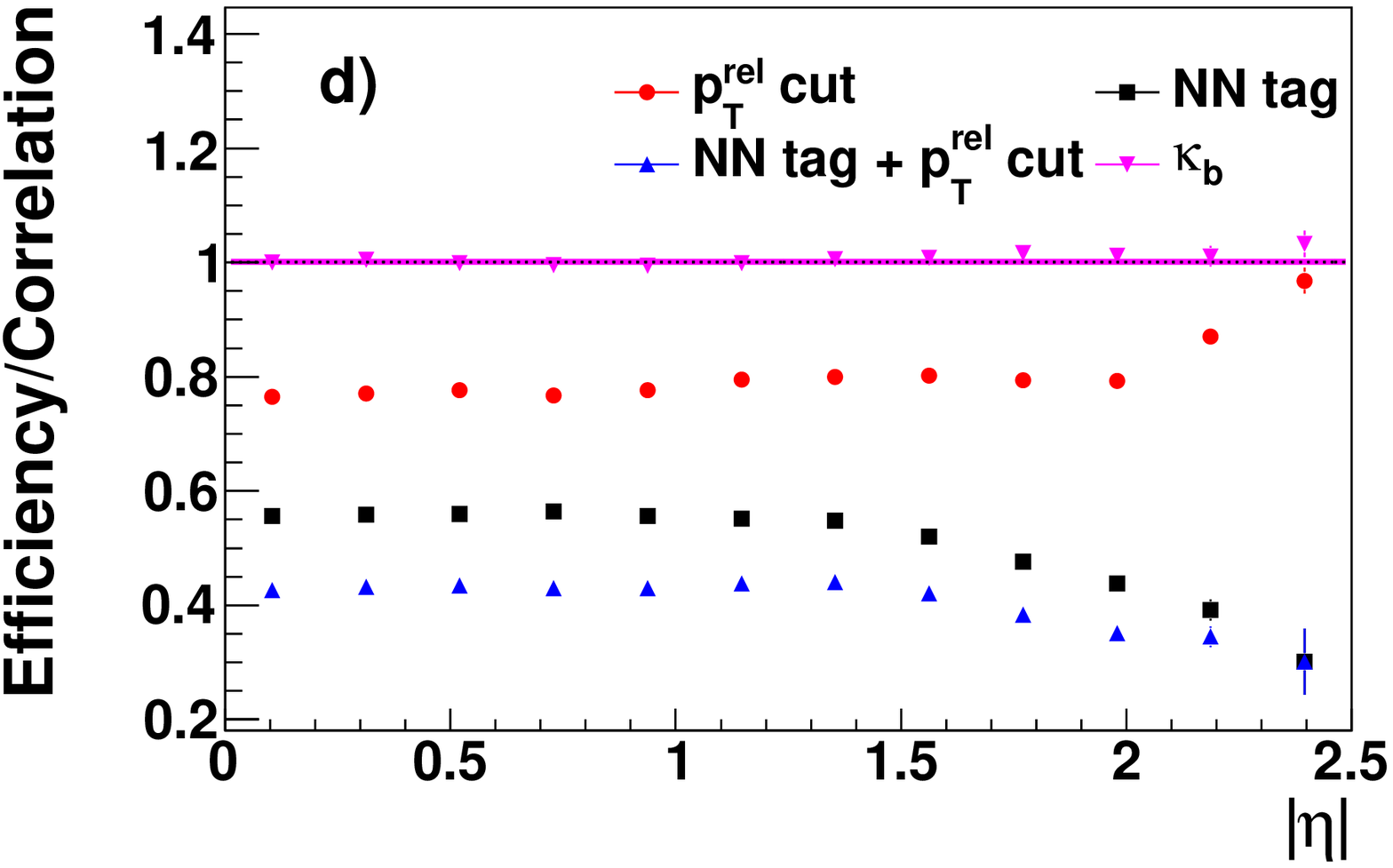}
  \end{center}
  \caption{\label{fig:kappa_b} (a): the efficiency of the \ptrel\ requirement (circles), 
    the L2 NN tagger (squares), the AND of the NN tag and \ptrel\ requirement (up
    triangles), and the correlation factor $\kappa_{b}$ (down triangles and
    fit), measured in the $\bquark\rightarrow\mu X$ MC sample in the jet \pt\ projection.
    (b): same in the $|\eta|$ projection. (c) and (d): same for the Tight NN
    tagger in the jet \pt\ and $|\eta|$ projection, respectively.
    The fit uncertainties are too small to be visible in this figure.}
\end{figure}

\begin{figure}[htbp]
  \begin{center}
    \includegraphics[width=2.67in]{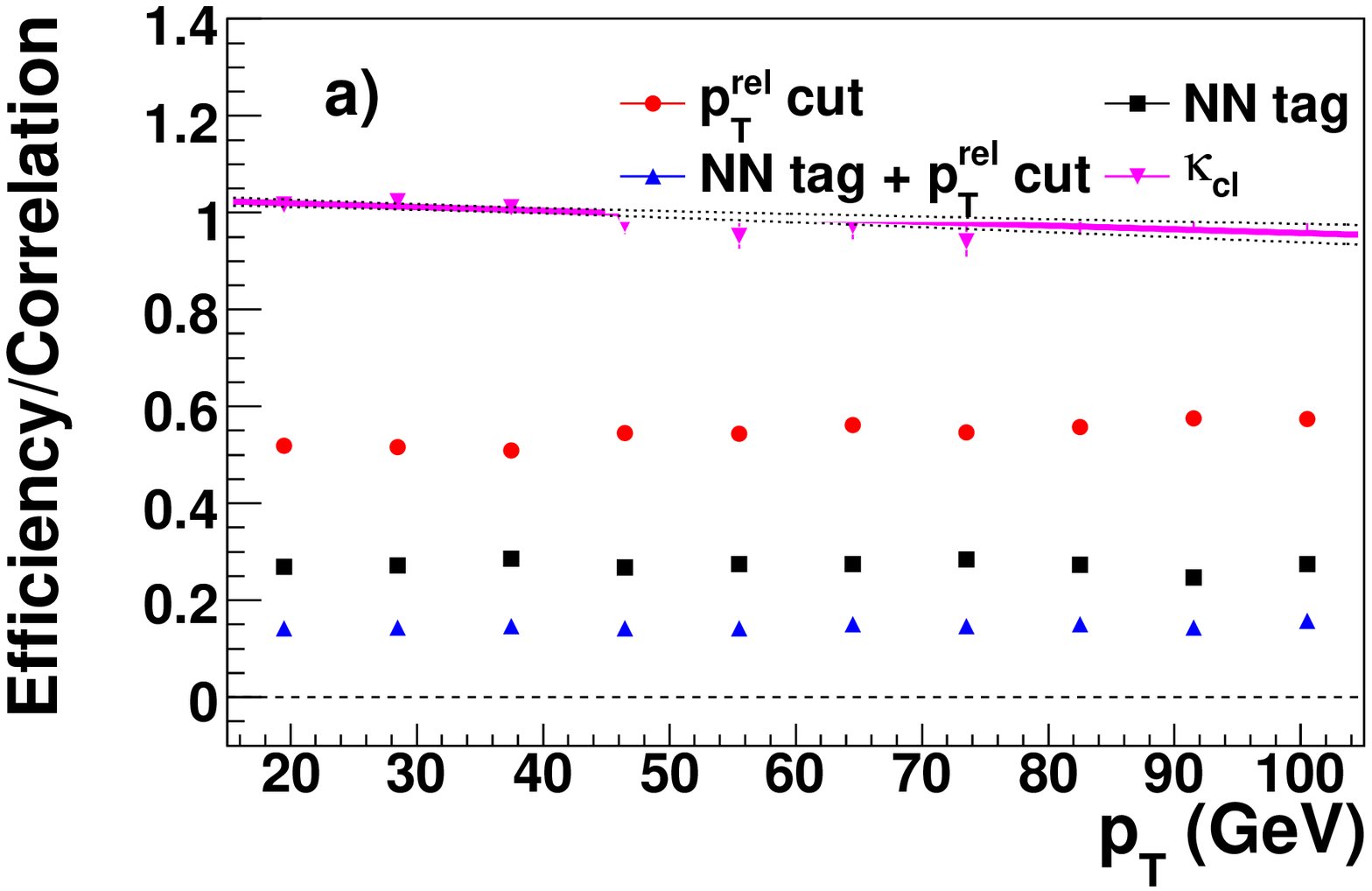}
    \includegraphics[width=2.67in]{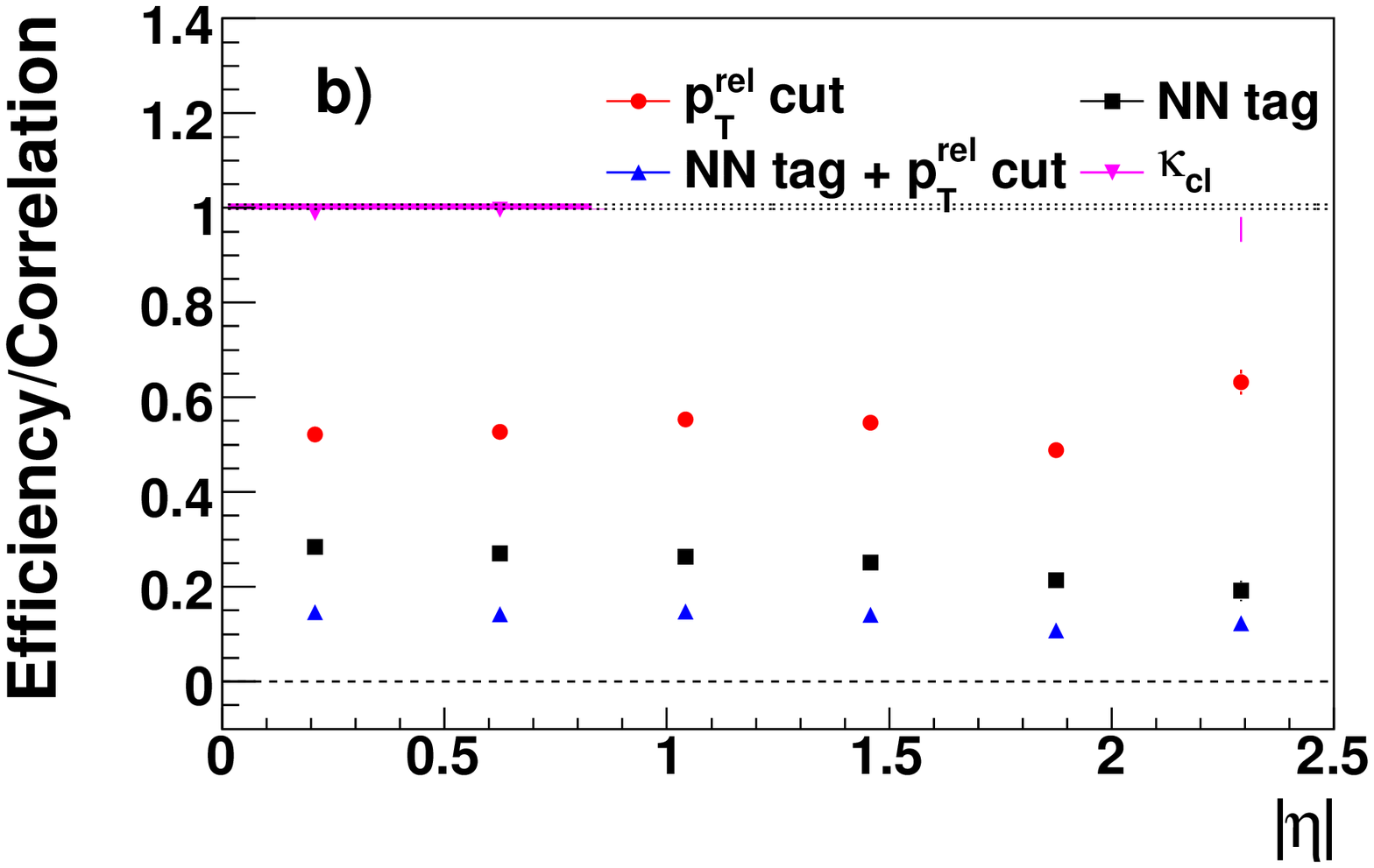}
    \includegraphics[width=2.67in]{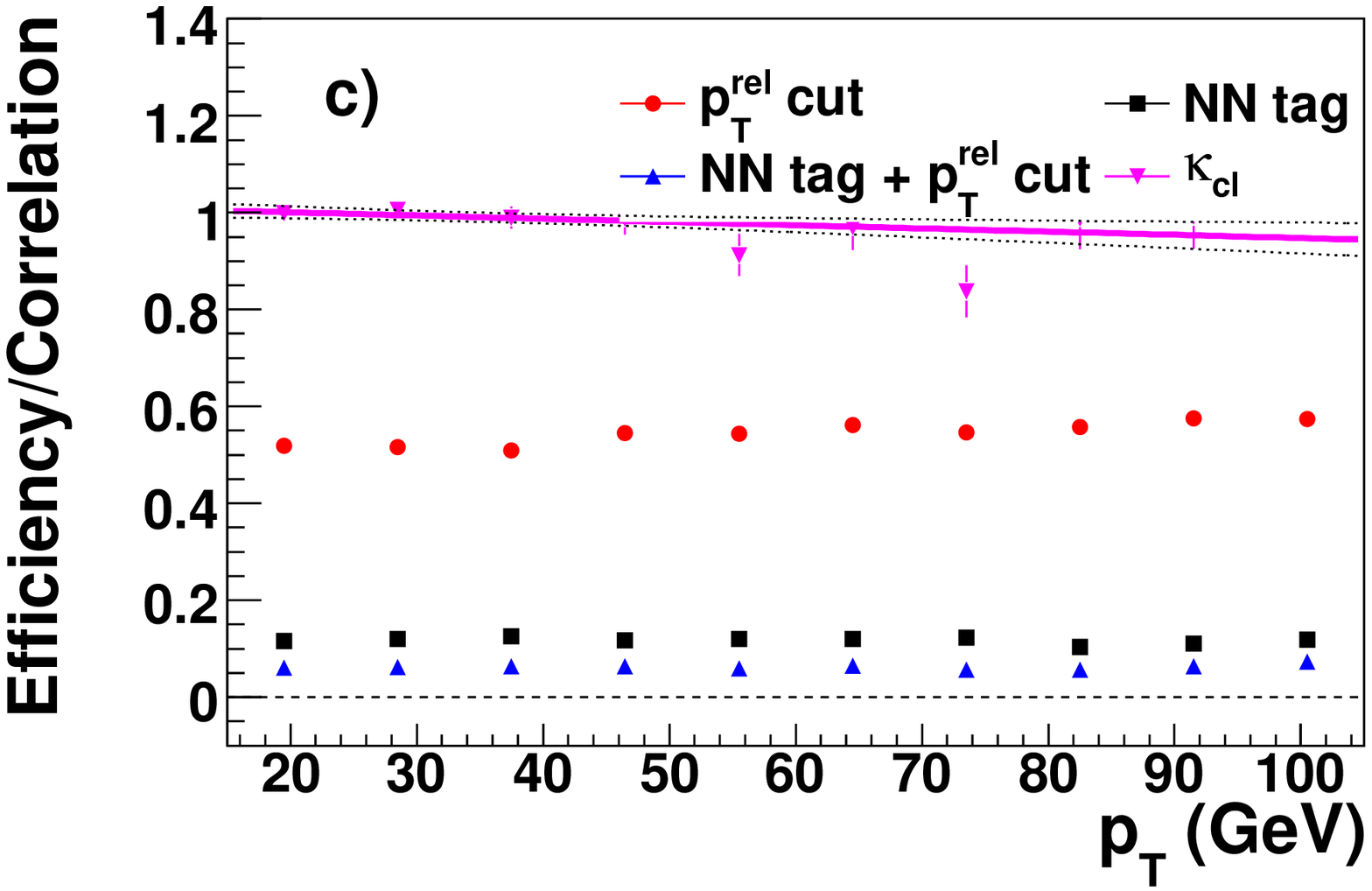}
    \includegraphics[width=2.67in]{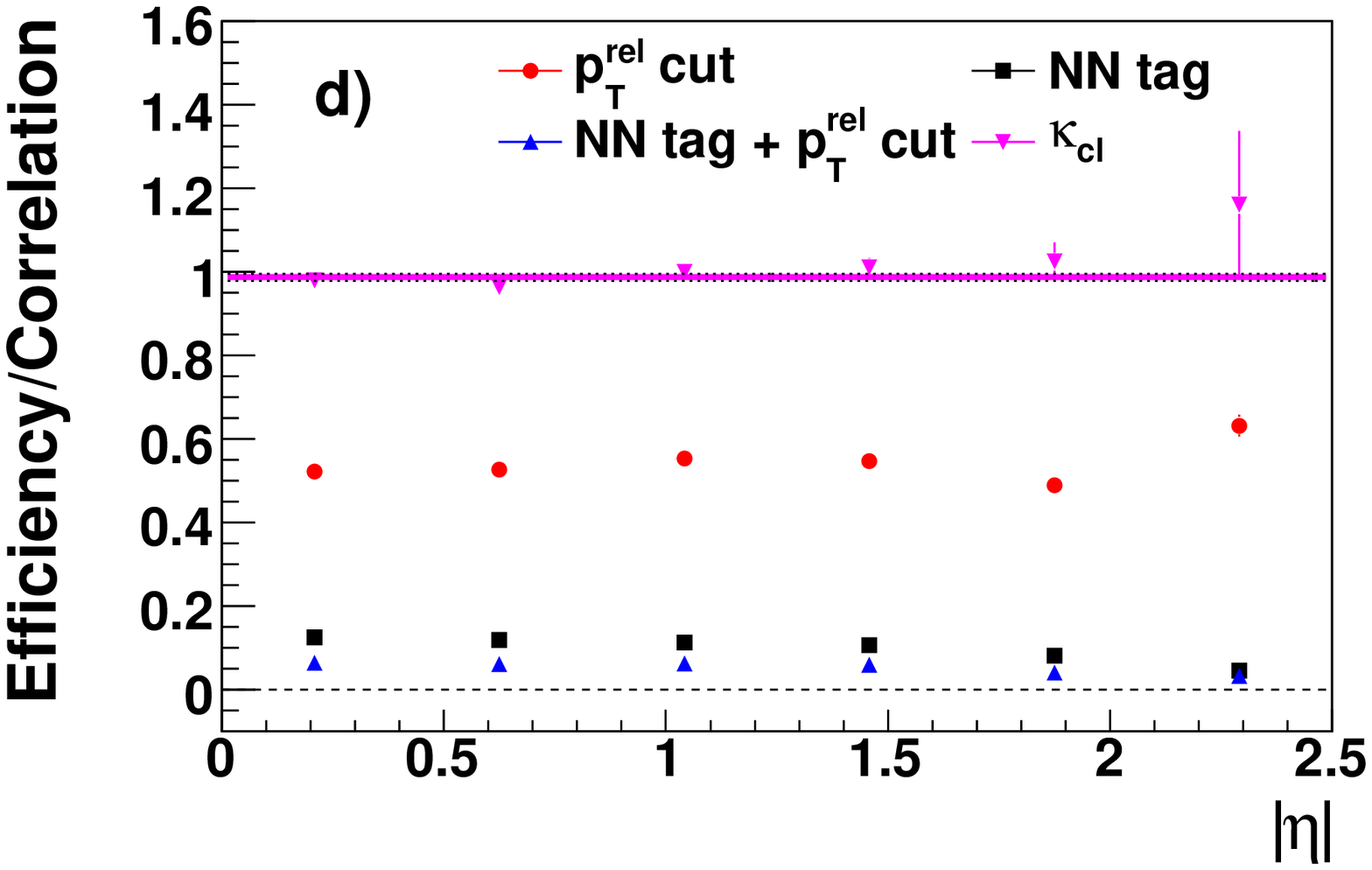}
  \end{center}
  \caption{\label{fig:kappa_cl} (a): the efficiency of the \ptrel\ requirement (circles),
    the L2 NN tagger (squares), the AND of the NN tag and \ptrel\ requirement (up
    triangles), and the correlation factor $\kappa_{cl}$ (down triangles and
    fit), measured 
    in the $\cquark l\rightarrow\mu X$ MC sample in the jet \pt\ projection.
    (b): same in the $|\eta|$ projection. (c) and (d): same for the Tight NN
    tagger in the jet \pt\ and $|\eta|$ projection, respectively.
    The dotted lines represent the fit uncertainties.}
\end{figure}

The requirement $\ptrel>0.5\GeVc$ is chosen in order to have a similar efficiency for
\cquark- and light-quark jets, so that the application of this requirement affects the
flavor composition of the background jets only modestly. Correction factors
$\kappa_{\bquark}$ and $\kappa_{cl}$ are determined for signal and background jets
by dividing the efficiency for jets satisfying both criteria by the product of the
efficiencies for jets satisfying the individual criteria.
They are shown in Figs.~\ref{fig:kappa_b} and \ref{fig:kappa_cl} for the L2 and
Tight operating points, respectively.

As indicated above, the application of the away side tag is not \emph{a priori}
expected to be correlated with the other two criteria. This hypothesis has been
verified explicitly in the simulation for the correlation with the \ptrel\ requirement.
The lifetime tagging requirements applied to both jets, however, could
be correlated by the fact that they involve the same primary vertex. The
corresponding correction factors are evaluated in the same way as
$\kappa_{\bquark}$ and $\kappa_{cl}$. They are denoted $\beta$ for \bquark\ jets
and $\alpha$ for background jets.

\begin{figure*}[htbp]
  \begin{center}
    \includegraphics[width=0.40\textwidth]{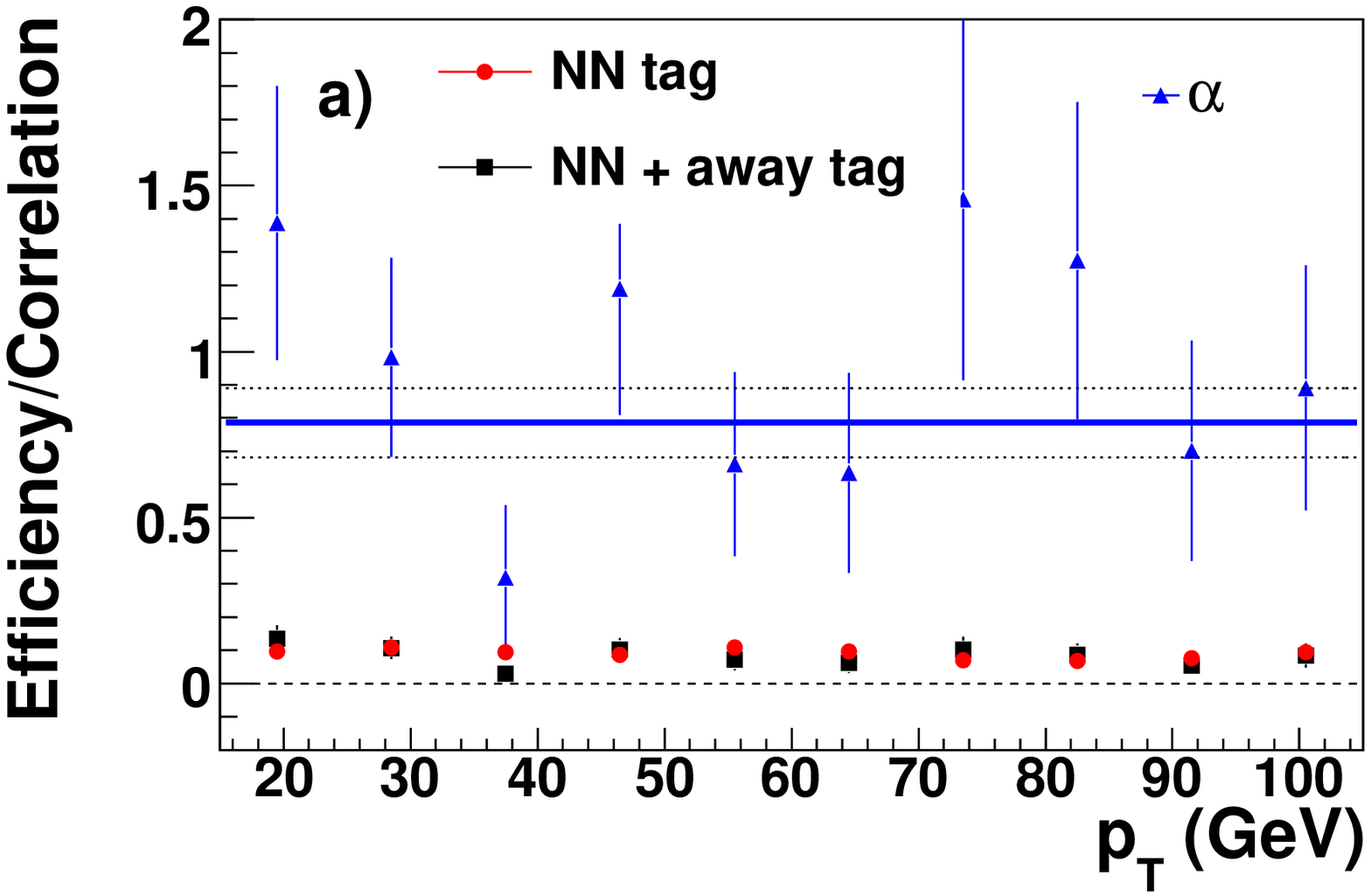}
    \includegraphics[width=0.40\textwidth]{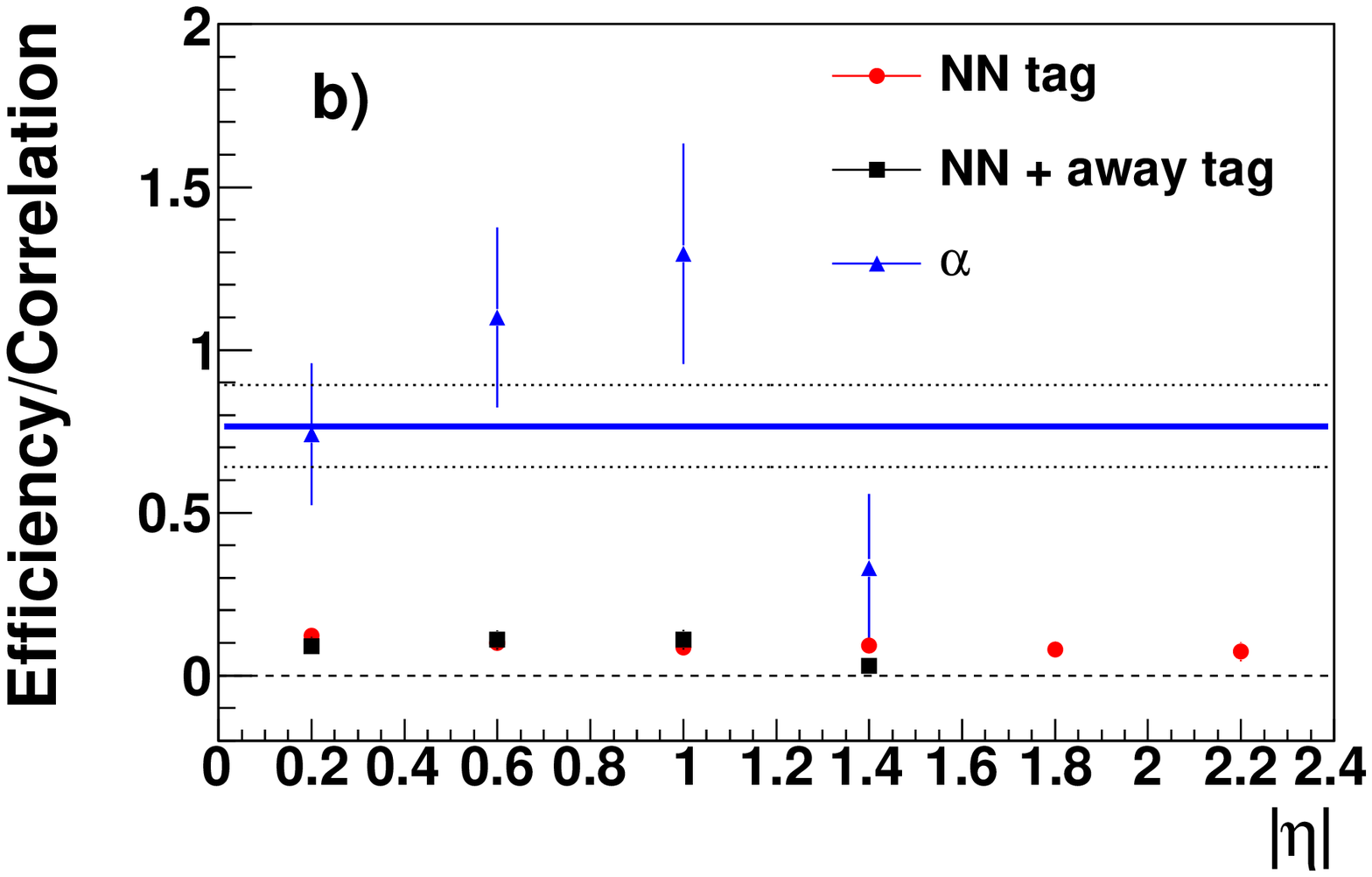}
    \includegraphics[width=0.40\textwidth]{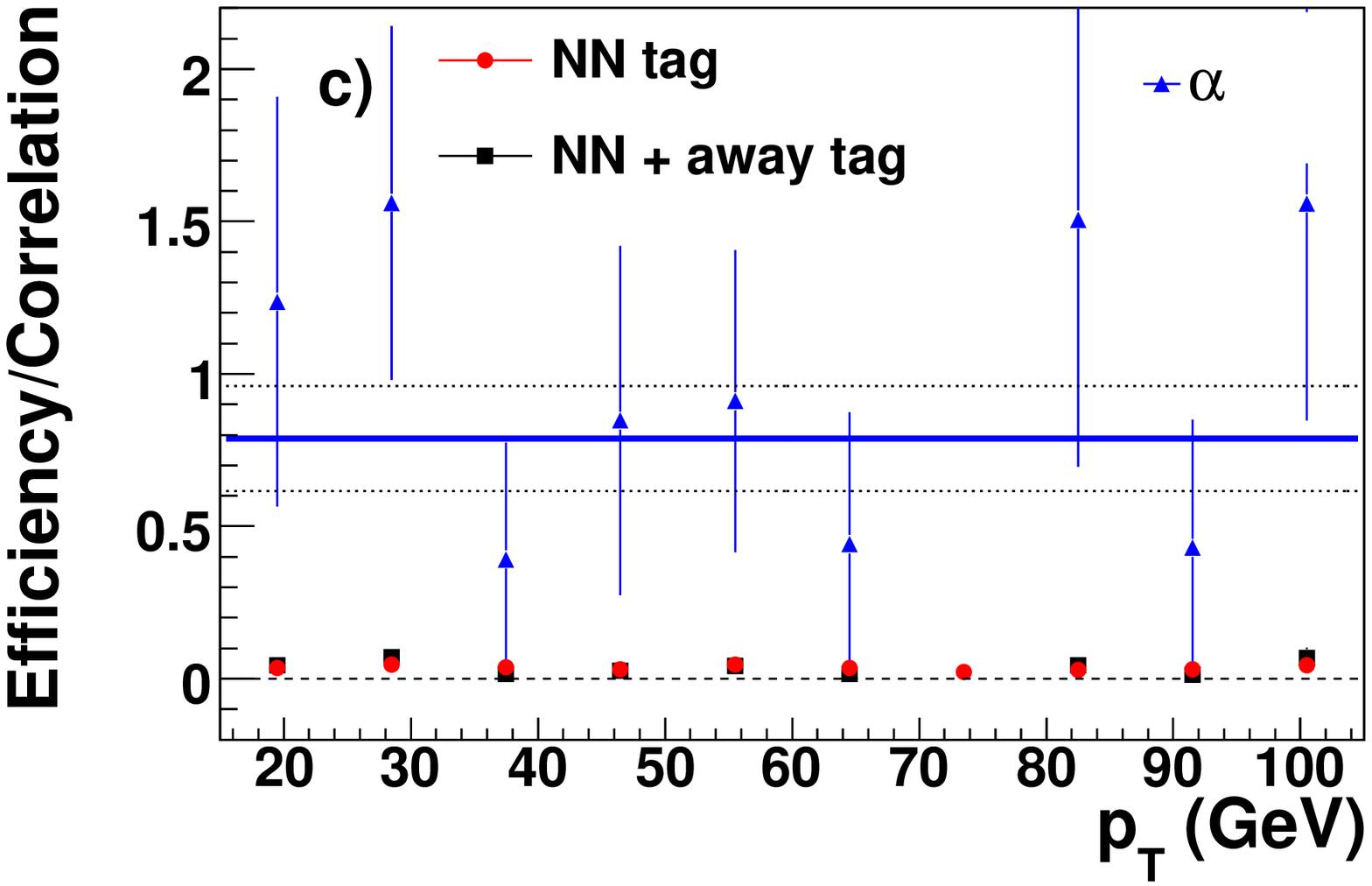}
    \includegraphics[width=0.40\textwidth]{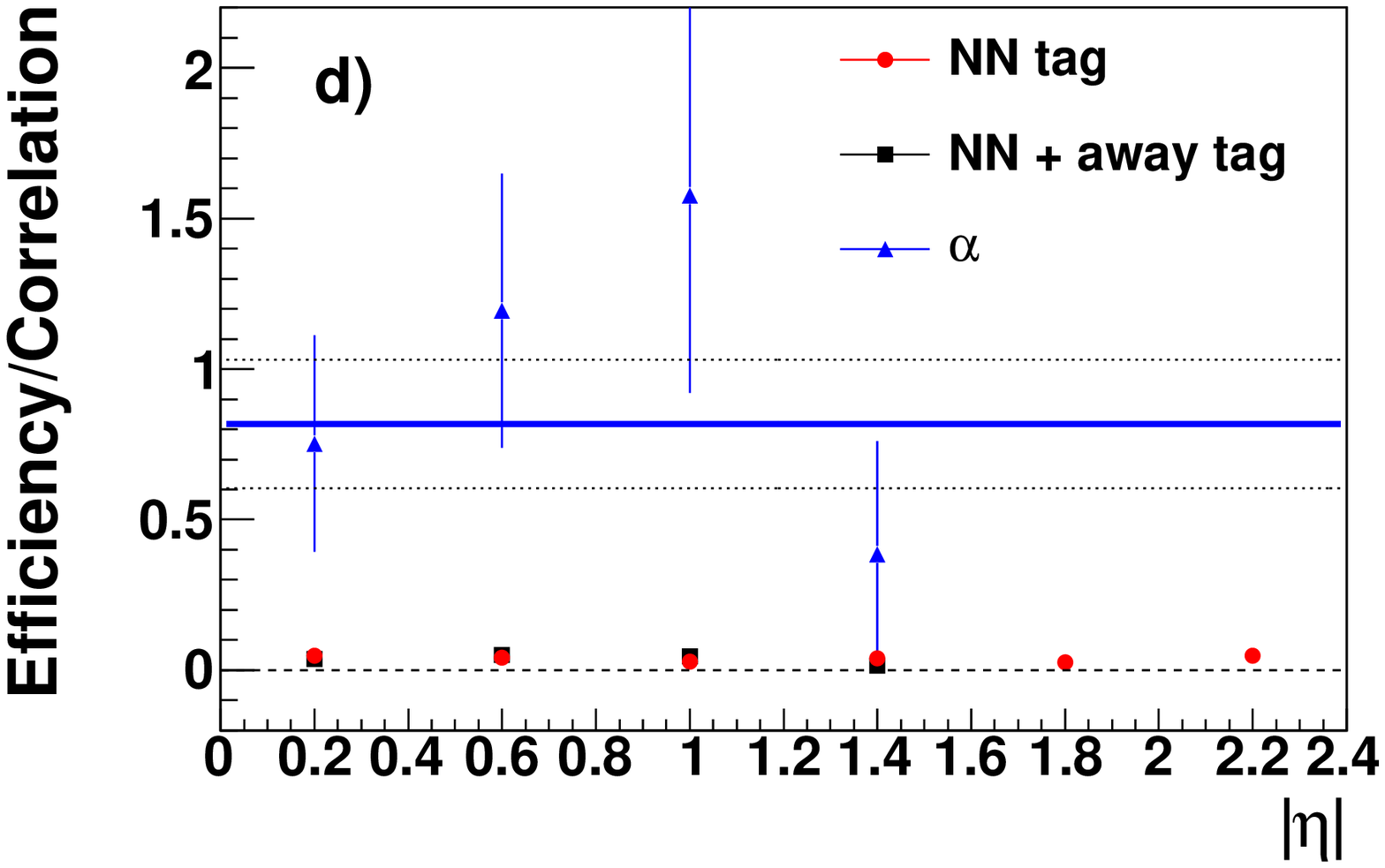}
  \end{center}
  \caption{\label{fig:alpha} (a): the L2 NN tagging efficiency (circles), the
    tagging efficiency after an away-tag requirement (squares), and
    their ratio, $\alpha$ (up triangles and fit) in the $cl\rightarrow\mu X$ MC
    sample as a function of jet \pt.
    (b): same as a function of $|\eta|$. (c) and (d): same for the Tight NN
    tagger as a function of jet \pt\ and $|\eta|$, respectively.
    The fit uncertainty on $\alpha$ is represented by the dotted lines.}
\end{figure*}

\begin{figure*}[htbp]
  \begin{center}
    \includegraphics[width=0.40\textwidth]{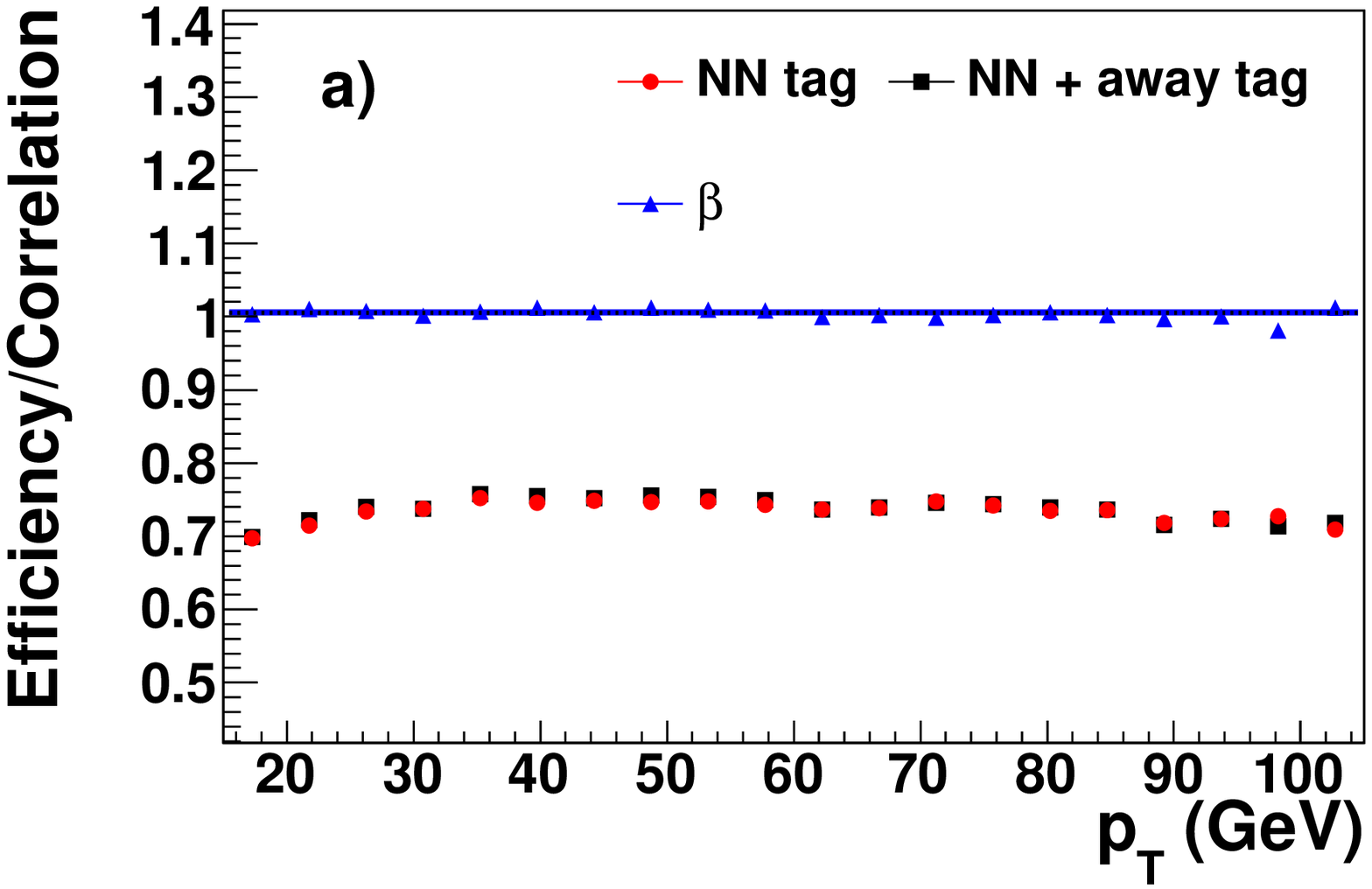}
    \includegraphics[width=0.40\textwidth]{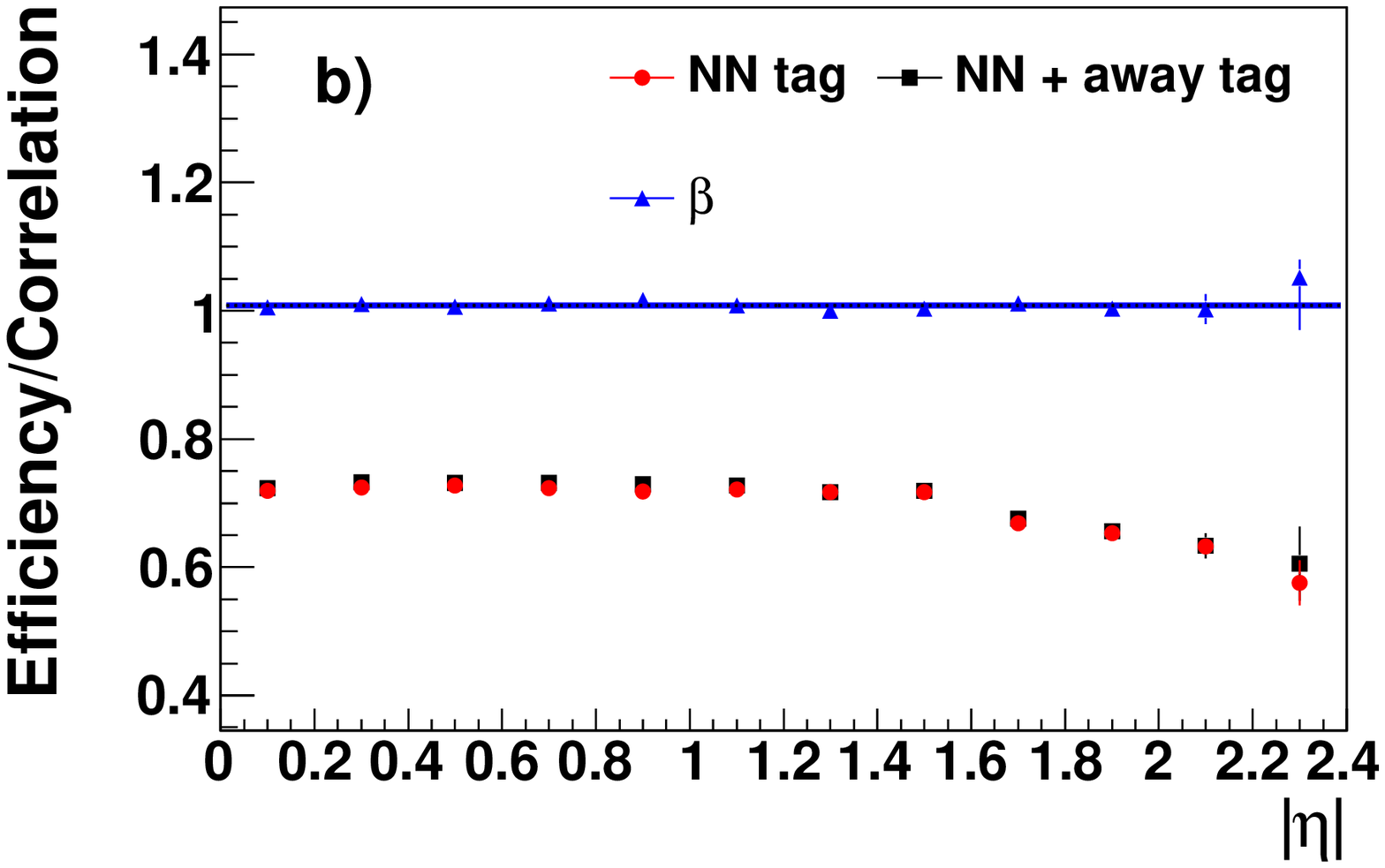}
    \includegraphics[width=0.40\textwidth]{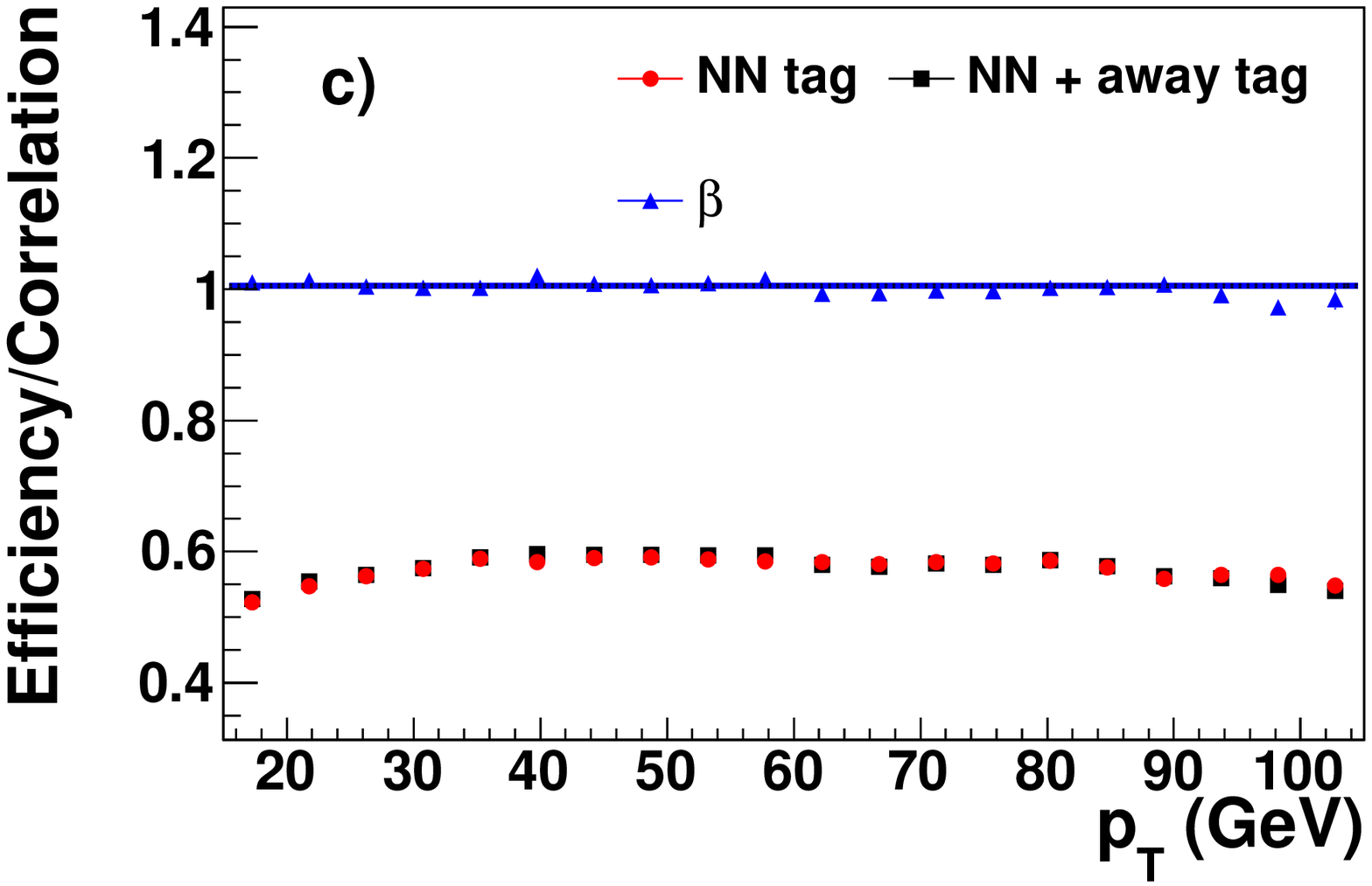}
    \includegraphics[width=0.40\textwidth]{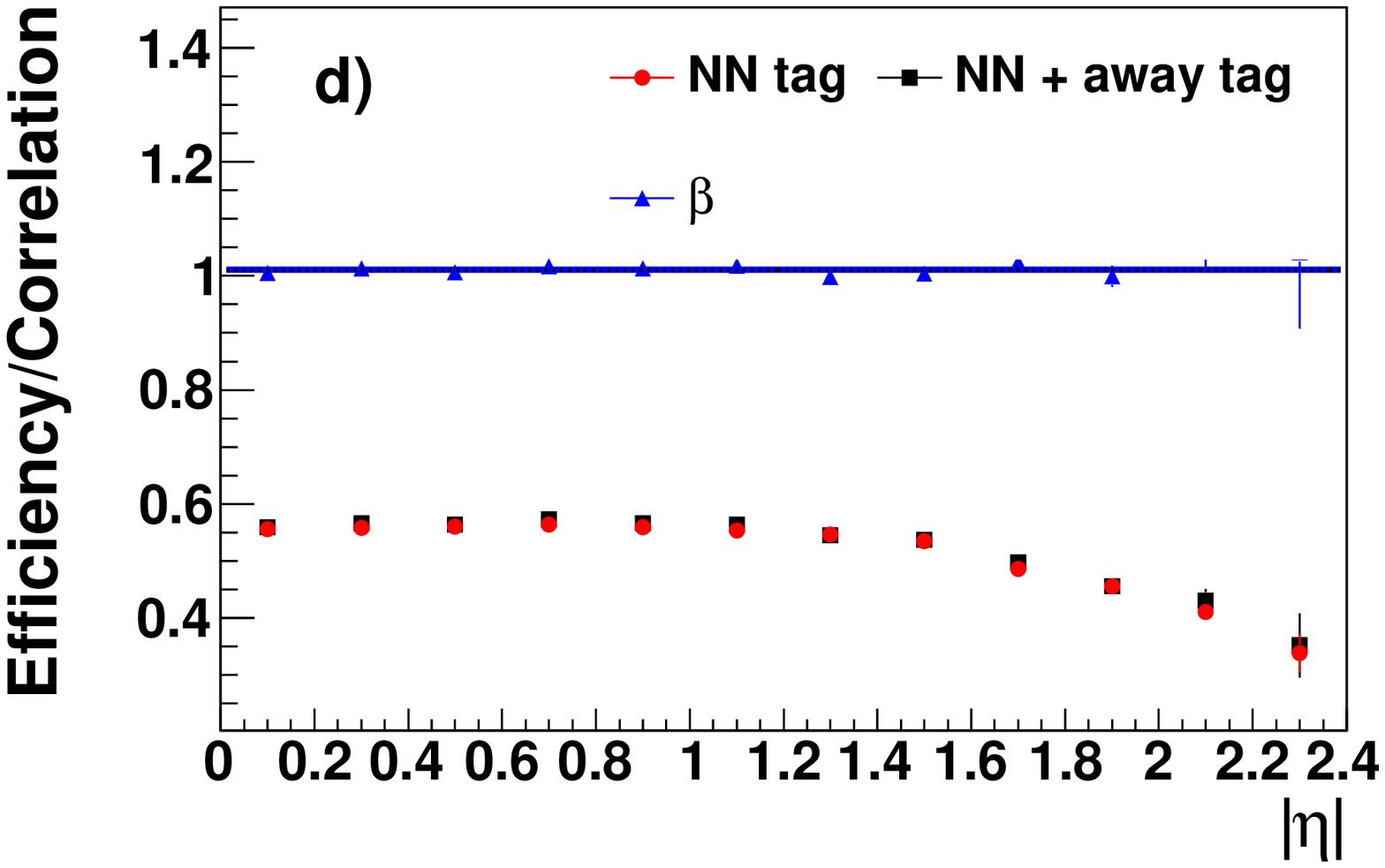}
  \end{center}
  \caption{\label{fig:beta} (a): the L2 NN tagging efficiency (circles), the
    tagging efficiency after an away-tag requirement (squares), and their ratio,
    $\beta$ (up triangles and fit) in the $\bquark\rightarrow\mu X$ MC sample as
    a function of jet \pt.
    (b): same as a function of $|\eta|$. (c) and (d): same for the Tight NN
    tagger as a function of jet \pt\ and $|\eta|$, respectively.
    The fit uncertainty on $\beta$ is too small to be visible in this figure.}
\end{figure*}

In addition to increasing the \bquark-jet fraction of the data sample, the
application of the away-side tag modifies the flavor composition 
of the background sample, as the charm tagging efficiency is expected to be
significantly higher than that for light-flavor jets. This causes a dependence
of $\alpha$ on the physics assumptions made in the MC event generators (in
which charm and light-flavor QCD samples are added weighted by their respective
production cross sections to yield a sample of $cl$ jets). Fortunately, it turns
out that the uncertainty on $\alpha$ affects the \bquark-tagging efficiency only
marginally. The factors $\alpha$ and $\beta$ are shown in
Figs.~\ref{fig:alpha} and \ref{fig:beta}, respectively, for the L2 and Tight
operating points.

\subsubsection{{\it SystemD} equations}
\label{sec:system8-equations}

Denoting the criteria used in {\it SystemD} as $l$ for the lifetime tagging
criterion, $m$ for the \ptrel\ requirement, and $b$ for the away-side tag, with
the notation for the correction factors as above, and with $f_{0}$ and $f_{1}$
of Eq.~\ref{eq:total} renamed to $f_{\bquark}$ and $f_{cl}$, respectively, the
final system to solve is therefore:
\begin{equation}
  \begin{array}{lclcl}
    f_{\bquark}&+&f_{cl} &=&1 \\
    f_{\bquark} \varepsilon_{\bquark}^{l} &+& f_{cl}
    \varepsilon_{cl}^{l}  &=& Q^{l} \\
    f_{\bquark} \varepsilon_{\bquark}^{m} &+& f_{cl}
    \varepsilon_{cl}^{m}  &=& Q^{m}\\
    f_{\bquark} \varepsilon_{\bquark}^{b} &+& f_{cl}
    \varepsilon_{cl}^{b}  &=& Q^{b}\\
    f_{\bquark} \kappa_{\bquark}\varepsilon_{\bquark}^{l}
    \varepsilon_{\bquark}^{m} &+& f_{cl} \kappa_{cl}
    \varepsilon_{cl}^{l}\varepsilon_{cl}^{m} &=& Q^{l,m}\\
    f_{\bquark} \varepsilon_{\bquark}^{m}\varepsilon_{\bquark}^{b} &+&
    f_{cl} \varepsilon_{cl}^{m}\varepsilon_{cl}^{b}
    &=& Q^{m,b}\\
    f_{\bquark} \beta\varepsilon_{\bquark}^{b}\varepsilon_{\bquark}^{l} &+&
    f_{cl} \alpha\varepsilon_{cl}^{b}
    \varepsilon_{cl}^{l} &=& Q^{b,l}\\
    f_{\bquark} \kappa_{\bquark}\beta\varepsilon_{\bquark}^{l}
    \varepsilon_{\bquark}^{m} \varepsilon_{\bquark}^{b} &+& f_{cl}
    \kappa_{cl}\alpha\varepsilon_{cl}^{l}
    \varepsilon_{cl}^{m}\varepsilon_{cl}^{b} &=& Q^{l,m,b}
  \end{array}
  \label{eq:system8}
\end{equation}

As already mentioned in Sec.~\ref{subsec:math}, the \textit{SystemD} method leads
to a set of \emph{nonlinear} equations, and two possible solutions exist for the
quantity of interest, the \bquark-tagging efficiency.
The ambiguity between these two solutions is resolved using the \emph{a priori}
knowledge that the efficiency of each selection criterion for \bquark\ jets should
be higher than for background jets:
$\varepsilon_{\bquark}^{l,m,b} > \varepsilon_{cl}^{l,m,b}$.

\subsection{Further corrections}
\label{sec:further-corrections}

The \bquark-tagging efficiency obtained with the {\it SystemD} method is valid
for jets with a semimuonic decay of the \bquark\ hadron.
To obtain the efficiency for \emph{inclusive} jets not biased by the requirement
of such a decay, a correction is determined 
using a sample of simulated $\bquark$ jets with \bquark\ hadrons
decaying inclusively or as $\bquark\rightarrow \mu X$.
The final efficiency is then defined as
\begin{equation}
  \label{eq:sfb}
  \varepsilon^{\mbox{\scriptsize data}}_{\bquark} =
  \frac{\varepsilon^{\mbox{\scriptsize data}}_{\bquark \rightarrow \mu X} \cdot
    \varepsilon^{\mbox{\scriptsize MC}}_{\bquark}} {\varepsilon^{\mbox{\scriptsize MC}}_{\bquark
      \rightarrow \mu X}} = \mbox{SF}_{\bquark} \cdot \varepsilon^{\mbox{\scriptsize
      MC}}_{\bquark} \; ,
\end{equation}
where
$\mbox{SF}_{\bquark} = \varepsilon^{\mbox{\scriptsize data}}_{\bquark \rightarrow \mu X}
/ \varepsilon^{\mbox{\scriptsize MC}}_{\bquark \rightarrow \mu X}$
is the data-to-simulation efficiency scale factor, and
$\varepsilon^{\mbox{\scriptsize data}}_{\bquark \rightarrow \mu X}$ is identical
to the quantity denoted as $\varepsilon^{t}_{\bquark}$ in Eq.~\ref{eq:system8}.
The tagging efficiency for \cquark-quark jets is not measured in data.
It is assumed that the data-to-simulation scale factor is identical
for \bquark\ and \cquark\ jets. The \cquark-jet tagging efficiency is then
derived from the simulation as
\begin{equation}
  \label{eq:sfc}
  \varepsilon^{\mbox{\scriptsize data}}_{\cquark} = \mbox{SF}_{\bquark} \cdot
  \varepsilon^{\mbox{\scriptsize MC}}_{\cquark} .
\end{equation}

\subsection{Tagging efficiency parametrization}
\label{nn:system8:efficiencies}

The tagging efficiencies are parametrized in terms of the \et\ and $\eta$ of the jets.
As the use of the \textit{SystemD} method requires high statistics to obtain
stable solutions, it is not possible to extract a proper 2D
parametrization. Instead, it is assumed that the dependence on these variables can
be factorized:
\begin{equation}
  \varepsilon(\et,\eta) = \frac{1}{\varepsilon_{\mbox{\scriptsize all}}}
  \cdot \varepsilon(\et) \cdot \varepsilon(|\eta|),
  \label{eq:efficiency-factorization}
\end{equation}
where $\varepsilon_{\mbox{\scriptsize all}}$ is the efficiency for the entire sample,
and
\begin{eqnarray}
  \label{eq:efficiency-parametrization}
 \varepsilon(\et) & = & \frac{c}{1 + ae^{-b\et}},\\
  \varepsilon(|\eta|) & = & d + e|\eta| + f|\eta|^{2} + g|\eta|^{3} + h|\eta|^{4}, \nonumber
\end{eqnarray}
where $a$ -- $h$ are fit parameters.

\begin{figure}[htbp]
  \begin{center}
    \includegraphics[width=2.67in]{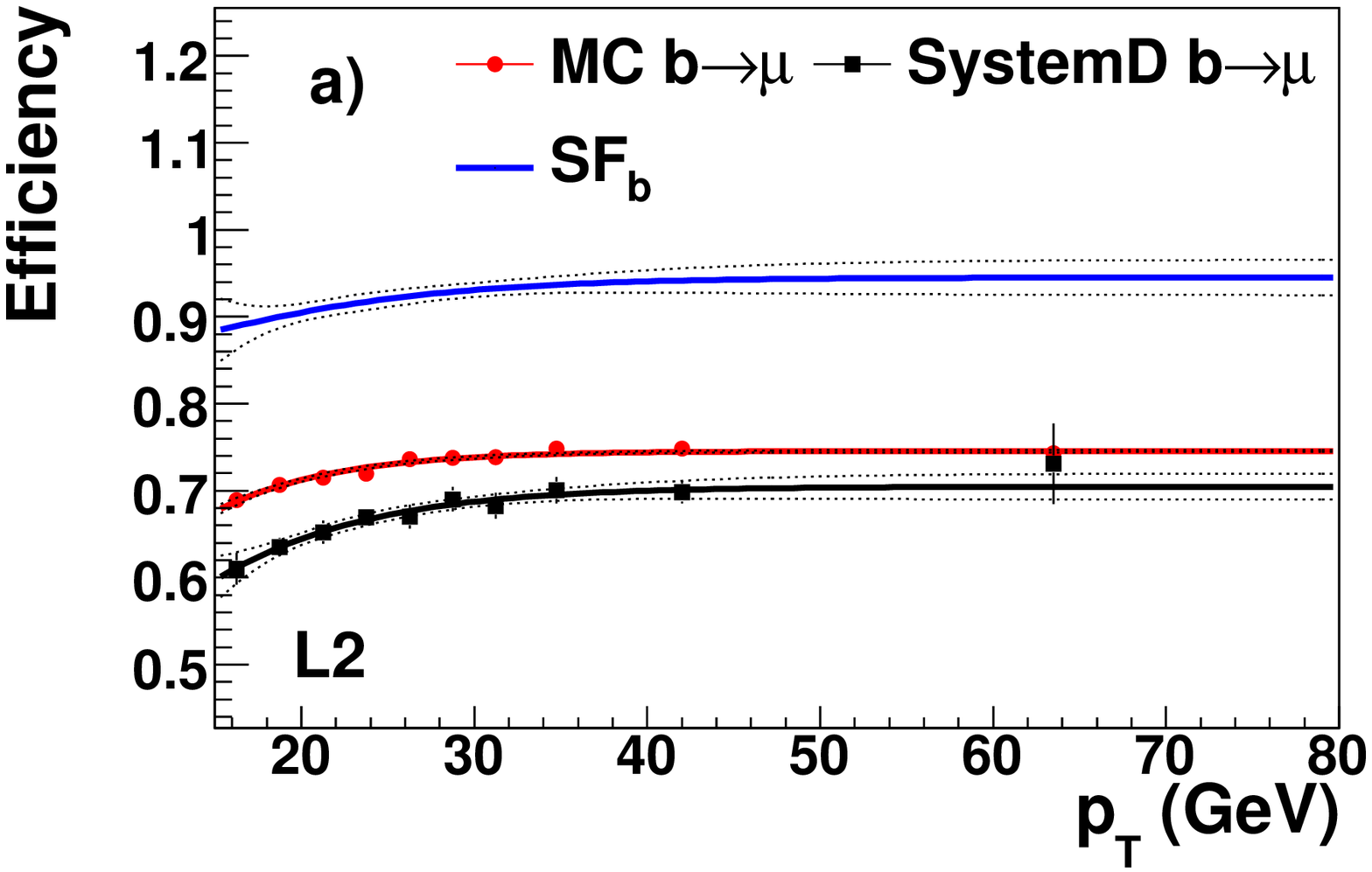}
    \includegraphics[width=2.67in]{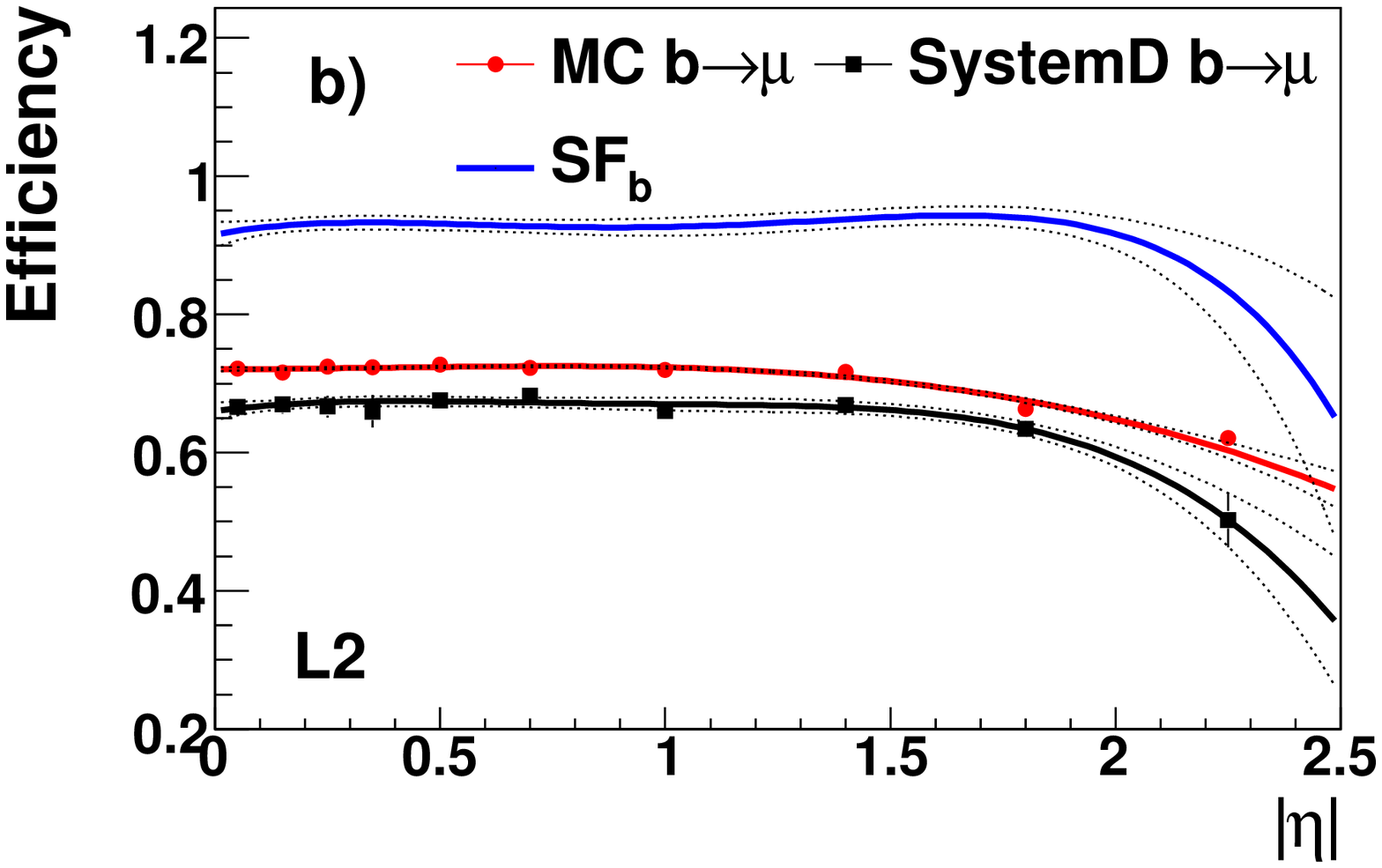}
    \includegraphics[width=2.67in]{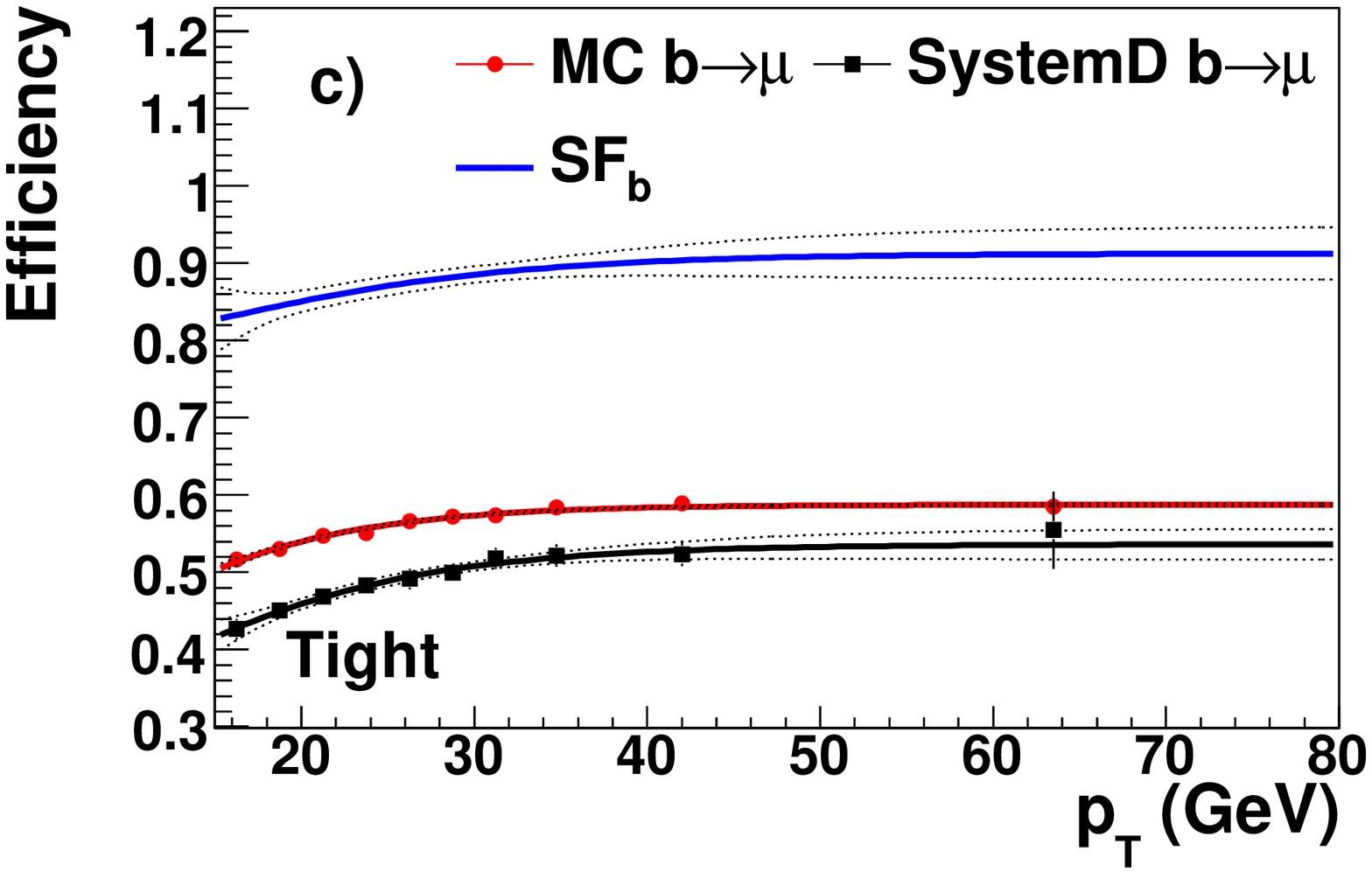}
    \includegraphics[width=2.67in]{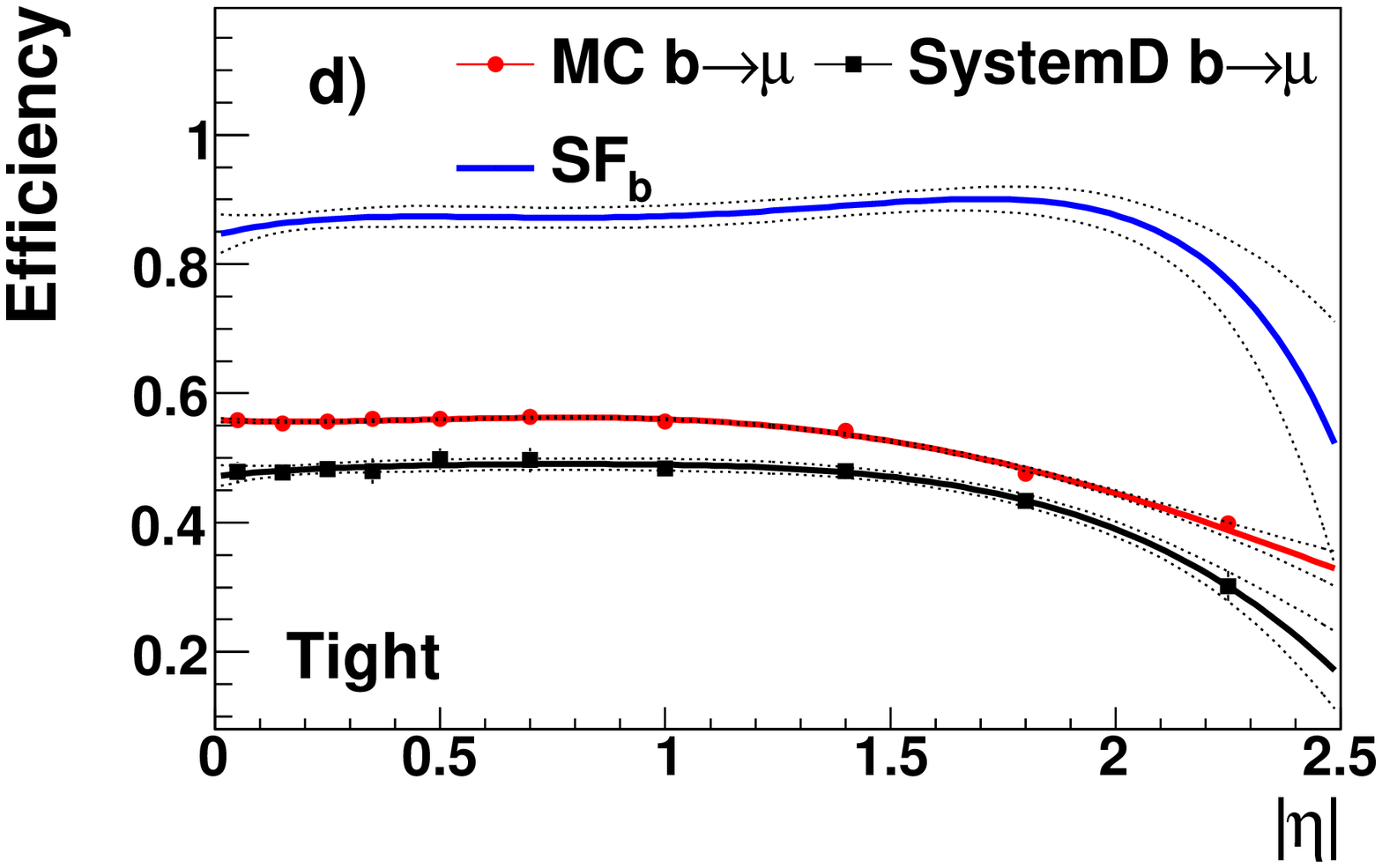}
  \end{center}
  \caption{(a): the L2 NN tagger scale factors (line) and
    the data (squares) and MC (circles) semimuonic \bquark-jet
    efficiencies as a function of jet \pt. (b): same as a function of jet
    $|\eta|$. (c) and (d): same for the Tight NN tagger, as a function of jet
    \pt\ and $|\eta|$, respectively. The functions used for the parametrization
    are outlined in the text and the dotted curves represent the $\pm1 \sigma$
    statistical uncertainty.}
  \label{fig:s8eff}
\end{figure}

\begin{figure}[htbp]
  \begin{center}
    \includegraphics[width=2.67in]{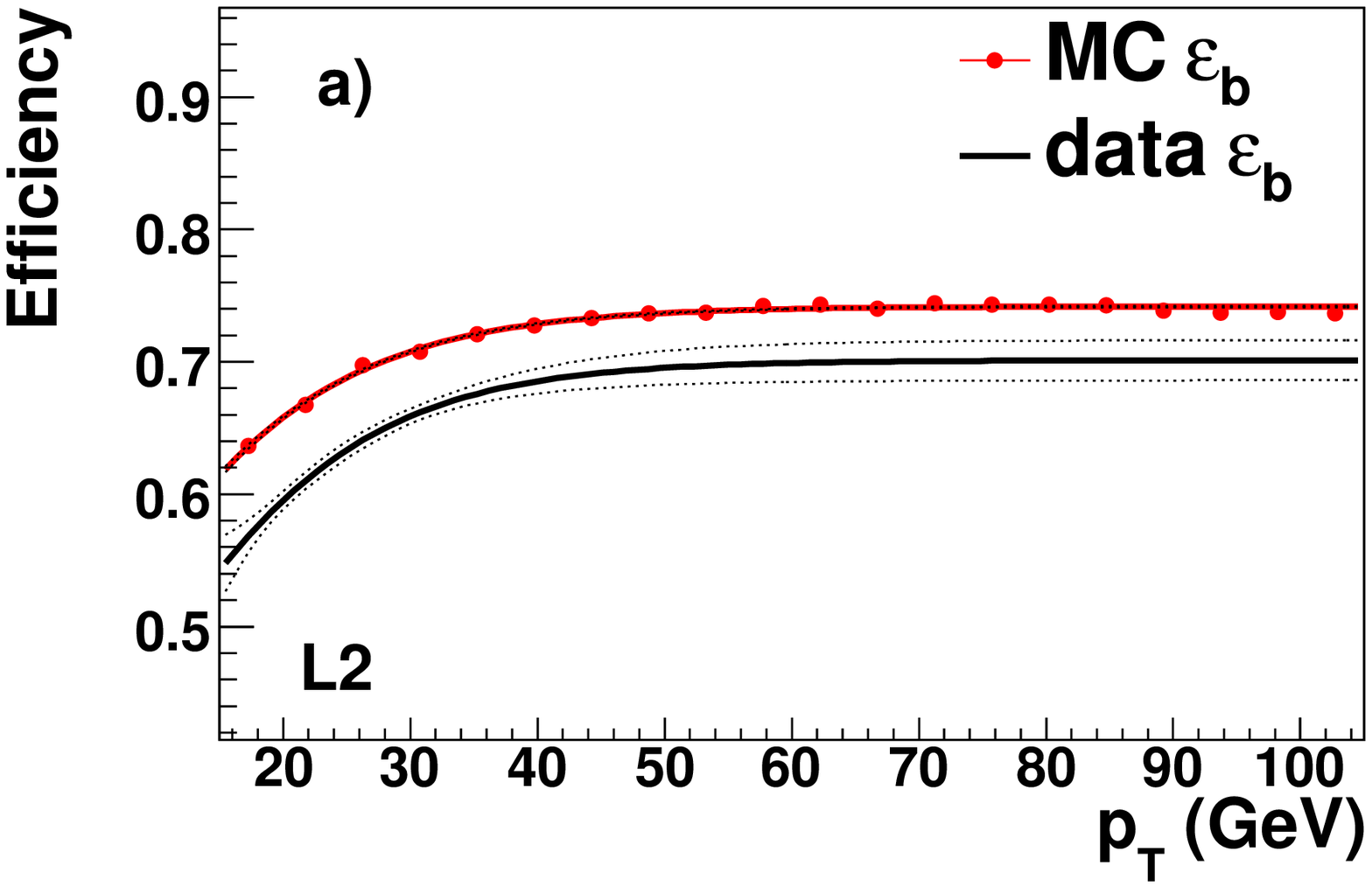}
    \includegraphics[width=2.67in]{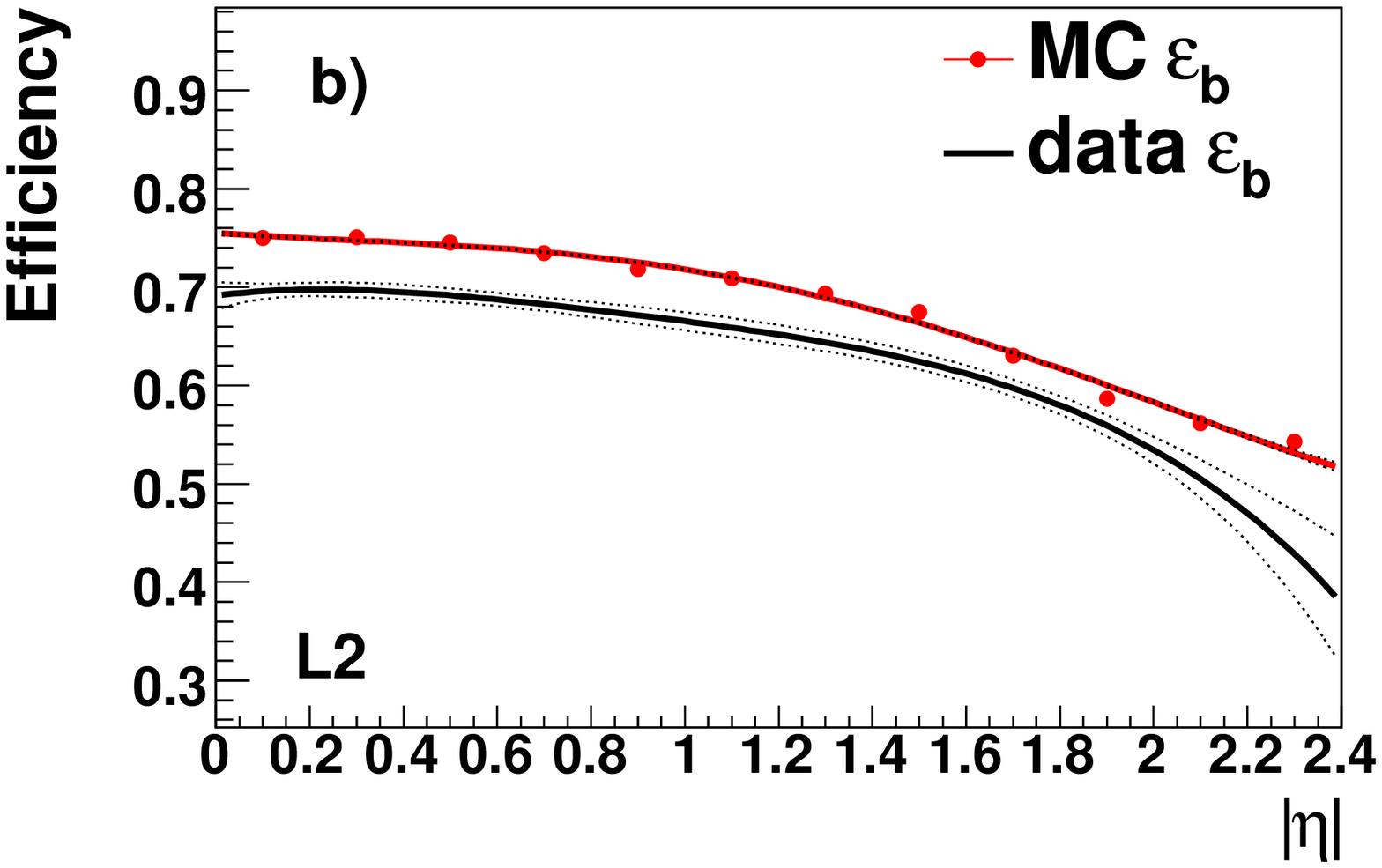}
    \includegraphics[width=2.67in]{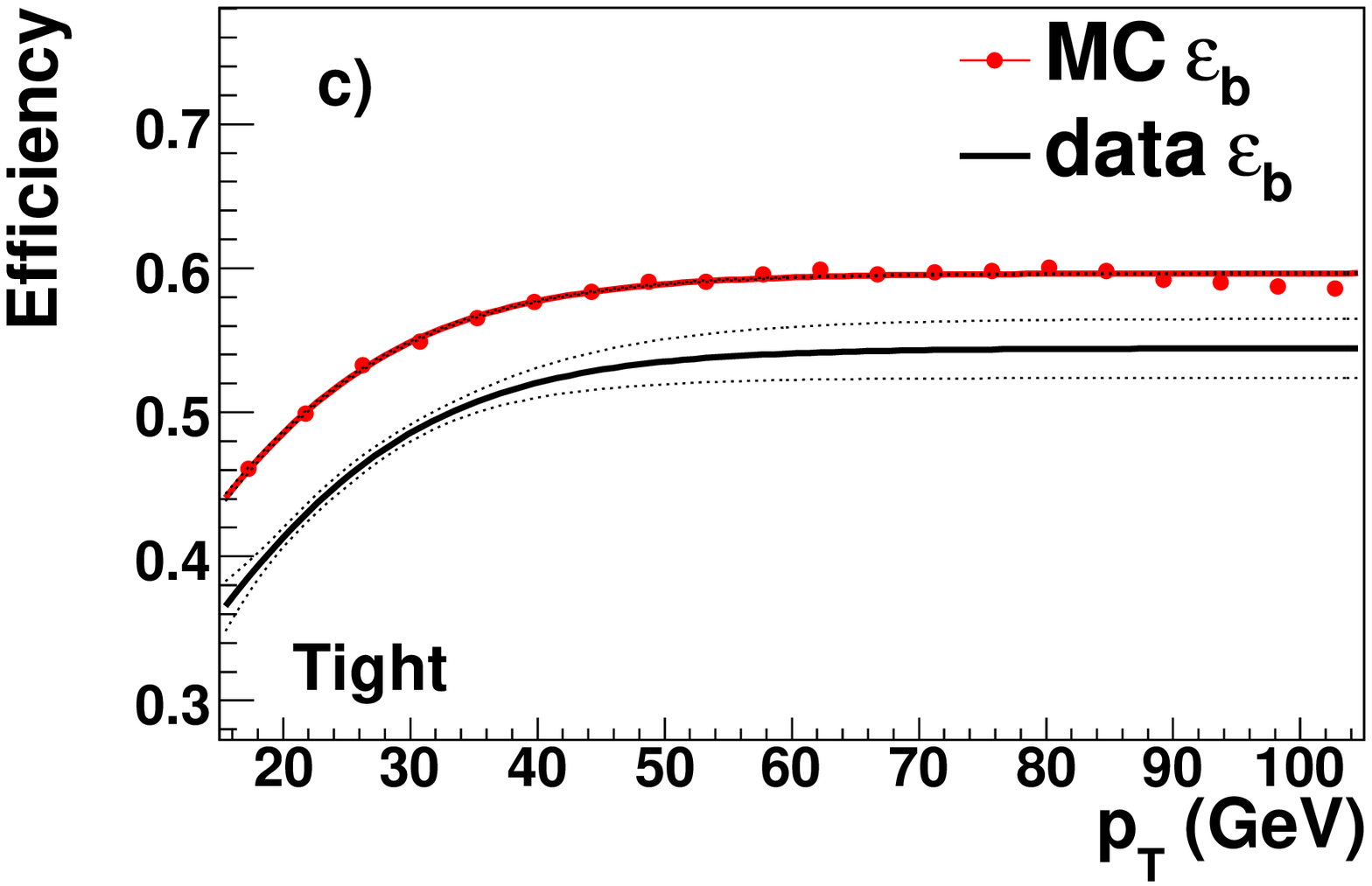}
    \includegraphics[width=2.67in]{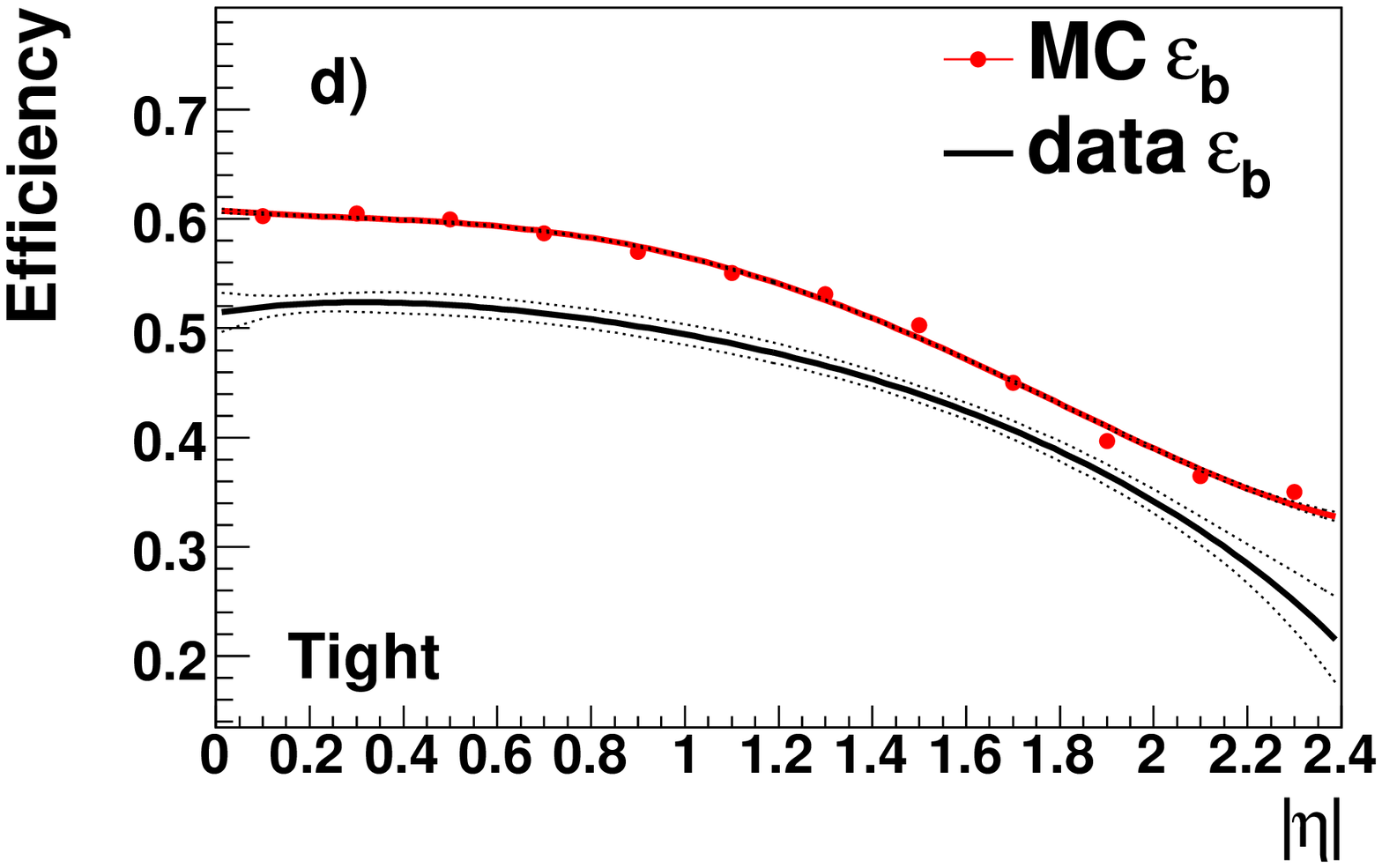}
  \end{center}
  \caption{(a): the L2 NN tagger inclusive \bquark-jet efficiency in both data
    (line) and MC (circles) as a function of jet \pt. (b): same as a
    function of jet $|\eta|$. (c) and (d): same for the Tight NN tagger, as a
    function of jet \pt\ and $|\eta|$, respectively.
    The dotted lines represent the fit uncertainty, which is almost
    entirely inherited from the uncertainty on the scale factor. The
    functions used for the parametrization are outlined in the text.}
  \label{fig:trf_b}
\end{figure}

\begin{figure}[htbp]
  \begin{center}
    \includegraphics[width=2.67in]{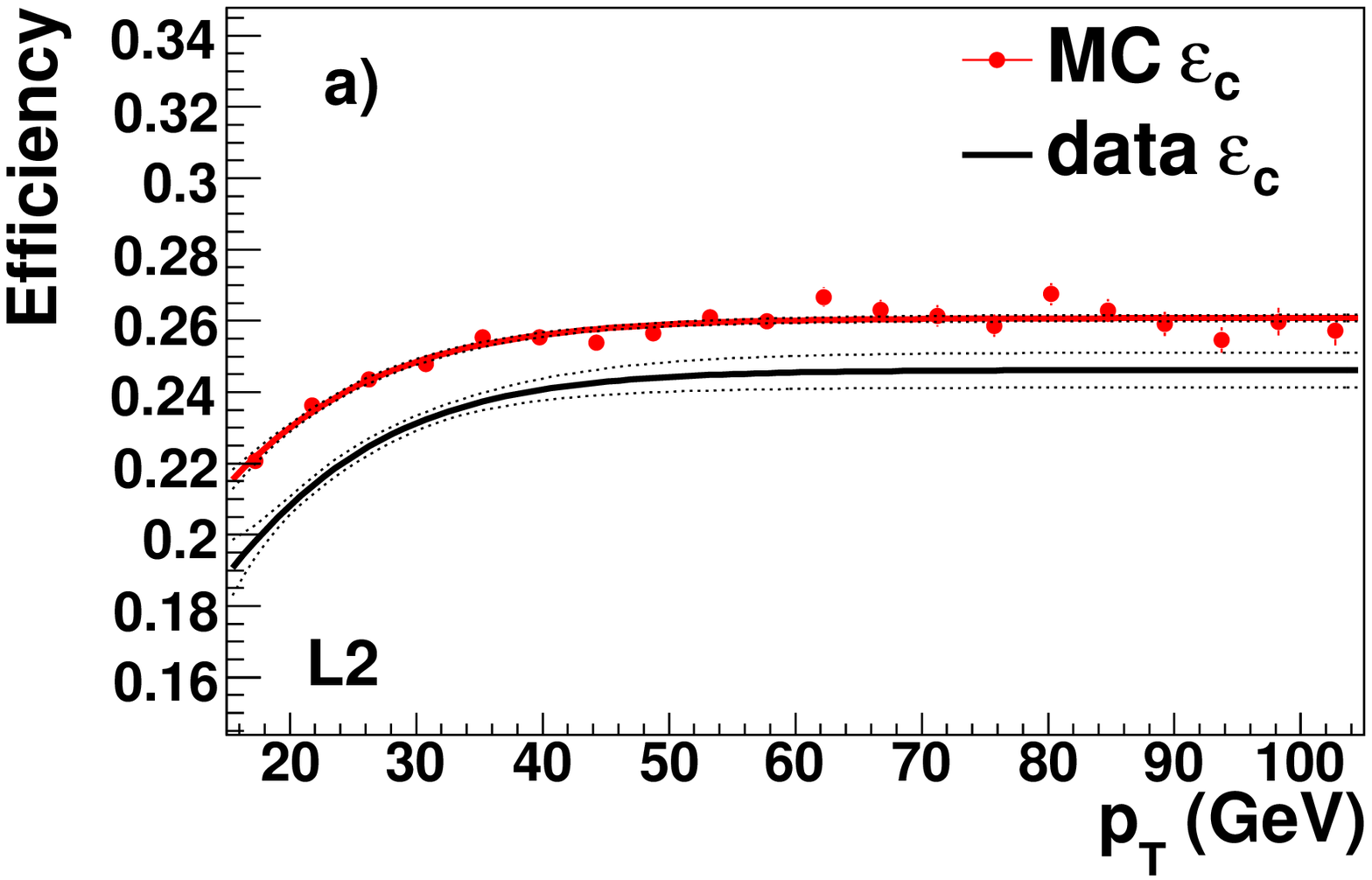}
    \includegraphics[width=2.67in]{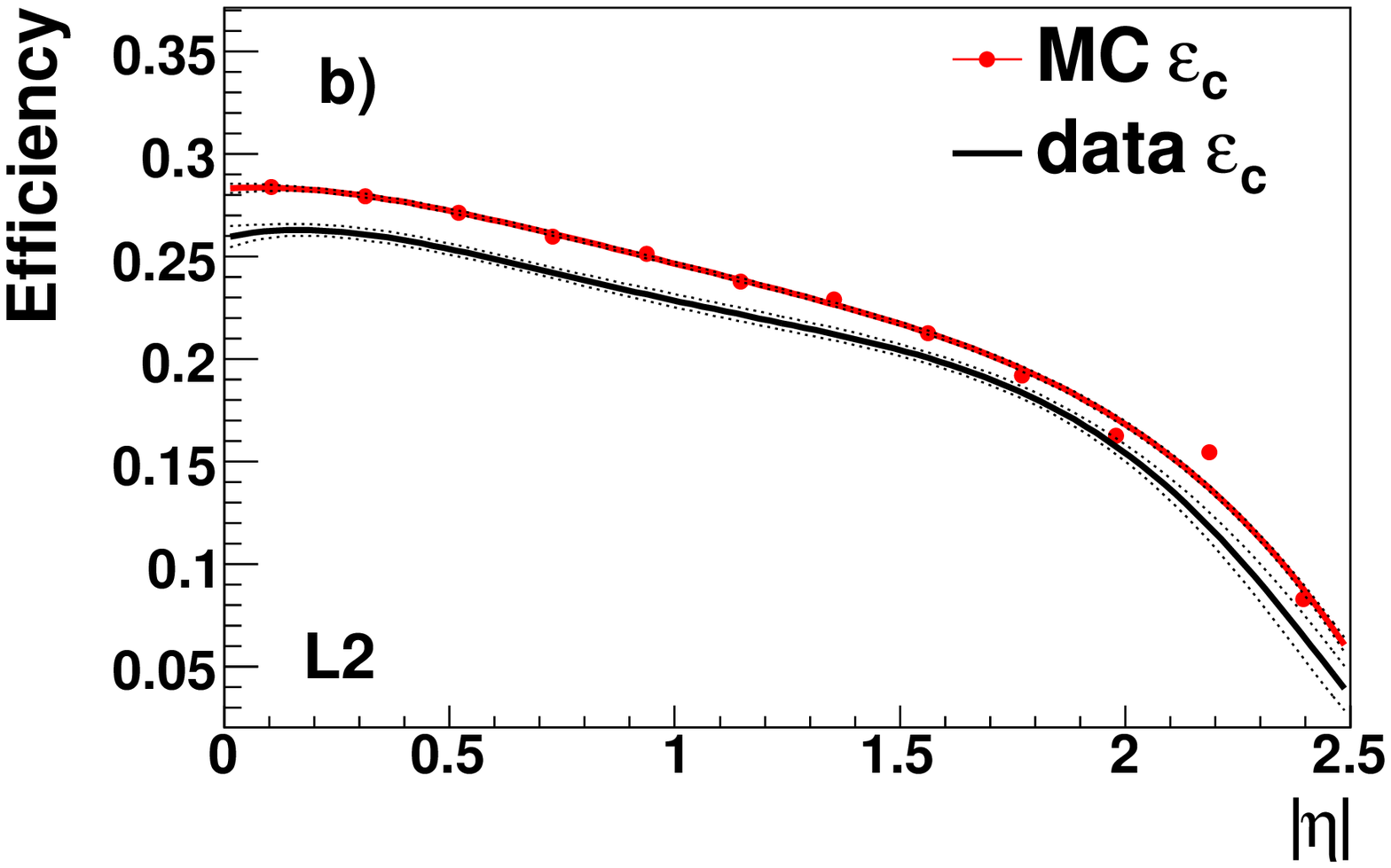}
    \includegraphics[width=2.67in]{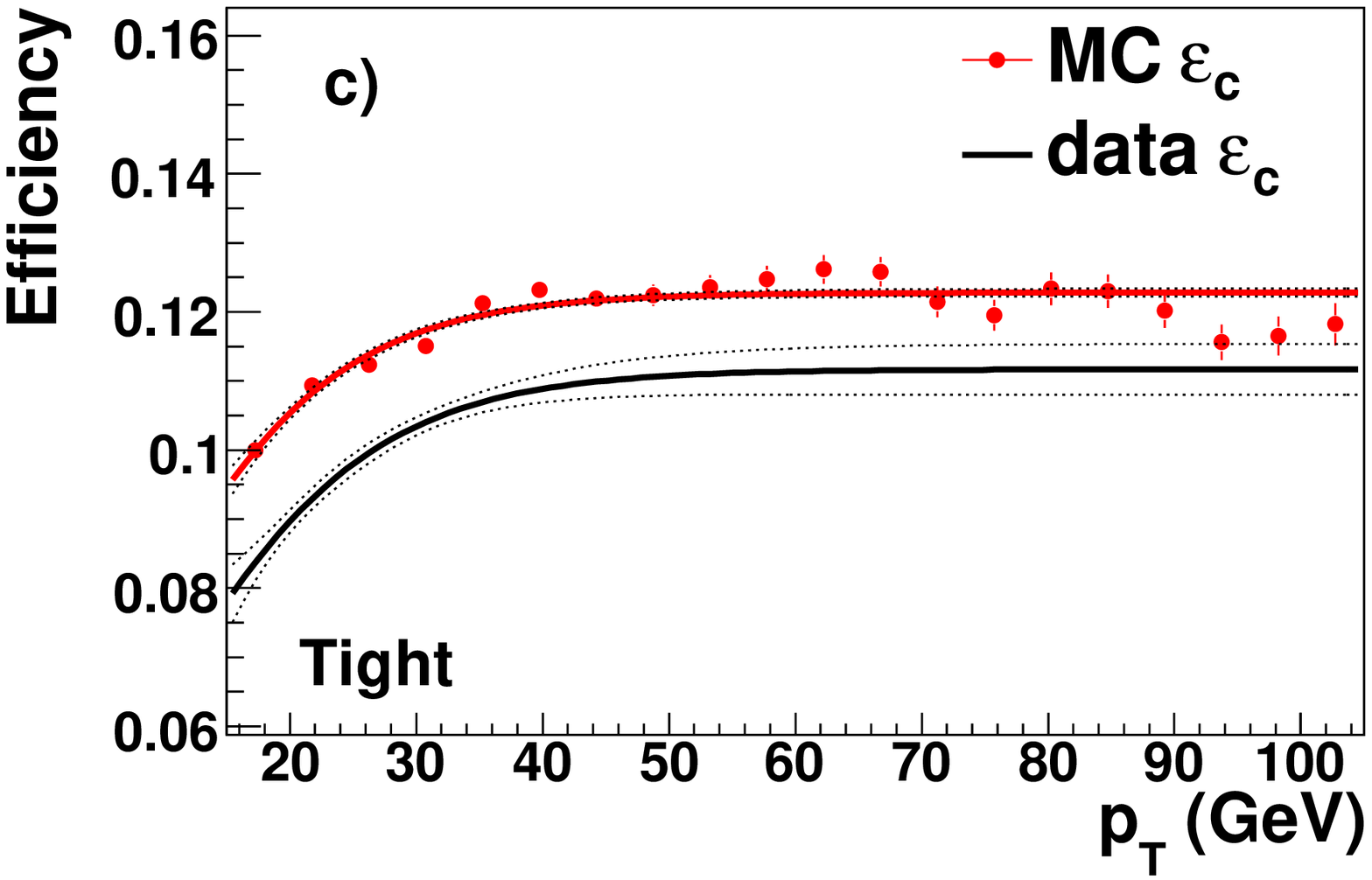}
    \includegraphics[width=2.67in]{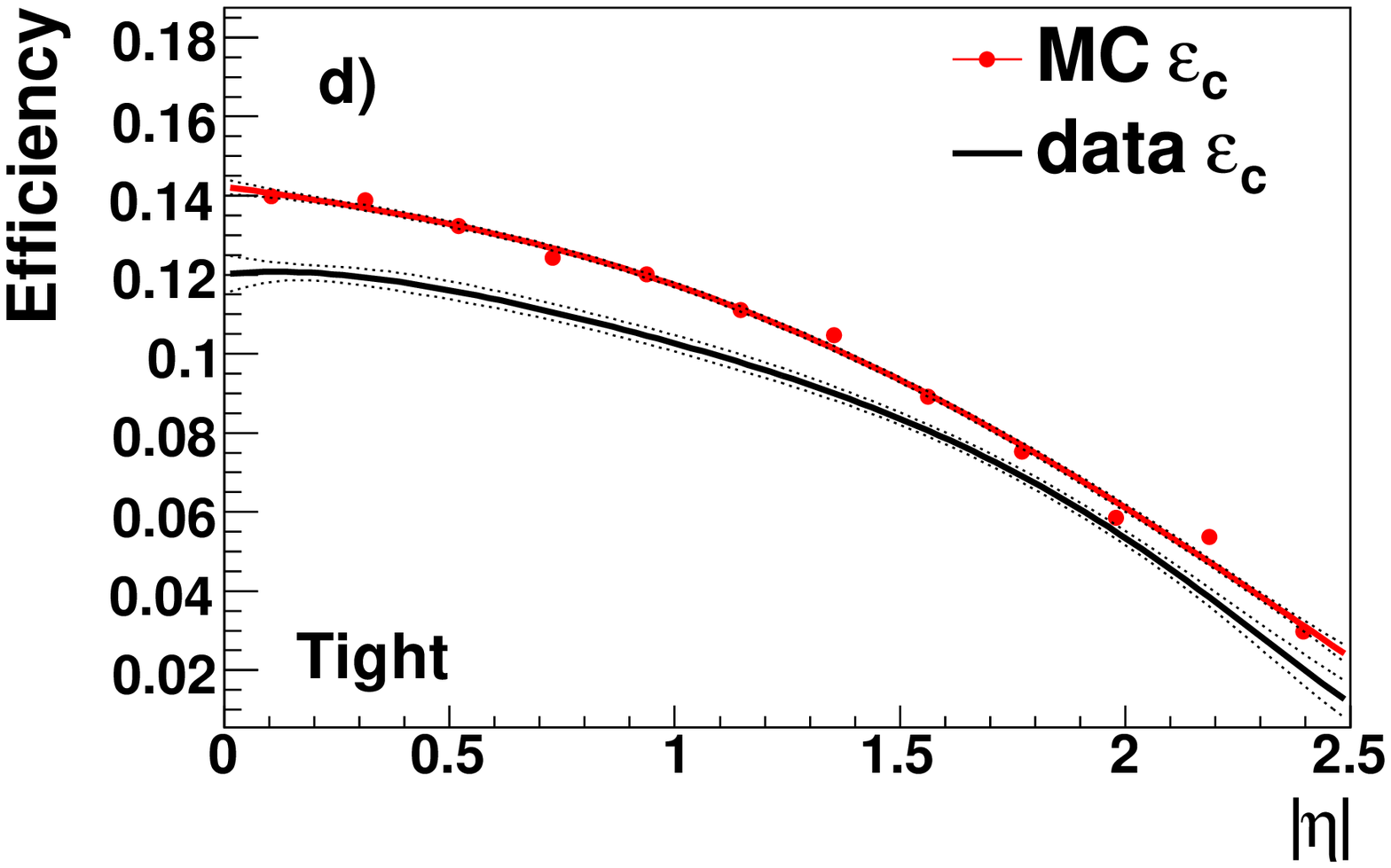}
  \end{center}
  \caption{(a): the L2 NN tagger inclusive \cquark-jet efficiency in both data
    (line) and MC (circles) as a function of jet \pt. (b): same as a
    function of jet $|\eta|$. (c) and (d): same for the Tight NN tagger, as a
    function of jet \pt\ and $|\eta|$, respectively.
    The dotted lines represent the fit uncertainty, which is almost
    entirely inherited from the uncertainty on the scale factor. The
    functions used for the parametrization are outlined in the text.}
  \label{fig:trf_c}
\end{figure}

The data \bquark-jet NN tagging efficiency calculated using the \textit{SystemD}
method is shown in Fig.~\ref{fig:s8eff}, along with the simulated semimuonic
\bquark-jet efficiency and $\mbox{SF}_{\bquark}$. The inclusive \bquark-jet
efficiencies, $\varepsilon_{\bquark}$, measured in data and in simulated events
are shown in Fig.~\ref{fig:trf_b}. The corresponding plots for $\cquark$ jets
are shown in Fig.~\ref{fig:trf_c}.

\subsection{Systematic uncertainties}
\label{system8:systematics}

Uncertainties on the resulting efficiencies arise from the following sources:
the \textit{SystemD} calculations (due to uncertainties on the correction
factors as well as limited data statistics), the dependence of the efficiencies on
the simulated samples, and possible imperfections in their chosen
parametrization including the assumption of factorization in
Eq.~\ref{eq:efficiency-factorization}. These uncertainties are discussed
below.

\subsubsection{\textit{SystemD} uncertainties}
\label{system8:s8syst}

The correction factors $\alpha$, $\beta$, $\kappa_{\bquark}$ and
$\kappa_{cl}$ (Sec.~\ref{sec:system8-application}) are evaluated using
simulated events, and have non-zero statistical uncertainties. The effect of these
is evaluated by repeating the \textit{SystemD} computations with the
parametrization of each individual factor shifted by its statistical
uncertainty, while all other correction factors are fixed to their nominal
values; the resulting changes in the computed efficiency are interpreted as
systematic uncertainties. 
The effect of the choice of minimum \ptrel\ requirement in the {\it SystemD}
calculations is evaluated by varying it between 0.3\GeVc\ and
0.7\GeVc.
The total relative systematic uncertainty associated with the {\it SystemD}
correction factors is estimated by adding the individual contributions in
quadrature, and varies between 1.3\% and 1.7\% for the different operating
points.
As an illustration, the results of this procedure when solving {\it SystemD} for
the entire sample are summarized in Table~\ref{tab:s8_systematics} for the NN L2
and Tight operating points.

For each bin in $\eta$ and \et, the {\it SystemD} systematic uncertainty for
that bin is added in quadrature with the statistical uncertainty resulting from
the {\it SystemD} fit.
This yields an overall uncertainty, referred to as ``statistical
uncertainty'' below, with which the efficiency is known for each
bin, and which is used in the fitting of the parametrized curves in \et\ and
$|\eta|$.

The relative combined statistical and systematic uncertainties as a function of
\et{} and $|\eta|$ are calculated by evaluating
\begin{equation}
  \Delta\varepsilon^{+}(\et,|\eta|) =
  \frac{\varepsilon^{+1 \sigma}(\et) \cdot \varepsilon^{+1\sigma}(|\eta|)}{%
    \varepsilon^{+1 \sigma}_{\mbox{\scriptsize all}}} -
  \frac{\varepsilon(\et) \cdot \varepsilon(|\eta|)}{\varepsilon_{\mbox{\scriptsize all}}},
\end{equation}
where the $+1\sigma$ quantities are the fluctuations upward by one standard
deviation of the quantities introduced in
Eq.~\ref{eq:efficiency-factorization}. This is also repeated with the downward
fluctuations, and the larger deviation is assigned as the uncertainty.

\begin{table}[htbp]
  \begin{center}
    \begin{tabular}{|l|c|c|}
      \hline
      & L2 & Tight \\
      \hline
      Efficiency & 65.9\% & 47.6\% \\
      \hline
      $\alpha$ & 0.0\% & 0.0\%\\
      $\beta$ & 0.2\% & 0.6\%\\
      $\kappa_b$ & 0.7\% & 1.2\%\\
      $\kappa_{cl}$ & 0.3\% & 0.2\%\\
      \ptrel & 1.0\% & 0.7\%\\
      \hline
      \emph{SystemD} Total & 1.3\% & 1.5\%\\
      \hline
    \end{tabular}
  \end{center}
  \caption{NN tagger efficiencies for the complete data sample, and relative
    systematic uncertainties originating from the \emph{SystemD} method. The
    total systematic uncertainty is determined by adding the individual
    uncertainties in quadrature.}
  \label{tab:s8_systematics}
\end{table}

\subsubsection{Efficiency parametrization and sample dependence uncertainty}

Both the parametrization and MC sample dependence systematic
uncertainties, which result from the use of efficiencies derived from generic combined
samples of simulated \bquark, \cquark, and muonic \bquark\ jets, are quantified
in one measurement. This is done by comparing the relative difference between
the actual and predicted numbers of tags in various bins in \et\ and $\eta$, and
for each of the simulated samples used to construct the efficiencies. This effectively
constitutes a closure test, and a total uncertainty is determined
from the spread of the relative differences.

In detail, the relative differences are calculated as a function of \et\ in three
$|\eta|$ ranges denoted CC ($|\eta|<1$), ICR ($1<|\eta|<1.8$), and EC
($|\eta|>1.8$).
The relative differences are histogrammed weighted by the number of actual
tags in the region. The RMS widths of the resulting distributions are used
to quantify the total uncertainty on each of the efficiencies.
The relative uncertainty determined by this method ranges from 1.2\% for the
loosest operating point to 3.5\% for the tightest operating point for the
inclusive \bquark-jet efficiency, and from 2.4\% to 4.0\% for the inclusive \cquark-jet efficiency.

\subsubsection{Total systematic uncertainty}

Total systematic uncertainties are assigned to all bins of \et\ and
$\eta$ for $\varepsilon_{\bquark}$, $\varepsilon_{\cquark}$, and SF$_{\bquark}$. They are
calculated as detailed below and are shown in Table~\ref{tab:trf_syst} for the
L2 and Tight NN operating points.

\begin{description}
 \item[\textnormal{SF}$_{\bquark}$:]  The closure test uncertainty for the $b\rightarrow\mu X$
efficiency.
\item[$\varepsilon_{\bquark}$:]  The SF systematic uncertainty added in quadrature with
  the closure test uncertainty for the \bquark-jet efficiency.
\item[$\varepsilon_{\cquark}$:]  The SF systematic uncertainty added in quadrature with
  the closure test uncertainty for the \cquark-jet efficiency.
\end{description}

The systematic uncertainties for $\varepsilon_{\bquark}$ range from $\pm1.9\%$ to
$\pm4.8\%$, for $\varepsilon_{\cquark}$ from $\pm2.8\%$ to $\pm5.2\%$, and for
$\mbox{SF}_{\bquark}$ from $\pm1.4\%$ to $\pm3.4\%$ for the loosest to tightest
working points.

\begin{table}[htbp]
  \begin{center}
    \begin{tabular}{|l|c|c|}
      \hline
      Uncertainty & L2 & Tight \\
      \hline
      MC $b\rightarrow \mu X$ & 2.4\% & 3.5\% \\
      MC $b$                                  & 1.8\% & 2.8\% \\
      MC $c$                                  & 2.9\% & 3.9\% \\
      \hline
      SF$_{\bquark}$                   & 2.4\% & 3.5\% \\
      $\varepsilon_{\bquark}$ & 3.0\% & 4.5\% \\
      $\varepsilon_{\cquark}$ & 3.8\% & 5.2\% \\
      \hline
    \end{tabular}
  \end{center}
  \caption{Total relative systematic uncertainties on the MC sample
    parametrizations, and their effect on $\varepsilon_{\bquark}$,
    $\varepsilon_{\cquark}$, and $\mbox{SF}_{\bquark}$.}
  \label{tab:trf_syst}
\end{table}

\begin{figure*}[htbp]
  \centering
  \includegraphics[width=0.45\textwidth]{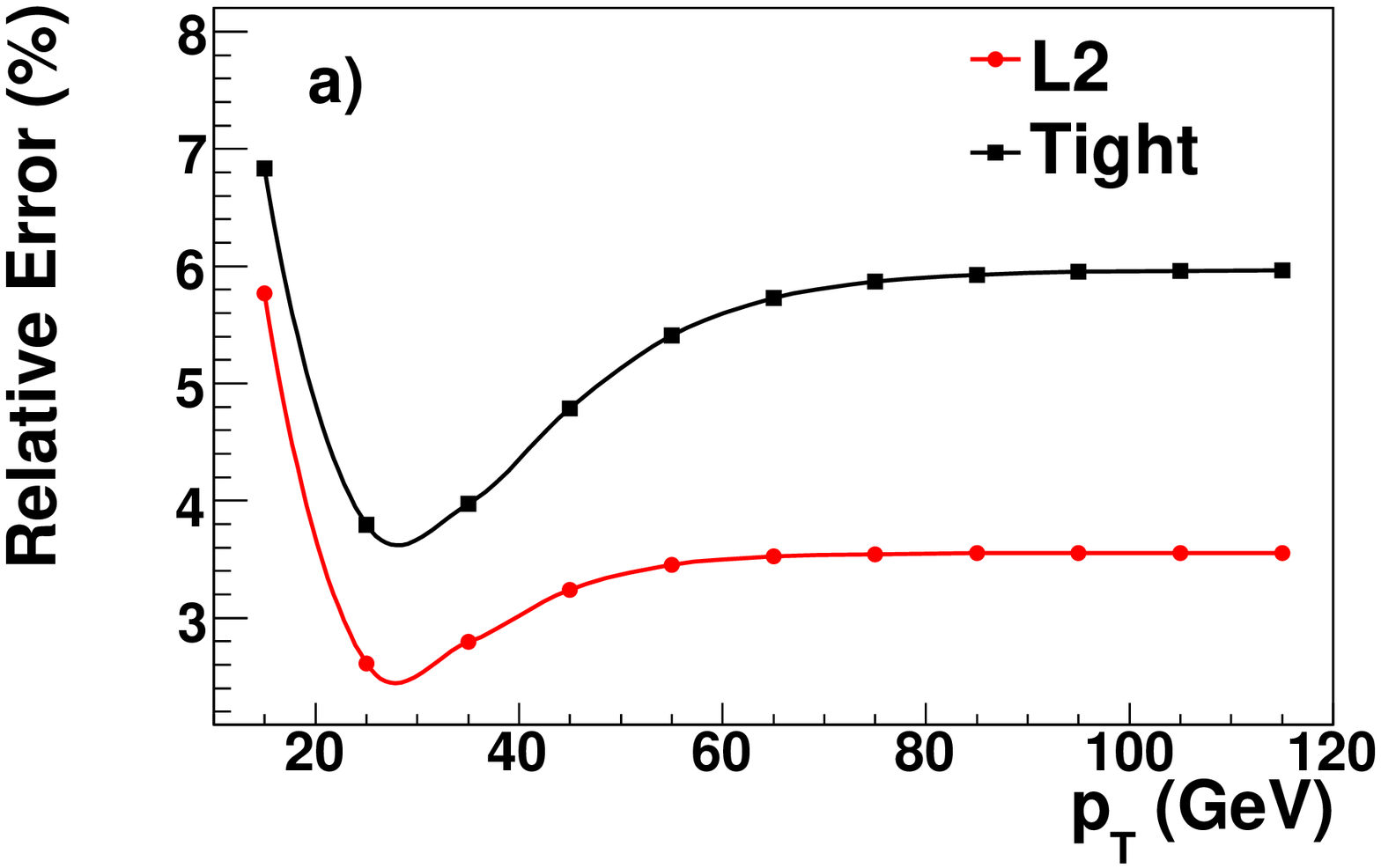}
  \includegraphics[width=0.45\textwidth]{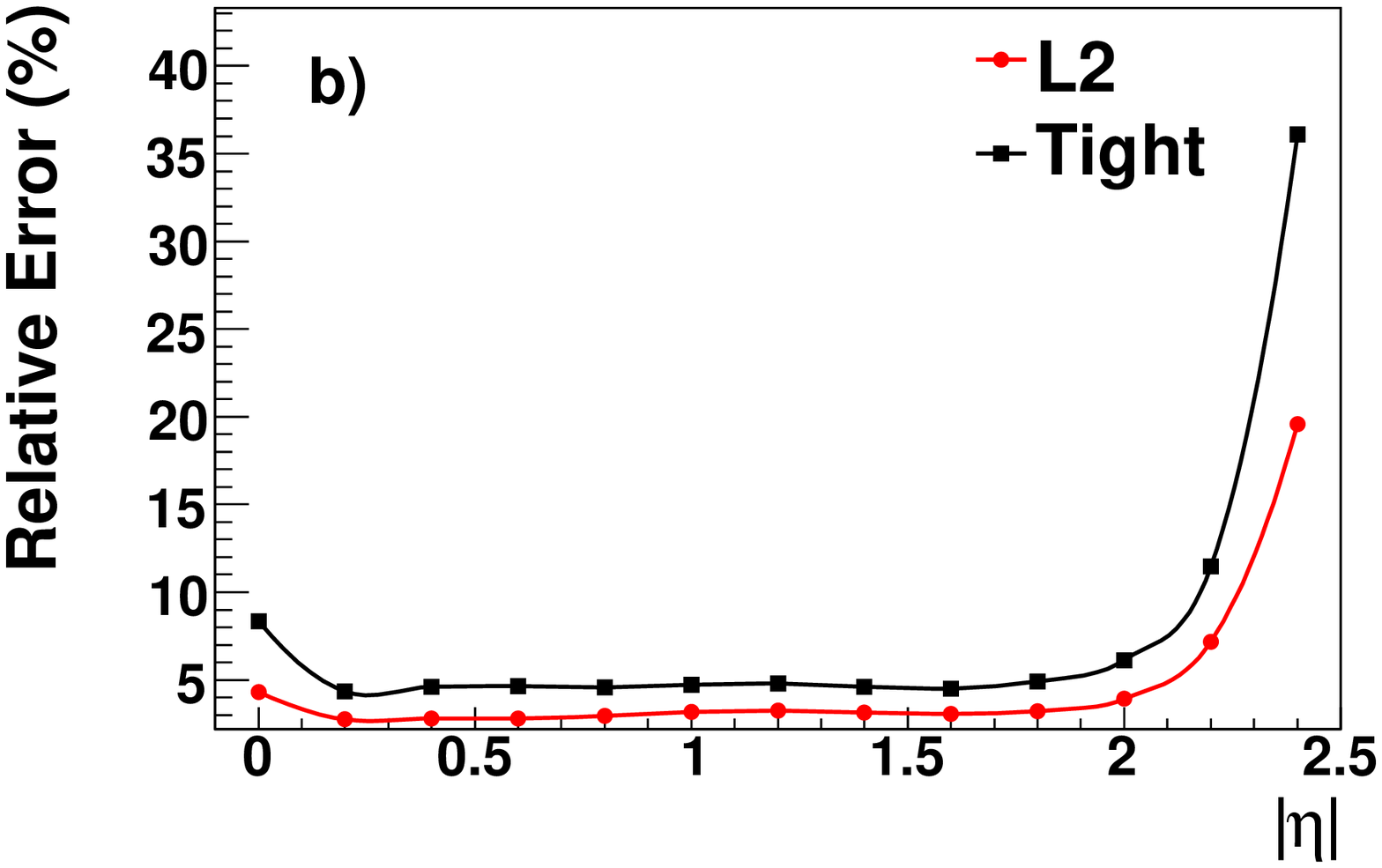}
  \includegraphics[width=0.45\textwidth]{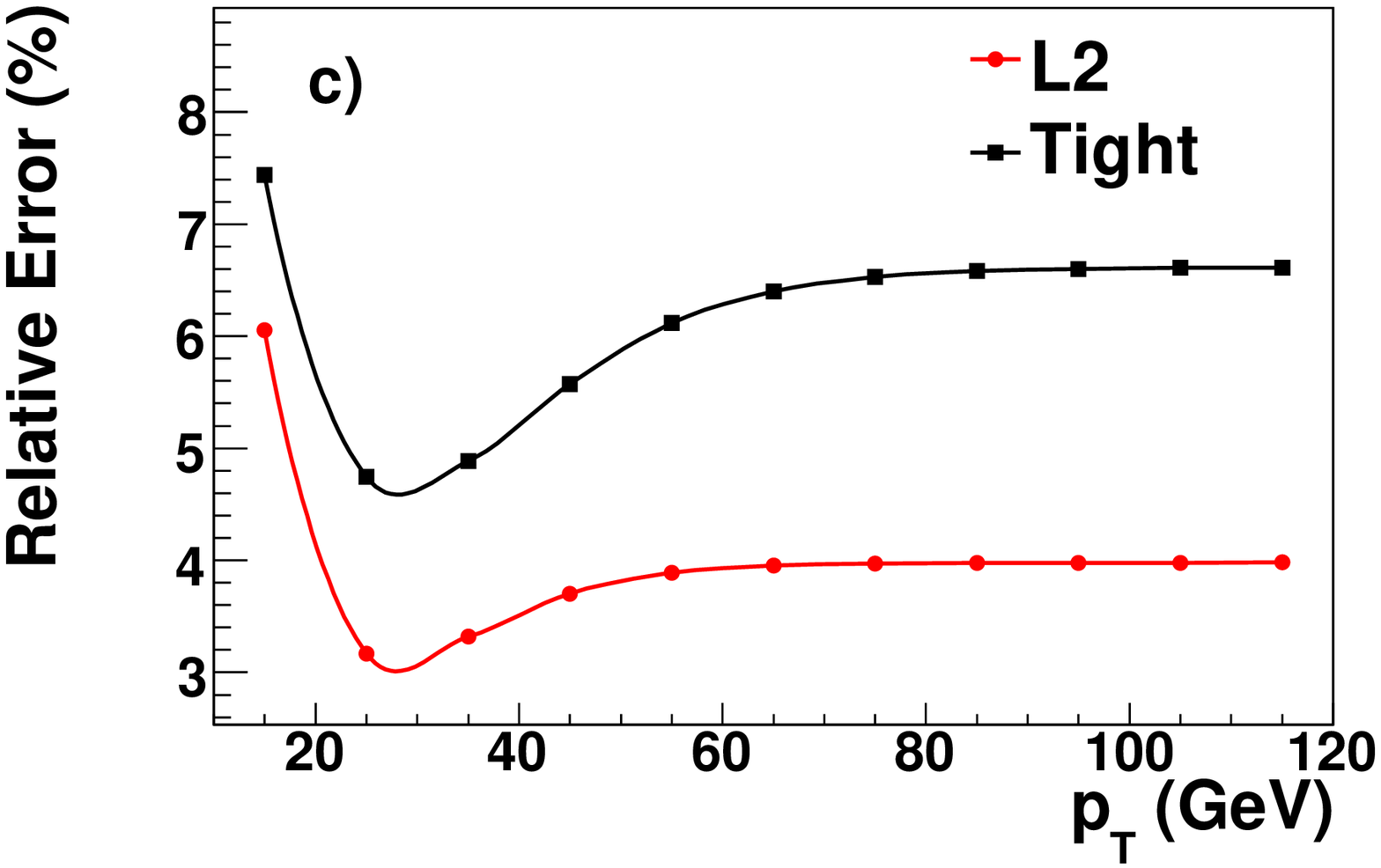}
  \includegraphics[width=0.45\textwidth]{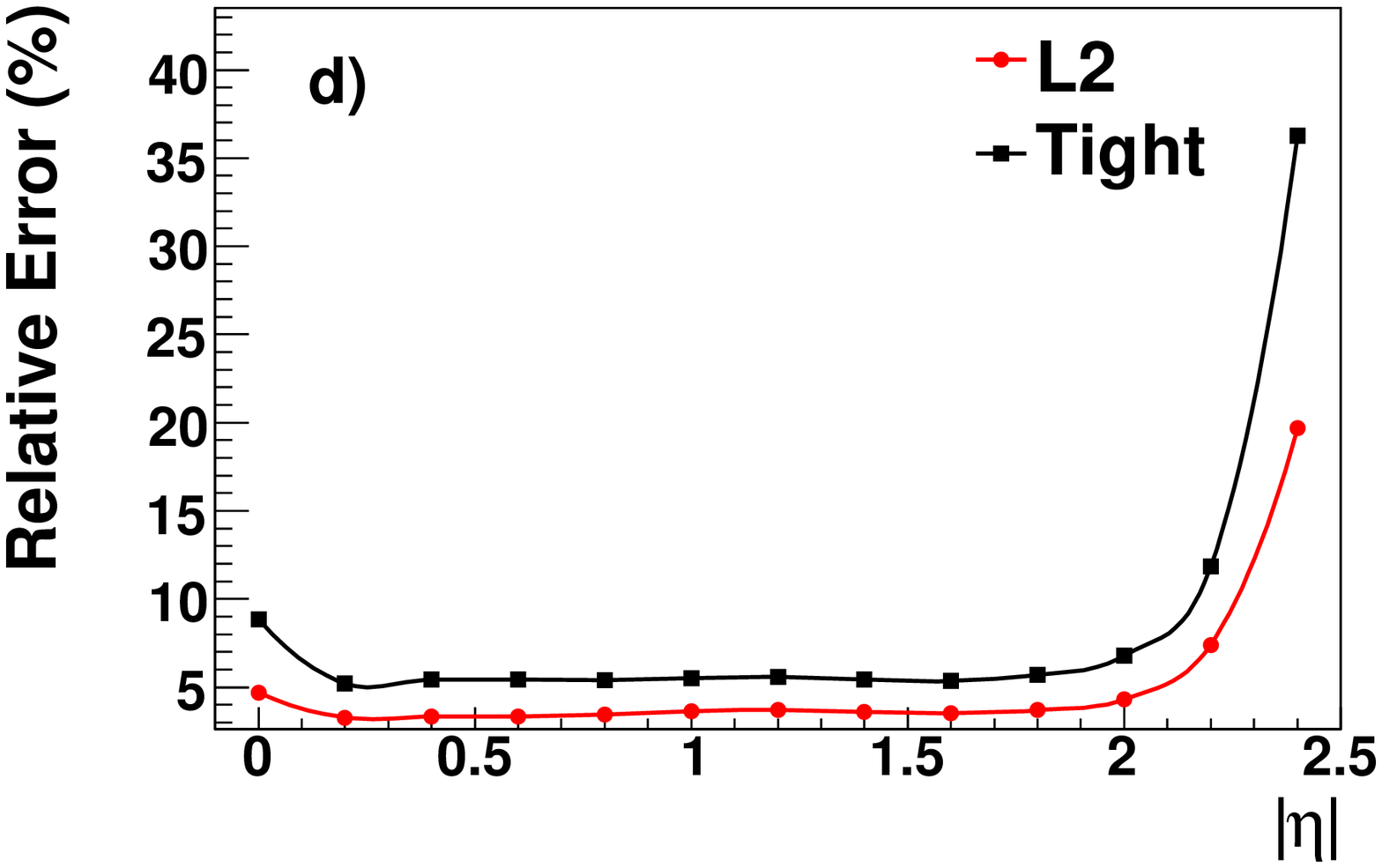}
  \includegraphics[width=0.45\textwidth]{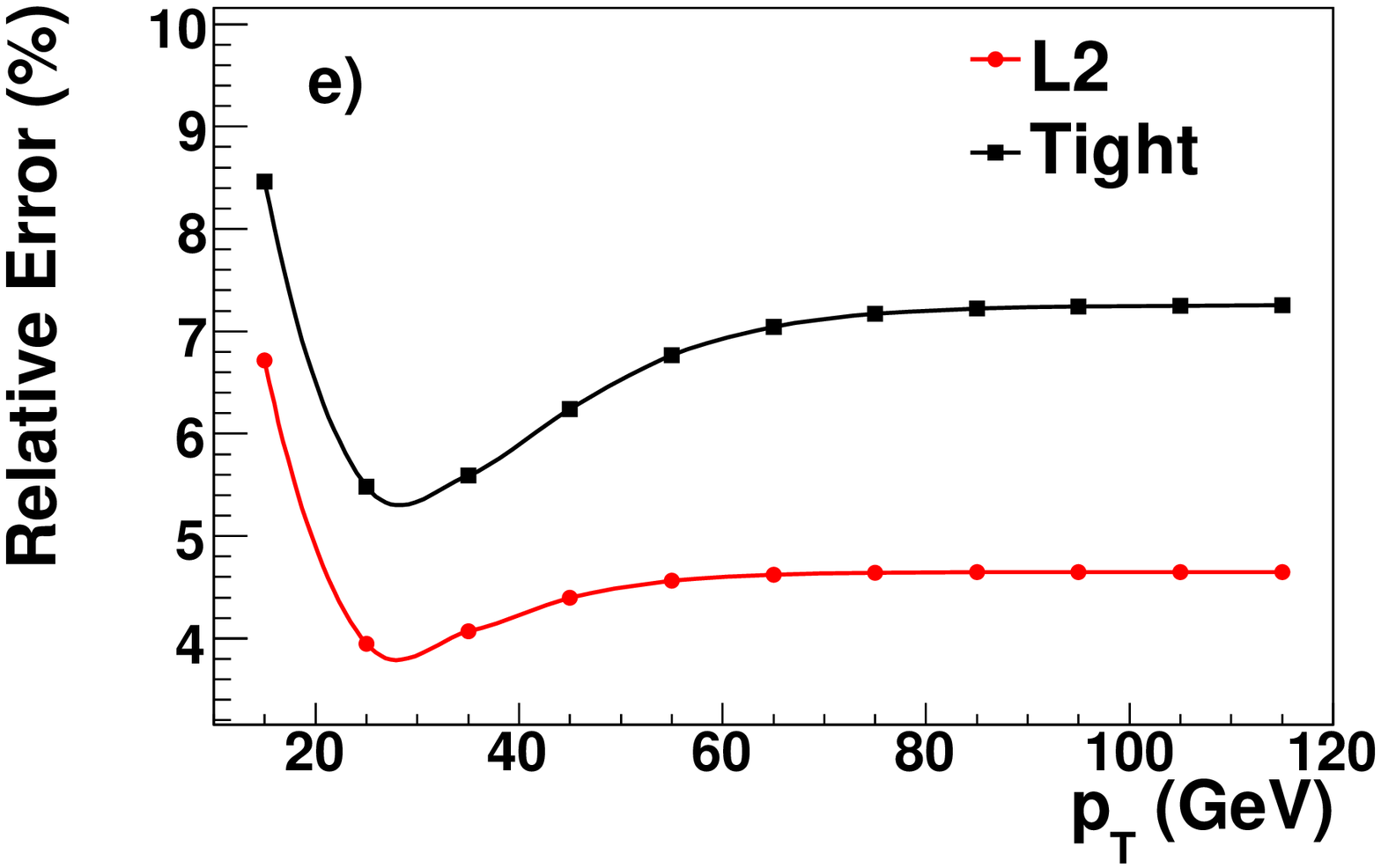}
  \includegraphics[width=0.45\textwidth]{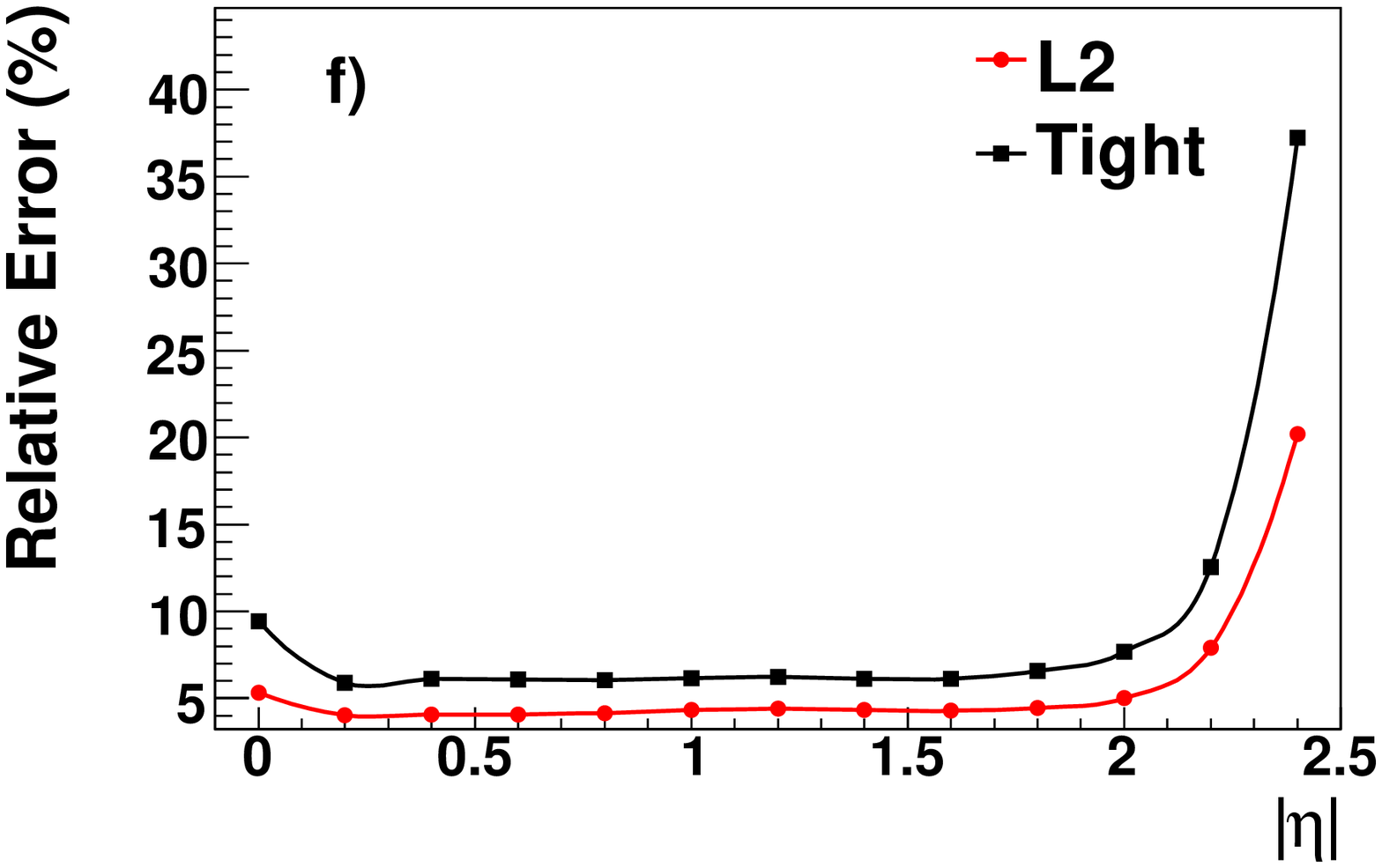}
  \caption{The total relative uncertainty (combined systematic and
    statistical) for the Scale Factor ($\mbox{SF}_{\bquark}$) (a, b),
    $\varepsilon_{\bquark}$ (c, d) and $\varepsilon_{\cquark}$ (e, f) as a function
    of \et\ (a, c, e) when $\eta = 1.2$, and $\eta$ (b, d, f) when
    $\et = 45\GeV$.}
  \label{fig:s8_total_errors}
\end{figure*}

The total uncertainties are computed by adding in quadrature the systematic
and the statistical uncertainties (which include the \emph{SystemD}
systematic uncertainties, as detailed in Sec.~\ref{system8:s8syst}).
They are shown for $\mbox{SF}_{\bquark}$, $\varepsilon_{\bquark}$, and
$\varepsilon_{\cquark}$ in Fig.~\ref{fig:s8_total_errors} for the L2 and Tight
operating points.
The relative uncertainty increases rapidly at high $\eta$ due to limited statistics
in that region and because the value of the scale factor drops rapidly for
$|\eta|>2$.

%% file: fakes.tex
\section{Fake Rate Determination}
\label{sec:fake_rate}

The determination of the light-flavor mistag rate (where ``light'' stands for
$uds$-quark or gluon jets) or \emph{fake rate} relies on the notion
that in the absence of long-lived particles such as $V^{0}$s (see
Sec.~\ref{sec:v0}), reconstructed high-impact parameter tracks or displaced
vertices reconstructed in light-flavor jets result from resolution and
misreconstruction effects. These effects are expected
to lead to tracks with negative impact parameters (see Sec.~\ref{sec:jlip} for
the impact parameter sign convention used) and displaced vertices with
negative decay lengths as often as to positive impact parameters and decay
lengths.
Barring incorrectly assigned negative impact parameter signs (which may occur
whenever a jet and a track are nearly aligned in azimuth, and which is important
for long-lived particles), using such negative impact parameter tracks and
negative decay length vertices should provide a reasonable estimate of the fake
rate.

\subsection{Data sample}
\label{sec:fake-sample}

\begin{figure*}[bth]
  \begin{center}
    \includegraphics[width=0.7\textwidth]{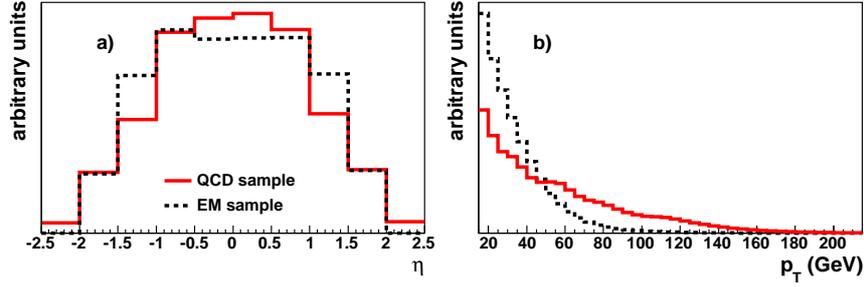}
  \end{center}
  \vspace{-0.5cm}
  \caption[]{Jet $\eta$ (a) and \et\ (b) distributions in the QCD and EM data
    samples.}
  \label{fig:datamc_multijet}
\end{figure*}

To minimize the impact of incorrectly attributed impact parameter signs, the
fake rate is determined in an inclusive jet sample with 
low heavy flavor content. Two samples are used for this purpose:
\begin{itemize}
\item A sample consisting of events selected by requiring at
 least one electron candidate with $\pt > 4 \GeVc$ and with low missing
  transverse energy, $\met < 10 \GeV$, and referred to below as the EM sample.
  As in Sec.~\ref{sec:system8-application},
  at least one trigger without lifetime bias is required. Most of
  the electron candidates are
  jets that deposit a large fraction of their energy in the EM section of the
  calorimeter. This may reduce the heavy flavor content of the sample, as
  the fraction of a jet's energy deposited through electromagnetic processes
  depends on the jet flavor. This bias is removed by only considering jets
  whose distance to the nearest identified EM cluster is $\Delta \mathcal{R} >
  0.4$. After these requirements, the EM sample contains 106 million taggable
  jets.
\item An inclusive jet sample (referred to in the following as the QCD sample)
  consisting of all events collected using jet triggers. It contains 154 million
  taggable jets. 
  For consistency, jets in the vicinity of identified EM
  clusters are not considered in this sample either. Since trigger requirements
  should not bias such jets in this sample, the effect of this removal should be
  small and will be evaluated below.
\end{itemize}
These two samples are combined (referred to as the COMB sample) for most studies;
their comparison allows a systematic uncertainty associated with the choice of a
particular sample to be estimated. The \pt\ and $\eta$ distributions in these
samples, after the taggability requirement, are compared in
Figure~\ref{fig:datamc_multijet}.
The main difference is an increased high-\et\ content in the QCD sample due to
trigger effects.

\subsection{Negative tag rate}
\label{sec:ntr}

The use of negative impact parameter tracks and negative decay length displaced
vertices is rather straightforward: the algorithms providing the NN input
variables listed in Sec.~\ref{sec:nn-variables} need only minor modifications
in order to provide ``negative'' equivalents of these variables, called
\emph{Negative Tag} (NT) results.
The NN output is then recomputed using the NT values rather than the
original ones, and the fraction of jets tagged in this modified fashion
represents the \emph{negative tag rate}. The NN input NT results are computed as
follows:
\begin{description}
\item[\textbf{CSIP}:] The $\mathcal{N}_{\mbox{\scriptsize CSIP}}$ variable is
  recalculated using tracks with negative instead of positive impact parameter
  significance to obtain the ``strong classifier'' numbers of tracks $N_{3s}$
  and $N_{2s}$ (see Sec.~\ref{sec:csip}).
\item[\textbf{JLIP}:] The Jet Lifetime Probability
  $\mathcal{P}_{\mbox{\scriptsize JLIP}}$ is recomputed using only tracks with
  negative rather than positive impact parameter significance.
\item[\textbf{SVT}:] In this case, no additional computation is necessary.
  Instead of the highest (positive) decay length significance, the most
  negative decay length significance displaced vertices (for both the
  SuperLoose and Loose algorithm versions discussed in
  Sec.~\ref{sec:svt}) are used to supply the SVT-related NN variables.
\end{description}

\begin{figure}[htbp]
  \centering
  \includegraphics[width=2.67in]{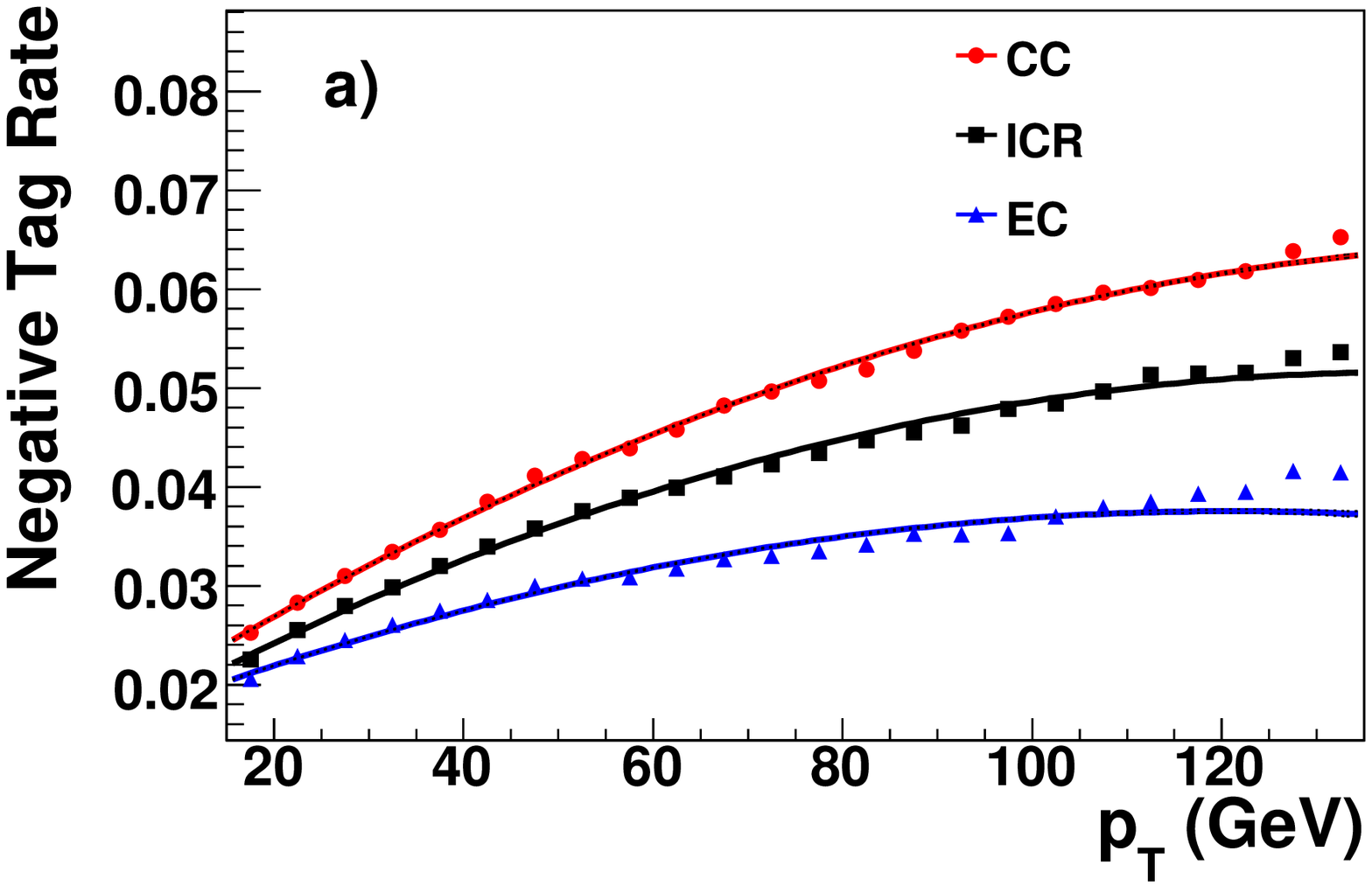}
  \includegraphics[width=2.67in]{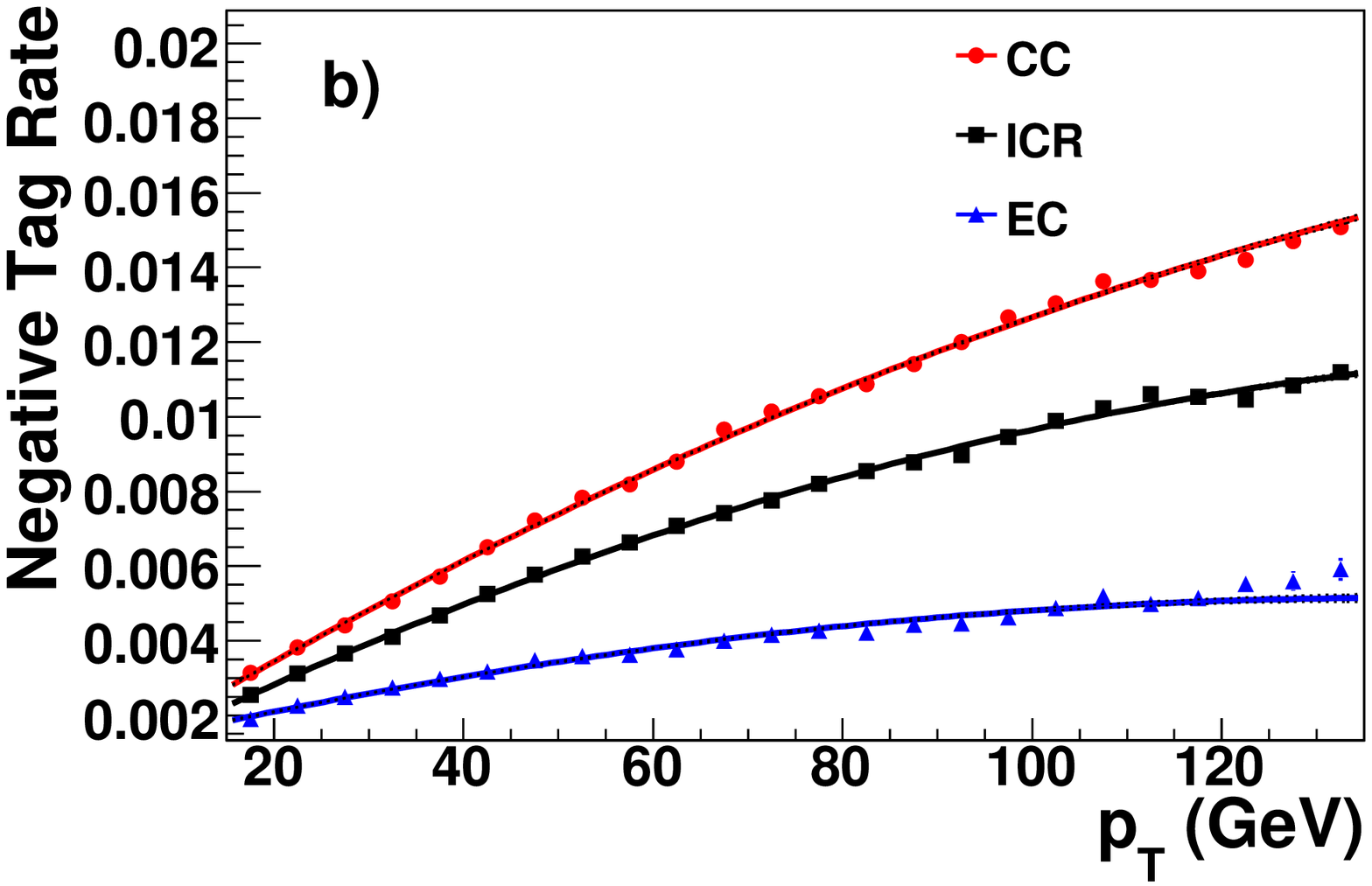}
  \caption{The NT rate parametrization for the COMB sample in the CC (circles),
    ICR (squares), and EC (triangles) for the L2 (a) and Tight (b)
    operating points. The negative tag rate is parametrized with a second order
    polynomial. The fit uncertainty is too small to be visible in this figure.}
  \label{fig:nt_rates}
\end{figure}

Like the \bquark-jet efficiency, the fake rate and negative tag rate are
parametrized as functions of a jet's kinematical (\et, $|\eta|$)
variables. However, in contrast to the efficiency, the data shows that the
dependence of the NT rate on jet \et\ and $|\eta|$ cannot be factorized into a
dependence on 
\et\ multiplied by a dependence on $|\eta|$. Instead, it is parametrized as a
function of jet \et\ in three regions: $0<|\eta|<1.0$ (CC), $1.0<|\eta|<1.8$
(ICR), and $1.8<|\eta|<2.5$ (EC). In each region, the \et\ dependence is
parametrized using a quadratic polynomial. The negative tag rate
parametrizations in the three $|\eta|$ regions are shown in
Fig.~\ref{fig:nt_rates} for the L2 and Tight operating points.

\subsection{Corrections}

The NT rate is not a perfect approximation of the fake rate.
Corrections for the following effects are applied:
\begin{itemize}
\item The presence of heavy flavor jets increases the NT rate, primarily due to
  tracks that originate from the decay of long-lived particles and are
  (mistakenly) assigned a negative impact parameter sign. As no method is
  available to estimate this effect on data, QCD events simulated using the
  \textsc{Pythia} event generator are used instead. This results in a correction factor
  $F_{\mbox{\scriptsize hf}} = \varepsilon^{\mbox{\scriptsize NT}}_{\mbox{\scriptsize QCD,light}} /
  \varepsilon^{\mbox{\scriptsize NT}}_{\mbox{\scriptsize QCD,all}}$, \emph{i.e.}, the ratio of
  NT rates with and without the presence of heavy flavor jets in
  these simulated events.
\item The $V^{0}$ removal algorithm (see Sec.~\ref{sec:v0}) is not fully
  efficient, so that some contribution from long-lived particles and photon
  conversions remains. Most of
  the resulting tracks will correctly be assigned 
  positive impact parameters, and the NT rate is affected less by their presence
  than the fake rate. This effect is estimated using the same simulated QCD
  events as for the $F_{\mbox{\scriptsize hf}}$ factor,  leading to a correction factor
  $F_{\mbox{\scriptsize lf}} = \varepsilon^{\mbox{\scriptsize PT}}_{\mbox{\scriptsize QCD,light}} /
  \varepsilon^{\mbox{\scriptsize NT}}_{\mbox{\scriptsize QCD,light}}$, \emph{i.e.}, the ratio of the
  positive- and negative-tag rates in the simulated light-flavor events.
\end{itemize}
Finally, the fake rate $\varepsilon_{\mbox{\scriptsize light}}$ is estimated as
the NT rate measured in data, $\varepsilon^{\mbox{\scriptsize
    NT}}_{\mbox{\scriptsize data}}$, corrected for the above effects: 
\begin{equation}
  \varepsilon_{\mbox{\scriptsize light}} =
  \varepsilon^{\mbox{\scriptsize NT}}_{\mbox{\scriptsize data}} \cdot
  F_{\mbox{\scriptsize hf}} \cdot F_{\mbox{\scriptsize lf}}.
\end{equation}

\begin{figure*}[htbp]
  \centering
  \includegraphics[width=0.45\textwidth]{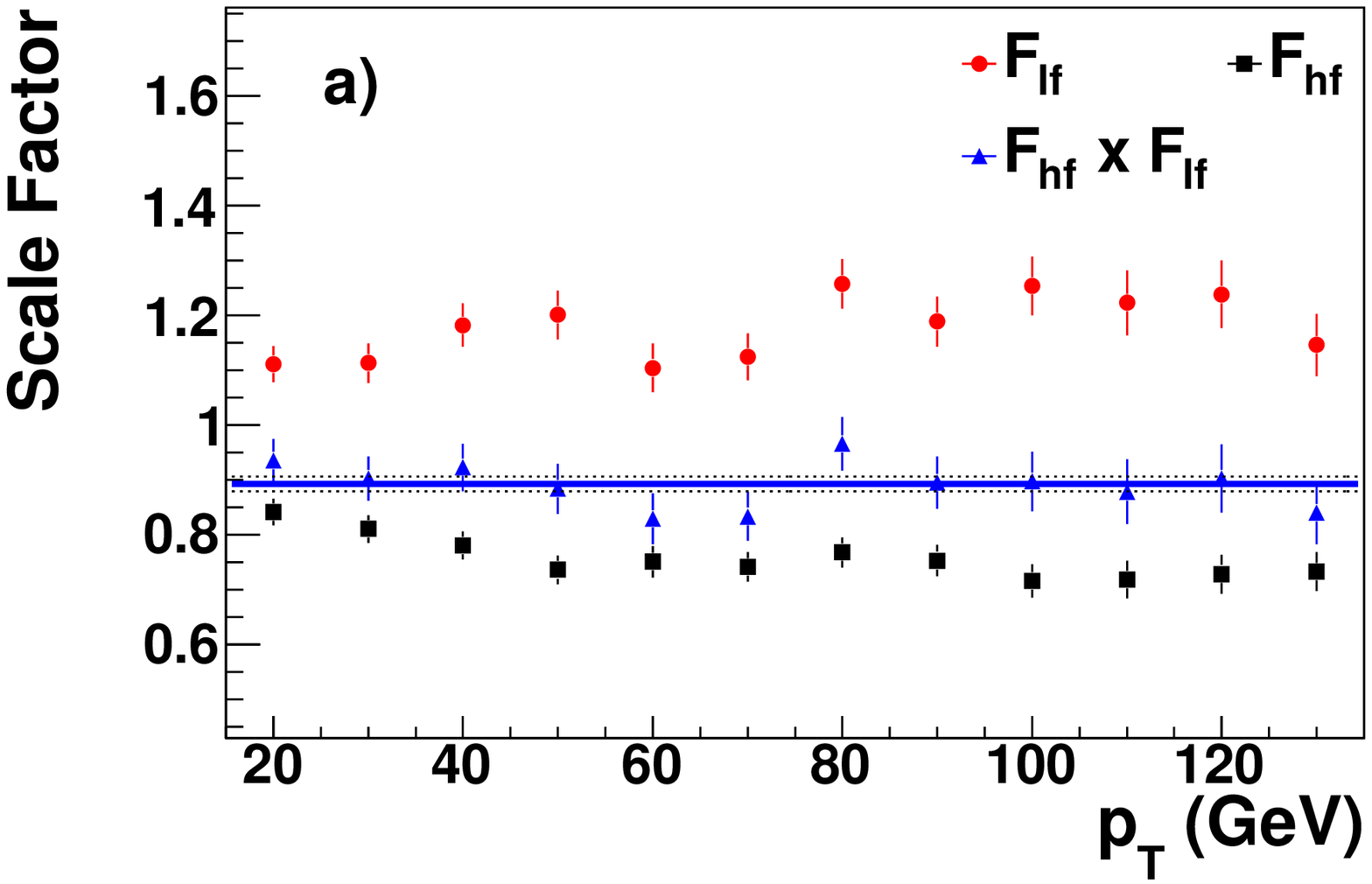}
  \includegraphics[width=0.45\textwidth]{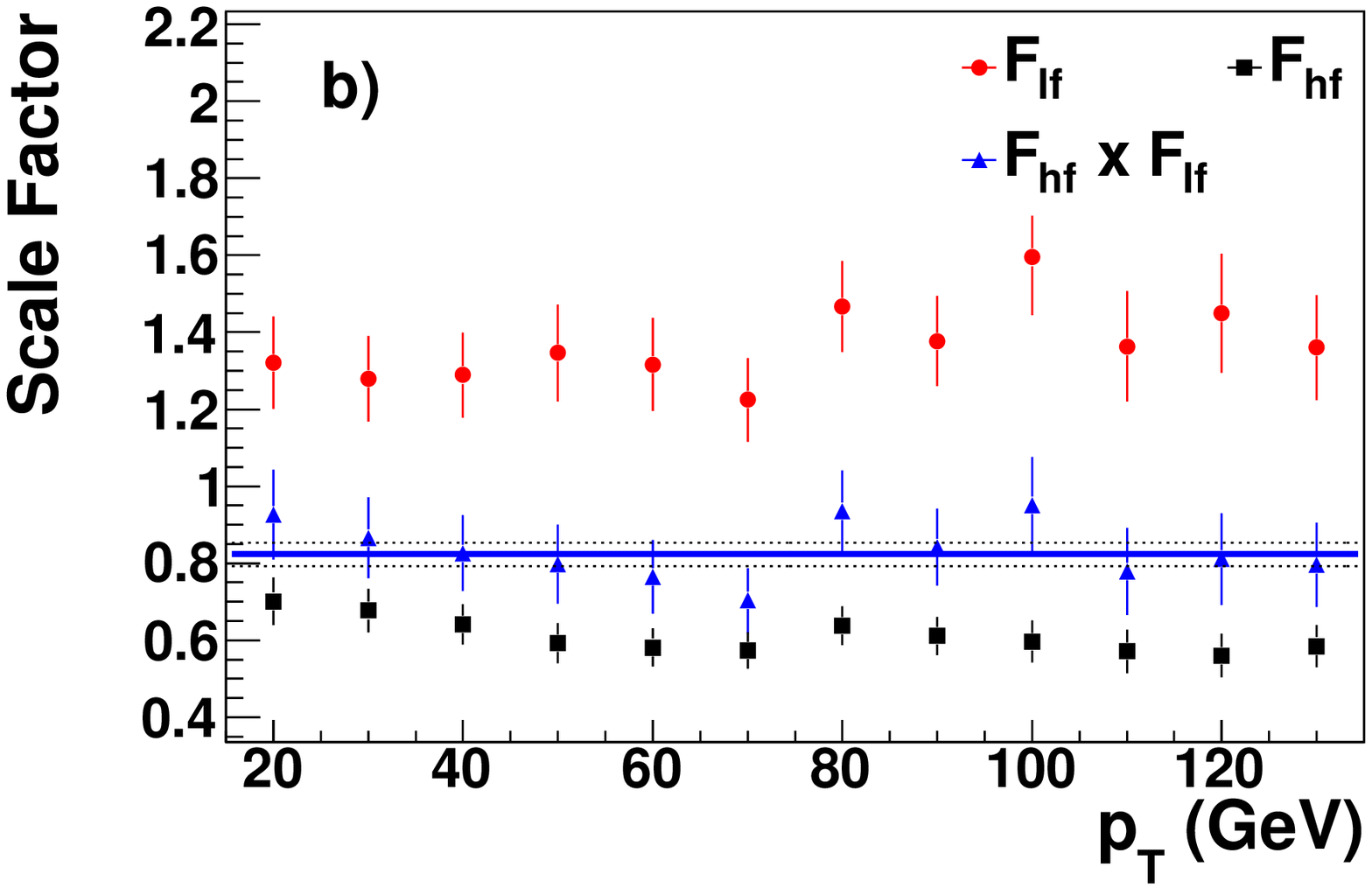}
  \includegraphics[width=0.45\textwidth]{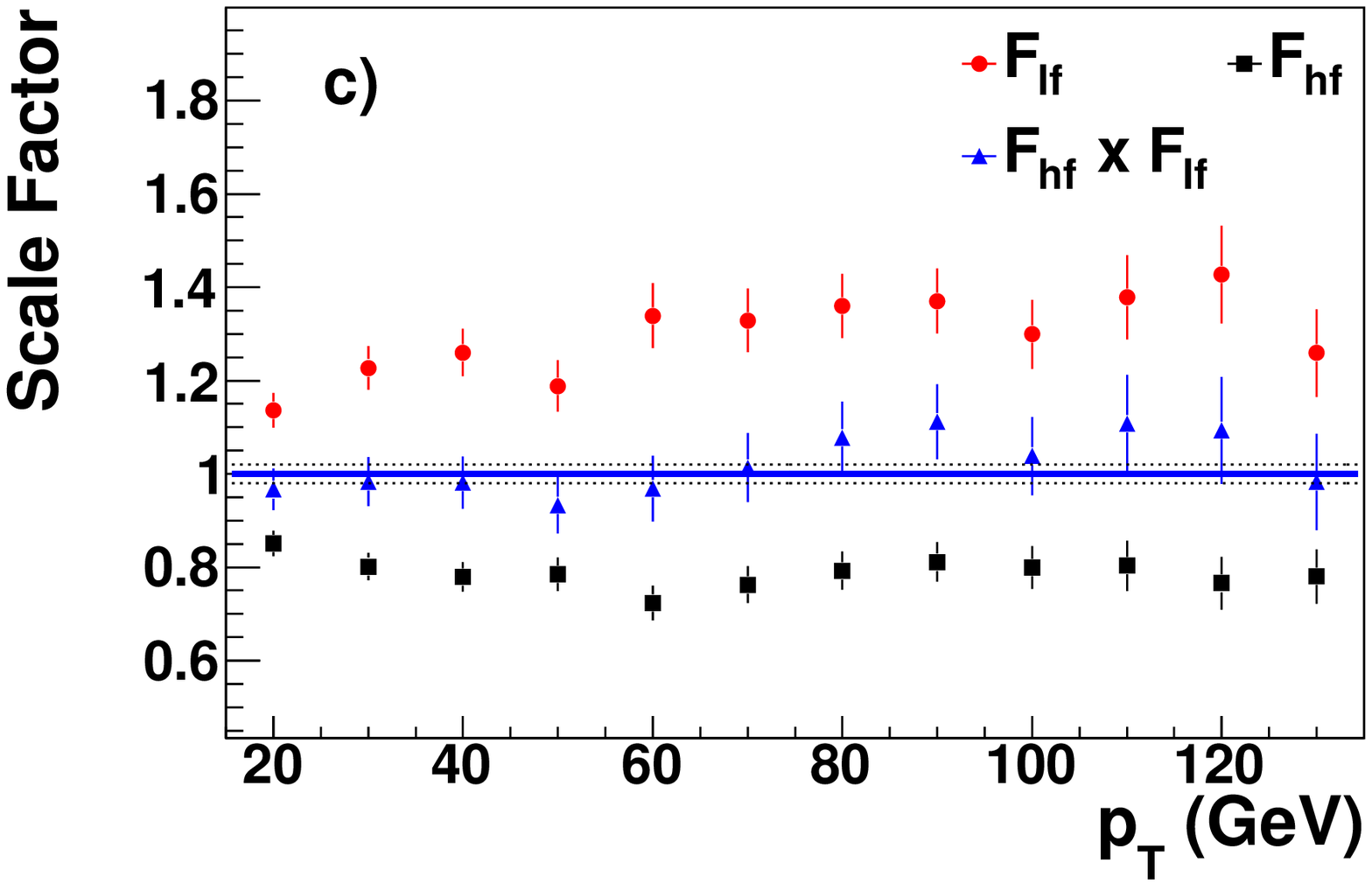}
  \includegraphics[width=0.45\textwidth]{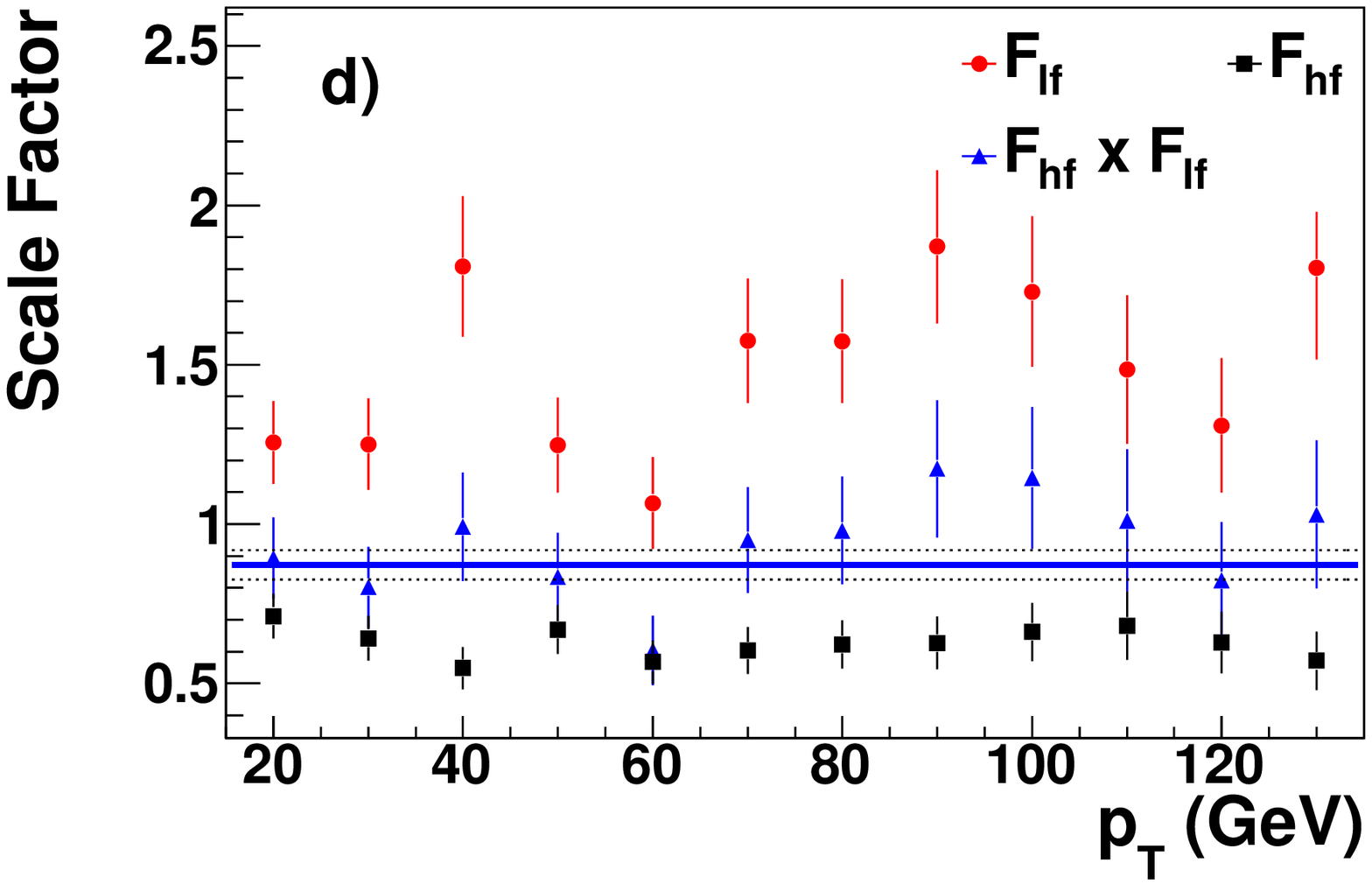}
  \includegraphics[width=0.45\textwidth]{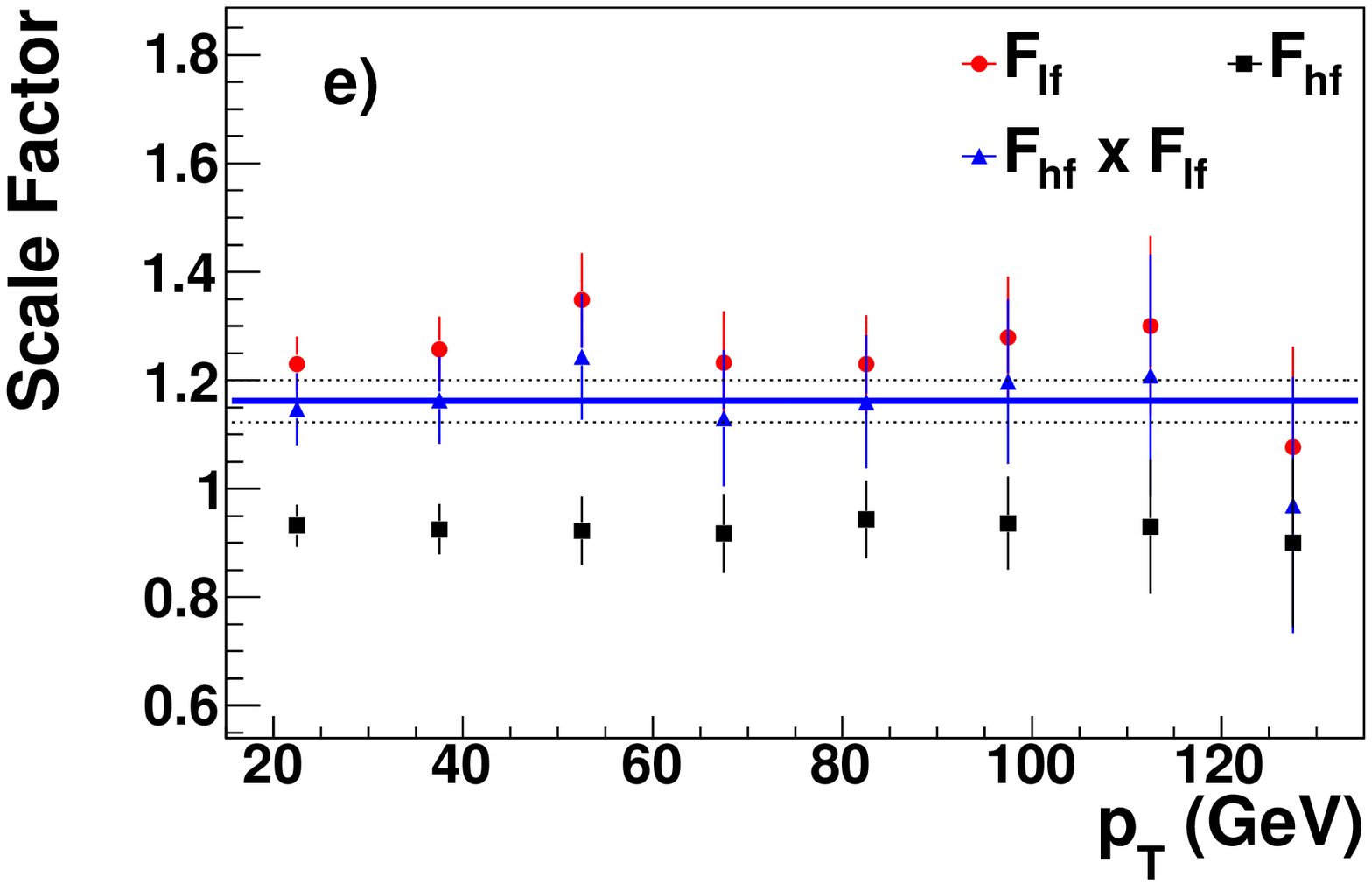}
  \includegraphics[width=0.45\textwidth]{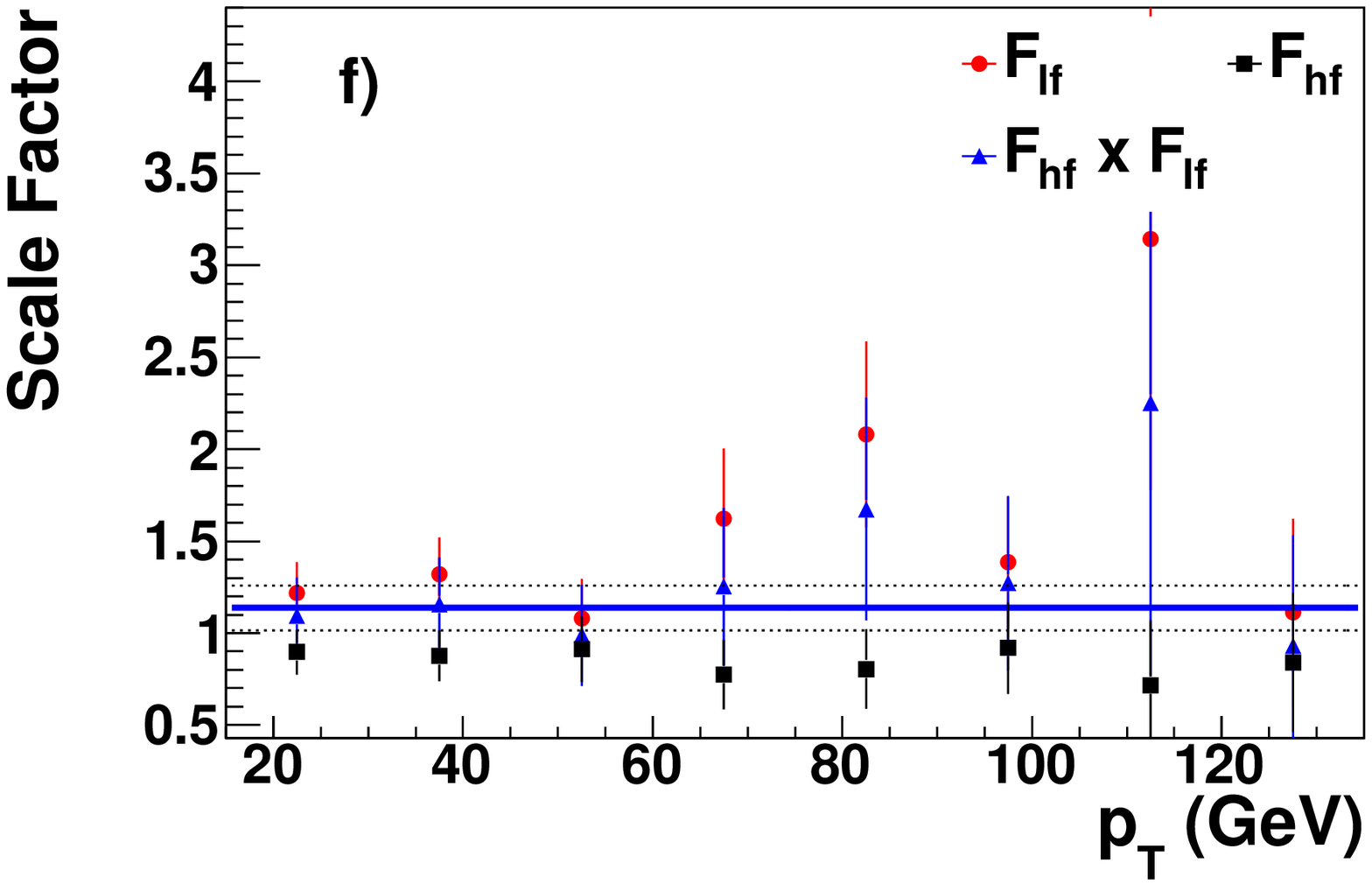}
  \caption{The light jet asymmetry correction, $F_{\mbox{\scriptsize lf}}$ (circles),
    heavy flavor correction, $F_{\mbox{\scriptsize hf}}$ (squares), and total
    negative tag correction (triangles) in the CC (a, b), ICR (c, d) and EC
    regions (e, f) for the L2 (a, c, e) and Tight (b, d, f) operating
    points. The solid and dotted lines indicate the fit of the total correction
    factor and its uncertainty, respectively.}
  \label{fig:mistag_corrections}
\end{figure*}

\begin{figure}[htbp]\centering
  \includegraphics[width=0.45\textwidth]{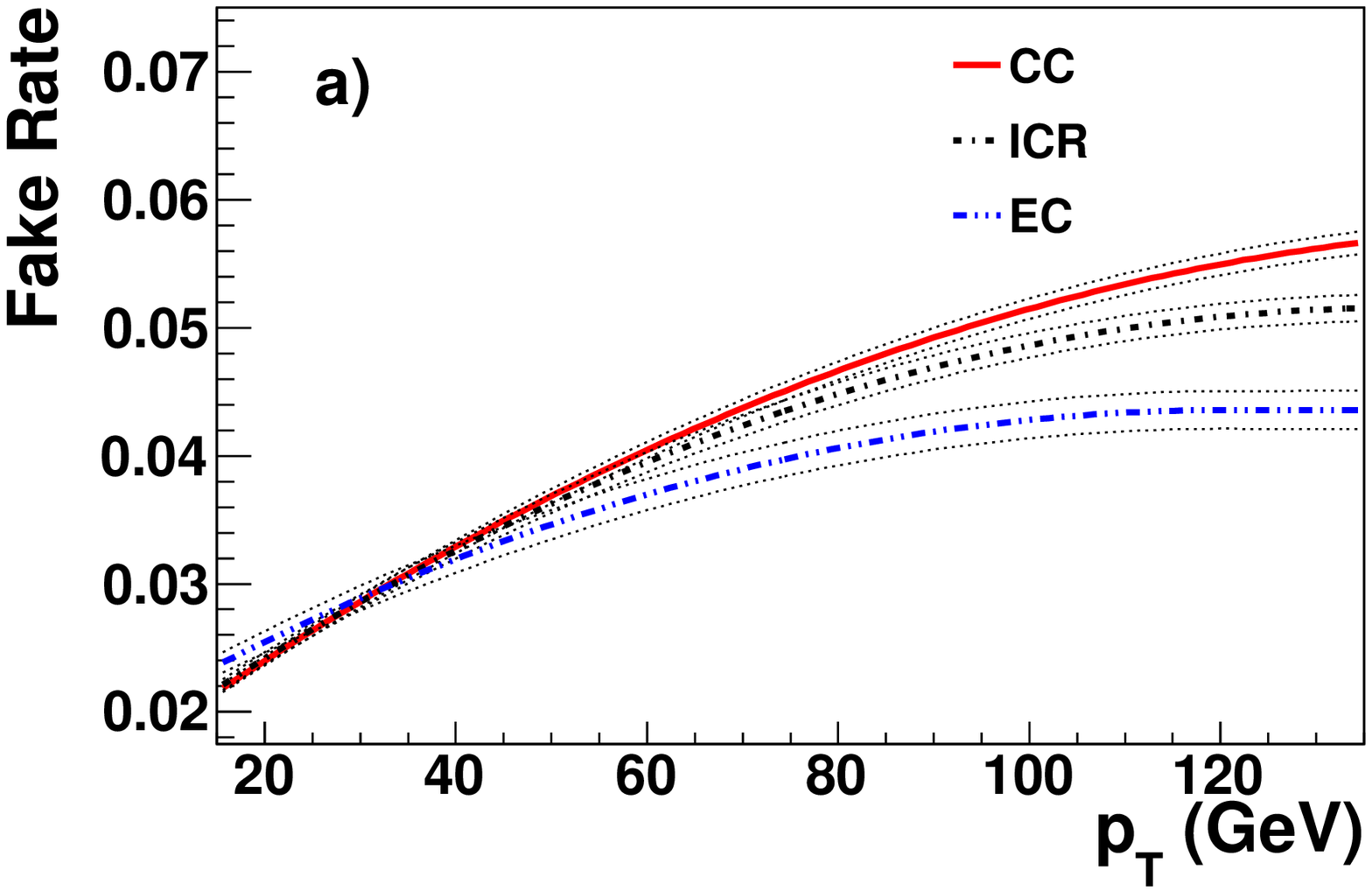}
  \includegraphics[width=0.45\textwidth]{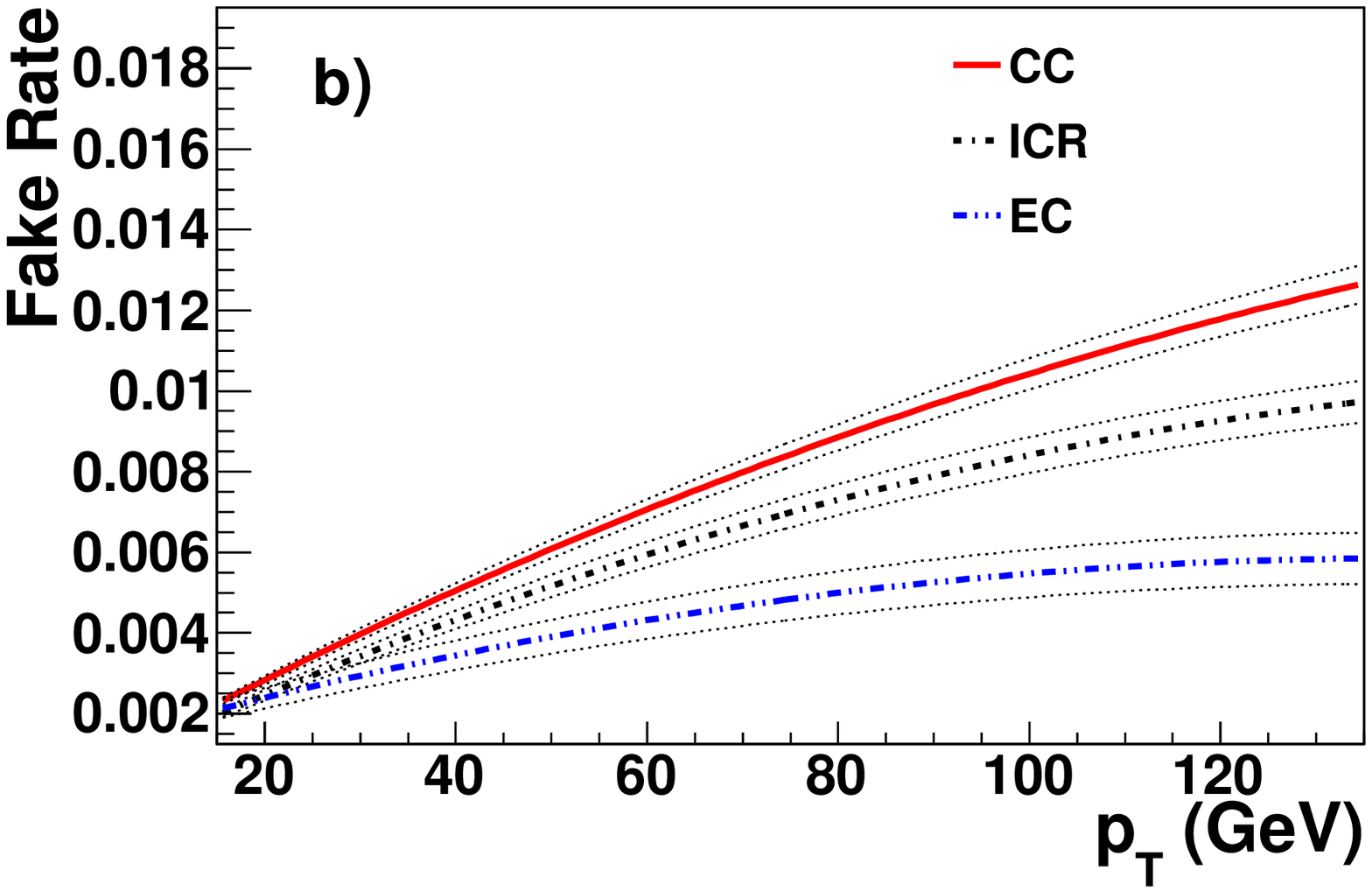}
  \caption{The estimated light quark tagging efficiency parametrized in the CC
    (continuous line), ICR (dot-dashed line) and EC (dot-dot-dashed line) for the
    L2 (a) and Tight (b) operating points. The dotted black lines represent
    the fit uncertainty.}
  \label{fig:mistag_pt}
\end{figure}

The jet \et\ dependences of $F_{\mbox{\scriptsize hf}}$ and $F_{\mbox{\scriptsize lf}}$
are shown in Fig.~\ref{fig:mistag_corrections} for the L2 and Tight operating
points. The estimated light quark tagging efficiencies
$\varepsilon_{\mbox{\scriptsize light}}$ for the L2 and Tight operating points are 
shown in Fig.~\ref{fig:mistag_pt}. Both the negative and positive tag rates for
light-flavor jets increase with increasing jet \et, for two reasons: (i) the
multiplicity of long-lived particles and their average decay length increase;
and (ii)
jets become more collimated, with the resulting higher
hit density leading to a larger number of wrongly reconstructed high-impact
parameter tracks.

\subsection{Systematic uncertainties}

\begin{figure*}[htbp]
  \centering
  \includegraphics[width=0.45\textwidth]{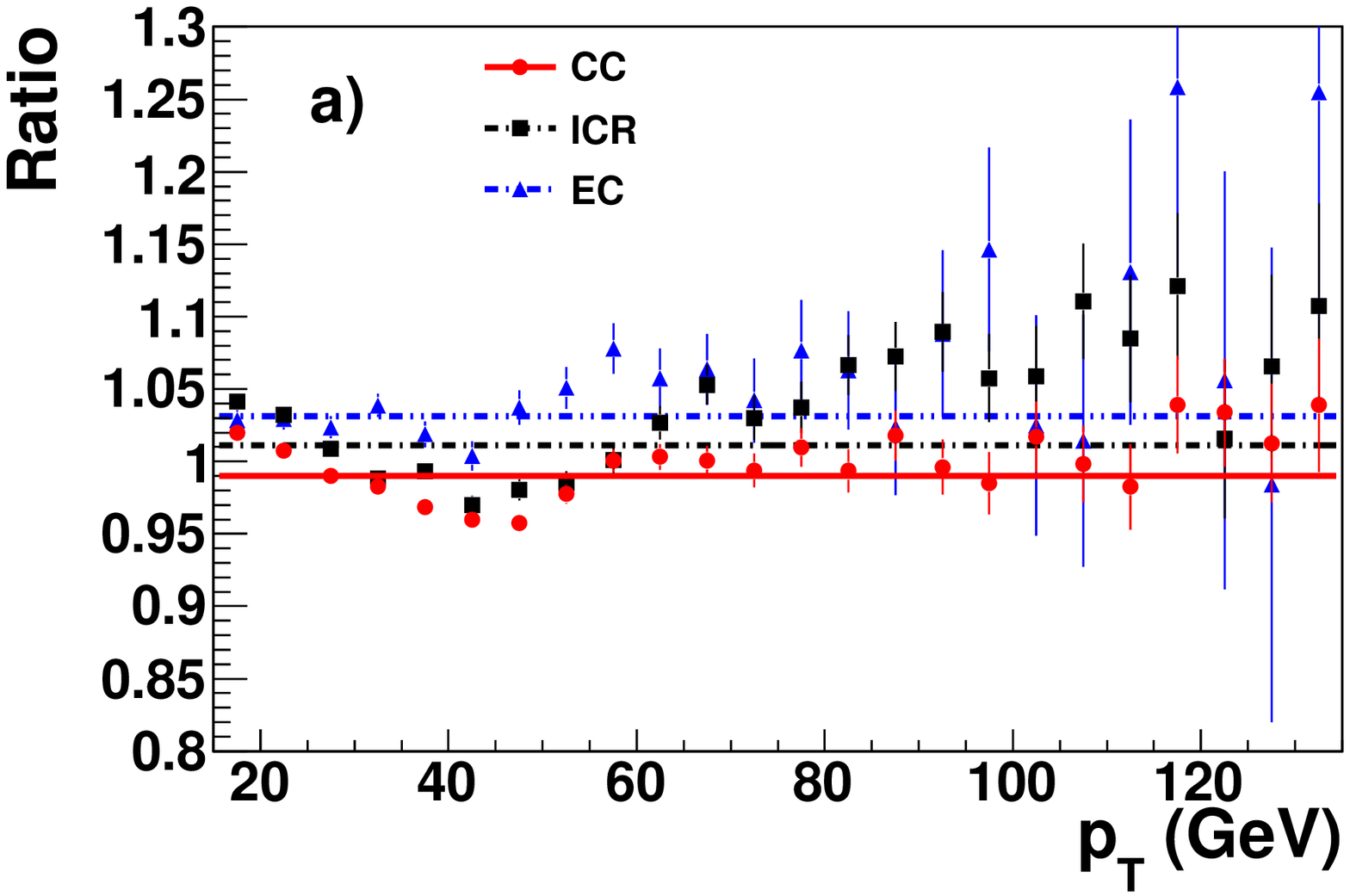}
  \includegraphics[width=0.45\textwidth]{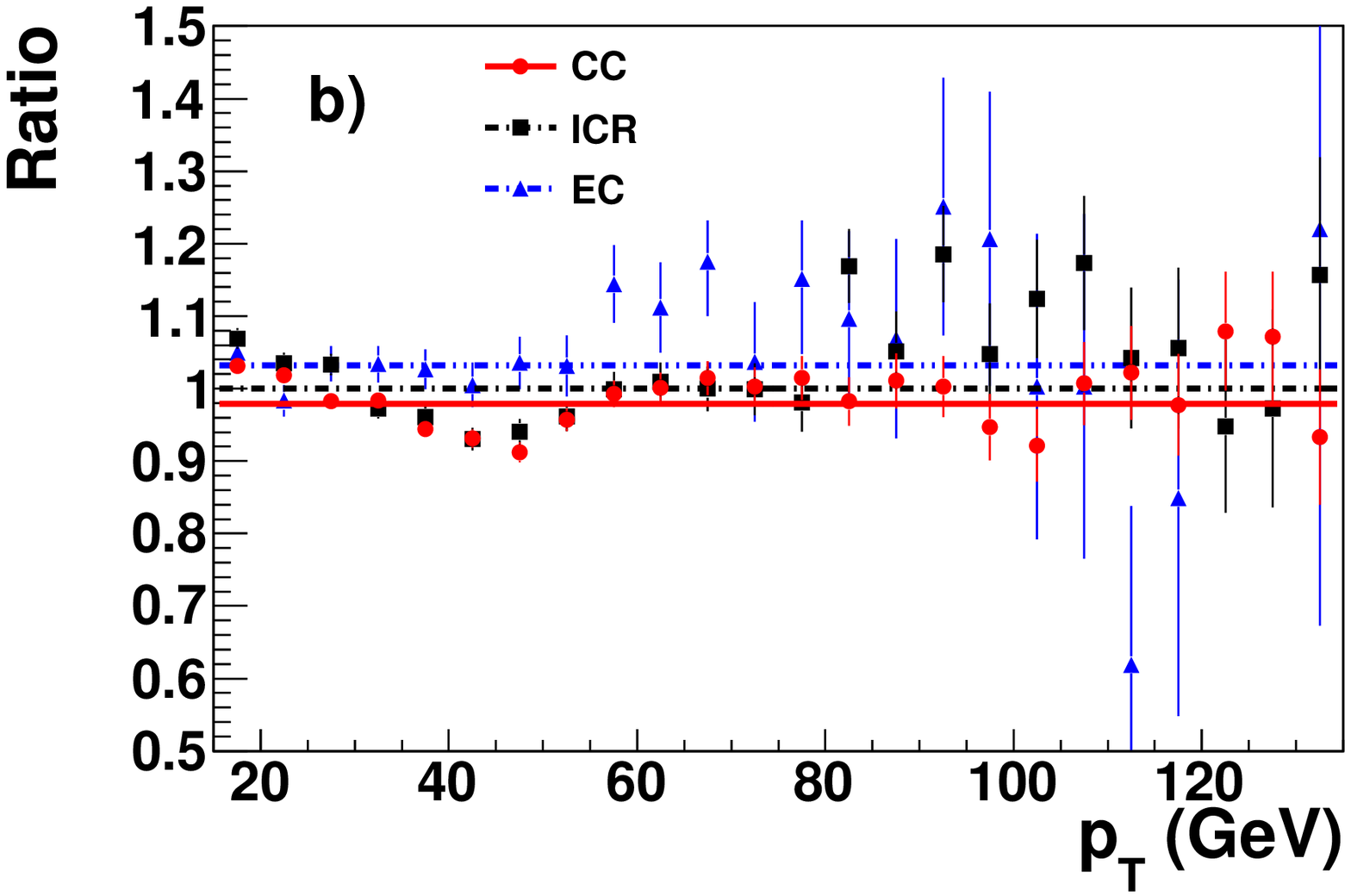}
  \caption{The ratio of the EM and QCD negative tag rates for the L2 (a) and
    Tight (b) operating points in the CC (circles, continuous line), ICR (squares, dot-dashed line), and
    EC regions (triangles, dot-dot-dashed line).}
\label{fig:nt_emqcd}
\end{figure*}

The use of a particular sample (in this case, the combination of the QCD and EM
multijet samples) to provide a ``universal'' estimate of the NT rate needs to be
validated. To this end, the ratio of the NT rates as measured in the separate
QCD and EM samples is determined as a function of the kinematical variables
and is shown in Fig.~\ref{fig:nt_emqcd} for the L2 and Tight operating points.

A corresponding systematic uncertainty is calculated from a constant fit to
the EM/QCD NT rate ratio. Half the difference between the fit value and
unity is taken as the systematic uncertainty, or if the ratio is
consistent with unity within the fit uncertainty scaled by
$\sqrt{\chi^{2}/N_{\mathrm{dof}}}$, this scaled fit uncertainty is taken as the
uncertainty. The relative uncertainty ranges from 0.2\% to 1.3\% in the CC, 0.3\% to
0.7\% in the ICR and 1.2\% to 3.1\% in the EC from the loosest to tightest operating
points.

In addition, the effect of removing jets in the vicinity of EM clusters in the
QCD sample needs to be taken 
into account. The NT rate in the QCD sample with the jets removed is slightly
lower than in the full QCD sample. The effect is small, ranging from 0.2\% for
the loosest to 1\% for the tightest operating point, and does not depend
on jet \et. A systematic uncertainty is assigned in the same way as that for the
difference between the EM and QCD samples, and ranges from 0.1\% to 0.6\% in the CC,
0.1\% to 0.3\% in the ICR, and 0.1\% to 0.7\% in the EC from the loosest to
tightest operating points.

To test the parametrization of the NT rate in the three $|\eta|$ regions a
comparison is made between the number of tags found by the tagger
and its prediction from the parametrized NT rate.
A systematic uncertainty is again calculated from a constant fit to the ratio of
the actual and predicted number of tags, following the same procedure as the
EM/QCD sample comparison. The systematic uncertainty ranges from 0.1\% to
0.3\% in the CC/ICR and from 0.1\% to 0.7\% in the EC from the loosest to tightest
operating points.

The $F_{\mbox{\scriptsize hf}}$ correction factor depends on the assumed \bquark- and
\cquark-fractions in the multijet data sample. In turn, these depend on the
cross sections for QCD heavy flavor production. 
To estimate the uncertainty on $F_{\mbox{\scriptsize hf}}$, the fraction
of \bquark\ (\cquark) jets is varied from its default value of 2.6\% (4.6\%) by
50\% (relative). Given that the individual \bquark- and \cquark-production
mechanisms are very similar, these fractions are varied coherently. The total
uncertainties from varying the fraction of \bquark\ (\cquark) jets ranges from
2.8\% (1.6\%) for the loosest to 19.5\% (4.7\%) for the tightest operating point in the CC/ICR and
from 1.0\% (0.7\%) to 6.7\% (3.5\%) in the EC regions.

The total uncertainty on the fake tag rate is given by adding in quadrature the
systematic contributions (as discussed above) for the appropriate region to
the statistical uncertainty, estimated as the difference between the fake tag
rate central value and one standard deviation fit curves.
The dominant contribution is the systematic one. The combined relative
systematic uncertainty ranges from 5.9\% (for the loosest operating point) to
23.1\% (for the tightest operating point) in the CC region, from 4.8\% to 24.2\%
in the ICR region, and from 2.2\% to 10.0\% in the EC region.
A more detailed breakdown of the systematic uncertainties is shown in
Tables~\ref{tab:mistagSyst_L2} and \ref{tab:mistagSyst_Tight} for the L2 and
Tight operating points, respectively.

\begin{table}[h]
  \begin{center}
    \begin{tabular}{|l|ccc|}
      \hline
      Region  & CC    & ICR   & EC    \\
      \hline
      Parametrization   & 0.1\% & 0.1\% & 0.1\% \\
      EM/QCD            & 0.5\% & 0.5\% & 1.6\% \\
      EM veto           & 0.2\% & 0.1\% & 0.1\% \\
      \cquark\ fraction & 3.7\% & 3.3\% & 1.3\% \\
      \bquark\ fraction & 7.3\% & 6.4\% & 2.2\% \\
      \hline
      Total   					& 11.0\% & 9.7\% & 3.8\% \\
      \hline
    \end{tabular}
  \end{center}
  \caption{Fake tag rate relative systematic uncertainties for the L2
    NN operating point.}
  \label{tab:mistagSyst_L2}
\end{table}

\begin{table}[h]
  \begin{center}
    \begin{tabular}{|l|ccc|}
      \hline
      Region  & CC    & ICR   & EC    \\
      \hline
      Parametrization   & 0.1\% & 0.2\% & 0.4\% \\
      EM/QCD            & 1.0\% & 0.5\% & 1.6\% \\
      EM veto           & 0.5\% & 0.2\% & 0.5\% \\
      \cquark\ fraction & 4.5\% & 4.5\% & 1.4\% \\
      \bquark\ fraction & 14.3\% & 13.8\% & 4.2\% \\
      \hline
      Total             & 18.8\% & 18.3\% & 5.9\% \\
      \hline
    \end{tabular}
  \end{center}
  \caption{Fake tag rate relative systematic uncertainties for the
    Tight NN operating point.}
  \label{tab:mistagSyst_Tight}
\end{table}


%% file: conclusion.tex
\section{Summary and Conclusion}
\label{sec:conclusion}

Several techniques to identify \bquark\ jets exploiting the long lifetime of
\bquark\ hadrons have been discussed in this article. Compared to the use of
individual \bquark-jet tagging algorithms, the combination of their results in an
artificial neural network leads to a considerable improvement in performance.

\begin{figure}
  \centering
  \includegraphics[width=0.48\textwidth]{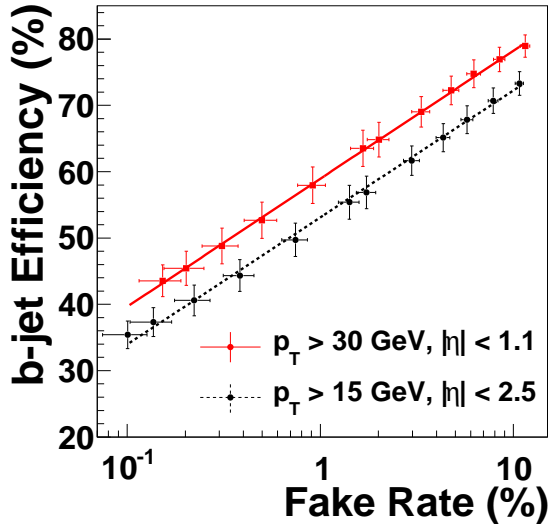}
  \caption{Data performance profile of the NN tagger as applicable to the
    kinematics of 
    $\PZz\rightarrow \bbbar$ and $\PZz\rightarrow \qqbar$ events 
    for all jets (dashed line) and for jets with $|\eta| < 1.1$ and
    $\et > 30\GeV$ (solid line). The vertical error bars on the plot
    represent the total uncertainty (statistical and systematic uncertainties
    added in quadrature) on the performance measurements.}
 \label{fig:data_performance}
\end{figure}

This performance needs to be calibrated using the actual collider data, as the
simulation cannot be expected to reproduce the performance of the detector in
all aspects relevant to \bquark-jet tagging. The calibration methods described
employ QCD jet samples, and therefore can make use of ample statistics at low
jet \et.
\begin{itemize}
\item Starting from the further requirement of a muon-jet association, the
  \textit{SystemD} method allows the determination of the \bquark-jet tagging
  efficiency, with minimal input from simulation, even in the presence of an
  \emph{a priori} unknown background.
\item The determination of the light-flavor mistag rate makes use of
  the fact that without such a muon requirement, the sample consists almost entirely
  of light-flavor jets. This method is limited by the knowledge on the remaining
  heavy-flavor content.
\end{itemize}
The resulting performance as measured using data, including full statistical and
systematic uncertainties, is shown in Fig.~\ref{fig:data_performance} for all
jets and for jets with $|\eta| < 1.1$ and $\et > 30\GeV$.

These tagging algorithms and calibration methods have been used in many
publications of \Dzero\ Run~IIa analyses. They are being refined to
make use of a new layer of silicon sensors installed at even smaller distance
from the beam line~\cite{Angstadt:2009ie}, as well as cope with the higher
instantaneous luminosities common in the Run~IIb data taking period that started
in June 2006. The calibration methods are not specific to \Dzero\ and could be
used at other experiments.


%% file: acknowledgement_paragraph_r2.tex
%
We thank the staffs at Fermilab and collaborating institutions, 
and acknowledge support from the 
DOE and NSF (USA);
CEA and CNRS/IN2P3 (France);
FASI, Rosatom and RFBR (Russia);
CNPq, FAPERJ, FAPESP and FUNDUNESP (Brazil);
DAE and DST (India);
Colciencias (Colombia);
CONACyT (Mexico);
KRF and KOSEF (Korea);
CONICET and UBACyT (Argentina);
FOM (The Netherlands);
STFC and the Royal Society (United Kingdom);
MSMT and GACR (Czech Republic);
CRC Program and NSERC (Canada);
BMBF and DFG (Germany);
SFI (Ireland);
The Swedish Research Council (Sweden);
and
CAS and CNSF (China).